\documentclass[a4paper,fleqn,usenatbib,useAMS]{mnras}

\usepackage{graphicx}	
\usepackage{amsmath}	
\usepackage{amssymb}	
\usepackage{multicol}        
\usepackage{bm}		
\usepackage{pdflscape}	
\usepackage{array}
\usepackage{geometry}
\usepackage{caption}
\usepackage{color}
\usepackage{ulem}

\DeclareCaptionFormat{cont}{#1 (cont.)#2#3\par}

\newcommand{\ha}{H$\alpha$}
\newcommand{\hb}{H$\beta$}

\newcommand{\heiiwr}{He\,{\sc ii}~$\lambda4686$}

\newcommand{\niiiwr}{N\,{\sc iii}~$\lambda\lambda4634,41$}
\newcommand{\nvwr}{N\,{\sc v}~$\lambda\lambda4603,20$}
\newcommand{\heineb}{He\,{\sc i}~$\lambda4713$}
\newcommand{\niineb}{[N\,{\sc ii}]~$\lambda5755$}
\newcommand{\feiii}{[Fe\,{\sc iii}]}
\newcommand{\civwrr}{C\,{\sc iv}~$\lambda\lambda5801,12$}
\newcommand{\civheir}{He\,{\sc i}~$\lambda5876$}
\newcommand{\ciiiwrb}{C\,{\sc iii}~$\lambda\lambda4647,66$}
\newcommand{\civwrb}{C\,{\sc iv}~$\lambda4658$}
\newcommand{\ciiiwrr}{C\,{\sc iii}~$\lambda5696$}
\newcommand{\ergs}{erg\,s$^{-1}$}

\newcommand{\oviwr}{O\,{\sc vi}~$\lambda\lambda3811,34$}
\newcommand{\hei}{He\,{\sc i}}
\newcommand{\heii}{He\,{\sc ii}}
\newcommand{\ciii}{C\,{\sc iii}}
\newcommand{\civ}{C\,{\sc iv}}
\newcommand{\nii}{[N\,{\sc ii}]}
\newcommand{\niii}{N\,{\sc iii}}

\newcommand{\nv}{N\,{\sc v}}

\newcommand{\fei}{[Fe\,{\sc i}]}

\newcommand{\oiii}{[O\,{\sc iii}]}
\newcommand{\oii}{[O\,{\sc ii}]}
\newcommand{\sii}{[S\,{\sc ii}]}
\newcommand{\siii}{[S\,{\sc iii}]}
\newcommand{\cliii}{[Cl\,{\sc iii}]}

\newcommand{\oiia}{[O\,{\sc ii]}~$\lambda3727$}
\newcommand{\oiiia}{[O\,{\sc iii]}~$\lambda4363$}
\newcommand{\oiiib}{[O\,{\sc iii]}~$\lambda4959$}
\newcommand{\oiiic}{[O\,{\sc iii]}~$\lambda5007$}

\newcommand{\feiiia}{[Fe\,{\sc iii}]~$\lambda4658$}
\newcommand{\oiib}{[O\,{\sc ii]}~$\lambda7320$}
\newcommand{\oiic}{[O\,{\sc ii]}~$\lambda7330$}

\newcommand{\kms}{\,km\,s$^{-1}$} 
\newcommand{\cmvol}{cm$^{3}$}

\usepackage[T1]{fontenc}
\usepackage{ae,aecompl}

\usepackage{txfonts}

\title[WR stars in the Antennae]
{Wolf--Rayet stars in the Antennae unveiled by MUSE}
\author[G\'{o}mez-Gonz\'{a}lez et al.]{V.~M.~A.\,G\'{o}mez-Gonz\'{a}lez$^{1}$\thanks{E-mail:\,v.gomez@irya.unam.mx, mau.gglez@gmail.com}, Y.~D.\,Mayya$^{2}$, J.~A.\,Toal\'{a}$^{1}$, S.~J.\,Arthur$^{1}$, J. Zaragoza-Cardiel$^{2,3}$ and \newauthor{M.~A.\,Guerrero$^{4}$}
\\
 $^{1}$Instituto de Radioastronom\'{i}a y Astrof\'{i}sica,
UNAM Campus Morelia, Apartado postal 3-72, 58090 Morelia, Michoac\'{a}n, Mexico\\
$^{2}$Instituto Nacional de Astrof{\'\i}sica, \'Optica y Electr\'onica,
Luis Enrique Erro 1, Tonantzintla 72840, Puebla, Mexico\\
$^{3}$Consejo Nacional de Ciencia y Tecnolog\'{i}a, Av. Insurgentes Sur 1582, 03940, Mexico City, Mexico\\
$^{4}$Instituto de Astrof\'{i}sica de Andaluc\'{i}a (IAA-CSIC), Glorieta de la Astronom\'{i}a S/N, 18008 Granada, Spain\\
}

\pubyear{2020}

\begin{document}
\label{firstpage}
\pagerange{\pageref{firstpage}--\pageref{lastpage}}
\maketitle

\begin{abstract}

We present the analysis of archival Very Large Telescope (VLT)
Multi Unit Spectroscopic Explorer (MUSE) observations of the
interacting galaxies NGC\,4038/39 (a.k.a. the Antennae) at a distance of 18.1~Mpc.
Up to 38 young star-forming complexes with evident contribution
from Wolf--Rayet (WR) stars are unveiled.
We use publicly available templates of Galactic WR stars in
conjunction with available photometric extinction measurements
to quantify and classify the WR population in each star-forming region,
on the basis of its nearly Solar oxygen abundance. The total estimated number of WR stars
in the Antennae is 4053 $\pm$ 84, of which there are 2021 $\pm$ 60~WNL and 2032 $\pm$ 59~WC-types.
Our analysis suggests a global WC to WN-type ratio of 1.01 $\pm$ 0.04,
which is consistent with the predictions of the single star evolutionary scenario
in the most recent {\sc BPASS} stellar population synthesis models.

\end{abstract}





\section{INTRODUCTION}

Most star formation occurs in groups and associations that were
once embedded in Giant Molecular Clouds \citep[][]{2003Lada}.
Observational evidence shows that star formation is enhanced in interacting
and merging galaxies \citep[see, e.g.,][]{2007Smith,2008Li,2015Knapen},
making them ideal environments
to study the formation and evolution of stars and,
in particular young massive stars \citep[][]{2009Whitmore}.
Super star clusters (SSCs), objects as compact and massive as
globular clusters (GC), though younger, are particularly abundant
in starburst and interacting galaxies.
The most massive and compact of these are the ones
expected to survive the disruptive effects of gravitational
shocks for a Hubble time, and hence are thought to be the
progenitors of GC \citep[][]{2010Portegies}.

Wolf--Rayet (WR) stars are considered descendants of O-type stars 
\citep[$M_\mathrm{ZAMS} \gtrsim 25$~M$_{\odot}$;]
[and references therein]{2007Crowther}.
For this reason, they are often used as indicators of young stellar
populations ($\sim$2--4~Myr) and to study the chemical enrichment of
its environments due to their characteristic strong stellar winds
enhanced with processed elements \citep[see, e.g.,][]{1992Maeder}.
They are also considered to be the most suitable candidates for core
collapse supernovae (SN) Ibc and long-duration soft-gamma ray burst
\citep[][]{2006WoosleyHeger,2006WoosleyBloom}.
There is a clear need to increase the sample of WR stars in the
local Universe, as that would raise the probability of a previously
classified WR exploding as a SN in a reasonable human life-time
\citep{2015Moffat}.
Up to now, not a single known WR star has exploded.

NGC\,4038/39, better known as the Antennae galaxies,
are located at a distance of
18.1~Mpc \citep[$m - M = 31.29$;][]{2016Riess},
making them the nearest and youngest pair of colliding galaxies at
an early stage of a merger \citep[][]{1994Whitmore}. 
Due to the significant number of star
forming zones and giant H\,{\sc ii} regions in the Antennae,
many multi-wavelength imaging \citep[e.g.,][]{2001Zhang, 2004Metz, 2018Matthews}
and spectroscopic studies \citep[e.g.,][]{2005Whitmore, 2018Weilbacher, 2019Gunawardhana}
have been conducted so far. 

Optical studies of the Antennae have revealed the presence of WR stars in the past.
\citet{2006Bastian} presented Very Large Telescope (VLT) VIMOS integral-field spectroscopy of
two northern fields in the Antennae and reported five SSCs with strong WR features.
Later, \citet{2009Bastian} presented Gemini GMOS observations and reported the discovery of
seven additional SSCs with broad WR signatures.
Nevertheless, a dedicated search for WR stars, specifying their number,
classification and distribution has not been performed yet.

Extragalactic spectroscopic observations, such as those obtained with 
the Multi Unit Spectroscopic Explorer (MUSE) at the VLT, are powerful
tools that can be used to unveil
the presence of WR stars in the nearby Universe.
It is important to note that at large distances most findings
of WR features correspond to unresolved populations of stars in
stellar clusters, rather than single stars.

In this paper we present the analysis of archival VLT MUSE observations
of the prototypical starburst/merging galaxy, the Antennae.
The MUSE data cubes were searched to identify tens of SSC complexes with
WR features, significantly increasing the sample of relatively nearby extragalactic WR stars
and SN candidates.
Publicly available templates of Galactic WR stars were used to quantify
and classify the WR population in each region.
We compare our observed results with the predictions of stellar population synthesis models.

This paper is organised as follows.
In Section~2, we describe the observations, technical details of
the data preparation and the search for WR features.
In Section~3, we present our classification and quantification scheme resulting from
the spectral analysis.
The discussion of our results is presented in Section~4, and the conclusions
are listed in Section~5.

\section{OBSERVATIONS}

\subsection{Public VLT MUSE data cubes}

MUSE is a panoramic integral-field spectrograph at the 8~m VLT
of the European Southern Observatory (ESO)\footnote{\url{http://muse-vlt.eu/science/}},
operating in the optical wavelength range from 4600 to 9300~\AA\,
with a spatial sampling	of 0.2 arcsec\,pixel$^{-1}$, a spectral sampling of 1.25~\AA\,pixel$^{-1}$ and
a spectral resolution $\sim$3~\AA\ \citep[][]{2010Bacon}.

Public MUSE data cubes covering the entire Antennae galaxies were
retrieved from the ESO Science Archive
Facility\footnote{\url{http://archive.eso.org/cms.html}}.
The data cubes were already flux
and wavelength calibrated, ready for scientific exploitation.
The data were obtained during three observing runs on 2015 April 22--24,
2015 May 11--23 and 2016 February 01 (PI: P.\,M.\,Weilbacher).
The observations correspond to a total of 23 data sets,
several of them covering the same regions in NGC\,4038/39,
which amounts to a total exposure time of $\approx$22.9~h.
Details of the observations have been presented in \citet{2018Weilbacher}
and \citet{2019Gunawardhana}.
A schematic view of all the field of views (FoV) of all the 
observations is shown
in Fig.~\ref{fig:datasets} {\it left panel} in Appendix~\ref{sec:datasets}.

\subsection{Search and distribution of WR features}

WR stars are spectroscopically identified by the presence of two broad features,
the blue bump (BB) at $\sim$4686\,\AA\, and the red bump (RB) at $\sim$5808\,\AA.
These features are the result of blends of several ionic transitions.
The BB contains one or several of the \heii, \niii, \nv, \ciii\ and \civ\ lines,
whereas the RB is made of \civ\ lines.
WR stars exhibiting any nitrogen line are classified as WN-type,
whereas those containing a carbon line are of WC-type (see details in Section 3).

We inspected the MUSE data cubes searching for spectral
WR features and detected their presence in four of these fields.
The outskirts of the Antennae did not show any hint of WR emission.
Consequently, in the present paper we only use the four observations
that cover the main regions in the Antennae (see Fig.~\ref{fig:datasets} 
{\it right panel}).
Details of these data cubes are listed in Table~\ref{tab:muse} 
in Appendix~\ref{sec:datasets}.

To identify WR stars in the Antennae, we used the data cubes to create
three different images simulating narrow-band filters,
each with 3~\AA\ of bandwidth,
centered on 4680~\AA\ (BB), 4800~\AA\ (continuum)
and 6596~\AA\ (\ha), taking into account the redshift of Antennae ($z\sim$0.005).
Then we used colour-composite images in order to highlight
any WR emission at 4686~\AA\ to facilitate its detection.
The resultant colour-composite images of the four
FoV presenting WR features are
shown in Fig.~\ref{fig:muse_fields} in Appendix~\ref{sec:datasets}.

We used the QFitsView\footnote{\url{http://www.mpe.mpg.de/~ott/dpuser/qfitsview.html}}
datacube file viewer \citep{2012Ott} to extract spectra
from all the identified candidates to confirm them as {\it bona fide} WR detections.
Subsequently, we followed a meticulous inspection by-eye of images
and spectra, simultaneously, in order to rule out probable WR presence
in places without an apparent excess in the BB band, and also to identify
any faint BB that did not show up in the colour images.
As a result, several weak-WR objects were included.

We note that not every emission feature detected in the BB is a WR finding.
The spectrum of a spurious WR candidate, probably a quasar at $z\sim0.264$,
was found in one of the data cubes. Its spectrum is presented in 
Fig.~\ref{fig:qso} of Appendix~\ref{sec:qso}.
These kinds of objects can mimic WR features when some of
its broad emission lines, at a given redshift,
coincide precisely with the wavelength of a WR feature.
This is uncommon, but happens \citep[see, e.g., figure~5 in][]{2012Neugent},
and care must be taken not to confuse
the nature of these objects when using the narrow-band filter
searching technique alone. The analysis of the spectral properties of this object
are out of the scope of the present work.

In order to quantify and classify the presence of WR stars in the Antennae
galaxies, we need spectra of each of the SSCs in which they are found. 
The main criterion to choose the size of the spectral extraction aperture 
was to maximise the intensity of the WR features.
Small apertures would underestimate the WR bump intensities whilst 
apertures larger than necessary could dilute these line profiles.

\begin{figure*}
\begin{center}
    \includegraphics[width=0.6\linewidth]{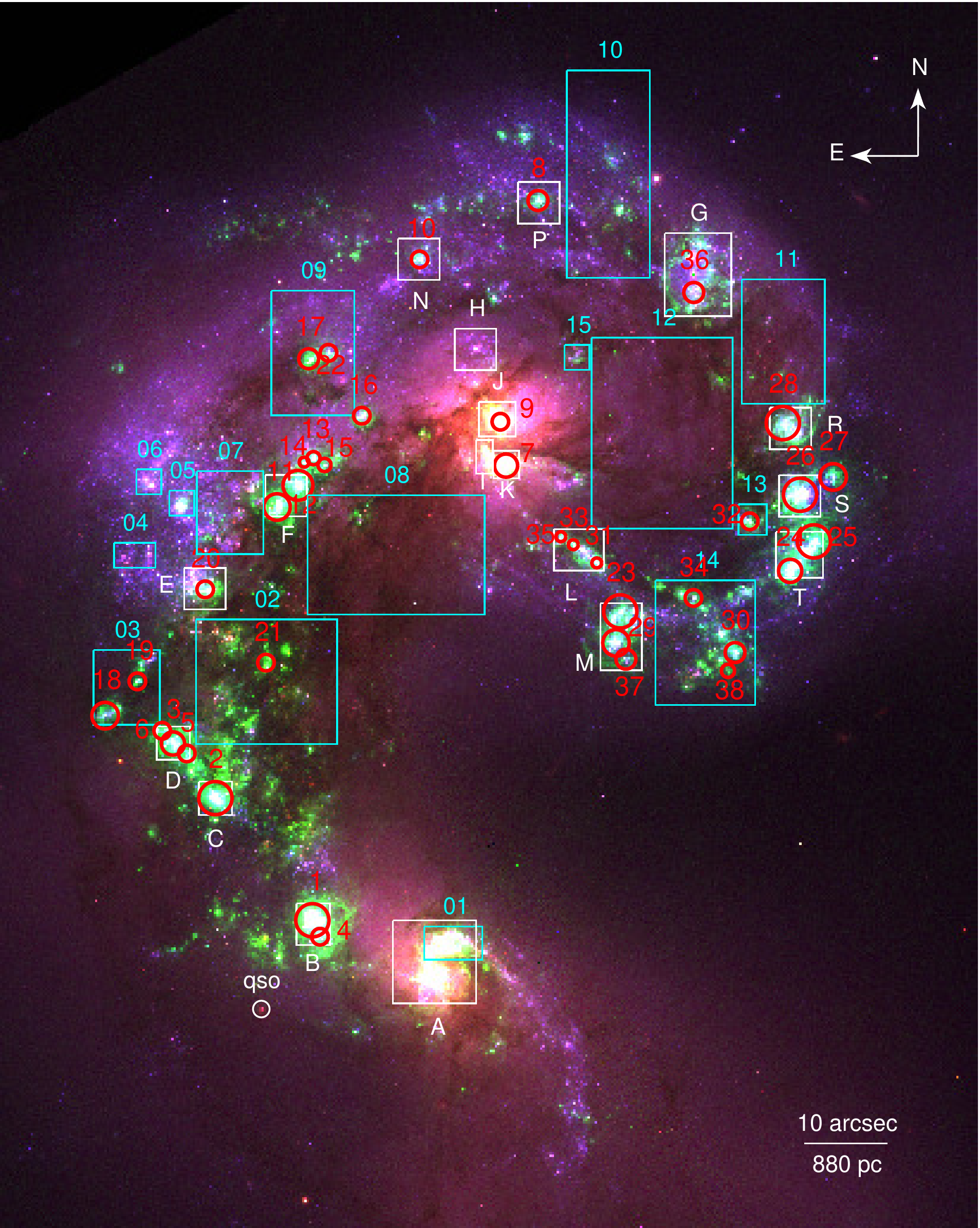}
\caption{Colour-composite {\it HST} ACS image of the Antennae galaxies.
Red, green and blue correspond to the F814W, F658W and F435W broad-band filters, respectively.
The locations of the 38 star-forming complexes with WR features are shown with red circles.
The white and cyan boxes correspond to the knots (letters: A--T)
and regions (number IDs: 01--15), respectively, studied by \citet{2010Whitmore},
who report the reddening information used in this study.
The extinction values and the angular and physical aperture 
sizes are listed in Table~\ref{tab:class}. The position of a detected 
quasar is denoted with a circle labeled as qso.} 
\label{fig:Antennae_RGB}
\end{center}
\end{figure*}

\begin{figure*}
\begin{center}
\includegraphics[width=1\linewidth]{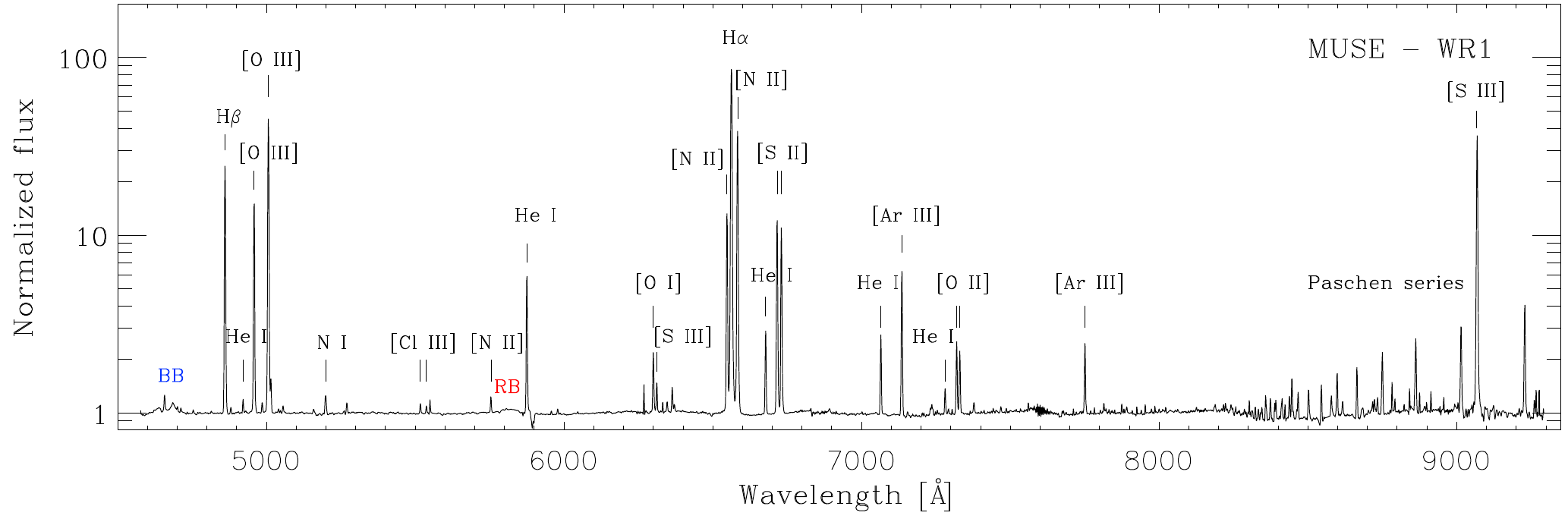}
\caption{VLT MUSE redshift corrected spectrum of WR\,1 in the Antennae.
The blue bump (BB) at 4686~\AA, and the red bump (RB)
at 5808~\AA, the most common optical WR features, are indicated,
as well as the narrow lines from the nebular environment.
The spectrum is shown normalized to the best fit continuum spectrum.}
\label{fig:MUSE}
\end{center}
\end{figure*}

We obtained a final sample of 38 spectra with confirmed WR features.
Their distribution is shown in the colour-composite {\it Hubble Space Telescope (HST)}
image presented in Fig.~\ref{fig:Antennae_RGB}.
Its locations are shown with red circles and labeled from 1 to 38.
As an example, we present in Fig.~\ref{fig:MUSE} the MUSE spectrum of WR\,1 
which displays the prominent BB and RB features. Other narrow emission lines indicated 
correspond to the nebular environment of the object.

\subsection{Public archival GTC OSIRIS spectra}

As described above, the VLT MUSE spectra are suitable for looking the 
so-called BB and RB. However, given the spectral coverage of MUSE,
we do not have access to another important WR feature, 
the violet bump (VB) at $\sim$3820~\AA. This broad emission is 
indispensable to reveal (or discard) the presence of 
Oxygen-type WR stars (WO).

\begin{figure*}
\begin{center}
\includegraphics[width=1\linewidth]{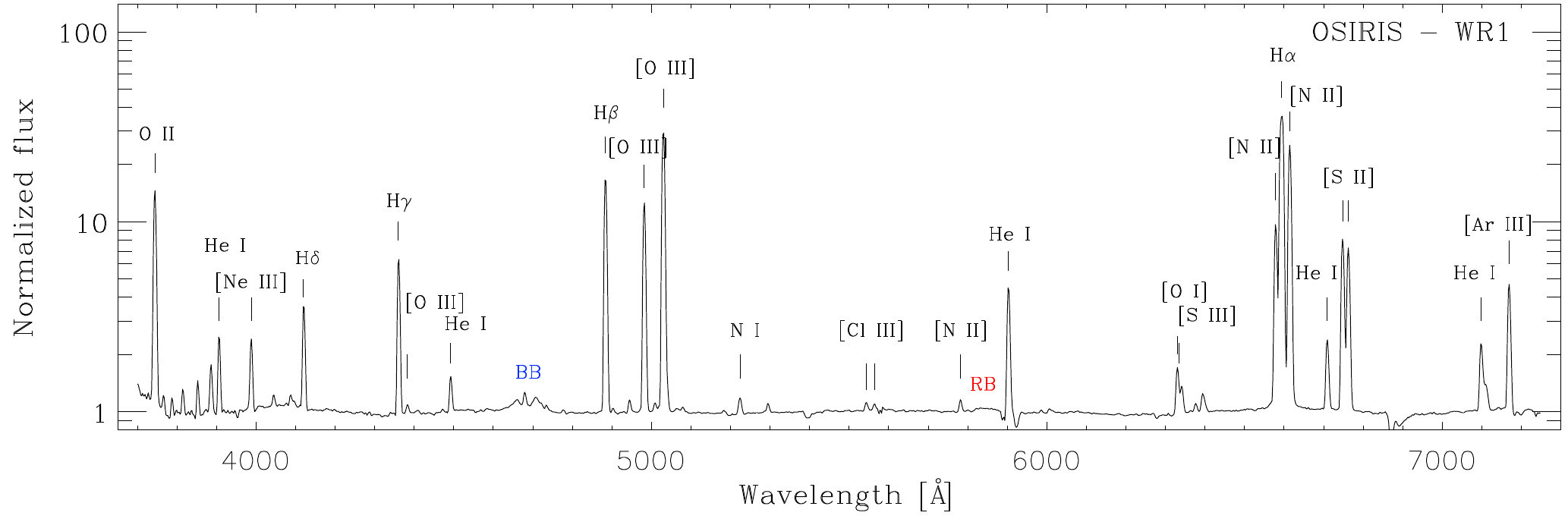}
\caption{GTC OSIRIS spectrum of WR\,1, complementary to that of MUSE.
The WR features BB and RB as well as some important nebular lines are identified.
The spectrum is shown normalized to the best fit continuum spectrum.}
\label{fig:OSIRIS}
\end{center}
\end{figure*}

Looking to cover this wavelength range,
we also retrieved available public archival observations in the
Antennae from the OSIRIS spectrograph in long-slit mode \citep{2016Cabrera}, at the 10.4~m
Gran Telescopio de Canarias
(GTC)\footnote{\url{http://gtc.sdc.cab.inta-csic.es/gtc/index.jsp}}.
The OSIRIS spectrum covers a spectral range from 3700 to 7500~\AA\, and is appropriate
to take a look in the VB band.
Unfortunately, only one of the identified WR features in the
Antennae has been detected by the OSIRIS observation, WR\,1.

These observations were carried out on 2013 January 12 in service mode
(PI: C.~M. Guti\'{e}rrez).
The total observing time was 2700~s which was split into three blocks of 900~s,
facilitating later removal of cosmic rays.
The settings were: R1000B grism, slitwidth
of 1~arcsec, position angle of $235.15^{\circ}$,
spectral resolution of $\sim$7~\AA\, and a CCD binning of 2$\times$2.
The observations have a spatial scale of 0.254 arcsec\,pixel$^{-1}$ and
a spectral sampling of $\sim$2\,\AA\,pixel$^{-1}$.
The ancillary files contained the spectrophotometric
standard GD140 (2.5 arcsec slitwidth),
$10\times$Bias (100 KHz),
$8\times$Flats (1.23 arcsec slitwidth) and 
the arc lamps: $2\times$HgAr, $2\times$Ne (1~arcsec slitwidth).
The night was clear of clouds, with Dark Moon,
and a seeing of 1.03 arcsec.
The reduction and extraction procedure we followed was similar to that
described in detail by \cite{2016Gomez}.

The resultant spectrum of WR\,1 covers a complementary bluer
wavelength range to that of MUSE and is shown in Fig.~\ref{fig:OSIRIS}.
With this valuable information, the presence of WO-type stars in this SSC 
can be discarded.

\subsection{Extinction correction}

In order to deredden our WR spectra, we used the results
presented by \citet{2010Whitmore},
who used {\it HST} ACS observations to produce $U-B$ versus $V-I$ colour-colour
diagrams to study the extinction across the Antennae galaxies.
These authors studied different star-forming knots and regions 
which we mark with white and cyan rectangles, respectively, in Fig.~\ref{fig:Antennae_RGB}.
It is worth mentioning that all of our WR detections lie within 
some of their analysed fields.

The $E(B-V)$ values were taken from \citet{2010Whitmore} and correspond to those
listed in their table~9 (column~8). With this information we determined
the visual extinction values ($A_\mathrm{V}$) 
for each WR listed in Table~\ref{tab:class}.
We use a standard total-to-selective extinction
of $R_\mathrm{V}=3.1$ and the equation $A_\mathrm{V}=3.1E(B-V)$.

The dereddened spectra of all the 38 WR sources in the Antennae galaxies obtained from
the MUSE observations are presented in Fig.~\ref{fig:spectra}. 
These spectra were fitted with template WR spectra to infer the number
and type of WR stars in each identified source following the method described below.

\section{Classification and number of WR}

\subsection{Galactic templates}

Template fitting has proven to be a very helpful tool to study extragalactic
WR stars in unresolved massive stellar populations 
\citep[e.g.,][]{2006Hadfield,2013Kehrig,2016Gomez,2020Gomez}.
The method consists in comparing the spectra with templates
obtained from averaging observed spectra of a number of individual classified WR stars
in the Galaxy and the Magellanic Clouds.
This methodology has been thoroughly described in \cite{2020Gomez}.
However, here we give a brief description of the quantification
and classification procedure of the 38 regions with WR features in the Antennae.

\begin{figure*}
\begin{center}
\includegraphics[width=0.33\linewidth]{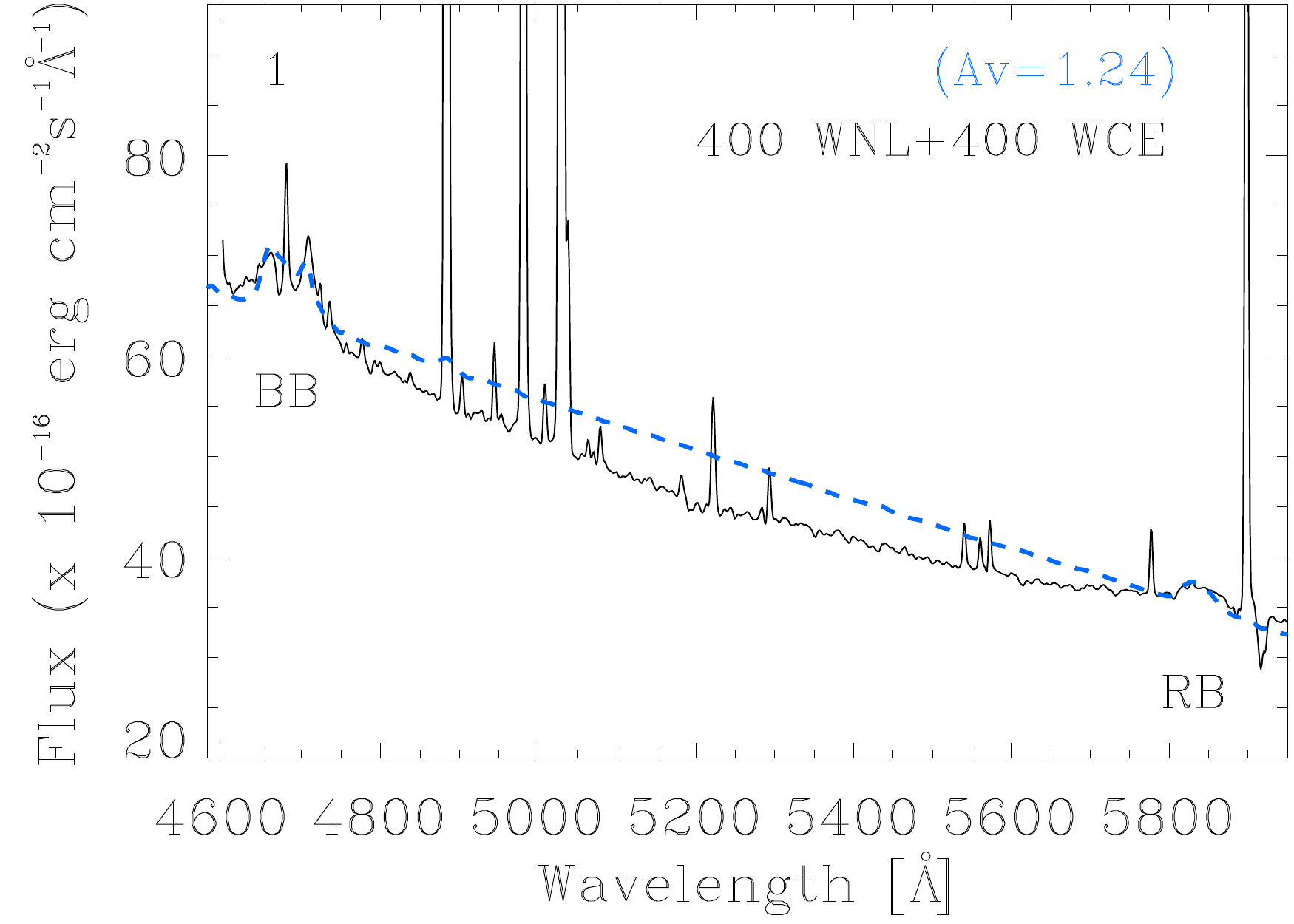}~
\includegraphics[width=0.33\linewidth]{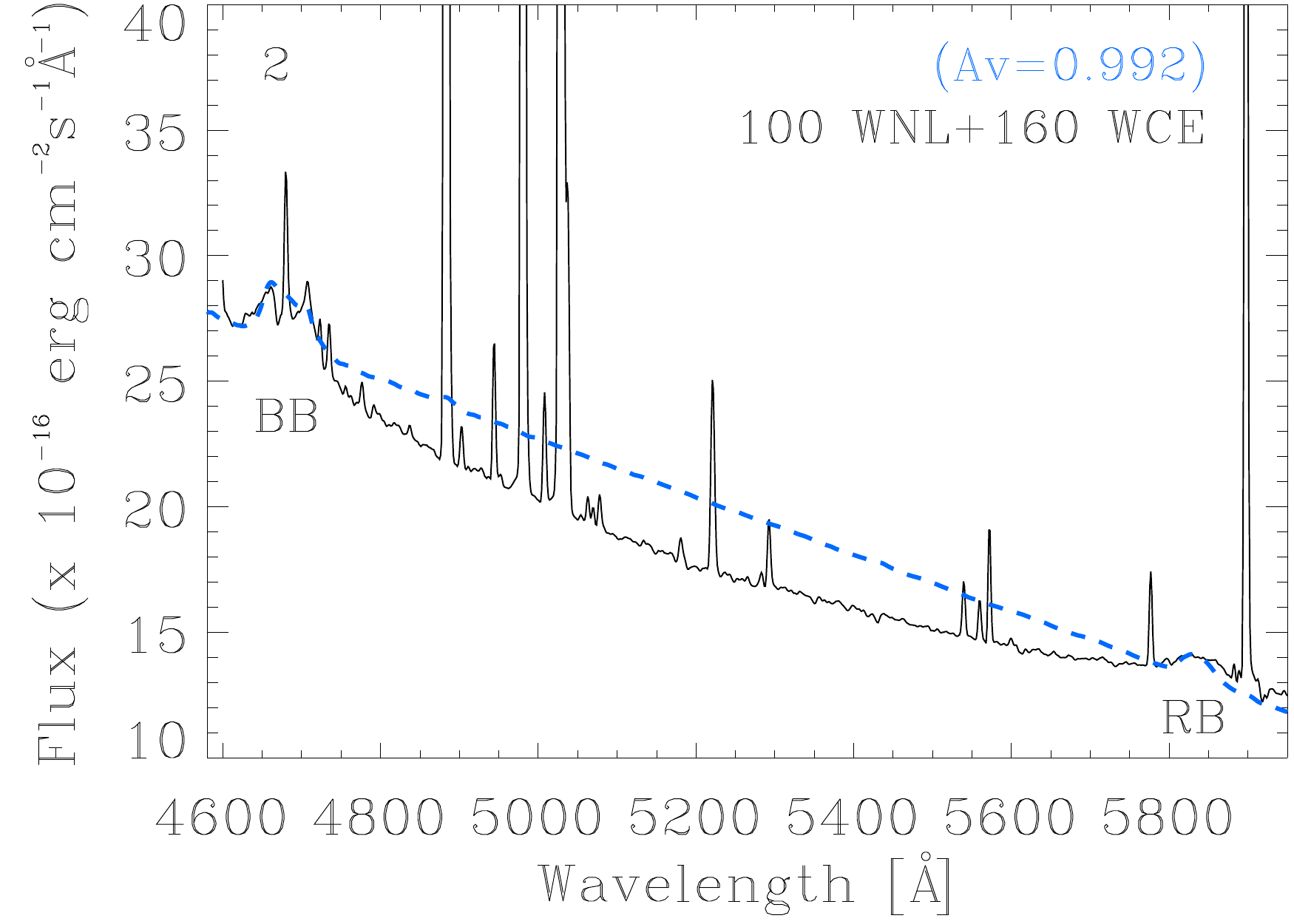}~
\includegraphics[width=0.33\linewidth]{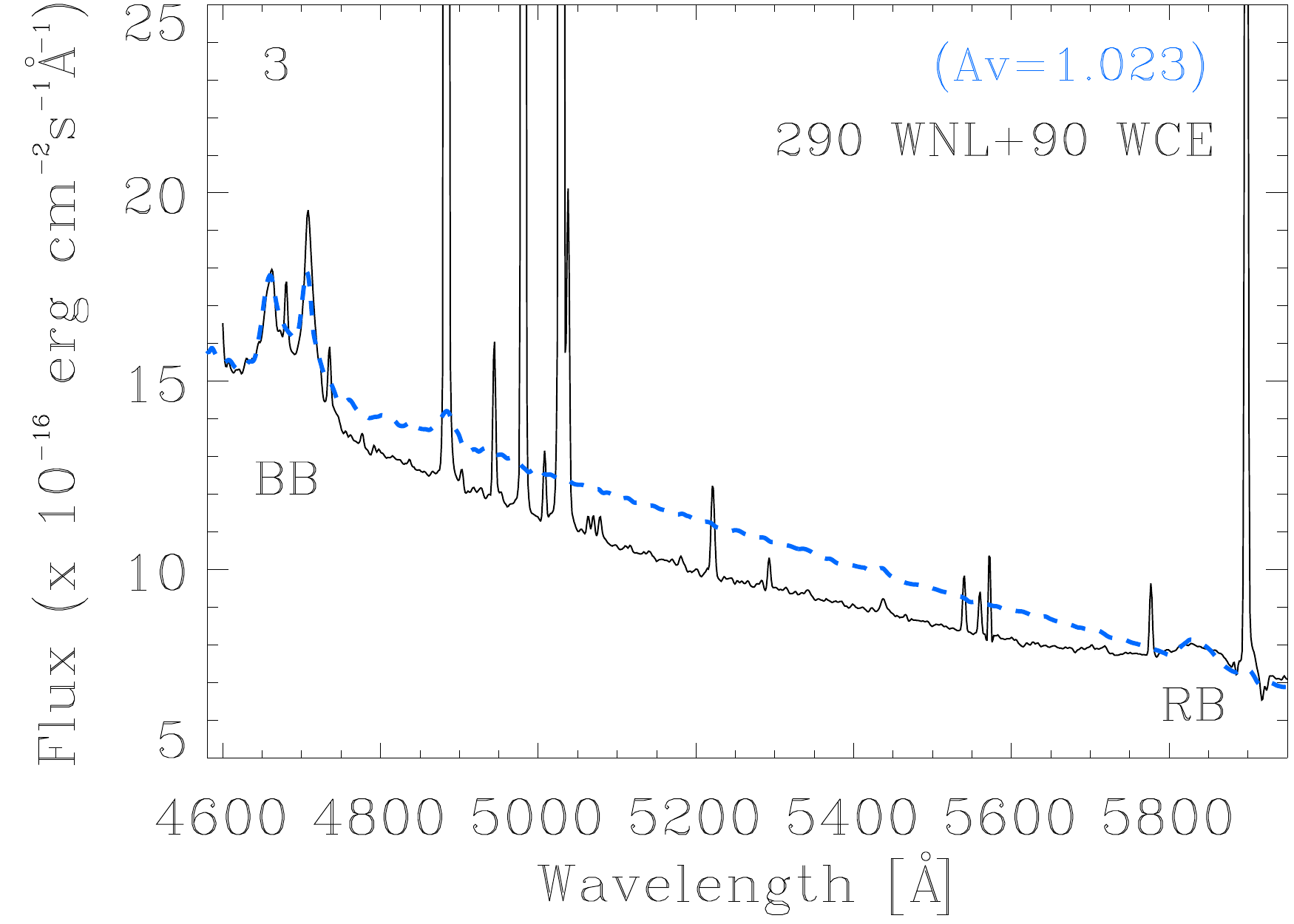}\\
\includegraphics[width=0.33\linewidth]{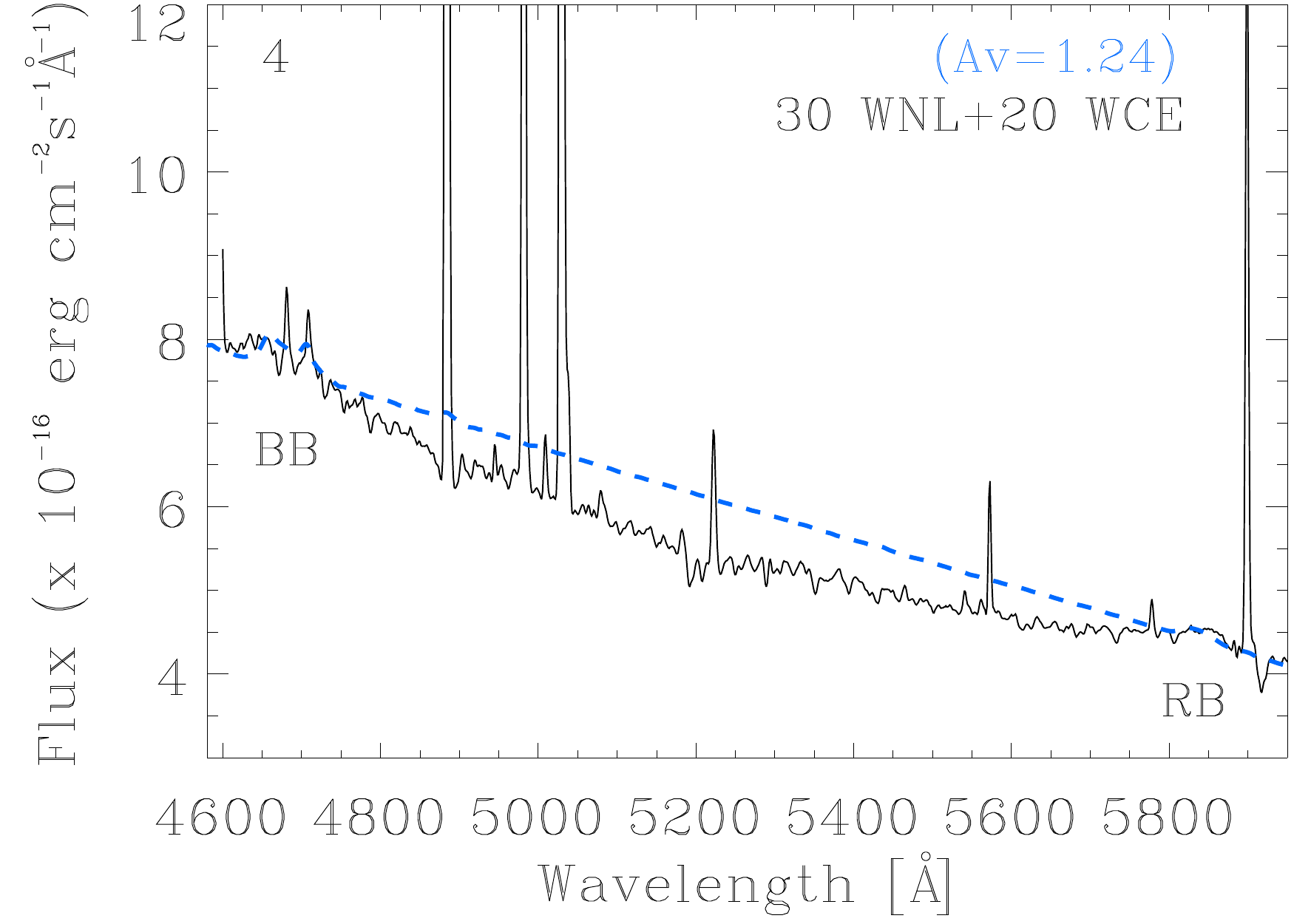}~
\includegraphics[width=0.33\linewidth]{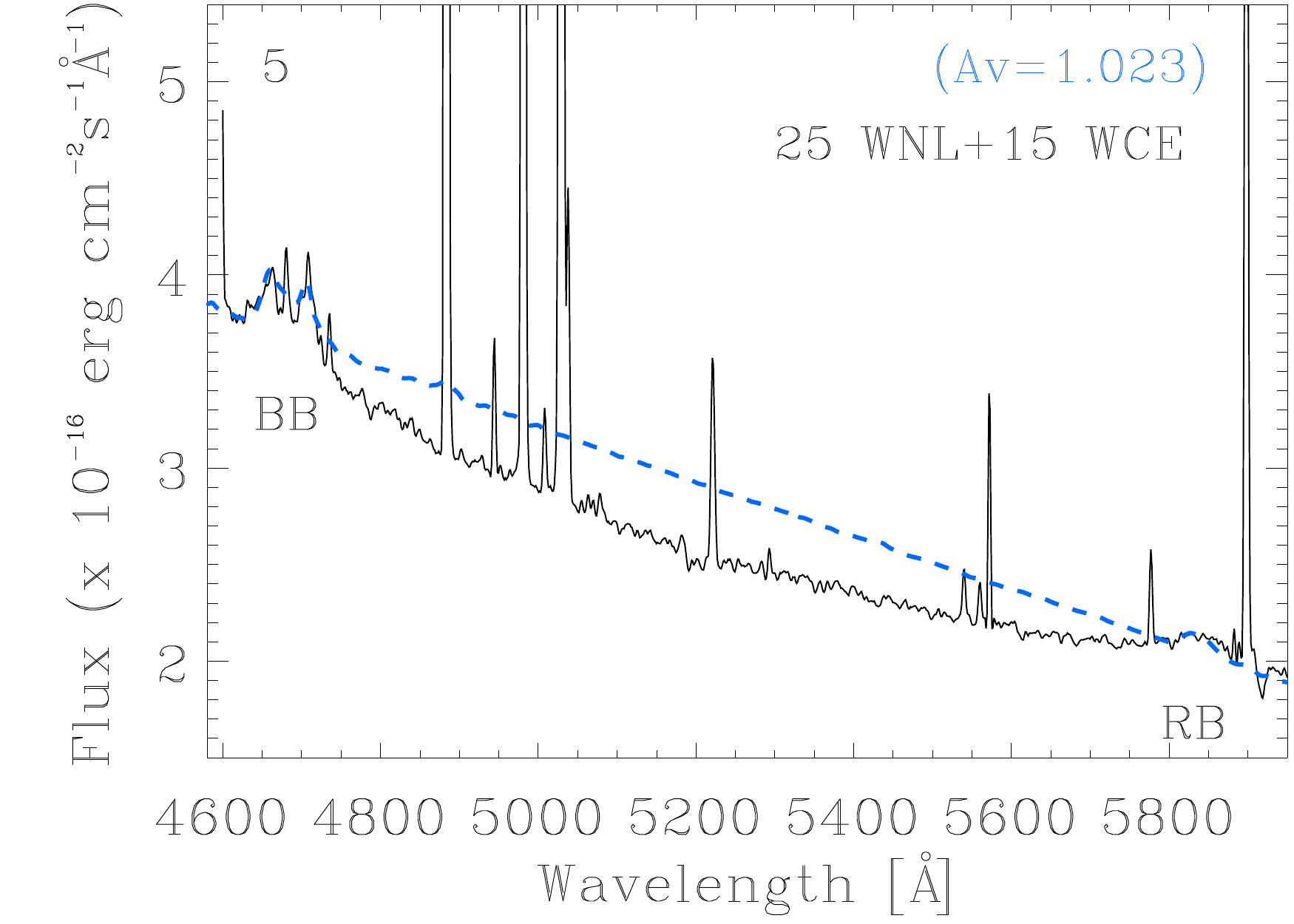}~
\includegraphics[width=0.33\linewidth]{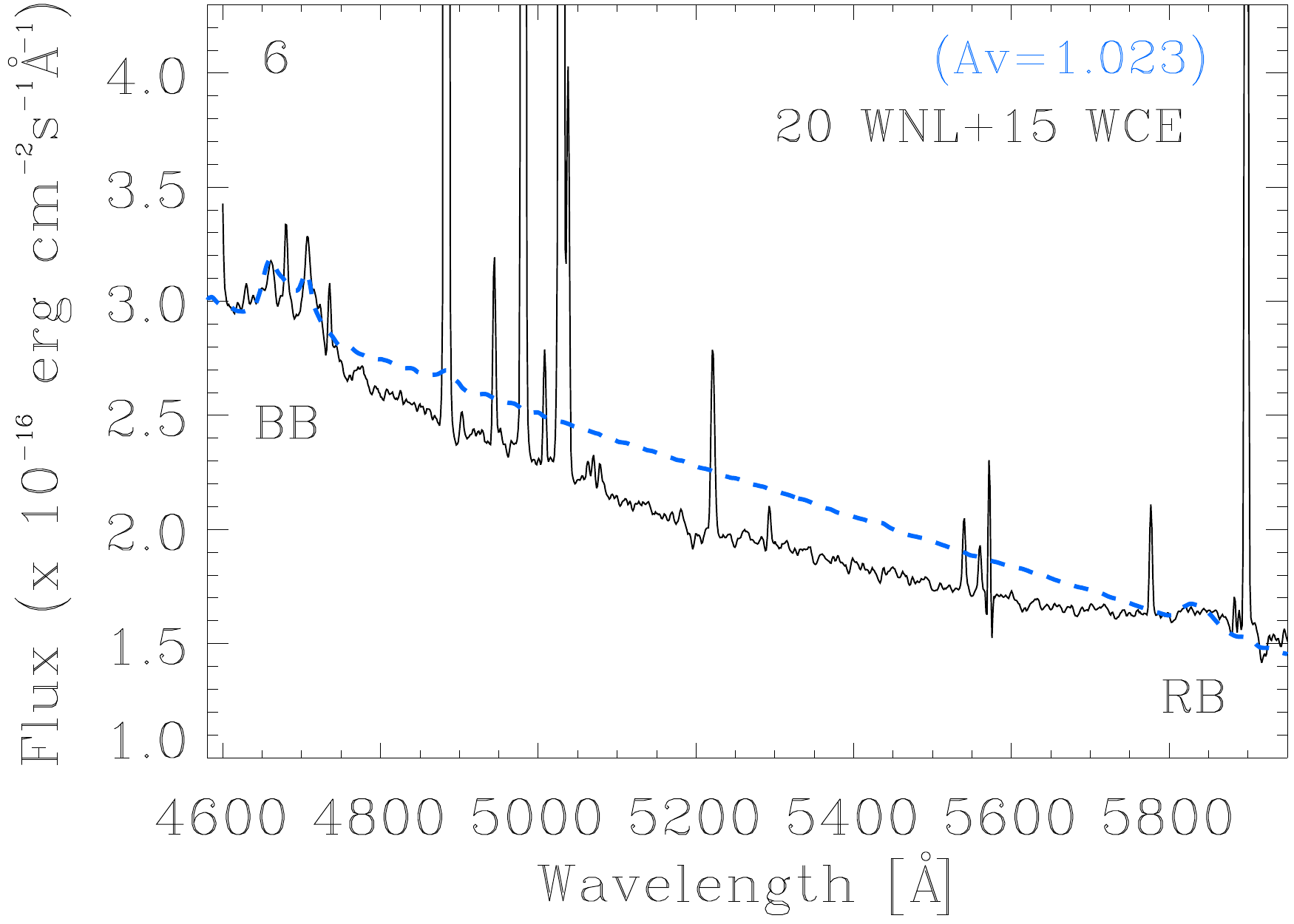}\\
\includegraphics[width=0.33\linewidth]{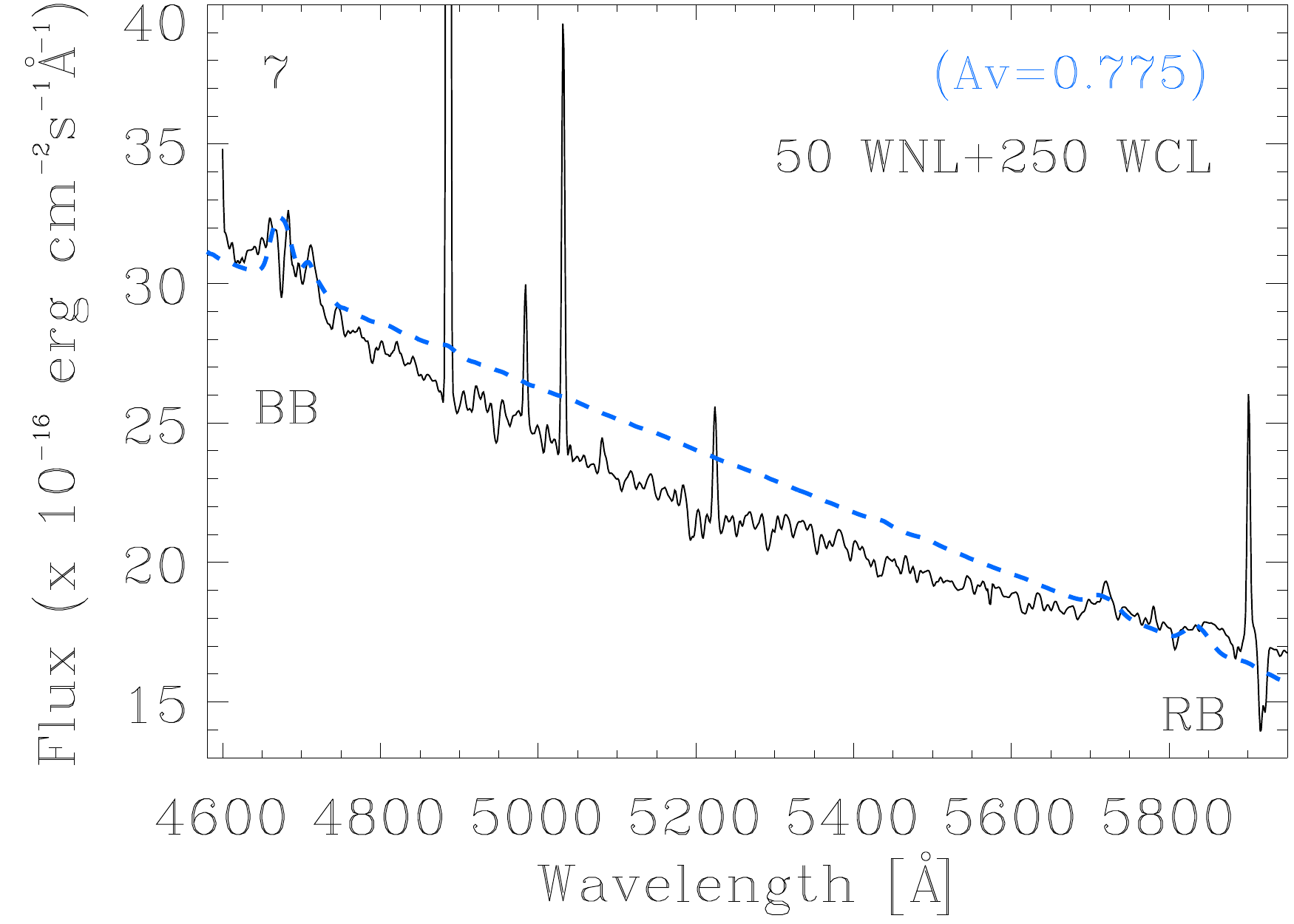}~
\includegraphics[width=0.33\linewidth]{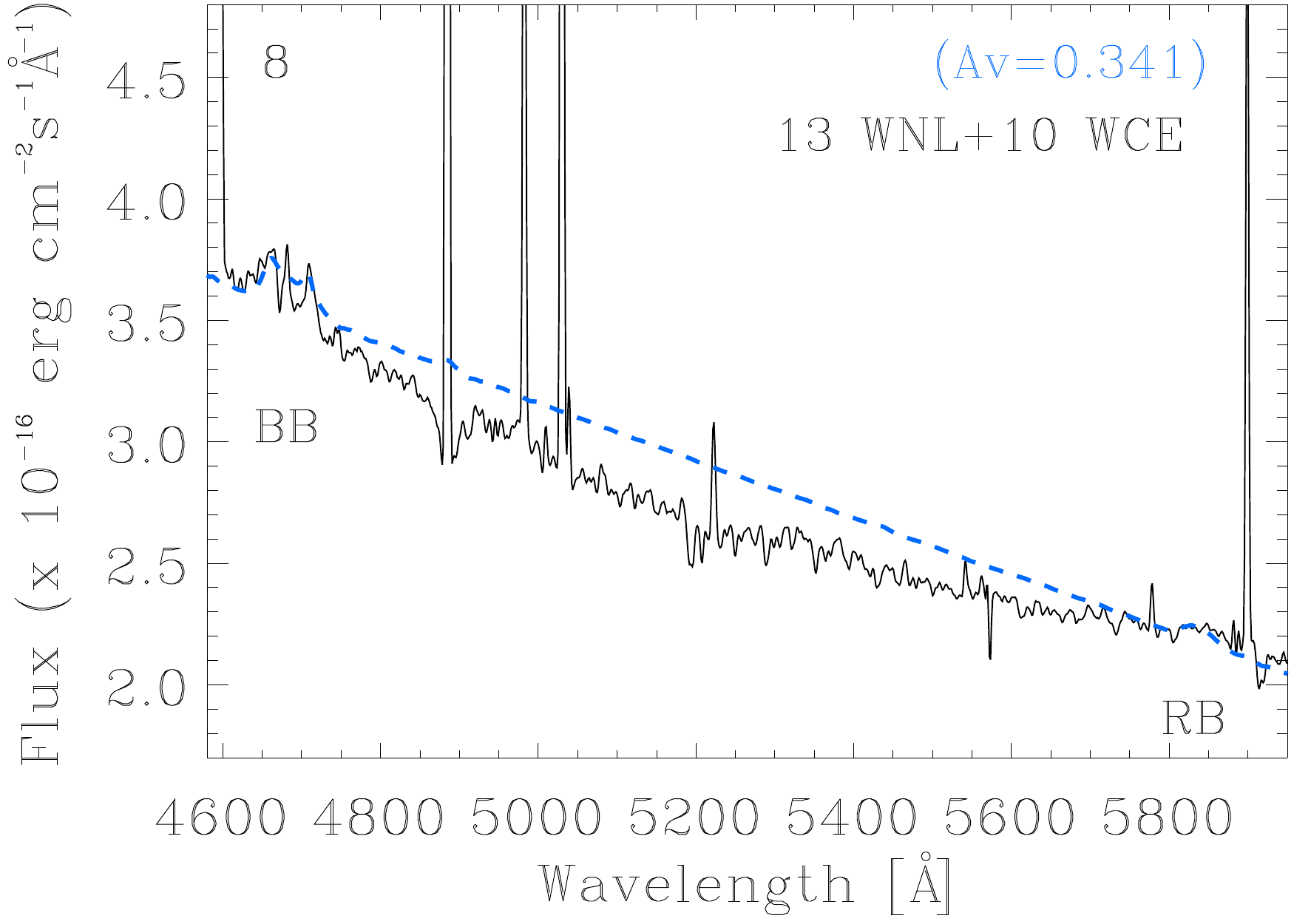}~
\includegraphics[width=0.33\linewidth]{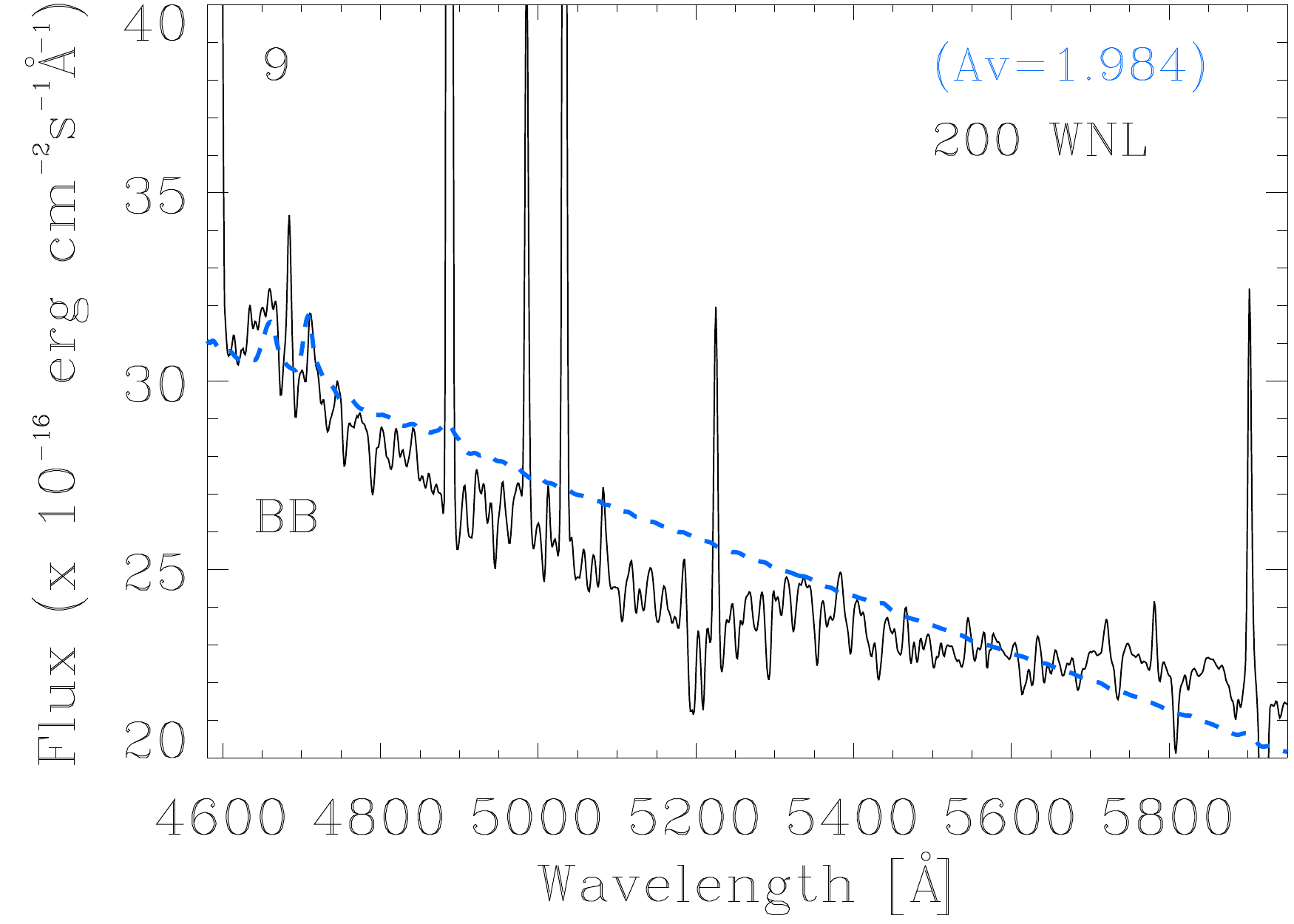}\\
\includegraphics[width=0.33\linewidth]{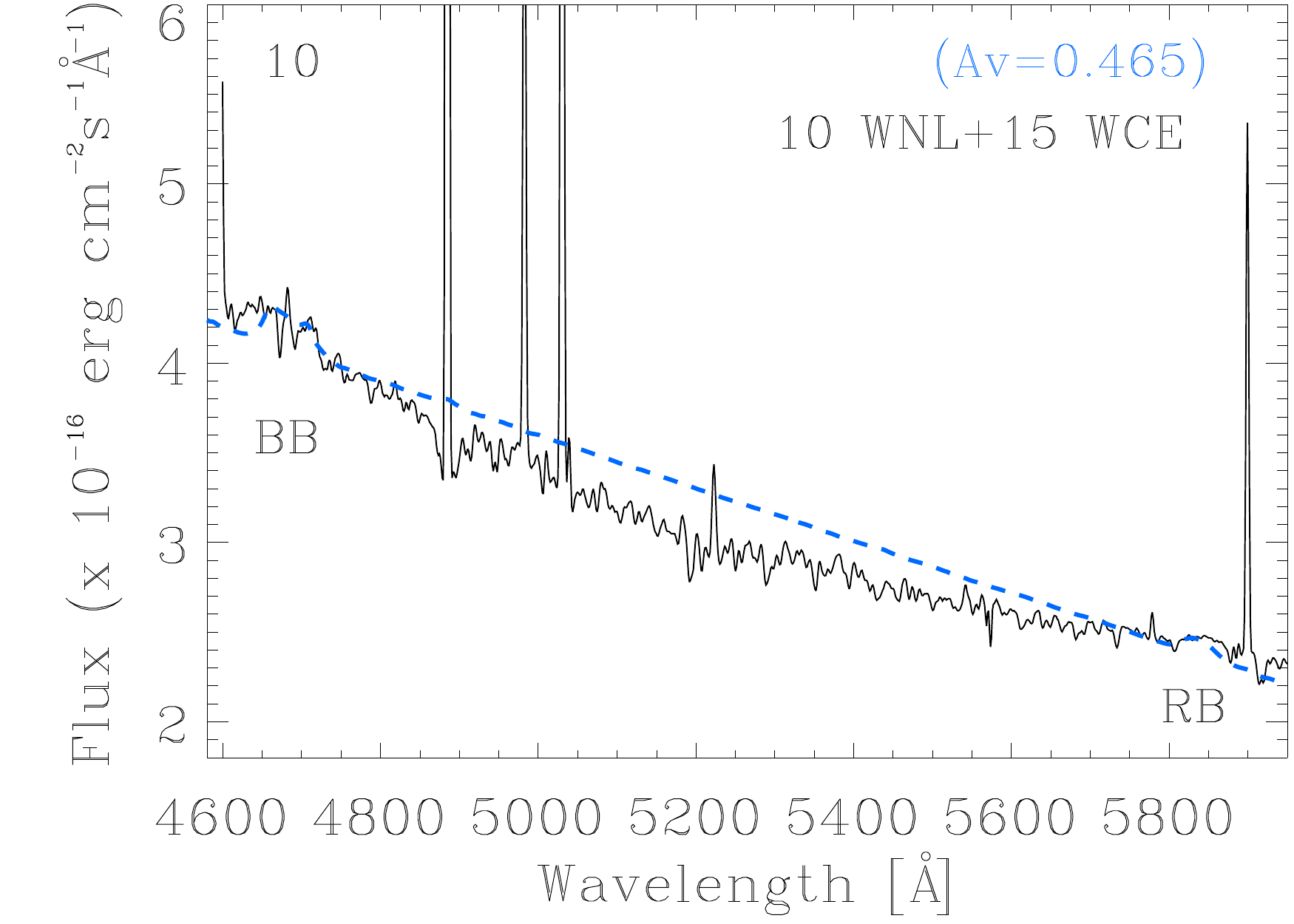}~
\includegraphics[width=0.33\linewidth]{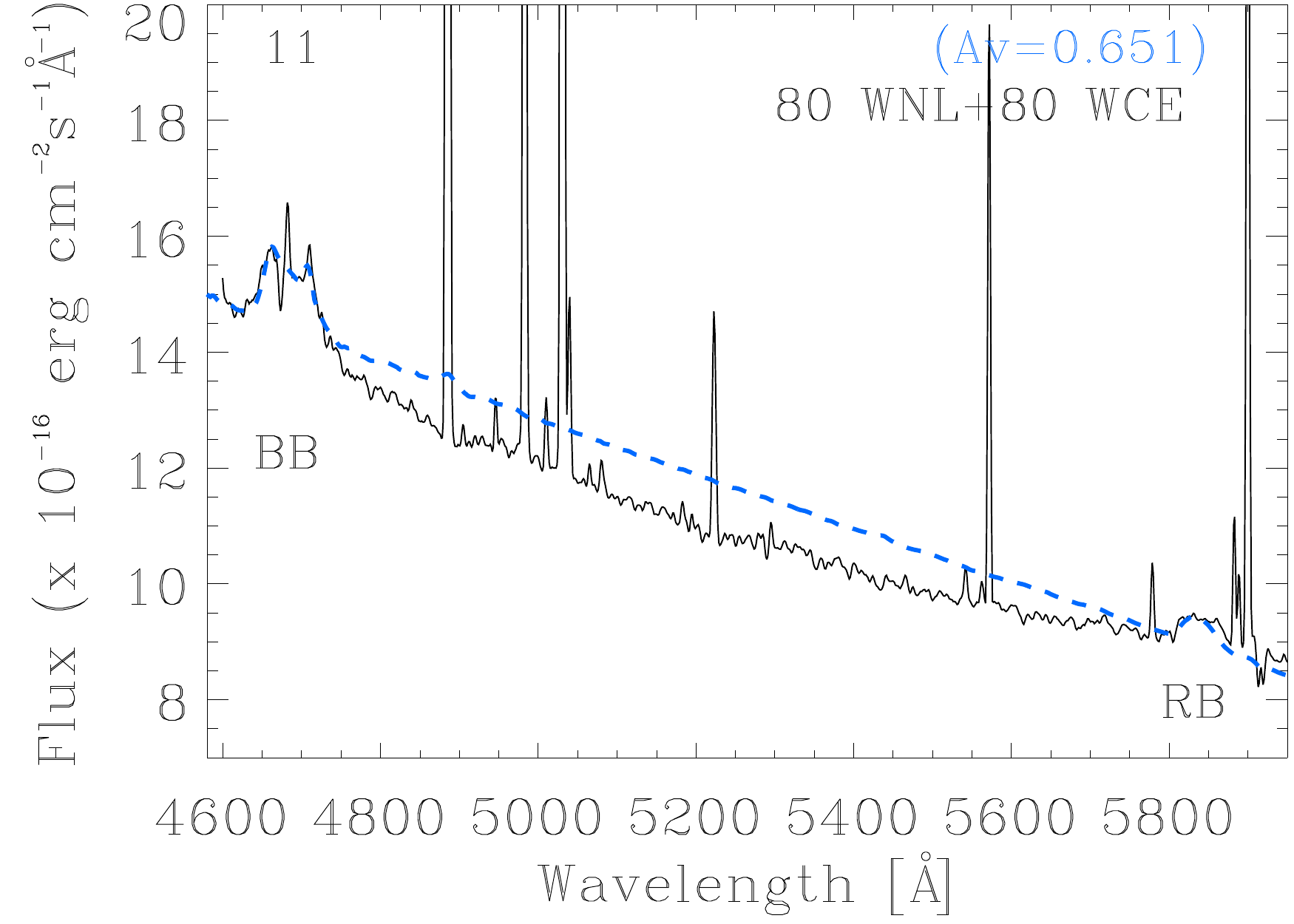}~
\includegraphics[width=0.33\linewidth]{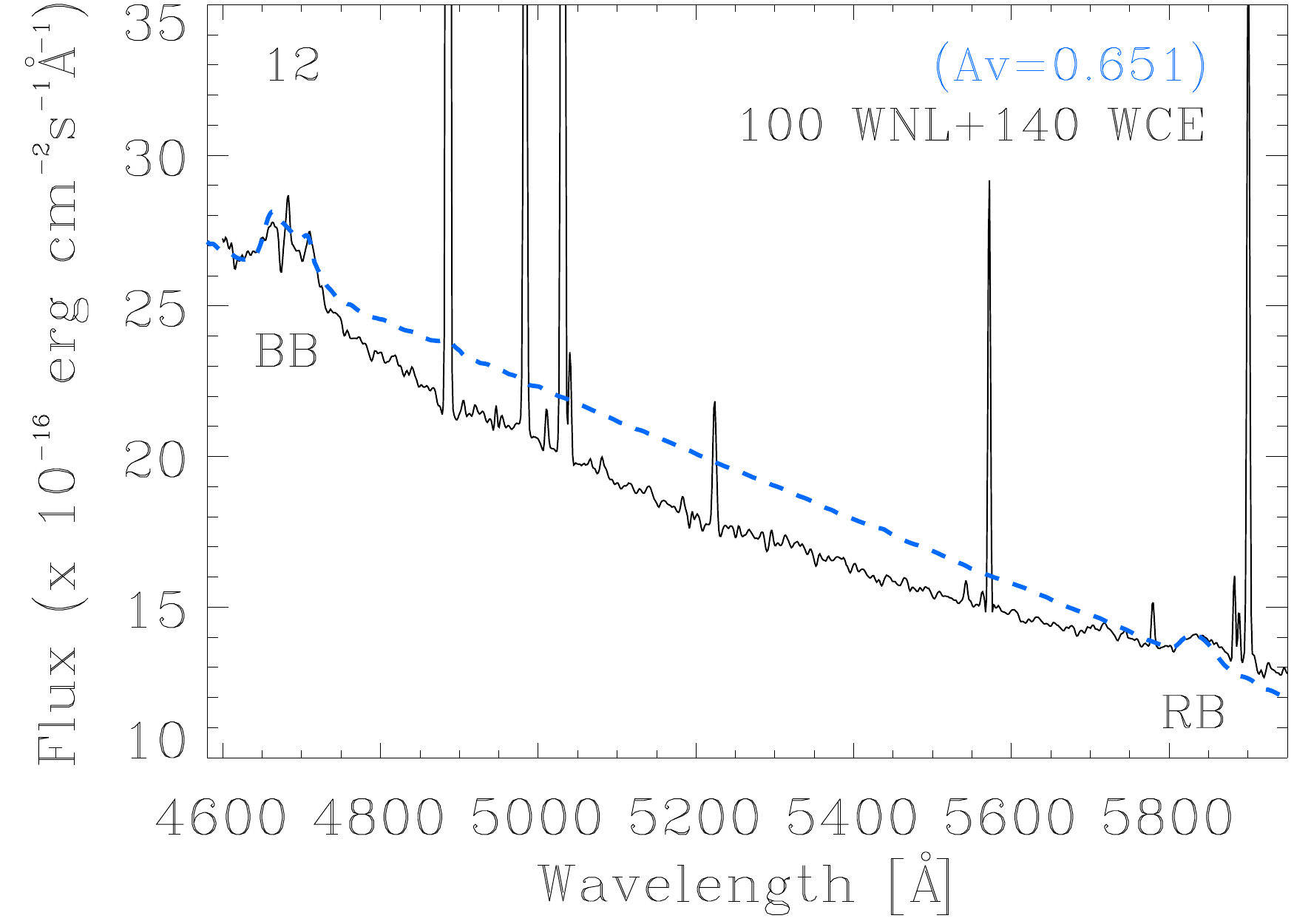}\\
\includegraphics[width=0.33\linewidth]{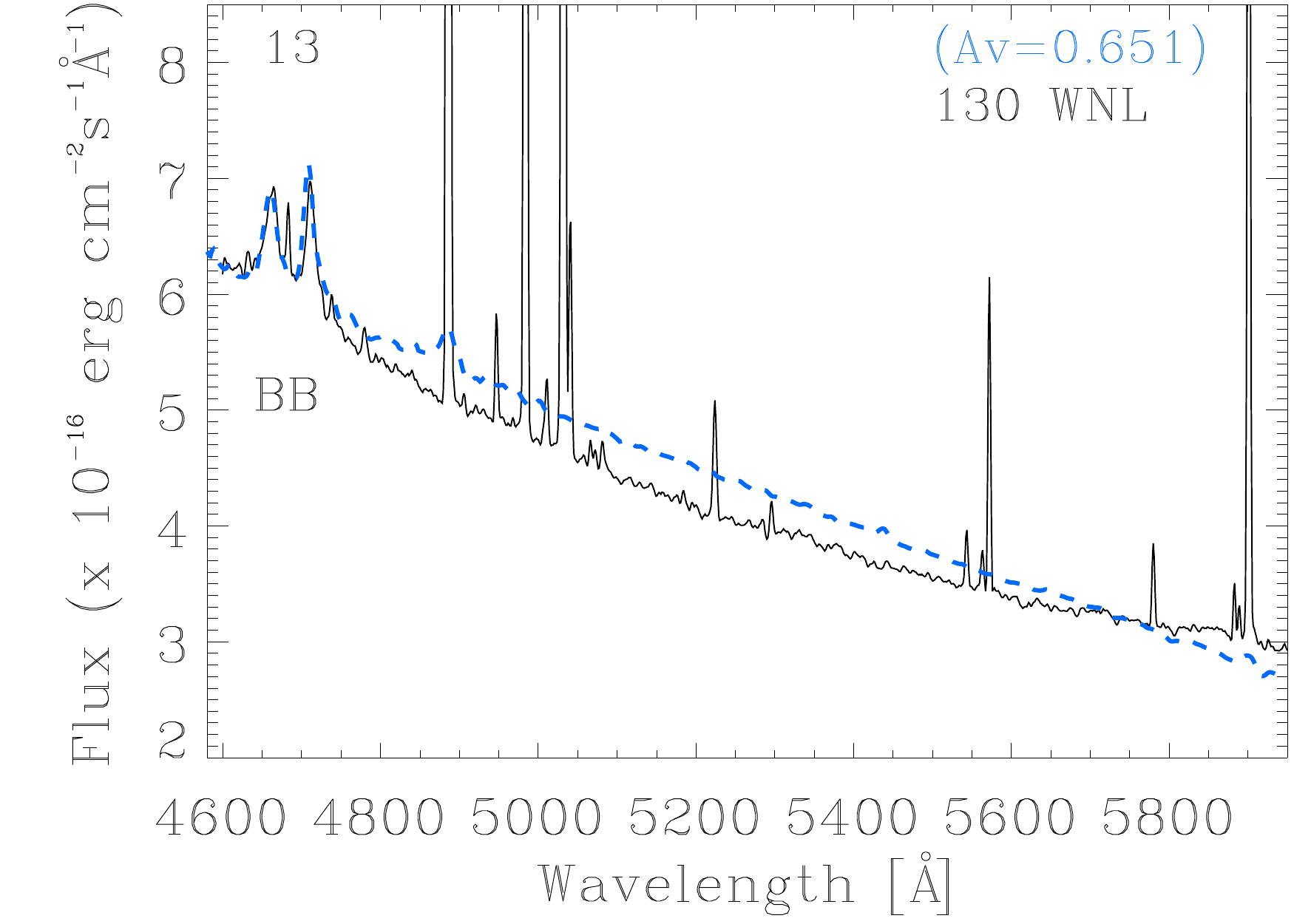}~
\includegraphics[width=0.33\linewidth]{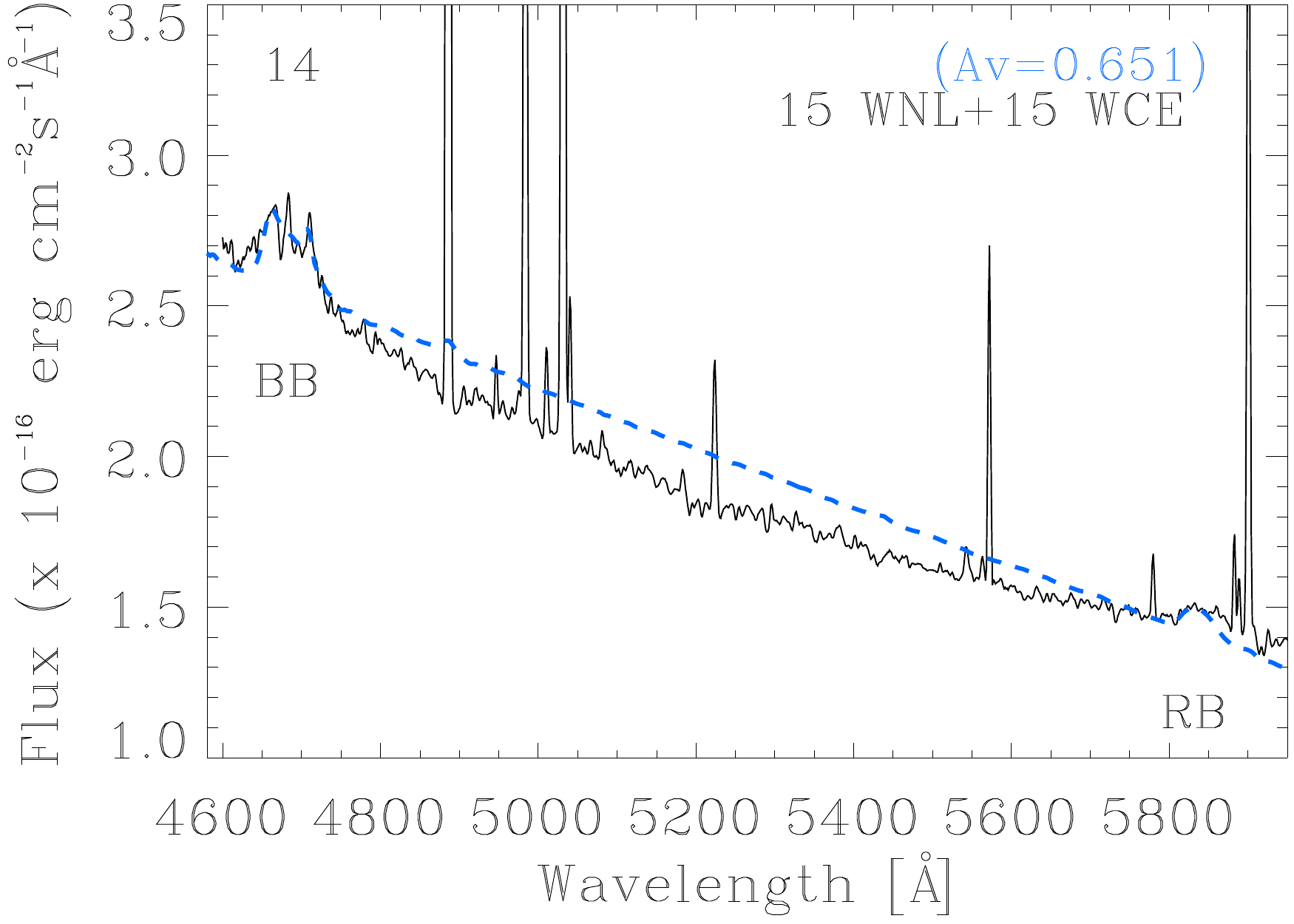}~
\includegraphics[width=0.33\linewidth]{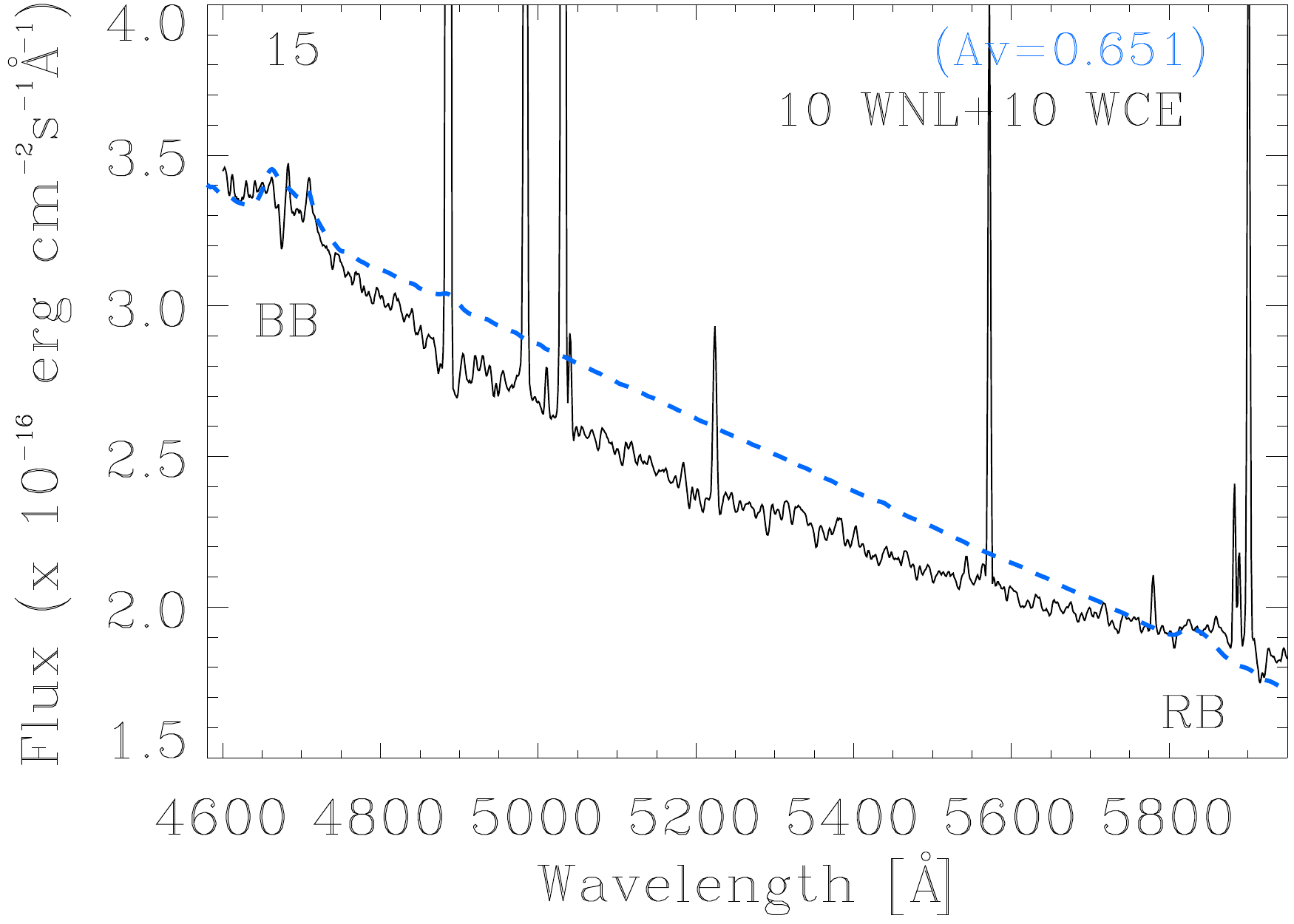}
\caption{
VLT MUSE spectra shown in the range of the BB and RB for the 38 star-forming complexes with WR features in the Antennae,
corrected by the extinction values reported in \citet{2010Whitmore} ({\it solid line}),
along with their best fit Galactic templates ({\it blue dashed line}).
The identification number of the WR region, the name of the templates and their multiplicative factors
required to match the observed bump strengths, as well of their $A_{\rm V}$ values,
are indicated.
The results are listed in Table~\ref{tab:class}.}
\label{fig:spectra}
\end{center}
\end{figure*}

\begin{figure*}
\begin{center}
\includegraphics[width=0.33\linewidth]{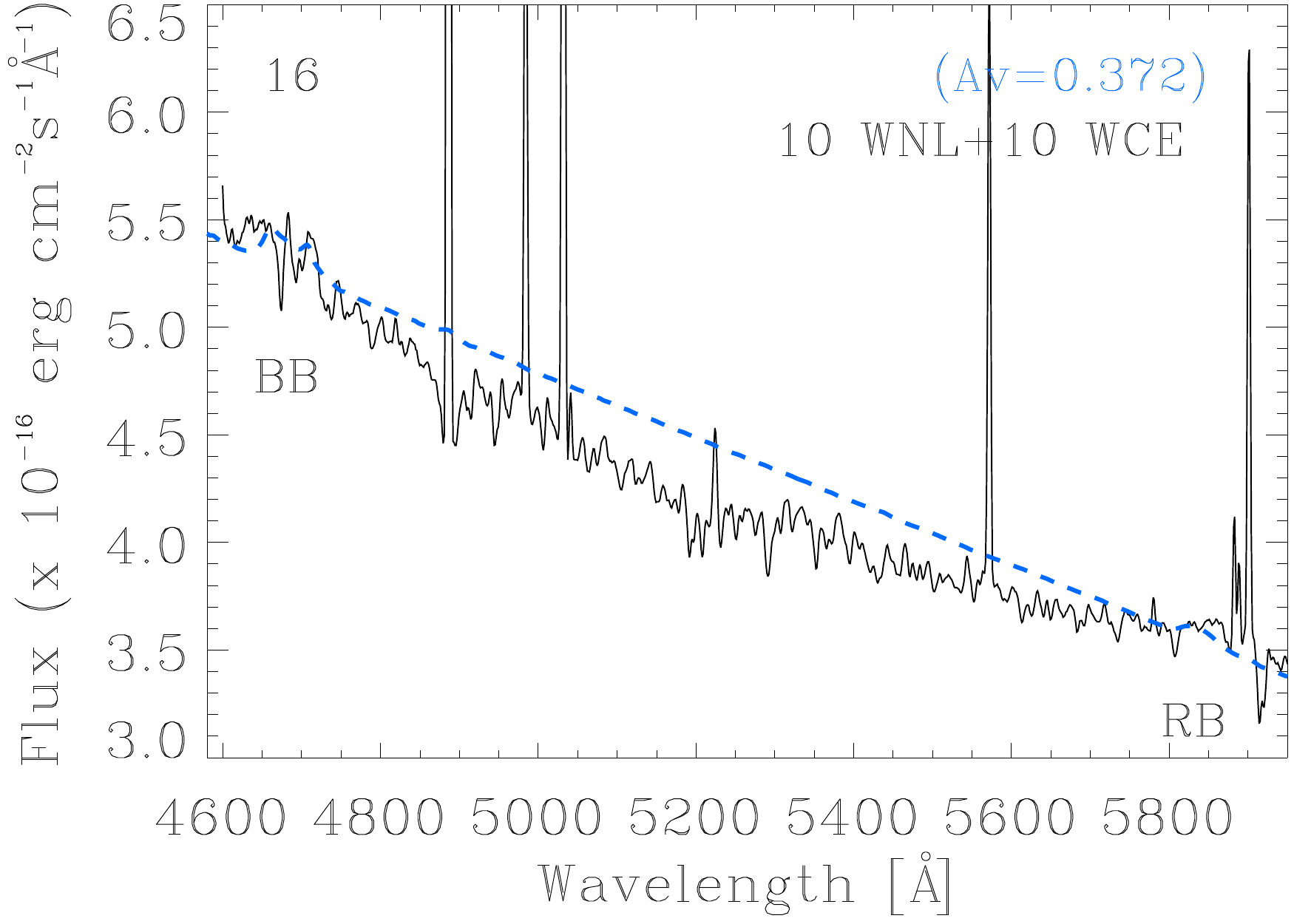}~
\includegraphics[width=0.33\linewidth]{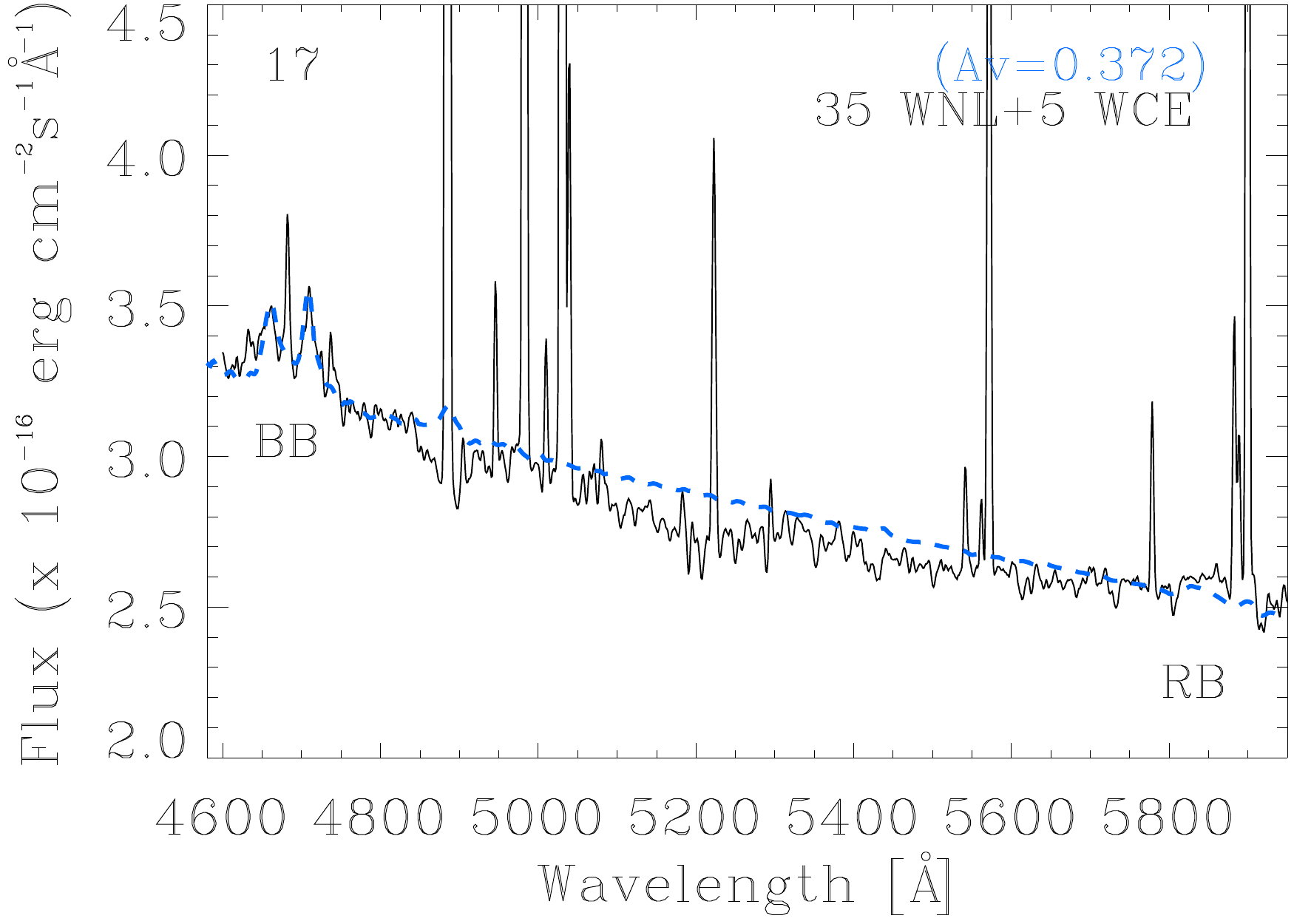}~
\includegraphics[width=0.33\linewidth]{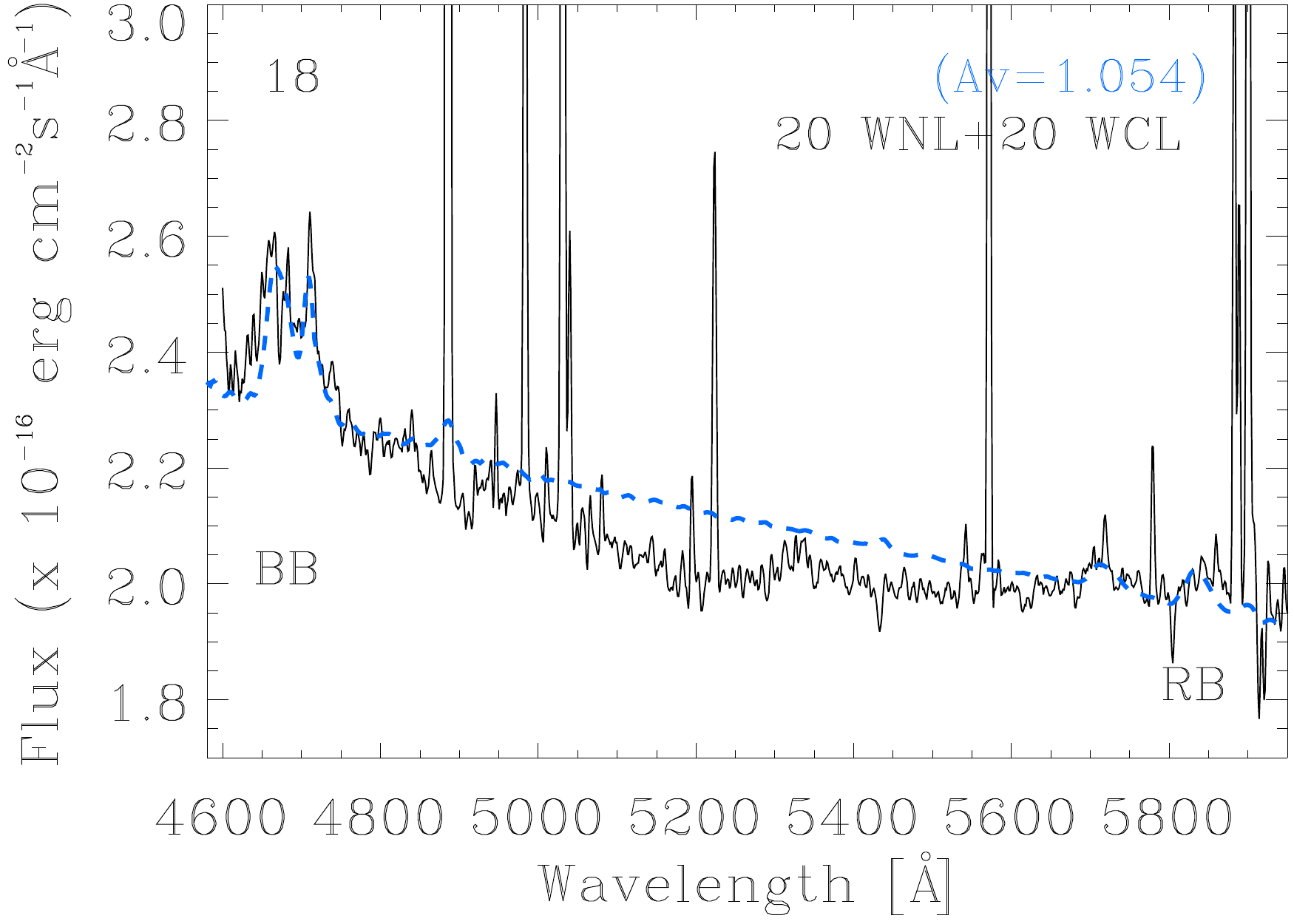}\\
\includegraphics[width=0.33\linewidth]{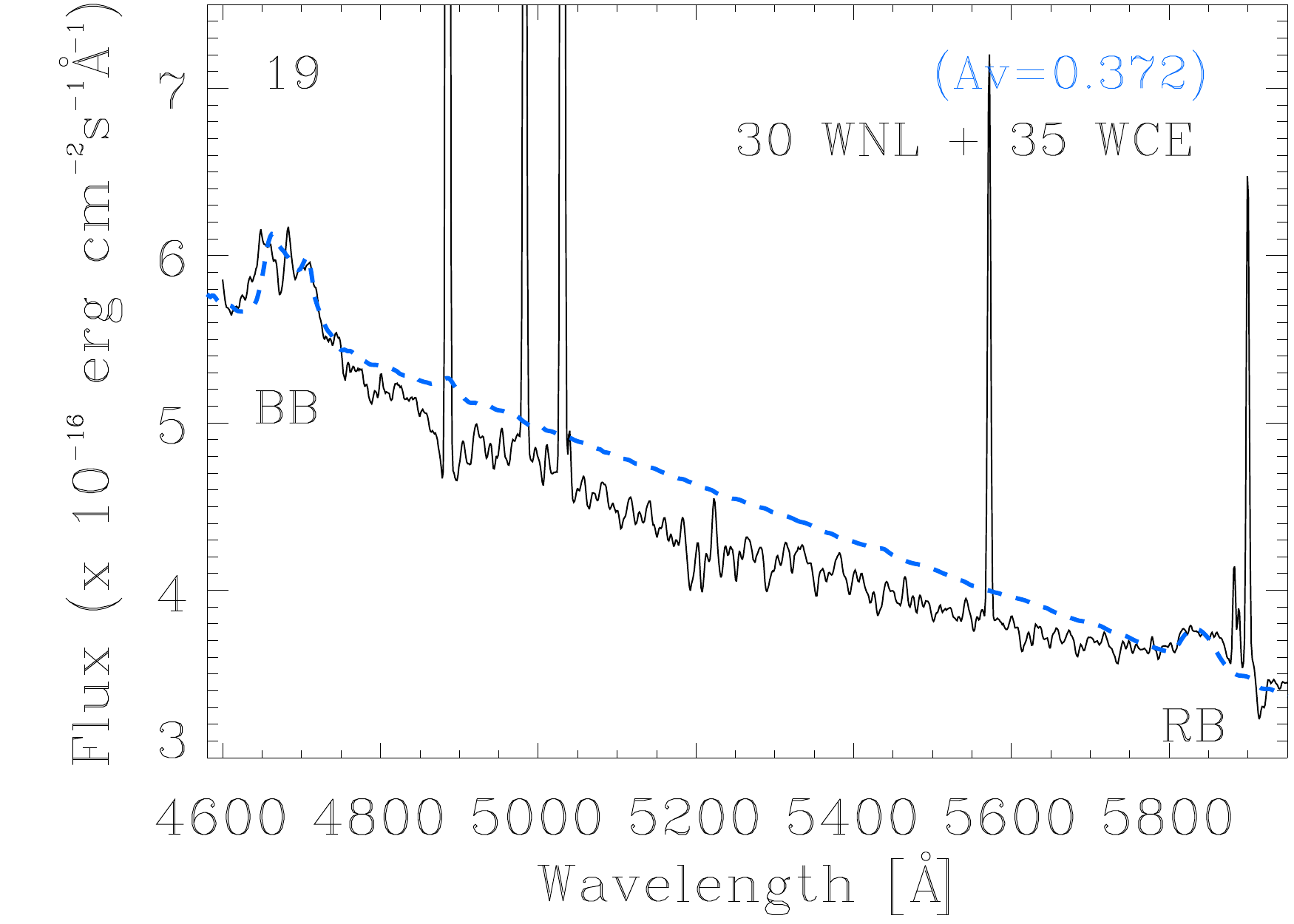}~
\includegraphics[width=0.33\linewidth]{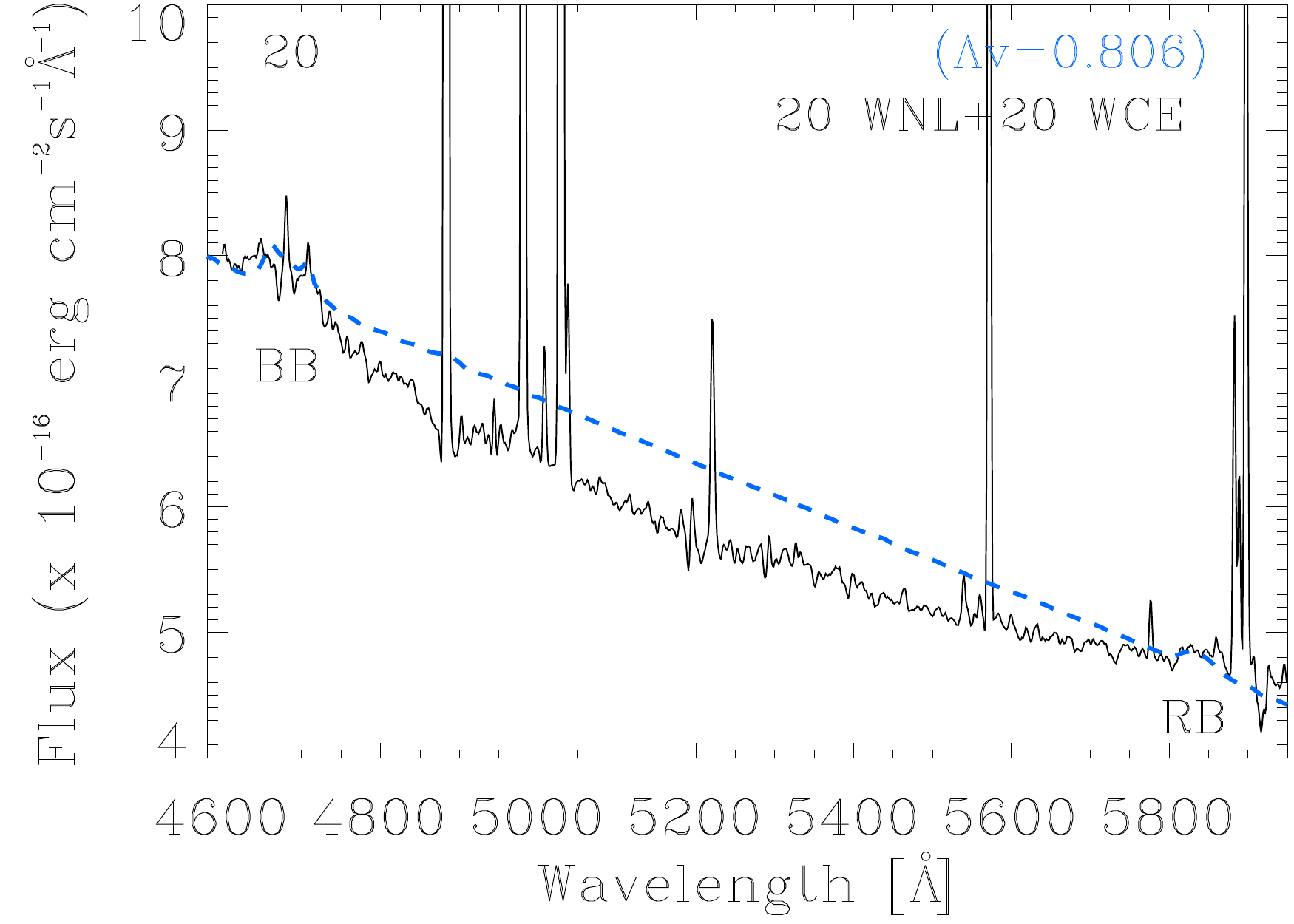}~
\includegraphics[width=0.33\linewidth]{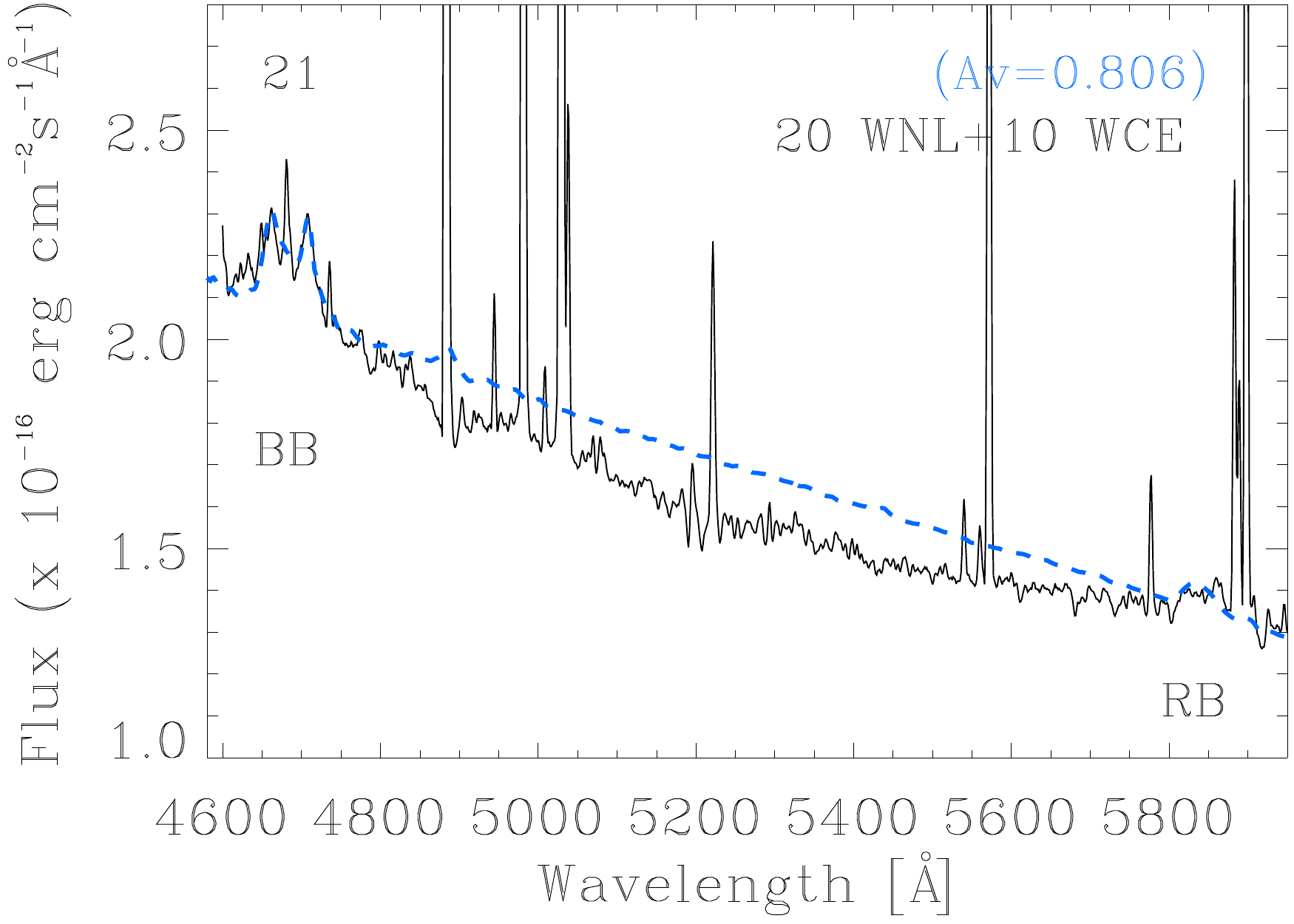}\\
\includegraphics[width=0.33\linewidth]{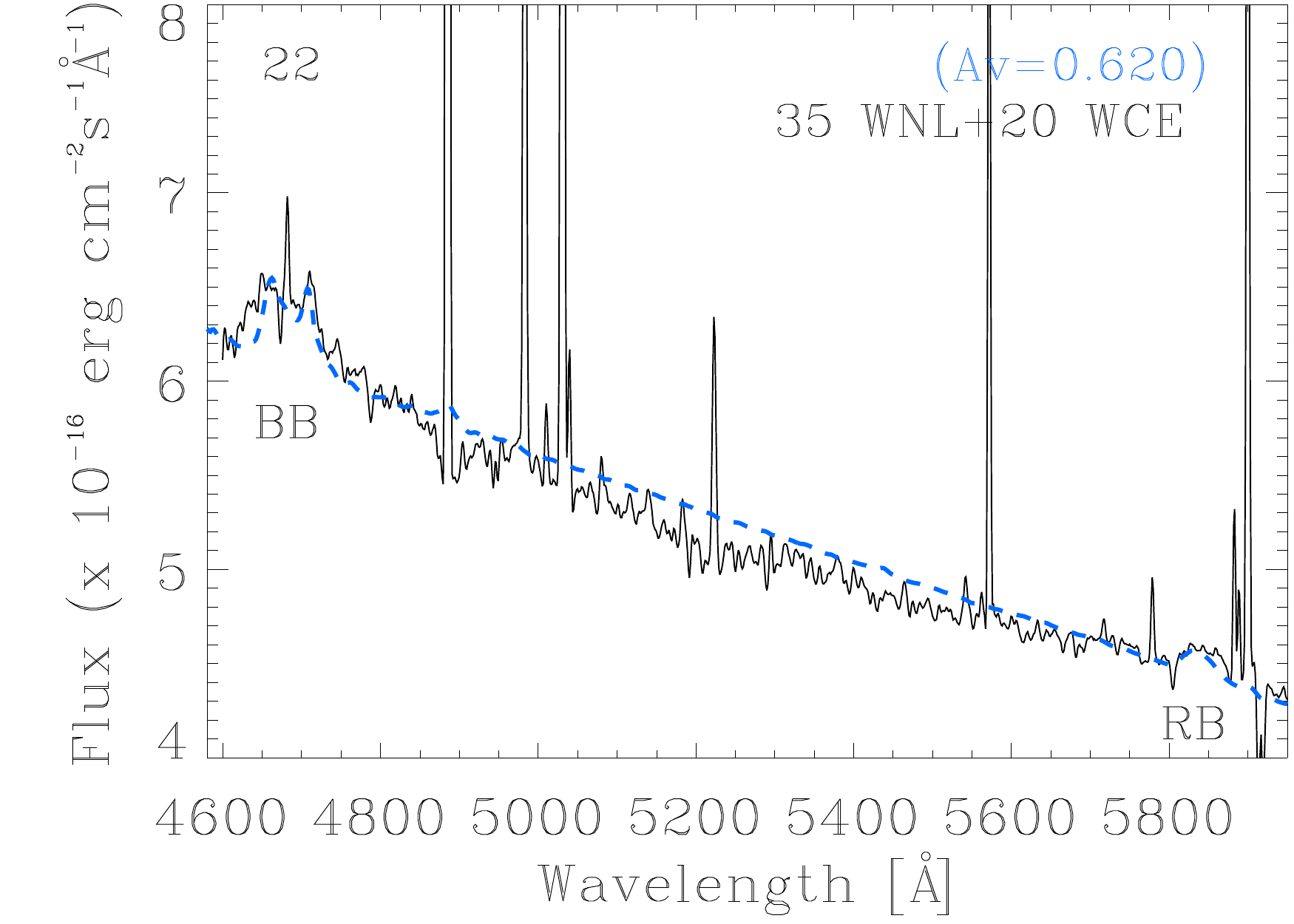}~
\includegraphics[width=0.33\linewidth]{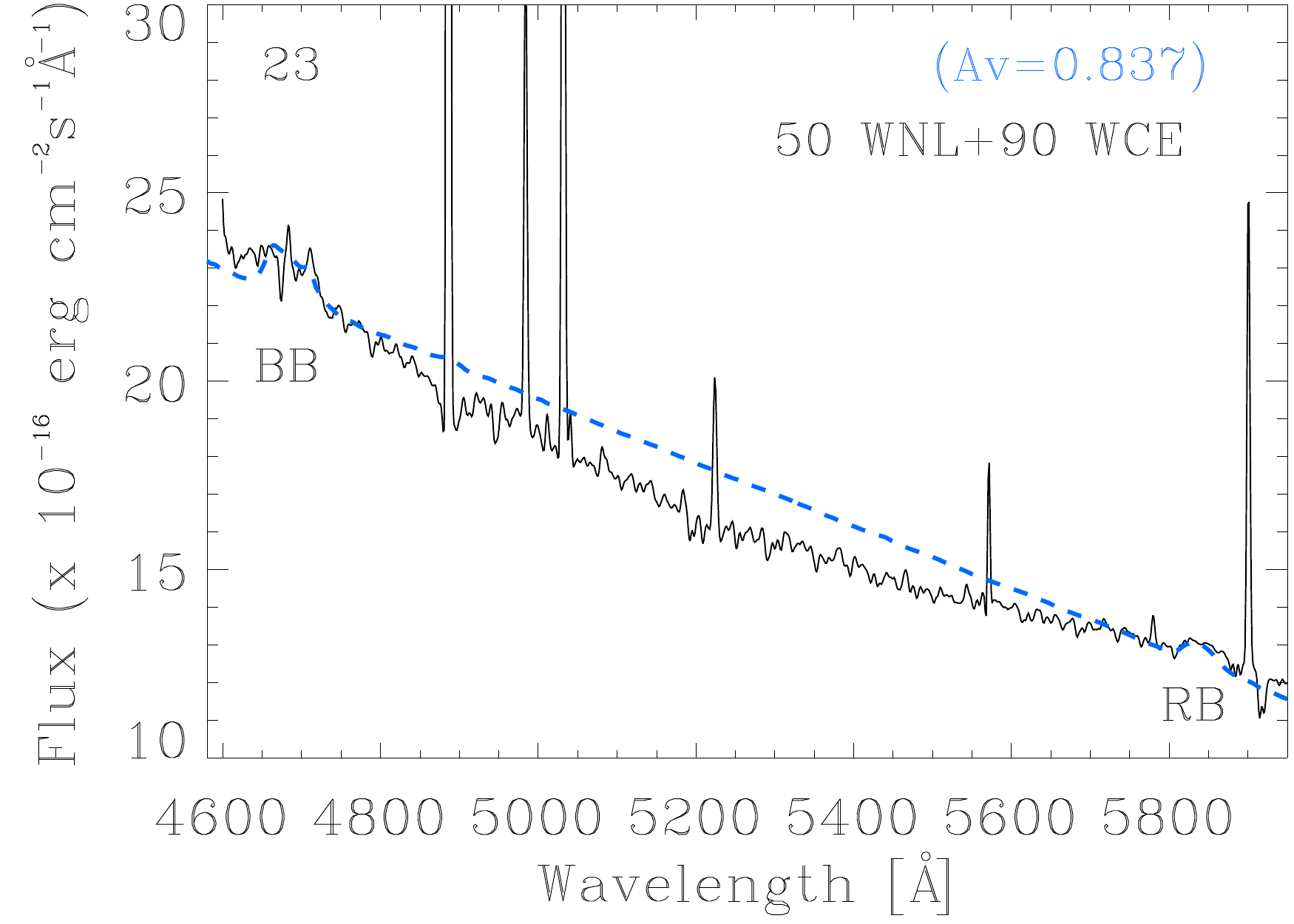}~
\includegraphics[width=0.33\linewidth]{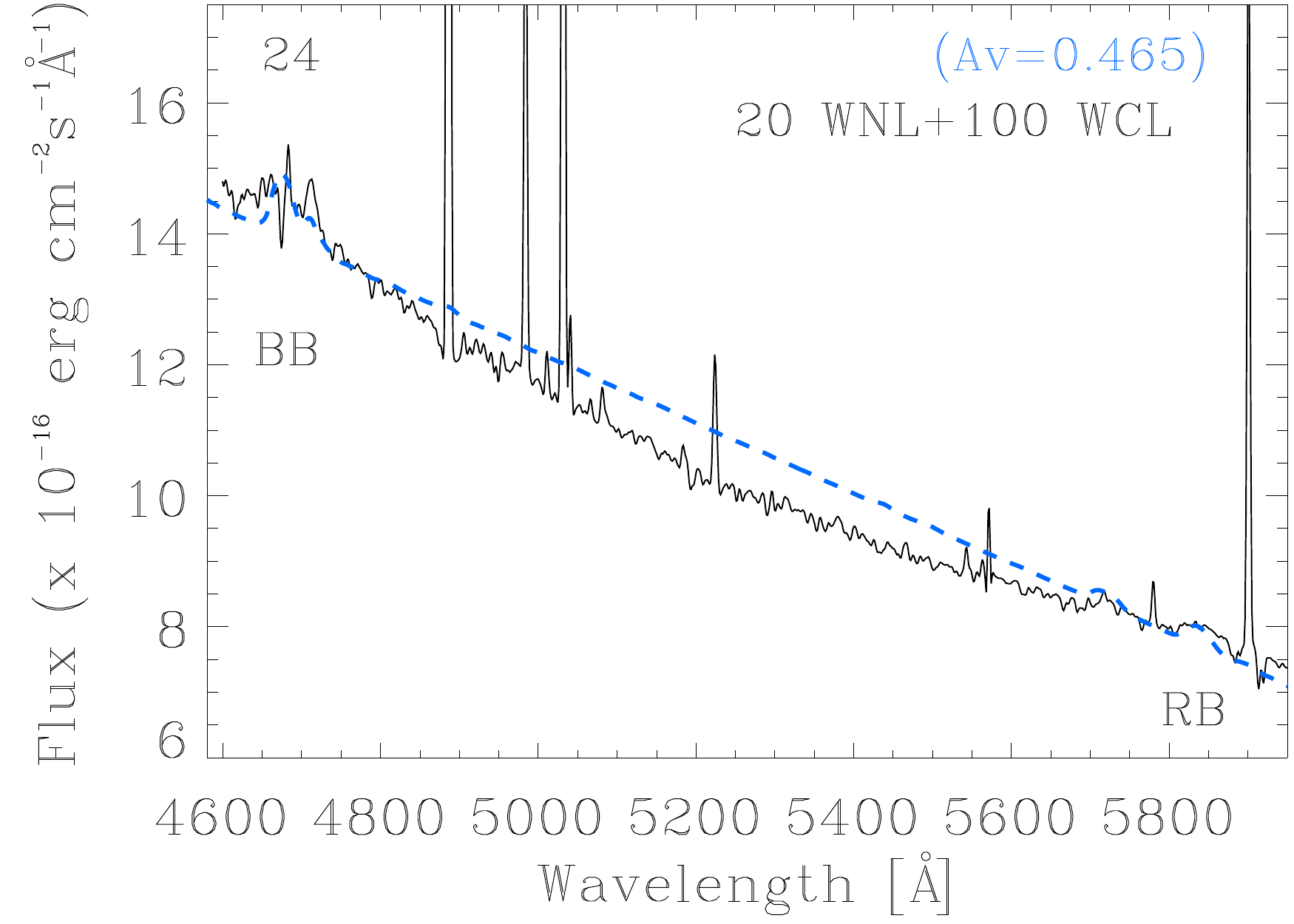}\\
\includegraphics[width=0.33\linewidth]{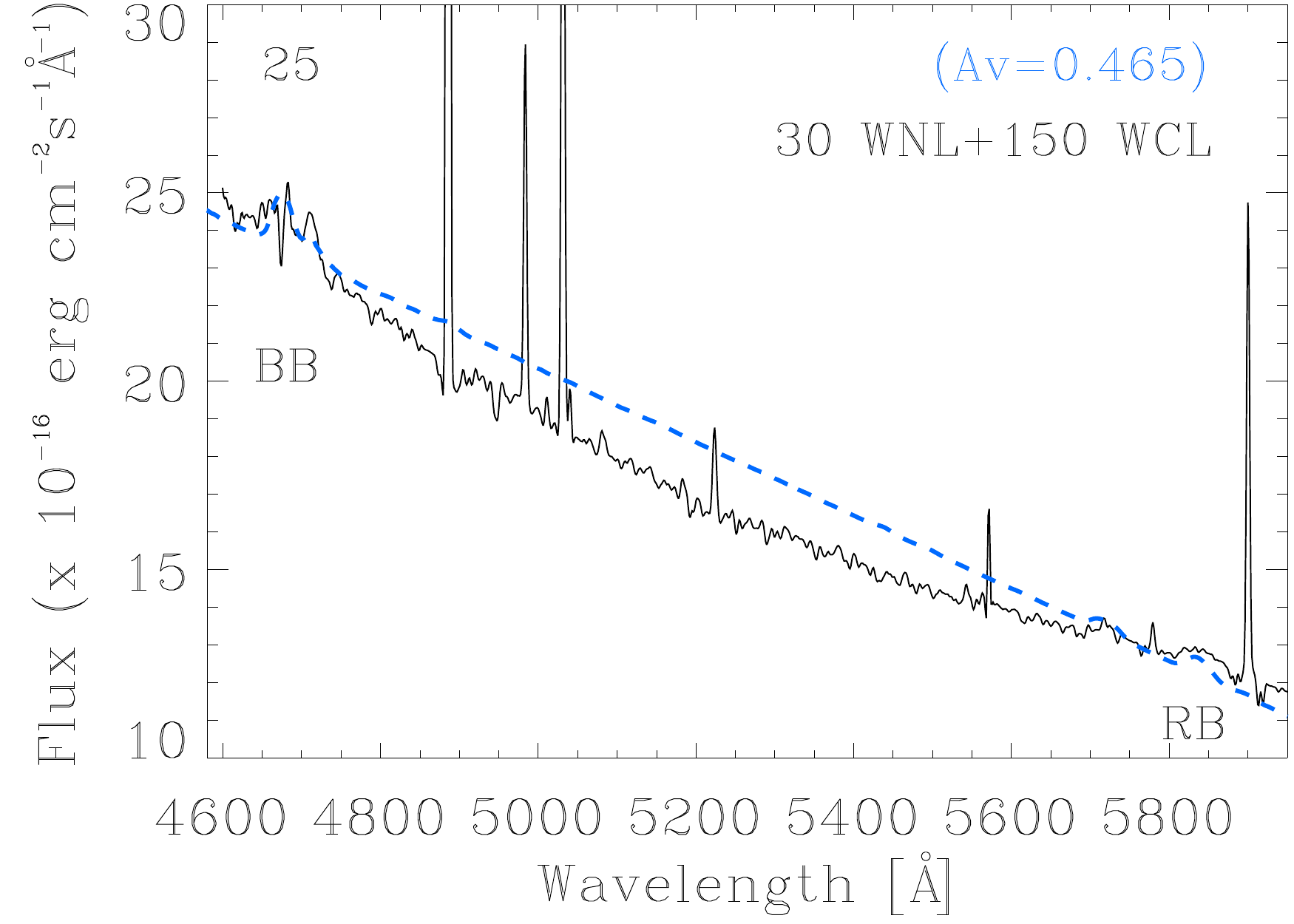}~
\includegraphics[width=0.33\linewidth]{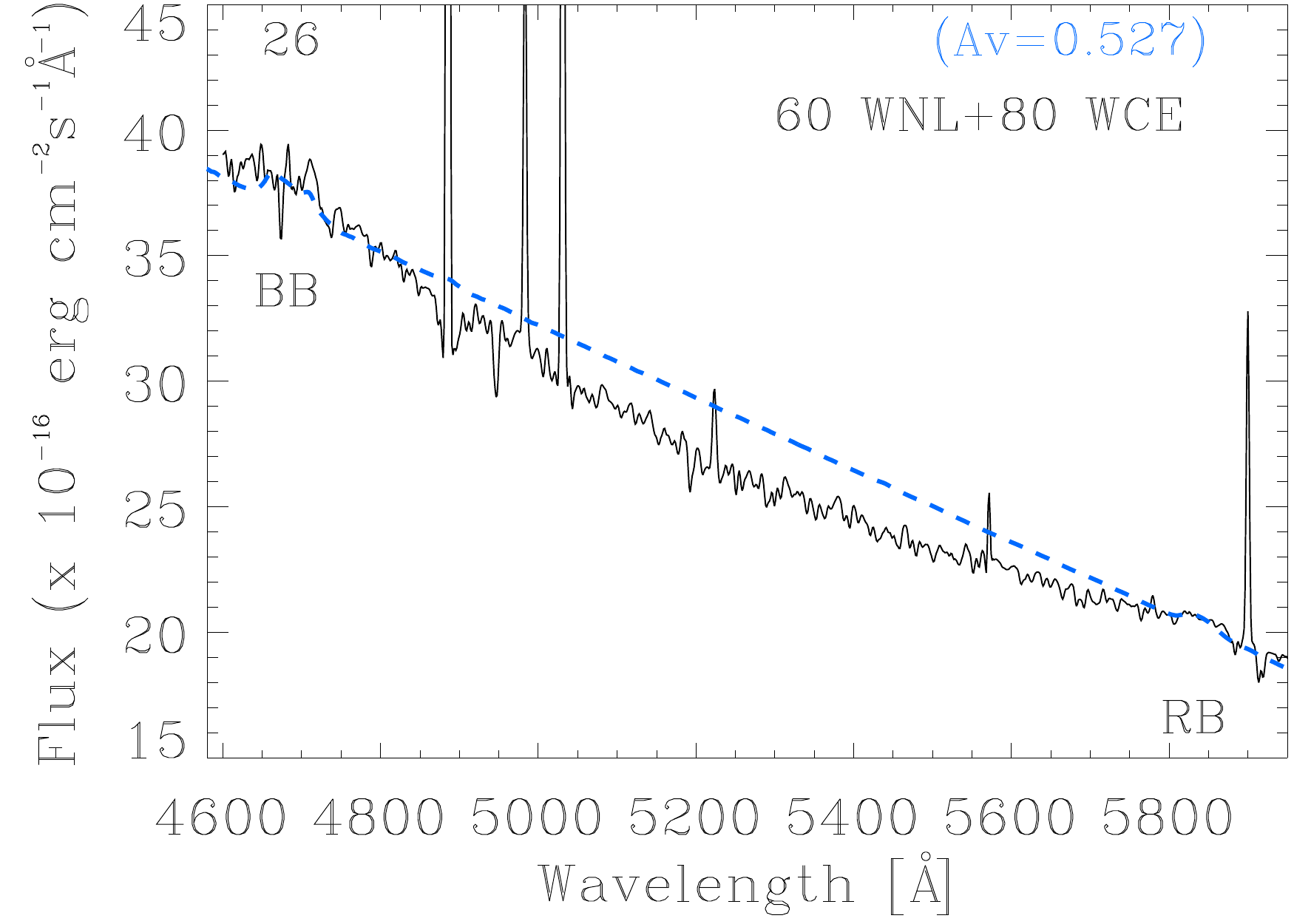}~
\includegraphics[width=0.33\linewidth]{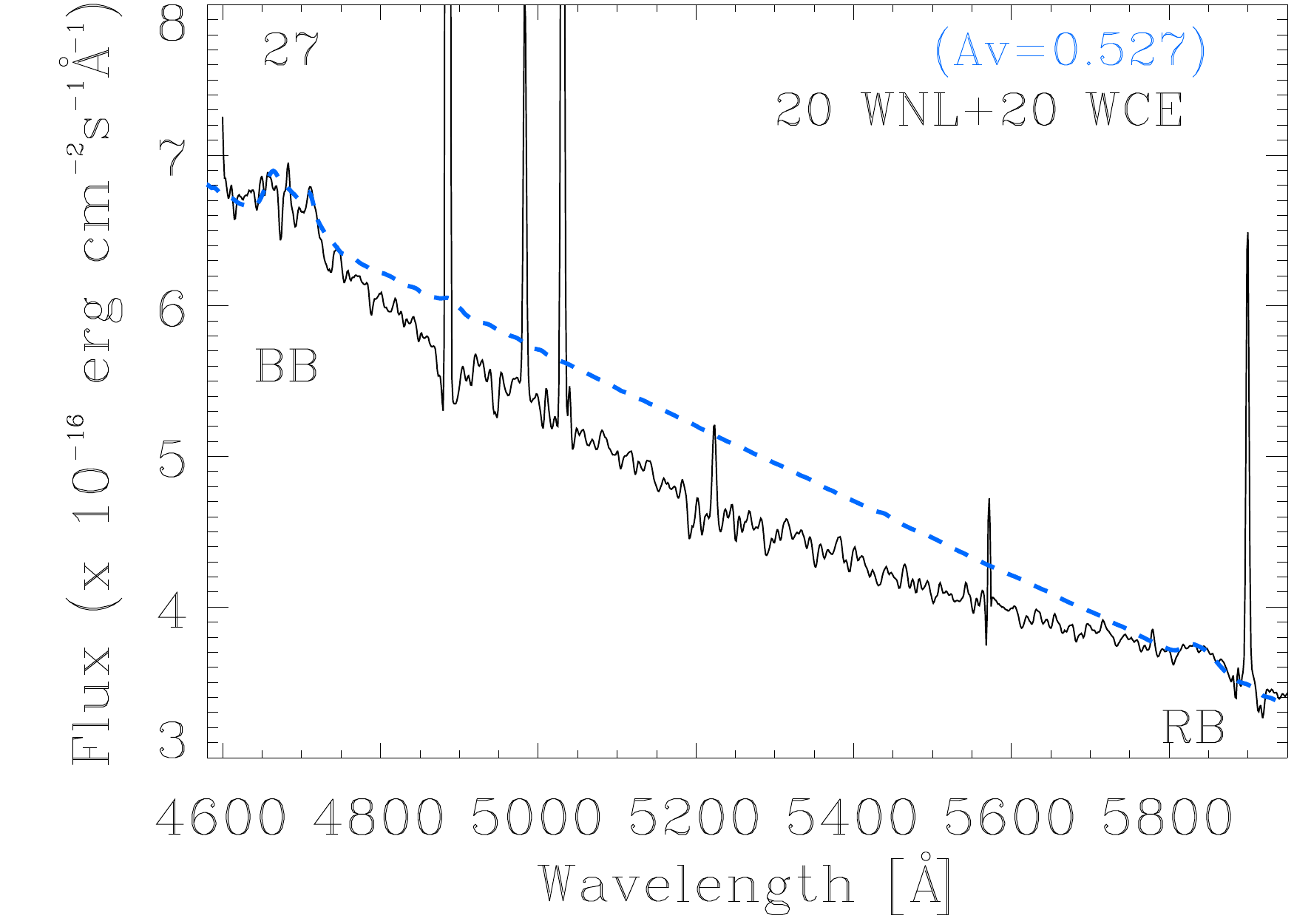}\\
\includegraphics[width=0.33\linewidth]{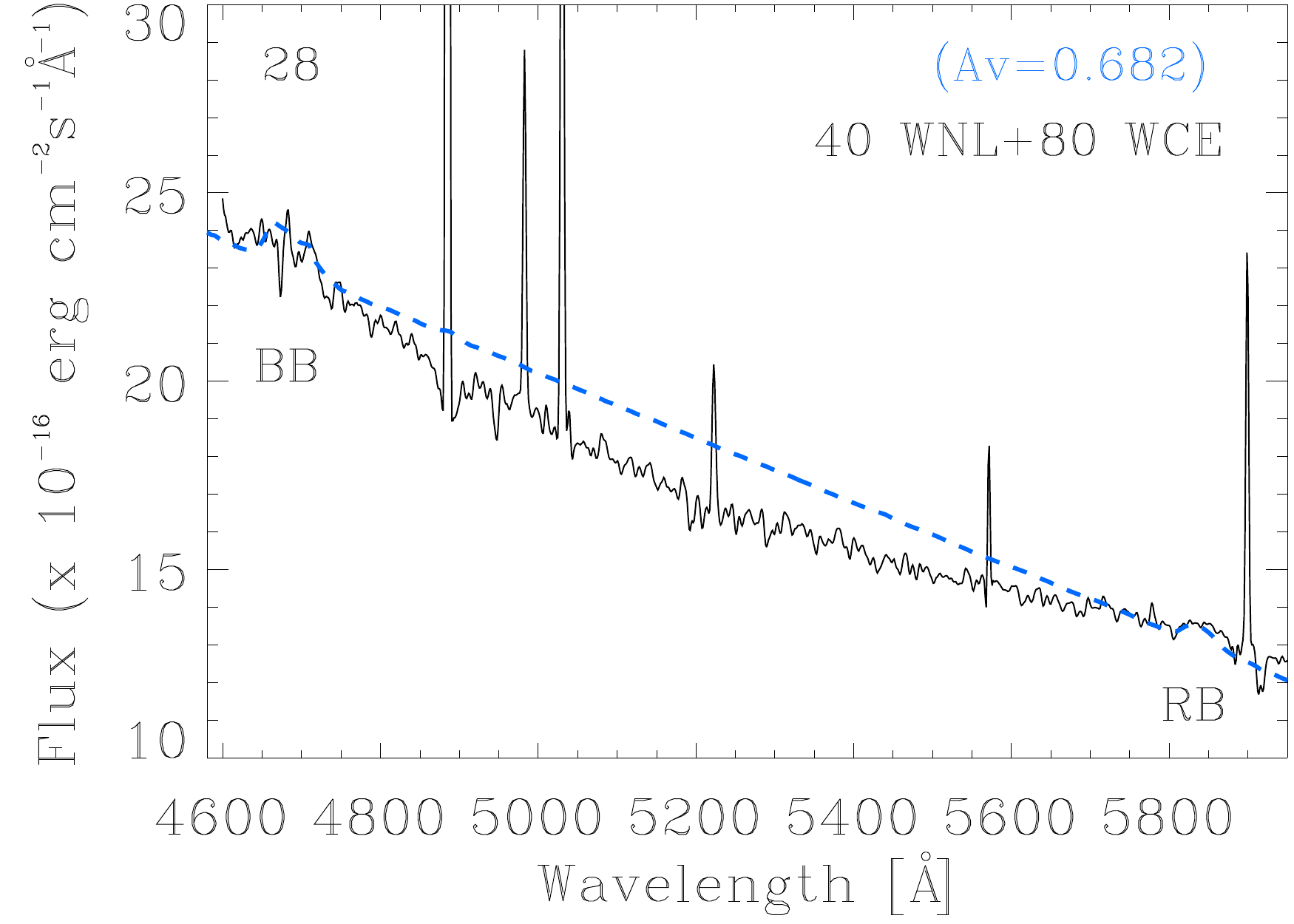}~
\includegraphics[width=0.33\linewidth]{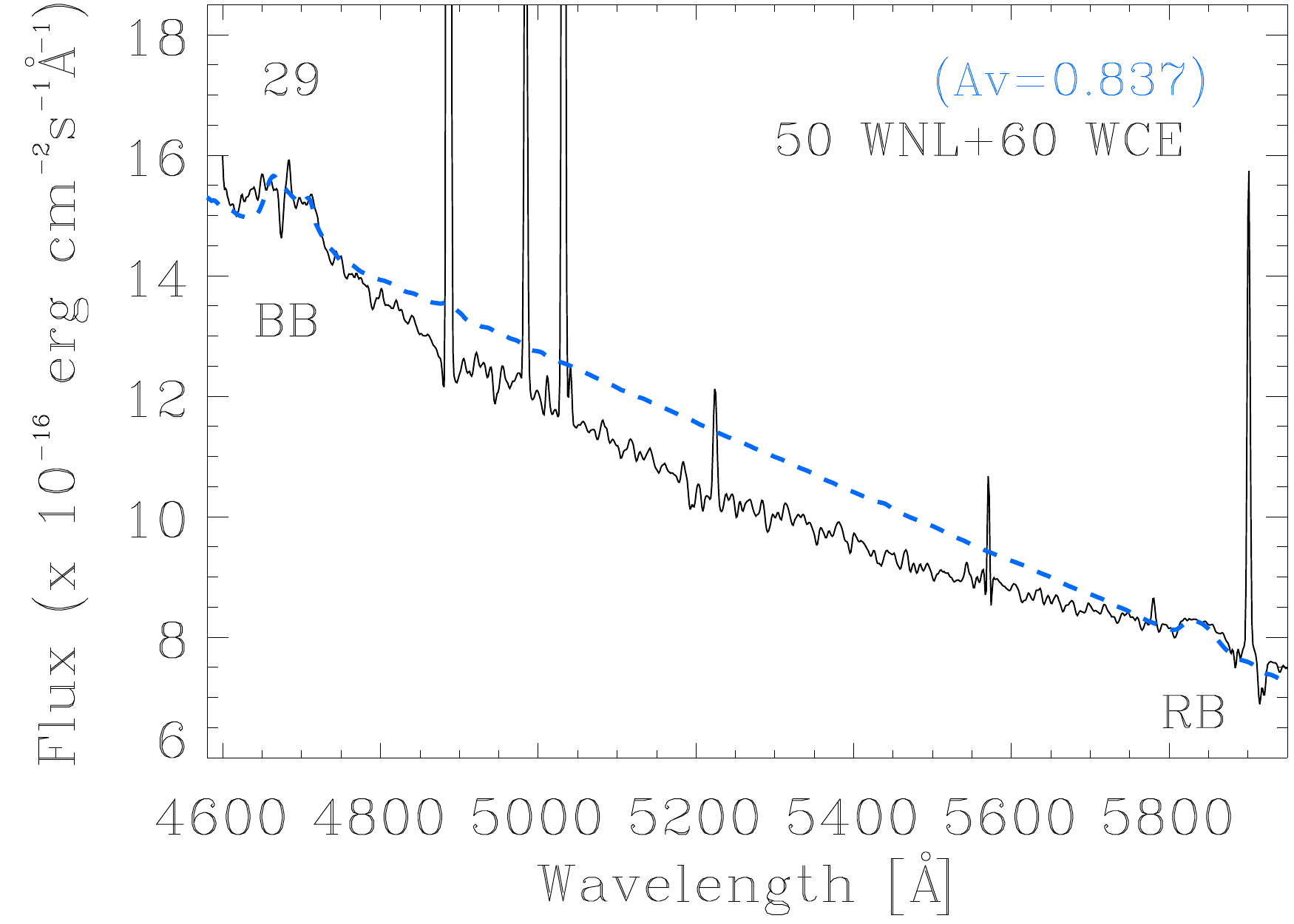}~
\includegraphics[width=0.33\linewidth]{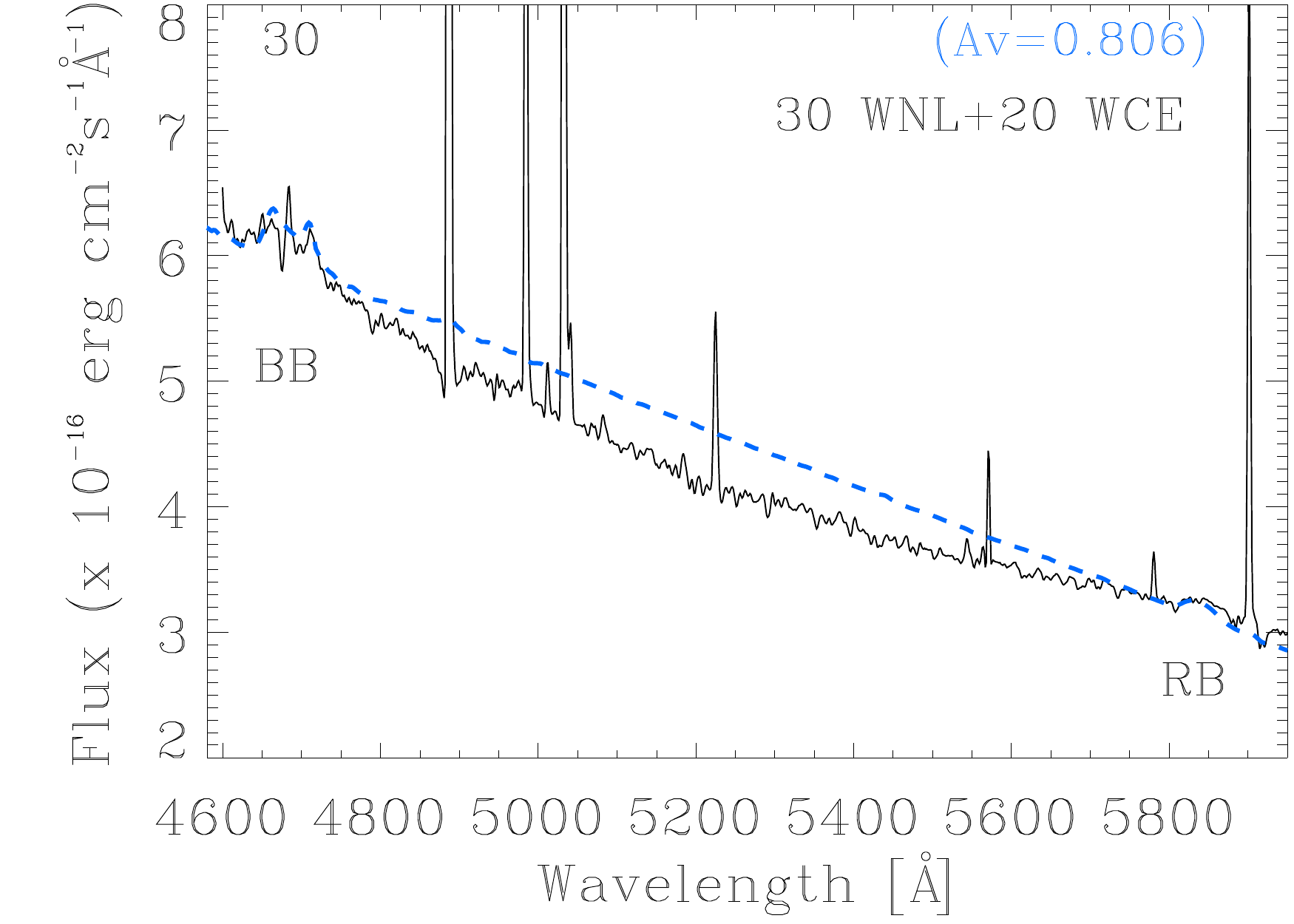}
\ContinuedFloat
\captionsetup{list=off,format=cont}
\caption{{\it -- continued}}
\end{center}
\end{figure*}

\begin{figure*}
\begin{center}
\includegraphics[width=0.33\linewidth]{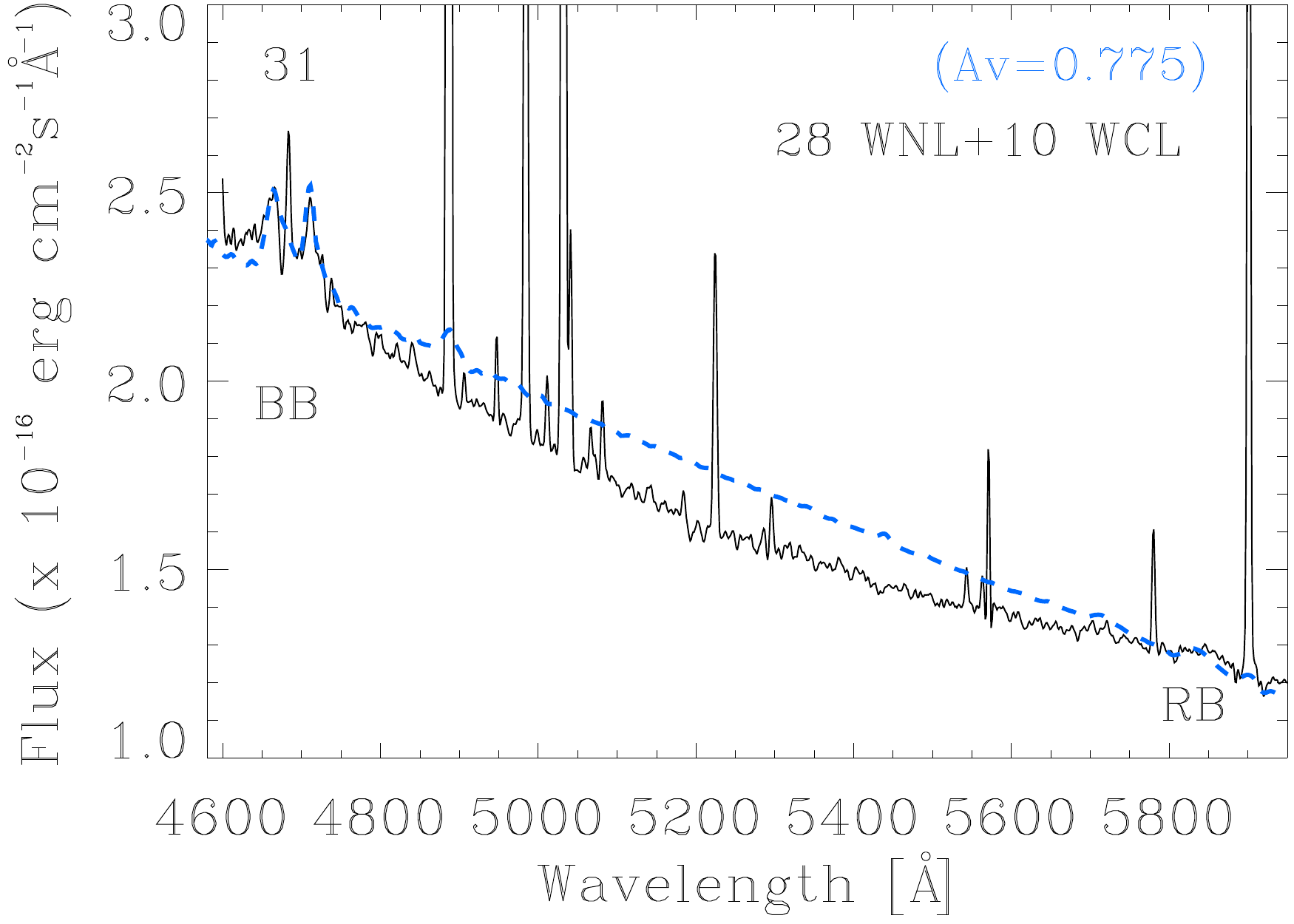}~
\includegraphics[width=0.33\linewidth]{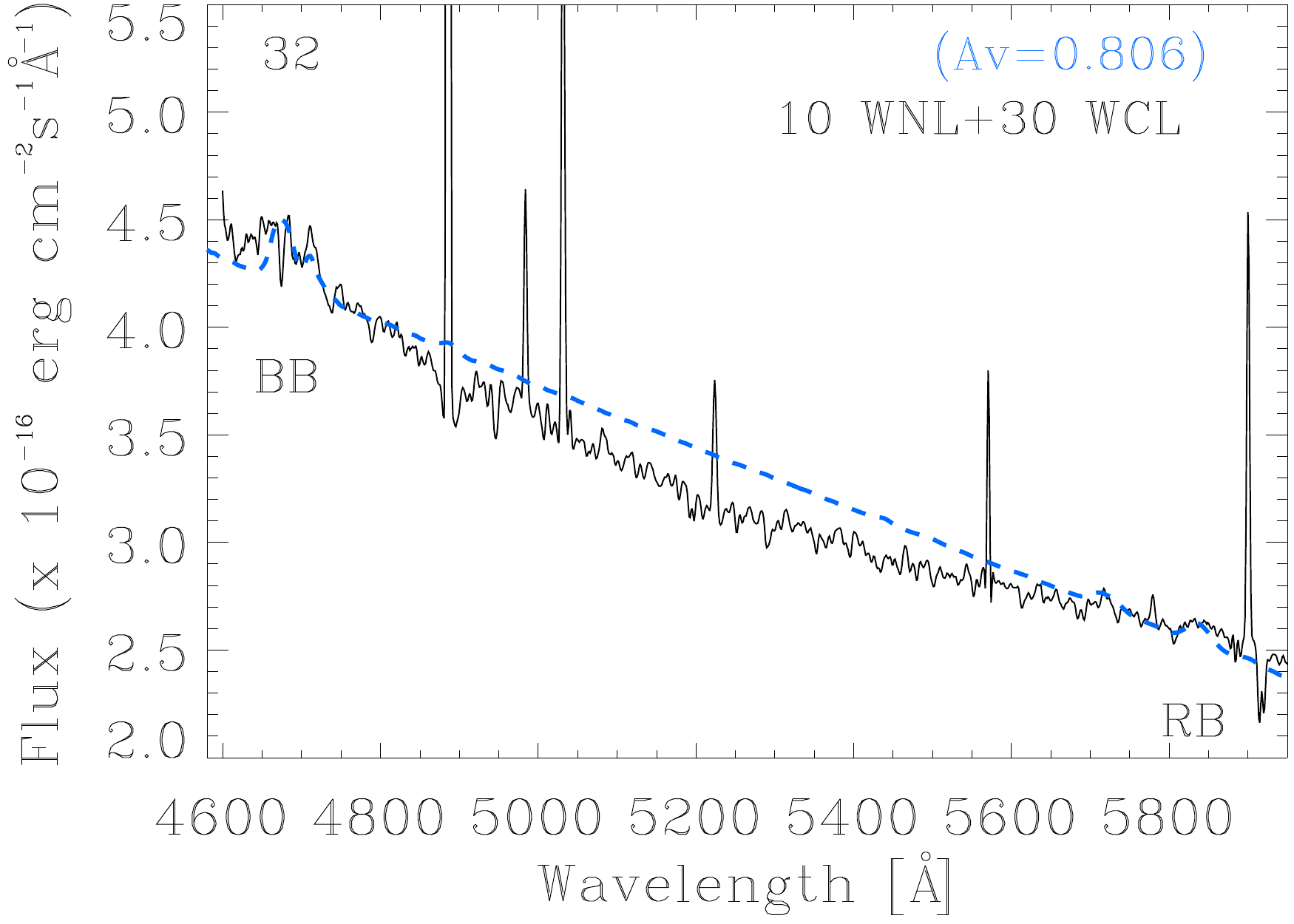}~
\includegraphics[width=0.33\linewidth]{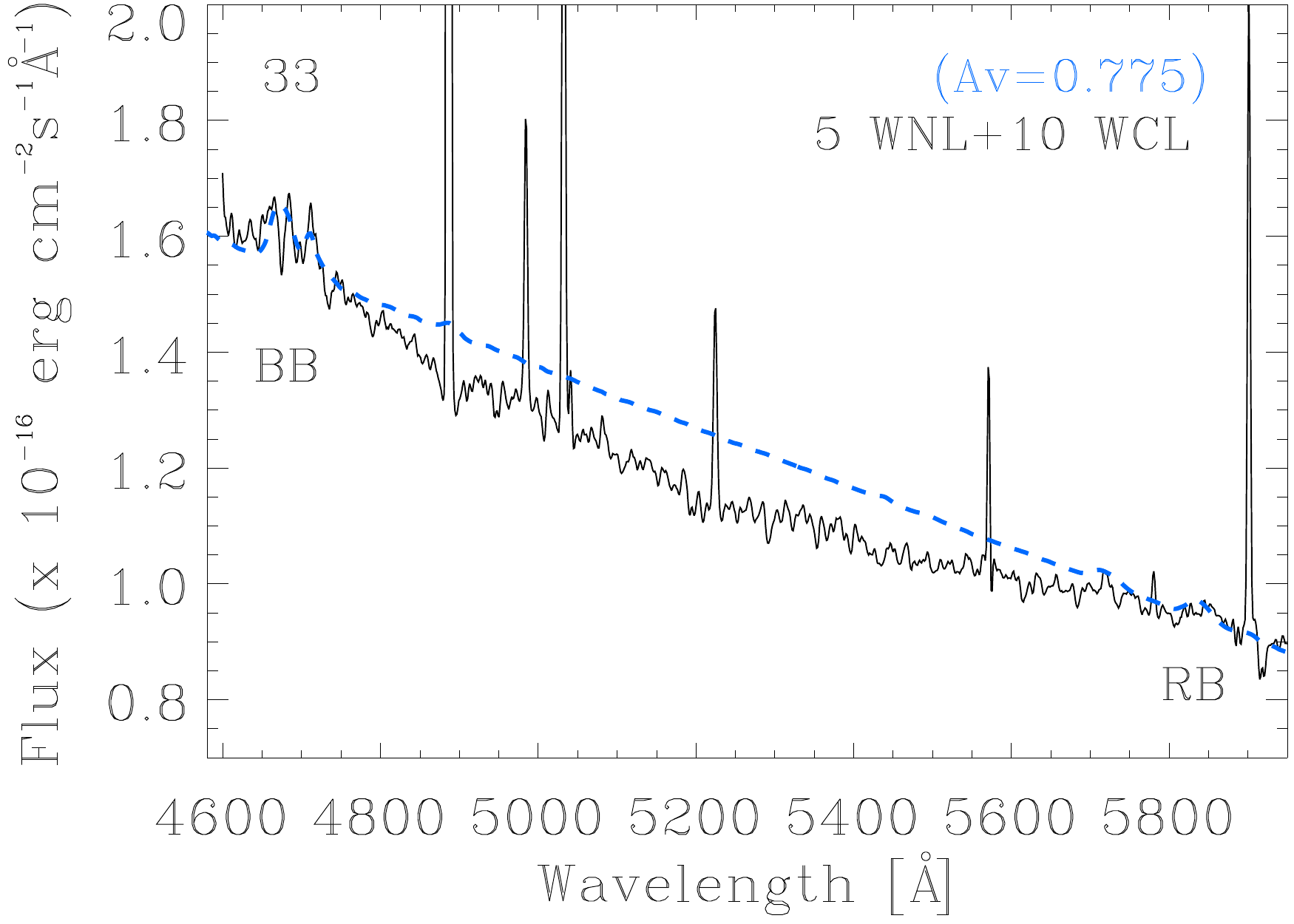}\\
\includegraphics[width=0.33\linewidth]{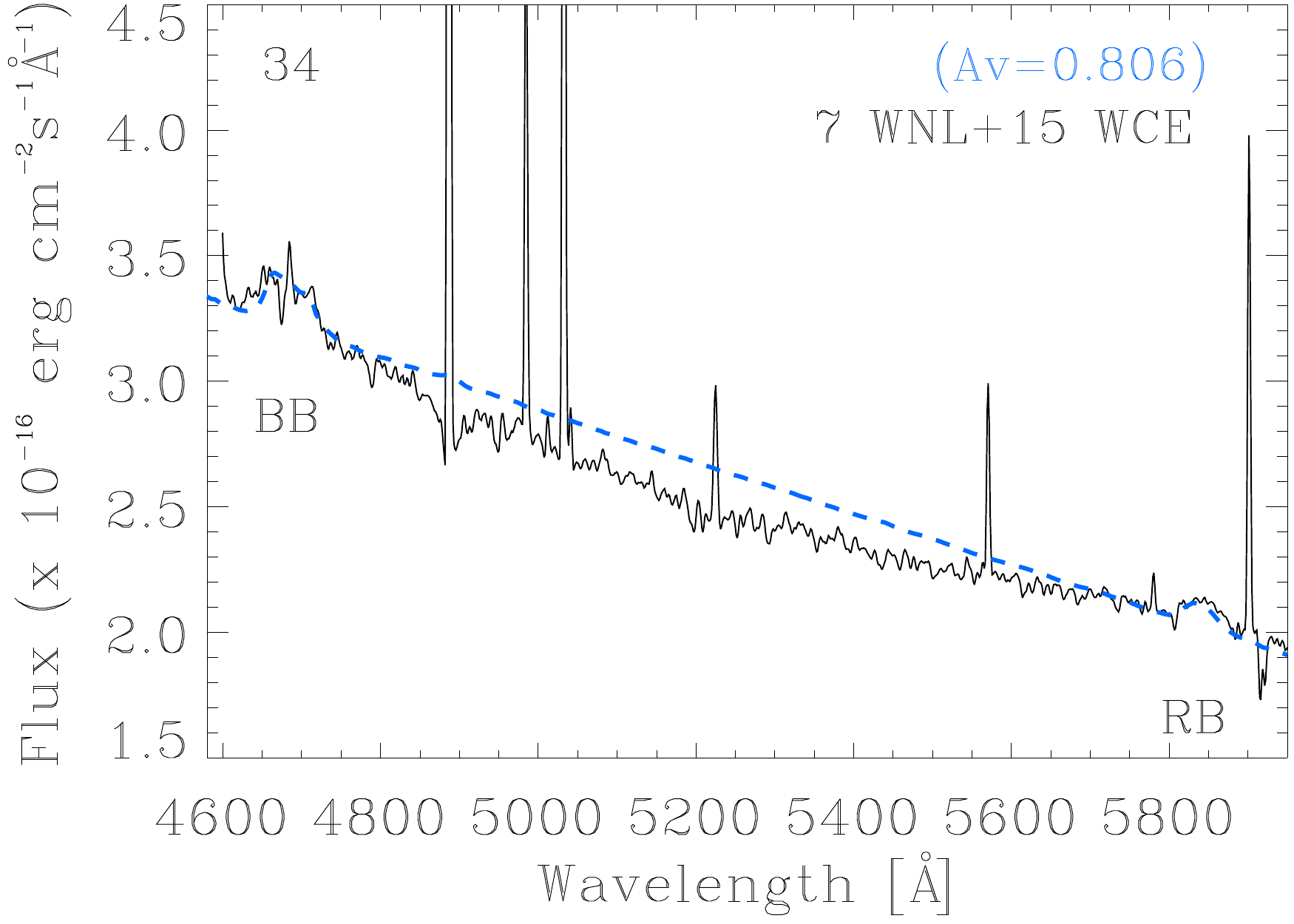}~
\includegraphics[width=0.33\linewidth]{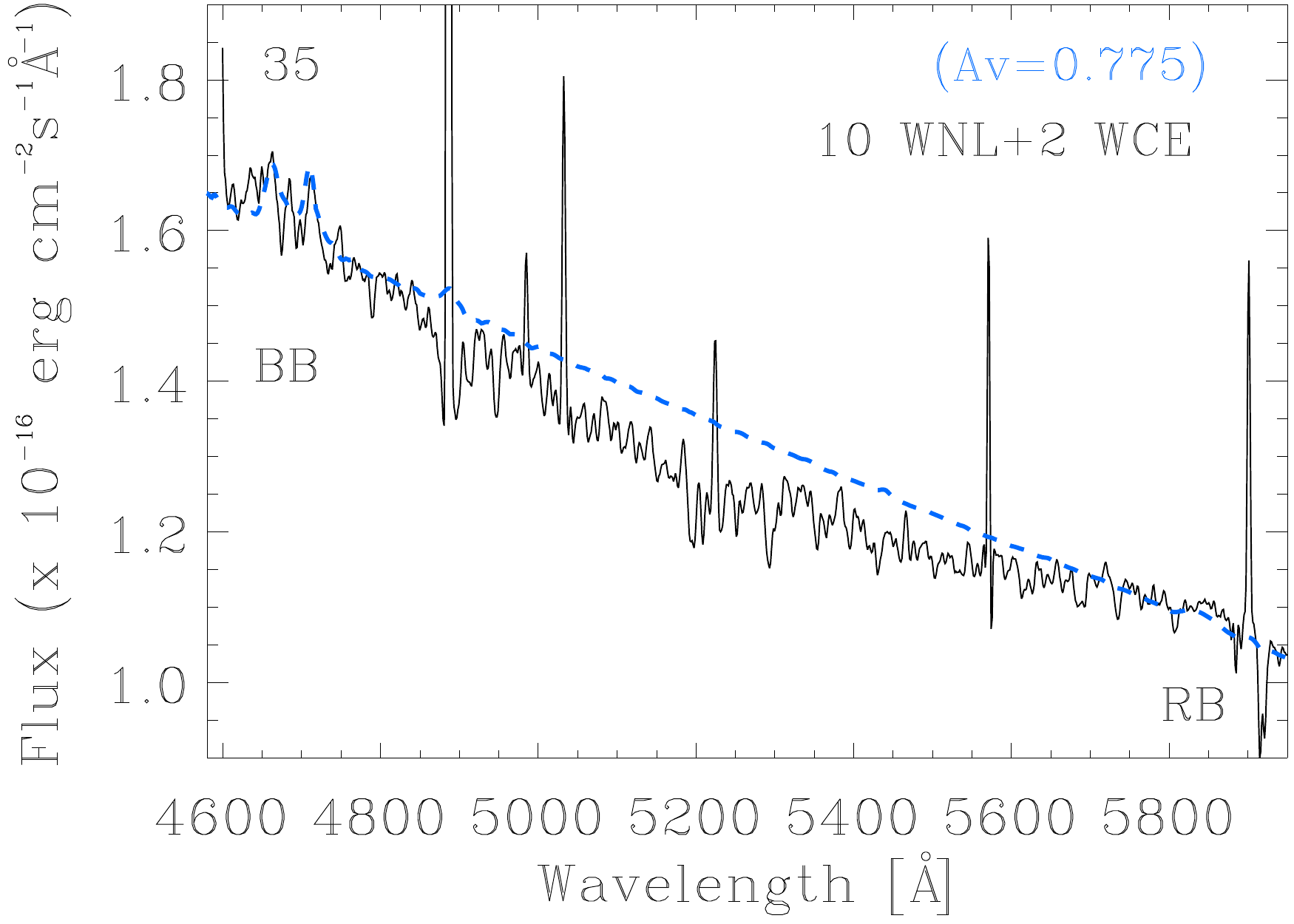}~
\includegraphics[width=0.33\linewidth]{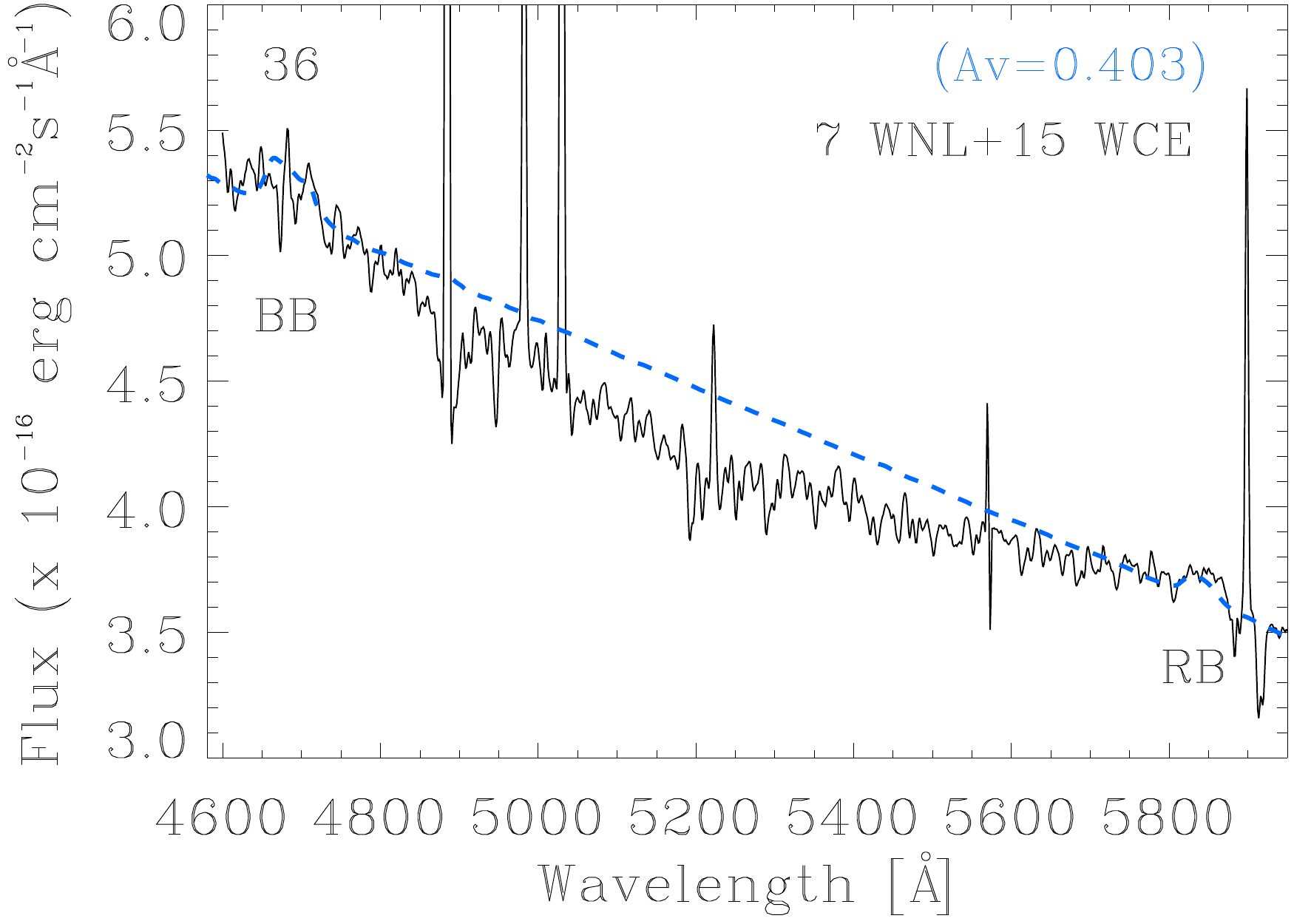}\\
\includegraphics[width=0.33\linewidth]{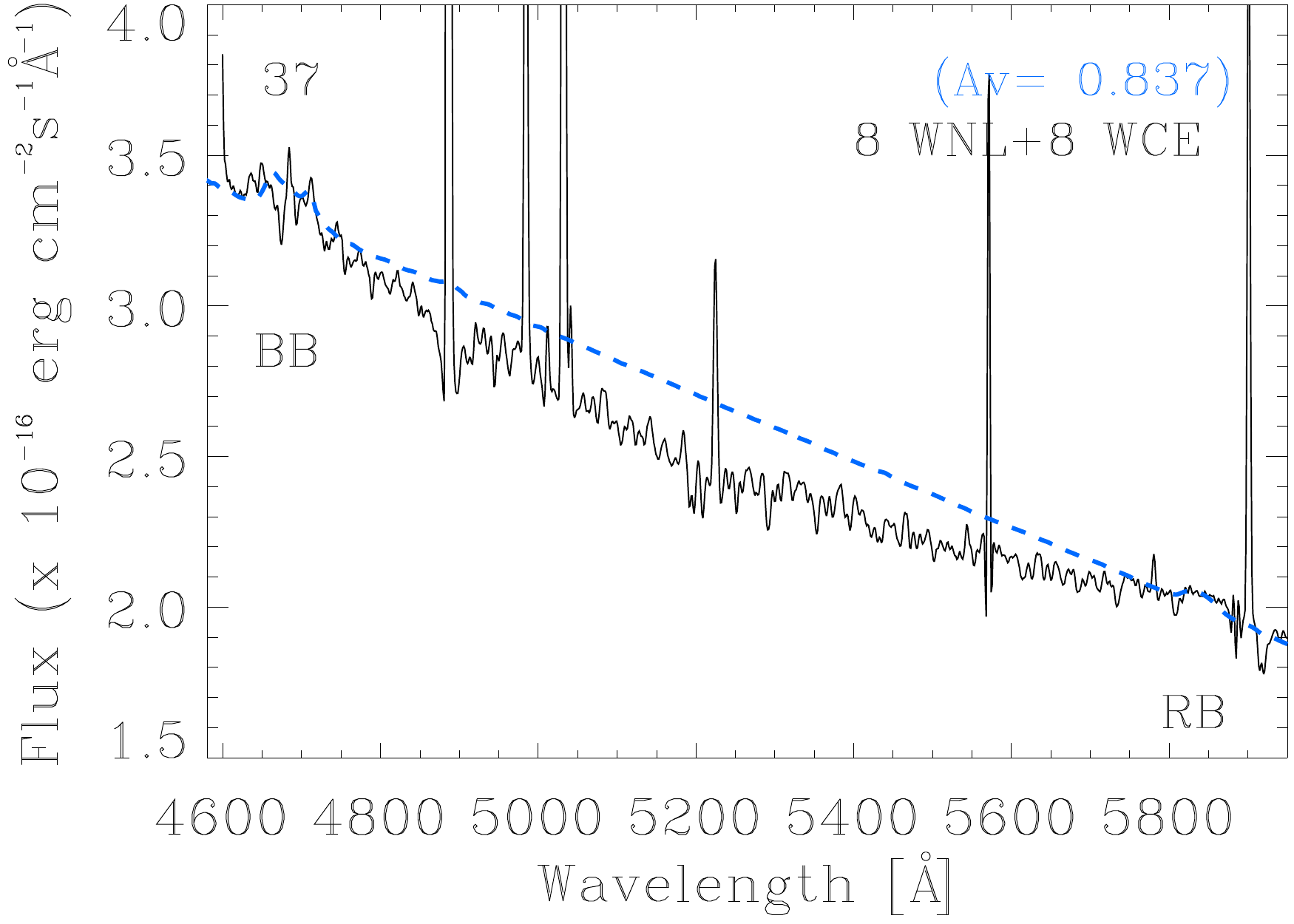}~
\includegraphics[width=0.33\linewidth]{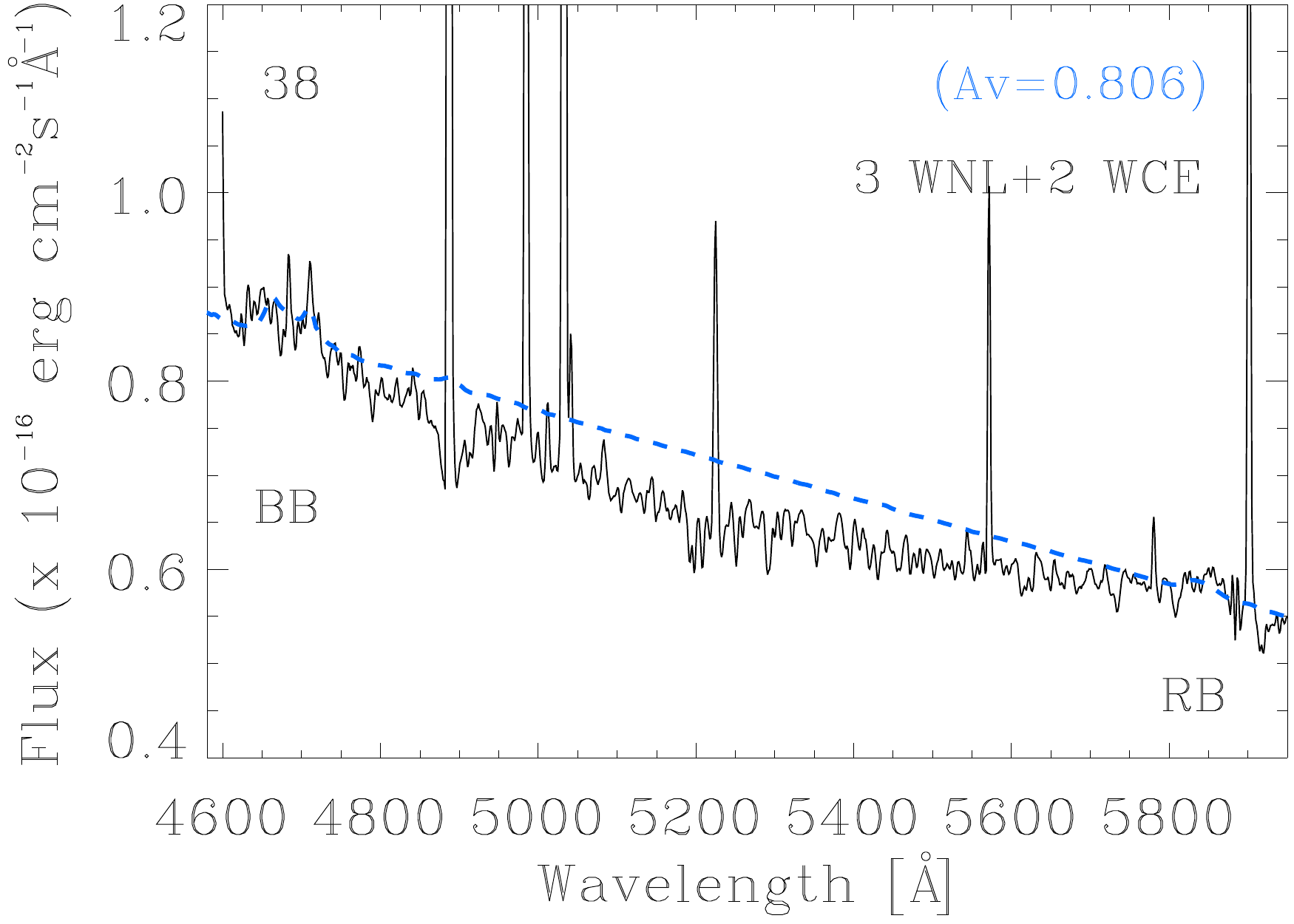}
\ContinuedFloat
\captionsetup{list=off,format=cont}
\caption{{\it -- continued}}
\end{center}
\end{figure*}

We use templates of individual Galactic WR stars,
publicly available from the personal website of
P.\,Crowther\footnote{\url{http://www.pacrowther.staff.shef.ac.uk/science.html}}.
Galactic templates are suitable for the Antennae
on the basis of its nearly Solar metallicity \citep[see][]{2009Bastian},
with a flat metallicity gradient \citep[see][]{2015Lardo}.
Furthermore, below we confirm that this is valid for our objects.

\begin{table*}
\small\addtolength{\tabcolsep}{-1.0pt}
\begin{center}
\caption{WR stars in star-forming complexes of the Antennae galaxies$^{\dagger}$.}
\begin{tabular}{ccccc|ccc|cccc}
\hline
ID & \multicolumn{4}{c}{Classification \&\ number} & WC/WN    & O  & WR/O & Zone& $A_{\rm V}$ & \multicolumn{2}{c}{$D_{\rm aper}$}\\ 
   & WNL          & WCE          & WCL          &  WR         &                 &             &                 &   &(mag)&($''$)&(pc)\\
(1)&  (2)         &  (3)         & (4)          &  (5)        & (6)             & (7)         & (8)             &(9)&(10)&(11) & (12) \\
\hline
 1 & 400 $\pm$ 40 & 400 $\pm$ 40 & $\cdots$     &800 $\pm$ 57 & 1.00 $\pm$ 0.14 &2196 $\pm$ 5 & 0.36 $\pm$ 0.03 & B& 1.2 & 4.0 & 351.0\\
 2 & 100 $\pm$ 10 & 160 $\pm$ 16 & $\cdots$     &260 $\pm$ 19 & 1.60 $\pm$ 0.23 &1566 $\pm$ 4 & 0.17 $\pm$ 0.01 & C& 1.0 & 4.0 & 351.0\\
 3 & 290 $\pm$ 29 &  90 $\pm$  9 & $\cdots$     &380 $\pm$ 30 & 0.31 $\pm$ 0.04 &1153 $\pm$ 3 & 0.33 $\pm$ 0.03 & D& 1.0 & 2.8 & 245.7\\
 4 &  30 $\pm$ 3  &  20 $\pm$  2 & $\cdots$     & 50  $\pm$ 4 & 0.67 $\pm$ 0.09 & 225 $\pm$ 2 & 0.22 $\pm$ 0.02 & B& 1.2 & 2.0 & 175.5\\
 5 &  25 $\pm$ 3  &  15 $\pm$  2 & $\cdots$     & 40  $\pm$ 3 & 0.60 $\pm$ 0.11 & 333 $\pm$ 1 & 0.12 $\pm$ 0.01 & D& 1.0 & 2.0 & 175.5\\
 6 &  20 $\pm$ 2  &  15 $\pm$  2 & $\cdots$     & 35  $\pm$ 3 & 0.75 $\pm$ 0.13 & 296 $\pm$ 1 & 0.12 $\pm$ 0.01 & D& 1.0 & 2.0 & 175.5\\
 7 &  50 $\pm$ 5  & $\cdots$     & 250 $\pm$ 25 &300 $\pm$ 25 & 5.00 $\pm$ 0.54 & 981 $\pm$ 9 & 0.31 $\pm$ 0.03 & K& 0.8 & 2.8 & 245.7\\
 8 &  13 $\pm$ 1  &  10 $\pm$  1 & $\cdots$     & 23 $\pm$  2 & 0.77 $\pm$ 0.10 & 233 $\pm$ 2 & 0.10 $\pm$ 0.01 &10& 0.3 & 2.4 & 210.6\\
 9 & 200 $\pm$ 20 & $\cdots$     & $\cdots$     &200 $\pm$ 20 & --              &1781 $\pm$ 25& 0.11 $\pm$ 0.01 & J& 2.0 & 2.0 & 175.5\\
10 &  10 $\pm$ 2  &  15 $\pm$  2 & $\cdots$     & 25 $\pm$  2 & 1.50 $\pm$ 0.36 & 142 $\pm$ 2 & 0.18 $\pm$ 0.01 & N& 0.5 & 2.0 & 175.5\\
11 &  80 $\pm$ 8  &  80 $\pm$  8 & $\cdots$     &160 $\pm$ 11 & 1.00 $\pm$ 0.11 &1444 $\pm$ 7 & 0.11 $\pm$ 0.01 & F& 0.7 & 3.2 & 280.8\\
12 & 100 $\pm$ 10 & 140 $\pm$ 14 & $\cdots$     &240 $\pm$ 17 & 1.40 $\pm$ 0.20 & 941 $\pm$ 6 & 0.26 $\pm$ 0.02 & F& 0.7 & 3.6 & 315.9\\
13 & 130 $\pm$ 13 & $\cdots$     & $\cdots$     &130 $\pm$ 13 & --              & 491 $\pm$ 1 & 0.26 $\pm$ 0.03 & F& 0.7 & 1.6 & 140.4\\
14 &  15 $\pm$ 2  &  15 $\pm$  2 & $\cdots$     & 30 $\pm$ 2  & 1.00 $\pm$ 0.19 & 142 $\pm$ 1 & 0.21 $\pm$ 0.01 & F& 0.7 & 1.2 & 105.3\\
15 &  10 $\pm$ 1  &  10 $\pm$  2 & $\cdots$     & 20 $\pm$ 2  & 1.00 $\pm$ 0.22 & 124 $\pm$ 1 & 0.16 $\pm$ 0.02 & F& 0.7 & 1.6 & 140.4\\
16 &  10 $\pm$ 1  &  10 $\pm$  1 & $\cdots$     & 20 $\pm$ 2  & 1.00 $\pm$ 0.14 & 167 $\pm$ 3 & 0.12 $\pm$ 0.01 &09& 0.4 & 2.0 & 175.5\\
17 &  35 $\pm$ 4  &   5 $\pm$  1 & $\cdots$     & 40 $\pm$ 4  & 0.14 $\pm$ 0.03 & 950 $\pm$ 7 & 0.04 $\pm$ 0.01 &09& 0.4 & 2.4 & 210.6\\
18 &  20 $\pm$ 2  & $\cdots$     &  20 $\pm$ 3  & 40 $\pm$ 4  & 1.00 $\pm$ 0.18 & 822 $\pm$ 6 & 0.05 $\pm$ 0.01 &03& 1.1 & 3.2 & 280.8\\
19 &  30 $\pm$ 3  &  35 $\pm$  4 & $\cdots$     & 65 $\pm$ 5  & 1.17 $\pm$ 0.18 & 131 $\pm$ 2 & 0.50 $\pm$ 0.04 &03& 0.4 & 2.0 & 175.5\\
20 &  20 $\pm$ 2  &  20 $\pm$  2 & $\cdots$     & 40 $\pm$ 3  & 1.00 $\pm$ 0.14 & 408 $\pm$ 4 & 0.10 $\pm$ 0.01 & B& 0.8 & 2.0 & 175.5\\
21 &  20 $\pm$ 2  &  10 $\pm$  1 & $\cdots$     & 30 $\pm$ 2  & 0.50 $\pm$ 0.07 & 198 $\pm$ 1 & 0.15 $\pm$ 0.01 &02& 0.8 & 2.0 & 175.5\\
22 &  35 $\pm$ 4  &  20 $\pm$  2 & $\cdots$     & 55 $\pm$ 4  & 0.57 $\pm$ 0.09 & 835 $\pm$ 8 & 0.07 $\pm$ 0.01 &09& 0.6 & 2.0 & 175.5\\
23 &  50 $\pm$ 6  &  90 $\pm$  9 & $\cdots$     &140 $\pm$ 11 & 1.80 $\pm$ 0.28 & 845 $\pm$ 10& 0.17 $\pm$ 0.01 & M& 0.8 & 4.0 & 351.0\\
24 &  20 $\pm$ 2  & $\cdots$     & 100 $\pm$ 10 &120 $\pm$ 10 & 5.00 $\pm$ 0.71 & 845 $\pm$ 6 & 0.14 $\pm$ 0.01 & T& 0.5 & 2.8 & 245.7\\
25 &  30 $\pm$ 3  & $\cdots$     & 150 $\pm$ 15 &180 $\pm$ 15 & 5.00 $\pm$ 0.71 & 856 $\pm$ 8 & 0.21 $\pm$ 0.02 & T& 0.5 & 4.0 & 351.0\\
26 &  60 $\pm$ 10 &  80 $\pm$  8 & $\cdots$     &140 $\pm$ 12 & 1.33 $\pm$ 0.26 & 937 $\pm$ 15& 0.15 $\pm$ 0.01 & S& 0.5 & 4.0 & 351.0\\
27 &  20 $\pm$ 2  &  20 $\pm$  2 & $\cdots$     & 40 $\pm$  3 & 1.00 $\pm$ 0.14 & 167 $\pm$ 2 & 0.24 $\pm$ 0.02 & S& 0.5 & 3.2 & 280.8\\
28 &  40 $\pm$ 6  &  80 $\pm$  8 & $\cdots$     &120 $\pm$ 10 & 2.00 $\pm$ 0.36 & 739 $\pm$ 10& 0.16 $\pm$ 0.01 & R& 0.7 & 4.0 & 351.0\\
29 &  50 $\pm$ 5  &  60 $\pm$  6 & $\cdots$     &110 $\pm$  8 & 1.20 $\pm$ 0.17 & 373 $\pm$ 3 & 0.29 $\pm$ 0.02 & M& 0.8 & 3.2 & 280.8\\
30 &  30 $\pm$ 3  &  20 $\pm$  2 & $\cdots$     & 50 $\pm$  4 & 0.67 $\pm$ 0.09 & 282 $\pm$ 2 & 0.18 $\pm$ 0.01 &14& 0.8 & 2.4 & 210.6\\
31 &  28 $\pm$ 3  & $\cdots$     &  10 $\pm$  1 & 38 $\pm$  3 & 0.36 $\pm$ 0.05 & 296 $\pm$ 1 & 0.13 $\pm$ 0.01 & L& 0.8 & 1.2 & 105.3\\
32 &  10 $\pm$ 1  & $\cdots$     &  30 $\pm$  3 & 40 $\pm$  3 & 3.00 $\pm$ 0.42 & 205 $\pm$ 2 & 0.20 $\pm$ 0.01 &13& 0.8 & 2.0 & 175.5\\
33 &   5 $\pm$ 1  & $\cdots$     &  10 $\pm$  1 & 15 $\pm$  1 & 2.00 $\pm$ 0.45 & 112 $\pm$ 1 & 0.13 $\pm$ 0.01 & L& 0.8 & 1.2 & 105.3\\
34 &   7 $\pm$ 1  &  15 $\pm$  2 & $\cdots$     & 22 $\pm$  2 & 2.14 $\pm$ 0.42 & 184 $\pm$ 2 & 0.12 $\pm$ 0.01 &14& 0.8 & 2.0 & 175.5\\
35 &  10 $\pm$ 1  &   2 $\pm$  1 & $\cdots$     & 12 $\pm$  1 & 0.20 $\pm$ 0.10 &  62 $\pm$ 1 & 0.19 $\pm$ 0.02 & L& 0.8 & 1.2 & 105.3\\
36 &   7 $\pm$ 1  &  15 $\pm$  2 & $\cdots$     & 22 $\pm$  2 & 2.14 $\pm$ 0.42 & 162 $\pm$ 3 & 0.14 $\pm$ 0.01 & G& 0.4 & 2.4 & 210.6\\
37 &   8 $\pm$ 1  &   8 $\pm$  1 & $\cdots$     & 16 $\pm$  1 & 1.00 $\pm$ 0.18 & 109 $\pm$ 1 & 0.15 $\pm$ 0.01 & M& 0.8 & 2.4 & 210.6\\
38 &   3 $\pm$ 1  &   2 $\pm$  1 & $\cdots$     &  5 $\pm$  1 & 0.67 $\pm$ 0.40 &  93 $\pm$ 1 & 0.05 $\pm$ 0.01 &14& 0.8 & 1.6 & 140.4\\
\hline
Total&2021  $\pm$ 60 &\multicolumn{2}{c}{2032  $\pm$ 59} &4053  $\pm$ 84& 1.01 $\pm$ 0.04&21834 $\pm$ 40& 0.186 $\pm$ 0.004&&&&\\
\hline
\end{tabular}
\\
{\it Notes}. Brief explanation of columns: $^{\dagger}$(1) Assigned number for star-forming
complex containing WR features;
(2) classification and number of WR stars obtained with Galactic templates;
the numbers represent the multiplicative factors for each template
corresponding to the number of WR stars of the given subtype; WNL;
(3) WCE;
(4) WCL-subtype;
(5) total number of WR stars;
(6) WC to WN-type ratio (WC/WN);
(7) number of estimated O-type stars (O);
(8) WR to O-type star ratio (WR/O);
(9) knots (letters: A--T) or regions (number IDs: 01--15) established in \citet{2010Whitmore};
(10) adopted extinction ($A_{\rm V}$) from zones in \citet{2010Whitmore};
(11) aperture diameter used in extracting spectra ({$D_{\rm aper}$}) in arcsec;
(12) {$D_{\rm aper}$} in pc.
The total numbers of WR stars, WR-types, O-type stars, and global fractions of
WC/WN and WR/O are indicated at the end of the respective column.
\label{tab:class}
\end{center}
\end{table*}

To start with, the template of a given WR subtype, scaled to the distance
of the Antennae, is superposed on the observed WR spectrum.
If the spectrum displays a RB, the fitting starts with a WC
template (WCE or WCL-subtype).
A multiplicative factor is used to match the intensity of the RB,
either with a WCE if it has \civwrr\ or WCL if it also shows \ciiiwrr.
We then proceed in a similar manner to the BB.
If the continuum levels at the BB are different,
we added a pseudo-continuum to the template spectrum.
If the observed BB shows an excess, especially on its
bluer edge, it suggests the presence of a nitrogen line.
We then found a multiplicative factor required to fit the BB
with a WN template, either with a WNL if it has \niiiwr\ or
WNE if it shows \nvwr.
The procedure is repeated until the observed bumps are well
matched by a single or a combination of templates.

Fig.~\ref{fig:spectra} shows the MUSE spectra of the 38 WR
locations in the Antennae, in the wavelength range covering 
the WR bumps (4600 to 5900 \AA\,),
corrected by the extinction reported in \citet[][]{2010Whitmore}.
The observed spectra are 
compared with the resultant best fit using Galactic WR templates.
Details of the fits are listed in Table~\ref{tab:class},
together with the adopted extinction,
estimated number of WR stars and their subtypes.
We took into account two possible sources of errors 
on the number of WR stars in each star-forming complex.
The primary source of error is that intrinsic to the template-fitting technique.
The goodness of the fit while using this technique is judged visually,
which is found to have an error of $\sim$10\%.
The secondary source of error is statistical in nature, which
was calculated as the 1-$\sigma$ deviation, $\sigma_l$, on each measured
flux of the WR feature using the expression \citep{Tresse1999}: \\
\begin{equation}
\sigma_l = \sigma_c D \sqrt{(2 N_{\rm pix} + \frac{EW}{D})},
\end{equation}
where $D$=1.24~\AA~pixel$^{-1}$ is the spectral dispersion, $\sigma_c$
is the mean
standard deviation per pixel of the continuum, $N_{\rm pix}$ is the number of
pixels covered by the feature, and EW is the equivalent width of the measured
feature. The percentage error on the BB flux is taken as the percentage
error on the number of WNL stars. The errors on the number of WCE and
WCL stars are
based on the percentage errors on the measured red bump and \ciiiwrr\
features, respectively. The error given in the table for each WR
source corresponds
to either the fitting error or the statistical error, whichever is larger.
For regions having more than $\sim$50 WR stars of a given type, the former error
dominates, whereas for the rest, the latter is the principal contributor.
The errors in columns 2, 3 and 4 are propagated to find the errors in columns
5, 6 and 8. The error on the number of O stars in column~7 is calculated based
on the error on the \hb\ flux. The errors on the total number of WR stars
in the Antennae were calculated by quadratically summing the errors of WR stars
in each complex.

This method allowed us to estimate a total number of 4053 $\pm$ 84 WR stars 
in the Antennae consisting of 2021 $\pm$ 60~WNL and 2032 $\pm$ 59~WC-types: 1462~WCE and 570~WCL-subtypes.
This corresponds to a global WC to WN-type ratio (WC/WN) of 1.01 $\pm$ 0.04, 
fifty per cent of nitrogen-types and fifty per cent of carbon-types,
with a WCL to WCE-type ratio (WCL/WCE)$\sim$40 per cent. 

Following \citet[][]{2010Lopez}, by
assuming that an O7V star produces an \hb\ luminosity of
$\sim$4.76$\times10^{36}$~\ergs \citep{1998Schaerer},
we can roughly estimate the number of O-stars per SSC with 
WR features by using their extinction-corrected \hb\
luminosity and dividing it by this number.
The estimated number of O-type stars and the WR to O-type star ratio (WR/O)
for each star-forming complex where WR sources were detected is given in Table~\ref{tab:class}.
In this way, the total number of O-type stars in the studied SSC regions 
with WR features resulted in 21834 $\pm$ 40, 
which gives a global WR/O ratio of 0.186 $\pm$ 0.004.
This ratio is an upper limit as we did not take into account the \hb\ flux
from complexes where we did not detect WR features for calculating
the number of O-type stars.

As mentioned in Section~2, the MUSE spectra do not cover
the wavelength range below 4600~\AA.
With the present data we are not able to say anything about
the presence or absence of \oviwr,
the seldom observed fingerprint of WO-type stars,
a short-lived final stage in the evolution of
massive stars \citep{2015Tramper}.
We note, however, that the OSIRIS spectrum of WR\,1,
with the highest number of WR stars in the Antennae (see Table~\ref{tab:class}),
clearly discards the presence of WO-type stars in this region.

\subsection{Spectral analysis of the WR features}

WR stars can also be analysed based on the presence
and strength of different emission lines that comprise the 
broad bumps in their spectra.
It is customary to relate the BB to a WN subtype, being
dominated by the \heiiwr\ and \niiiwr\, emission lines.
However, it can also be related to a WC subtype if there is a
contribution from the \civwrb\ line.
On the other hand, the RB is associated with the \civwrr\
line from the WC subtype.

\begin{figure*}
\begin{center}
\includegraphics[width=0.245\linewidth]{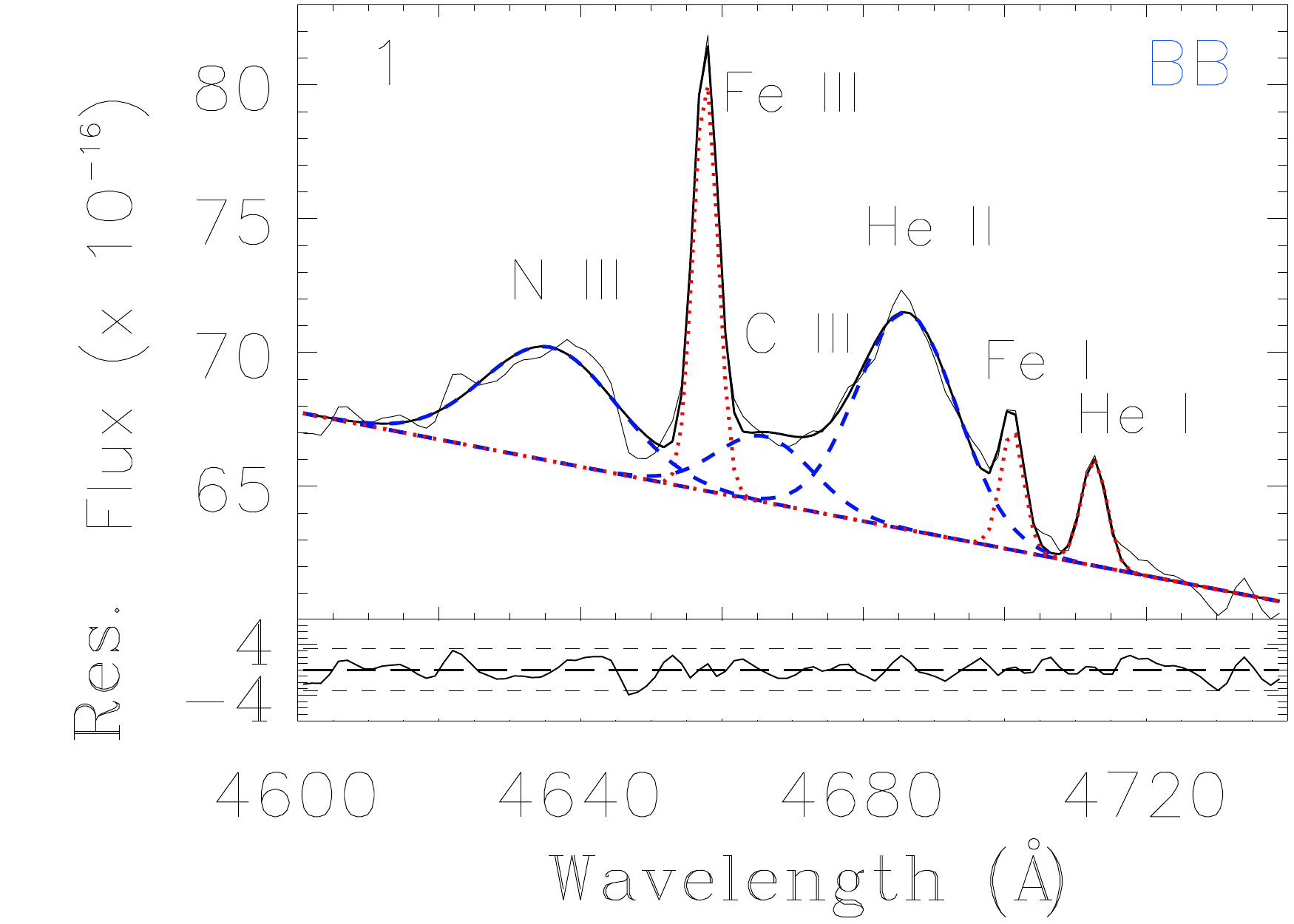}~
\includegraphics[width=0.245\linewidth]{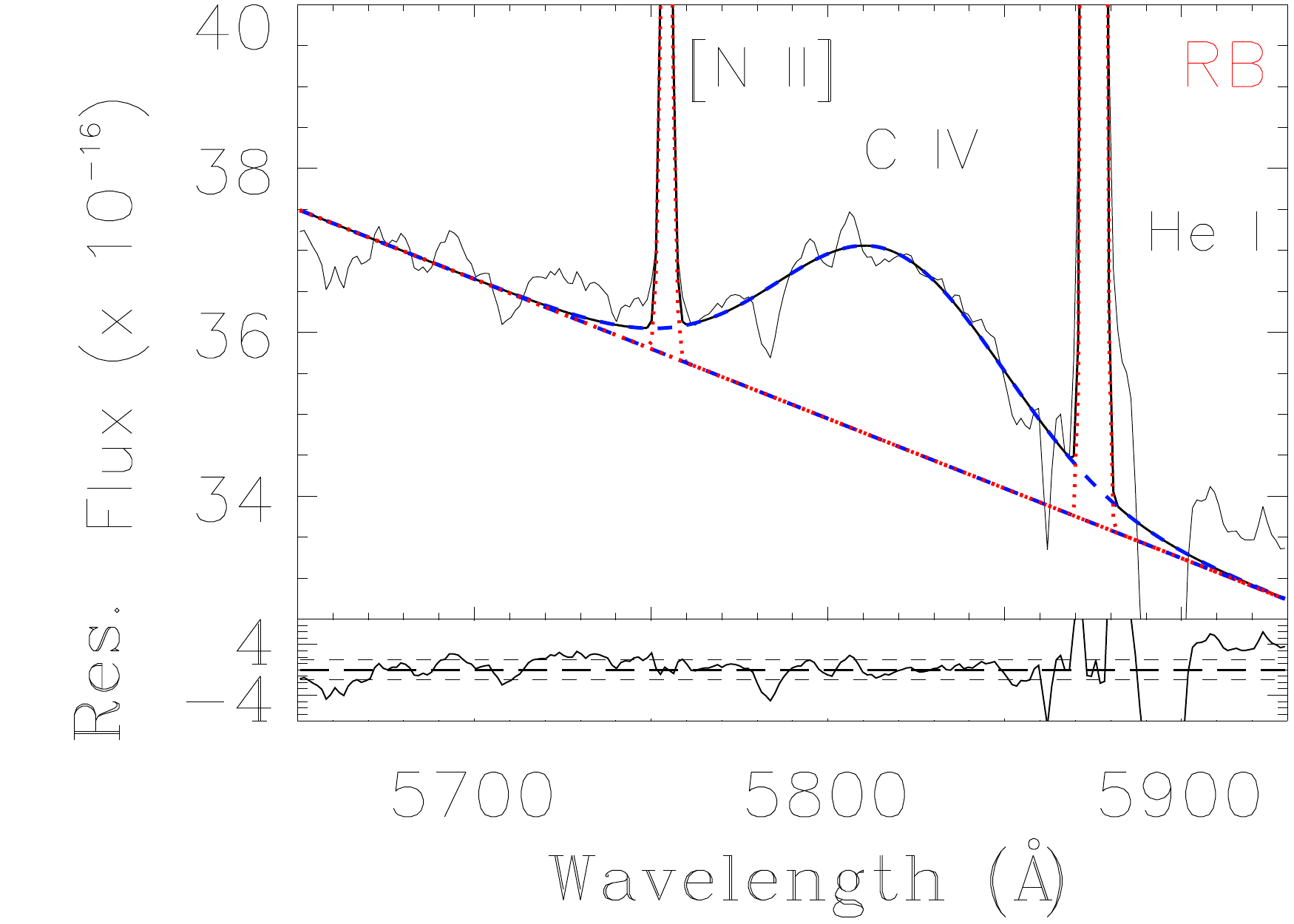}~
\includegraphics[width=0.245\linewidth]{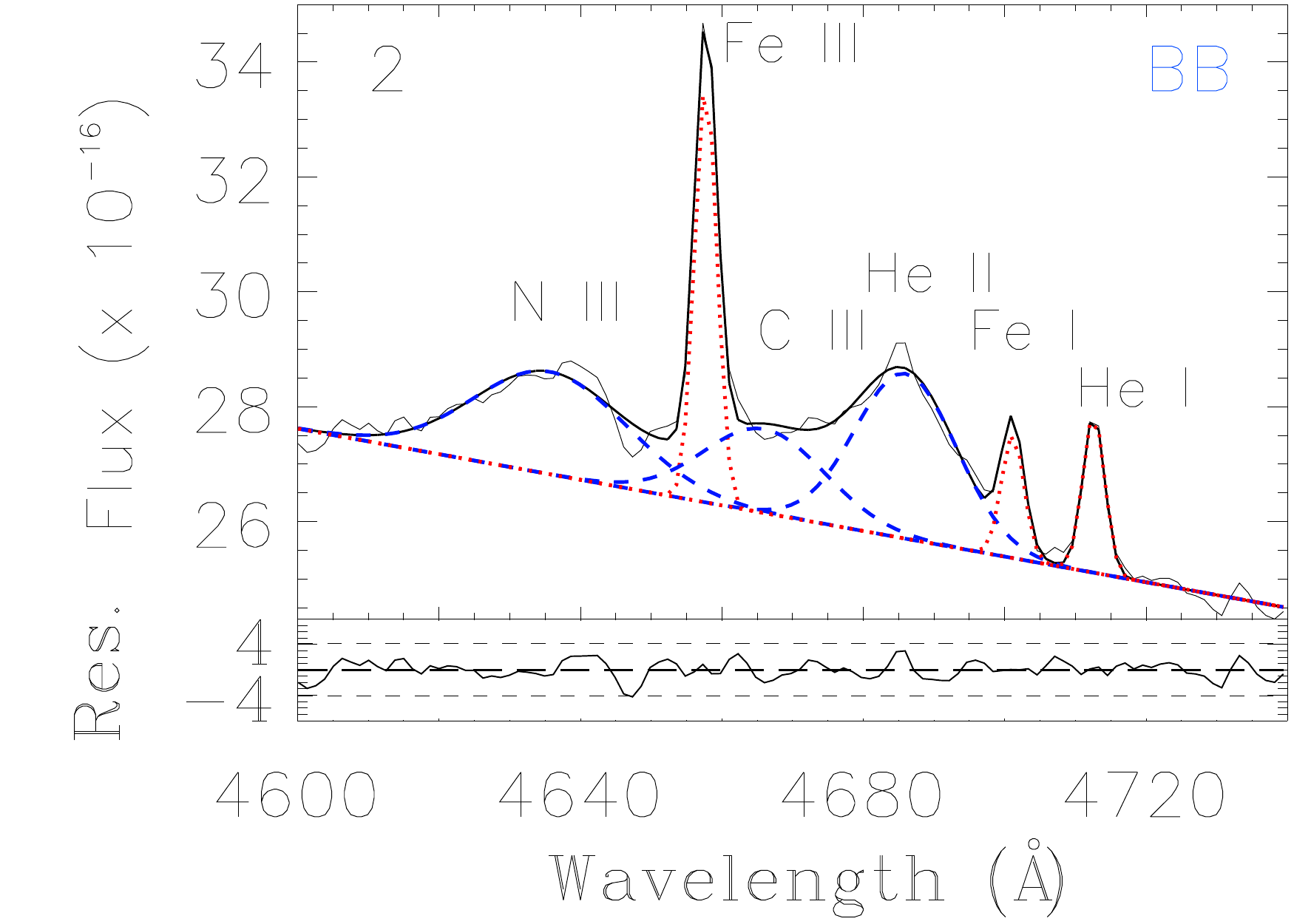}~
\includegraphics[width=0.245\linewidth]{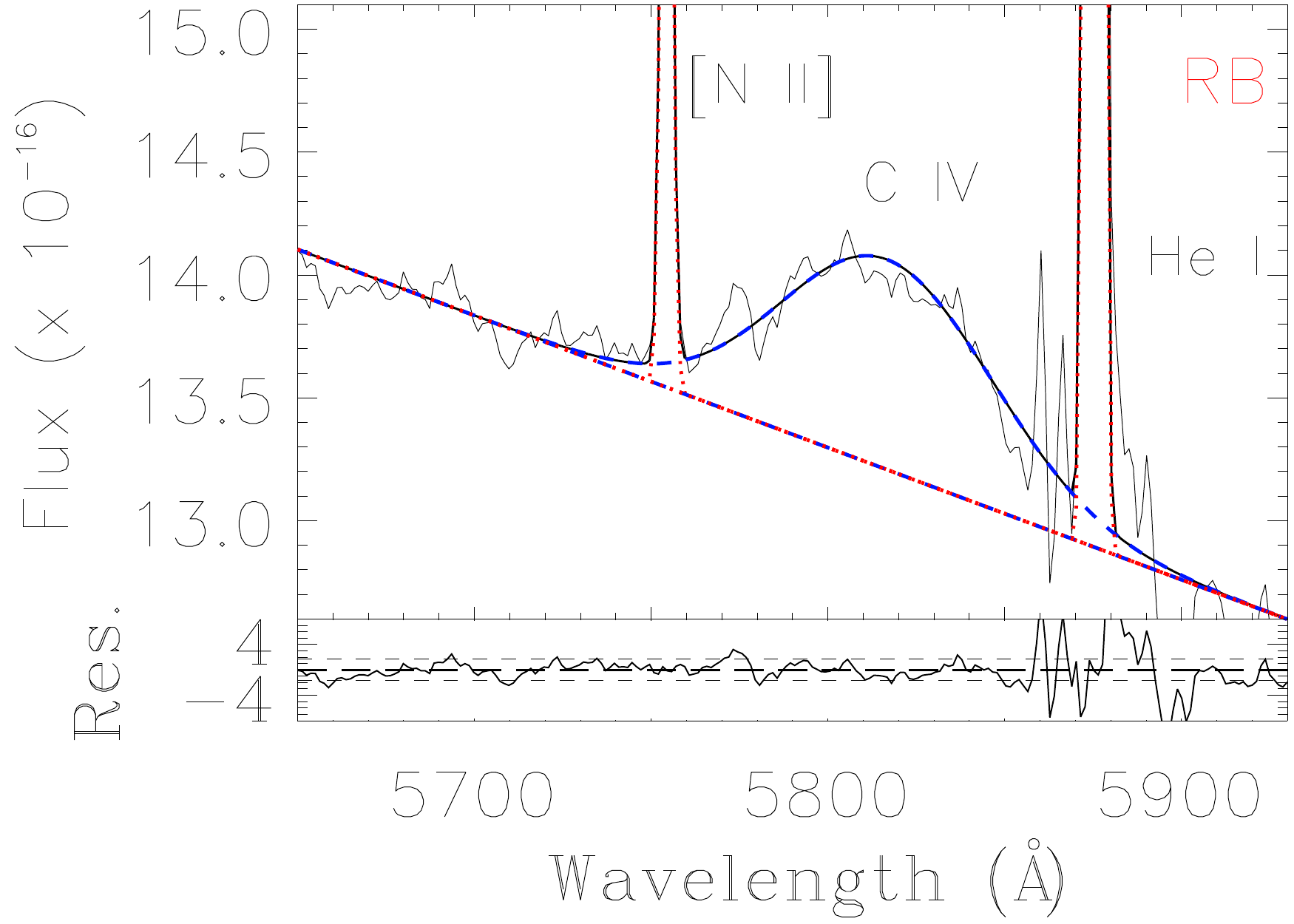}\\
\includegraphics[width=0.245\linewidth]{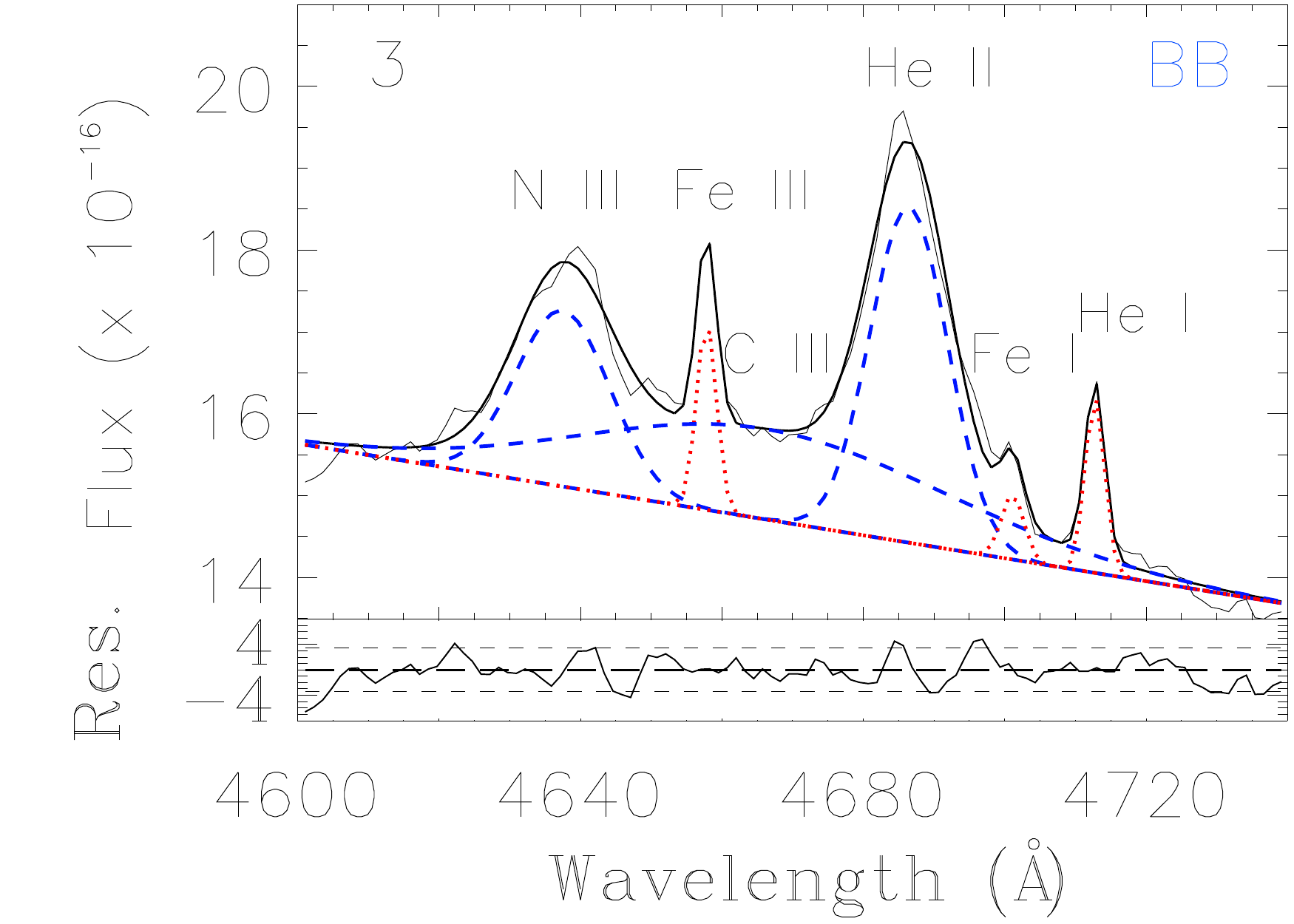}~
\includegraphics[width=0.245\linewidth]{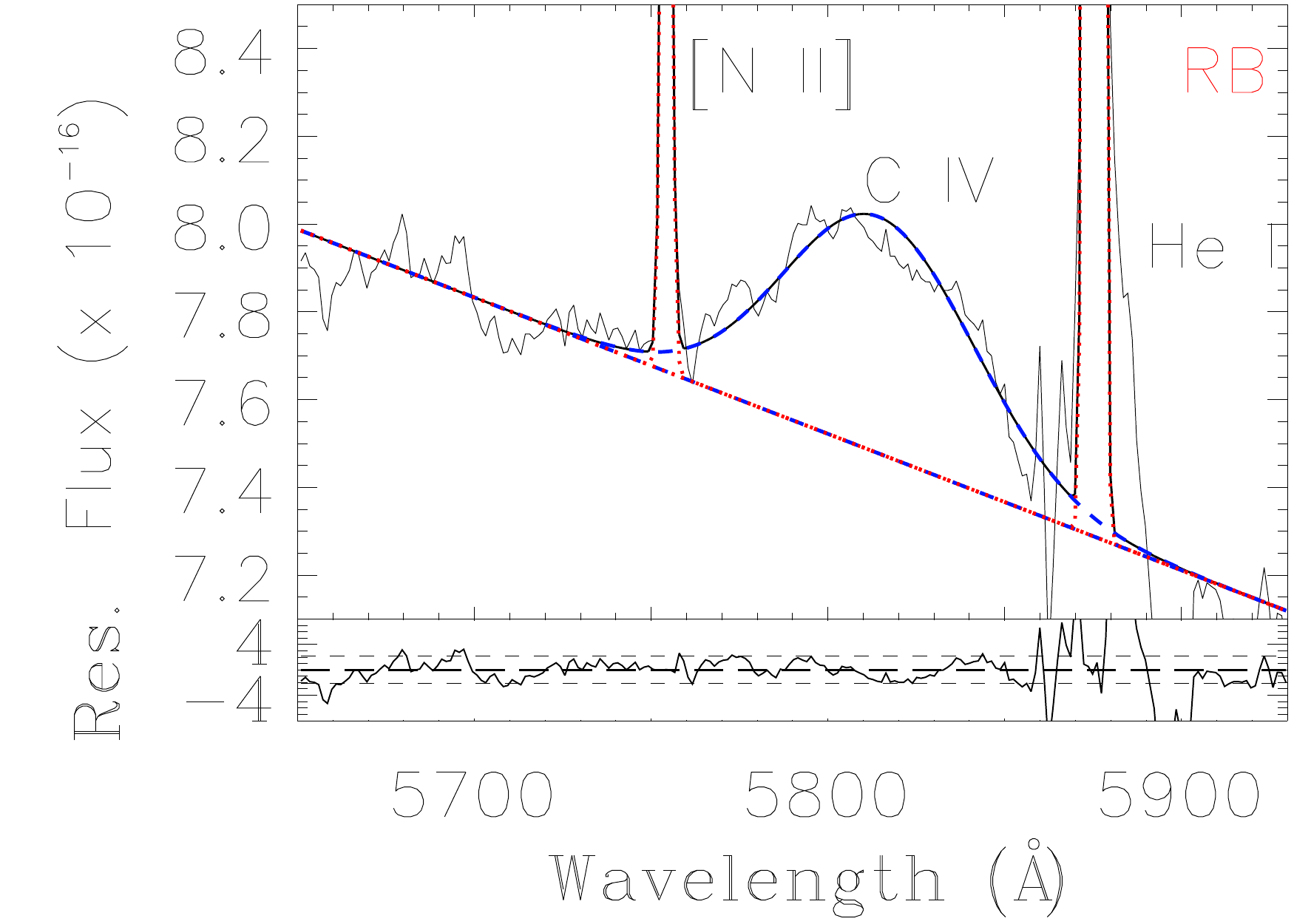}~
\includegraphics[width=0.245\linewidth]{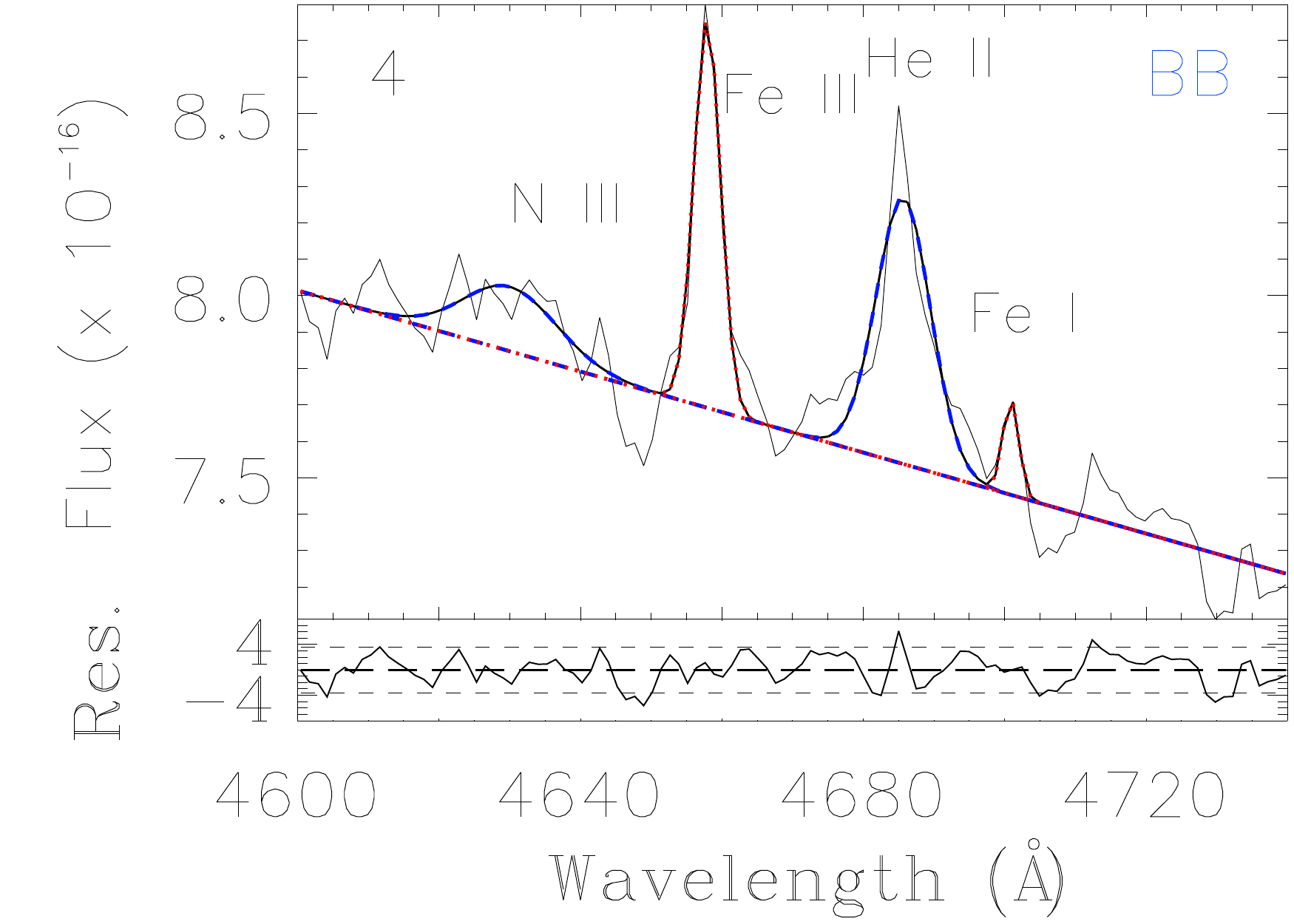}~
\includegraphics[width=0.245\linewidth]{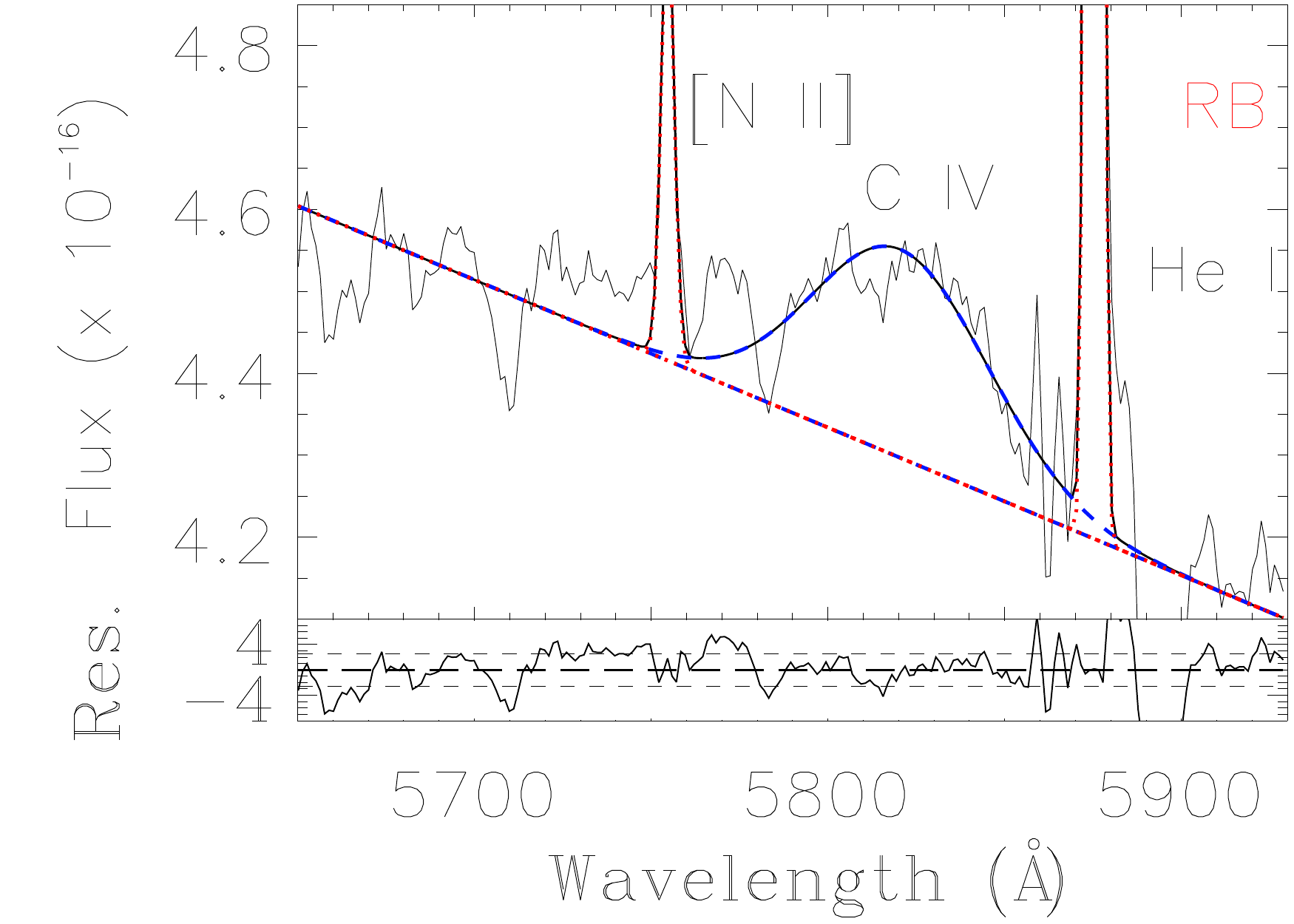}\\
\includegraphics[width=0.245\linewidth]{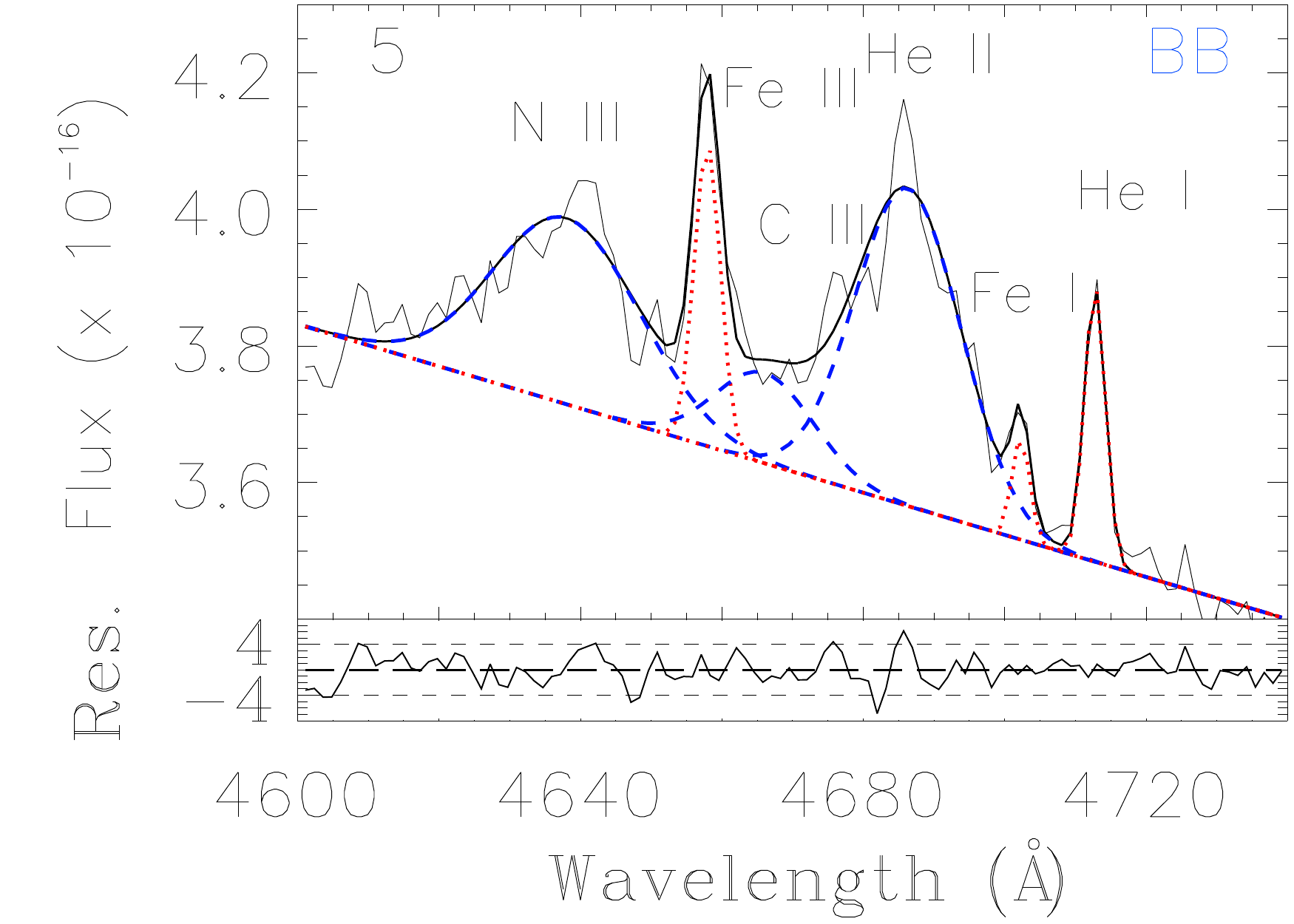}~
\includegraphics[width=0.245\linewidth]{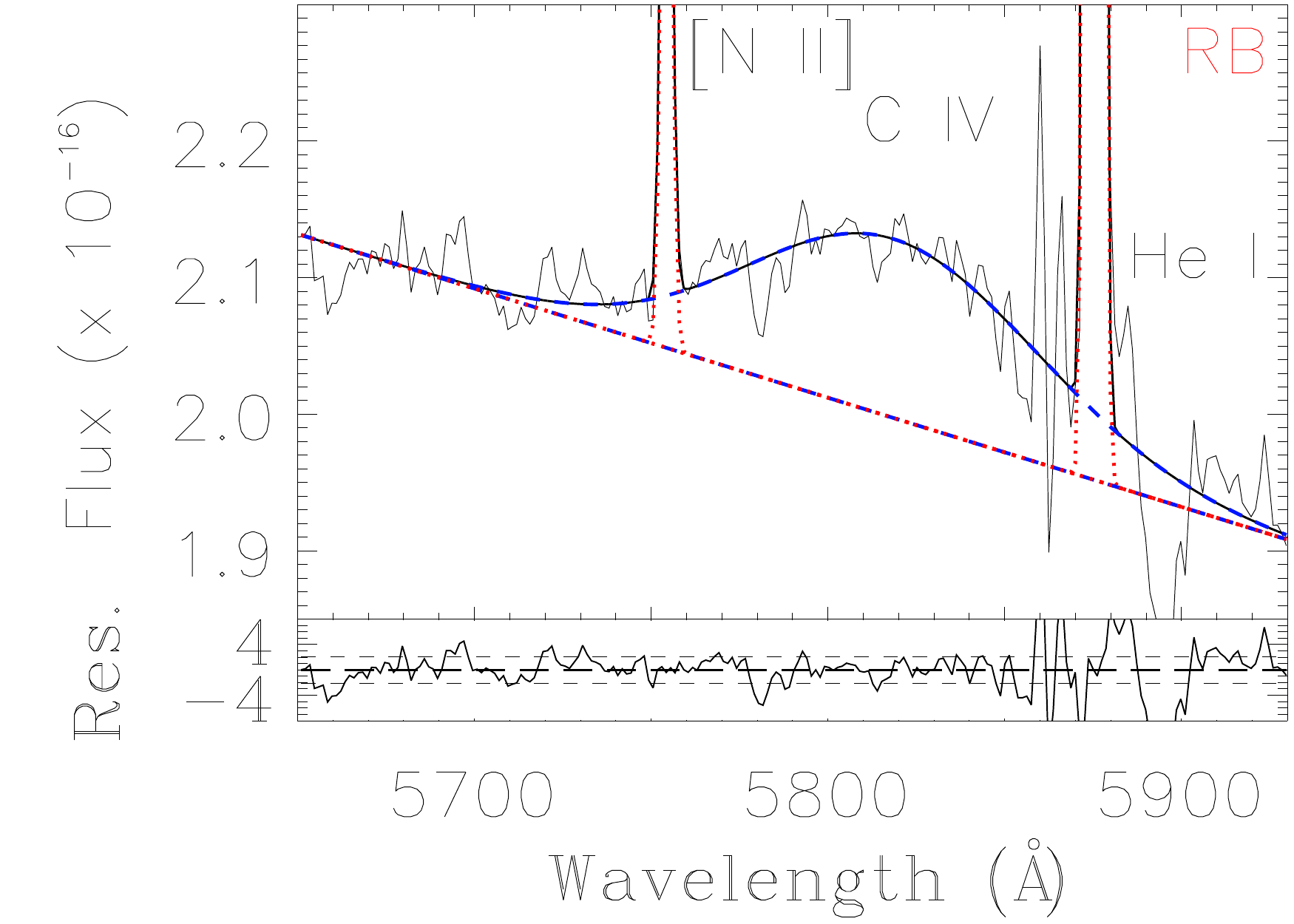}~
\includegraphics[width=0.245\linewidth]{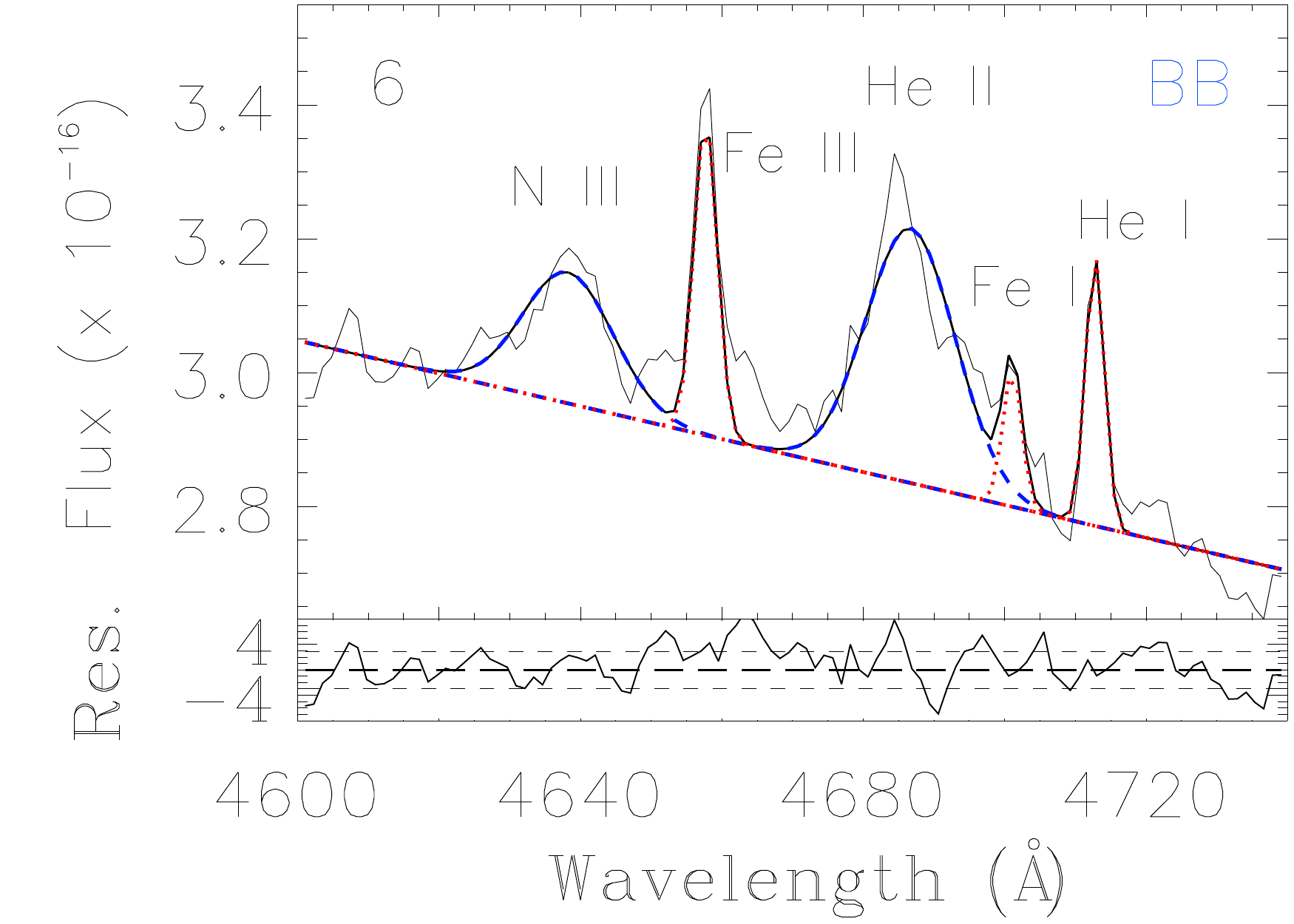}~
\includegraphics[width=0.245\linewidth]{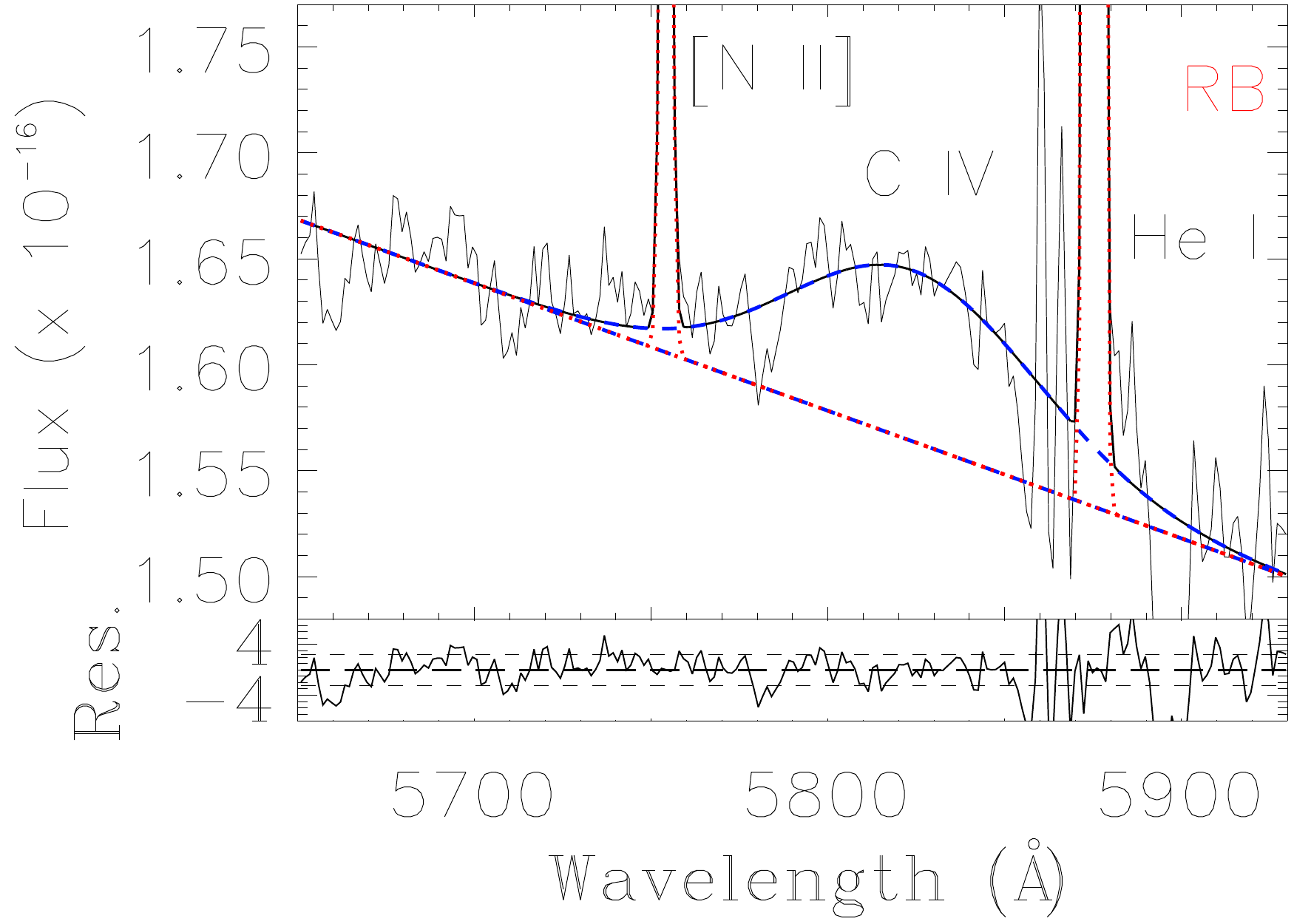}\\
\includegraphics[width=0.245\linewidth]{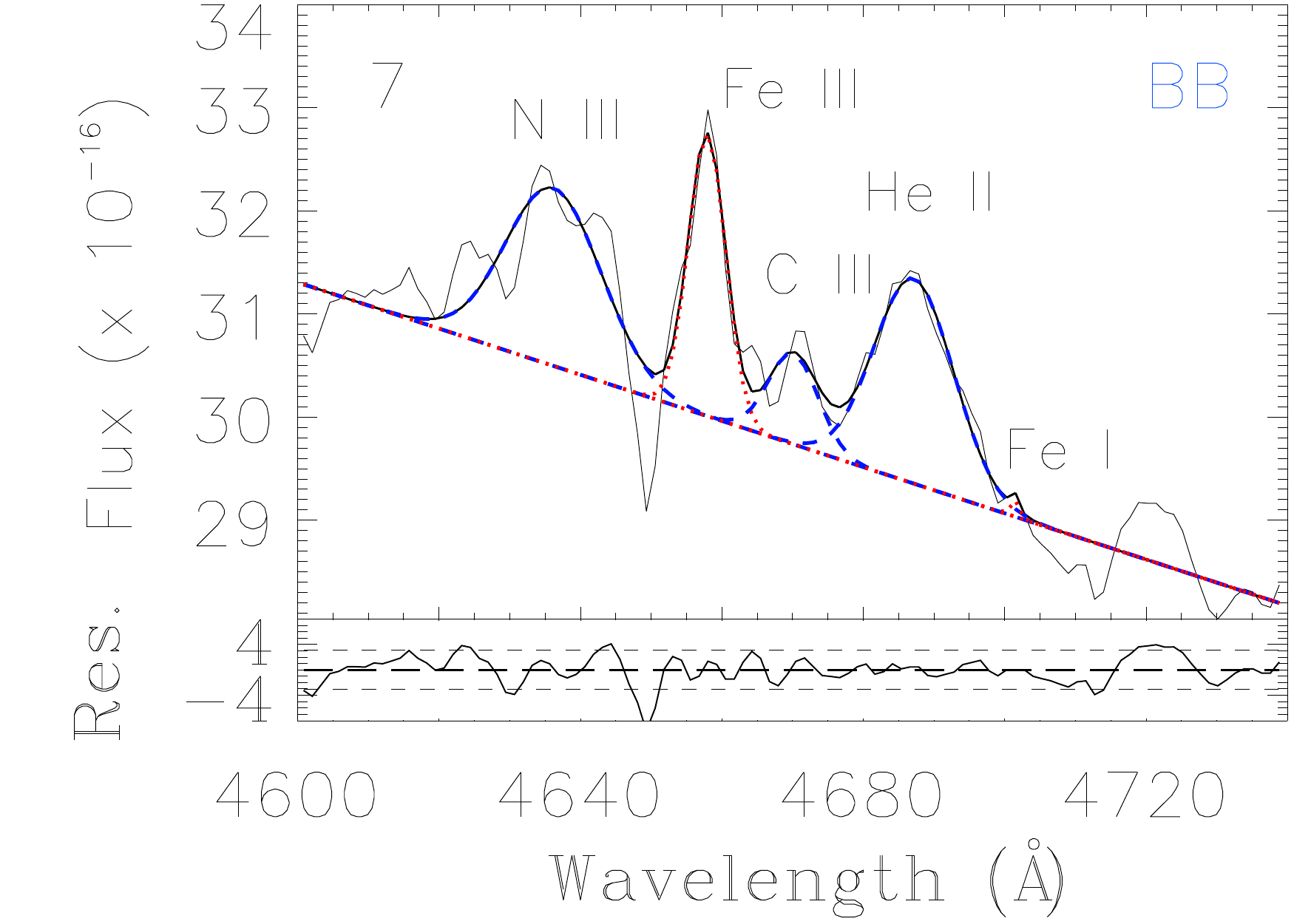}~
\includegraphics[width=0.245\linewidth]{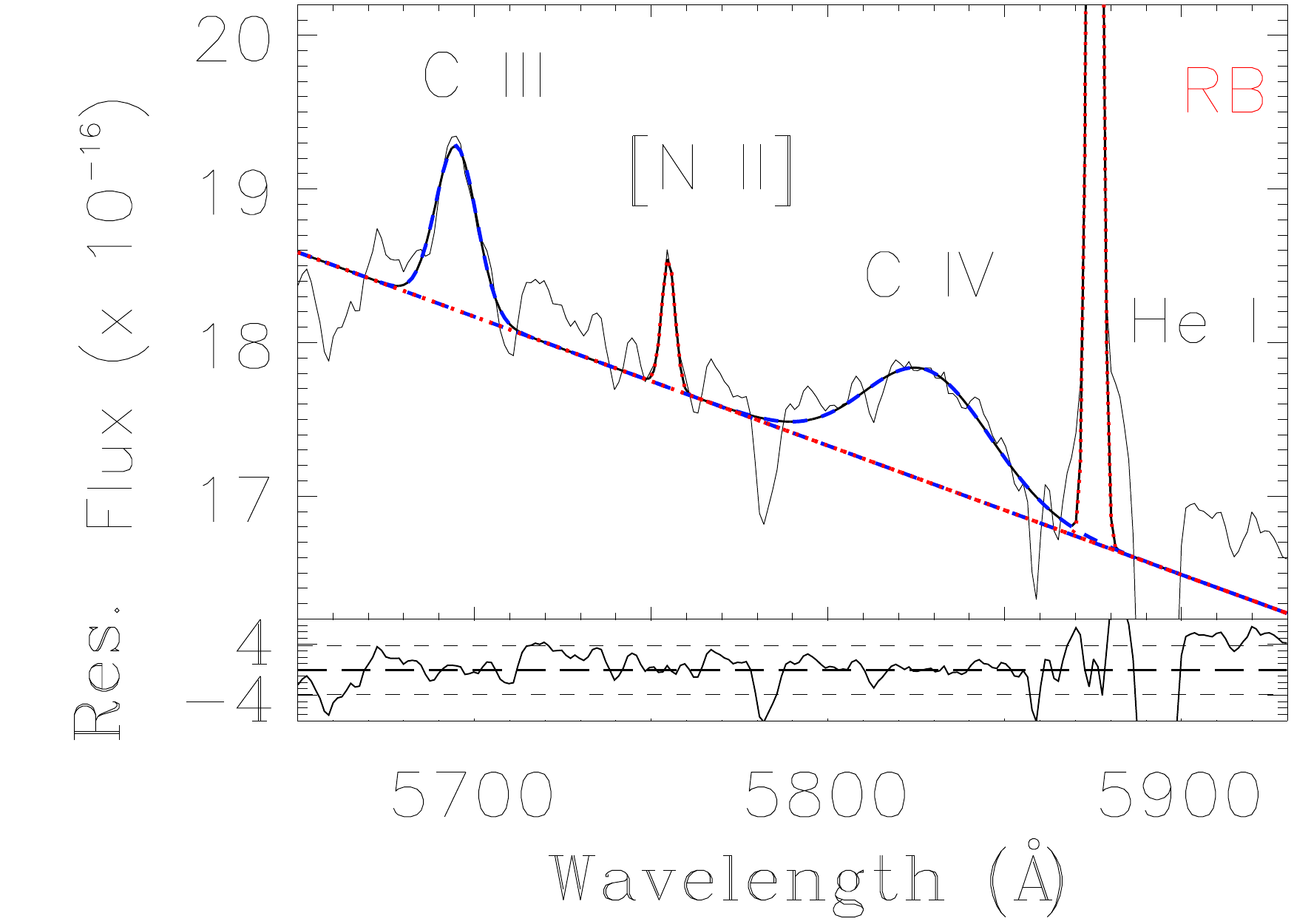}~
\includegraphics[width=0.245\linewidth]{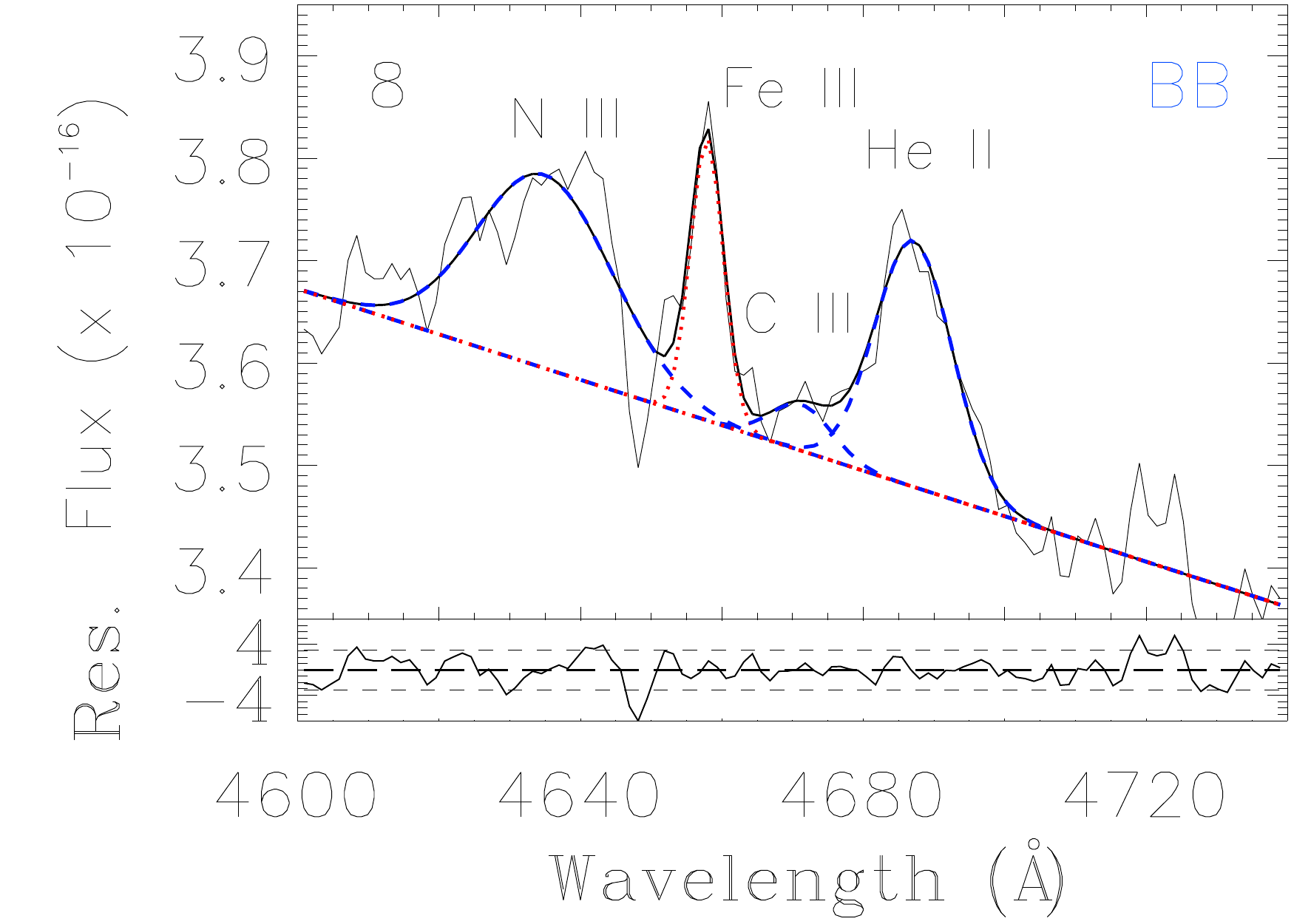}~
\includegraphics[width=0.245\linewidth]{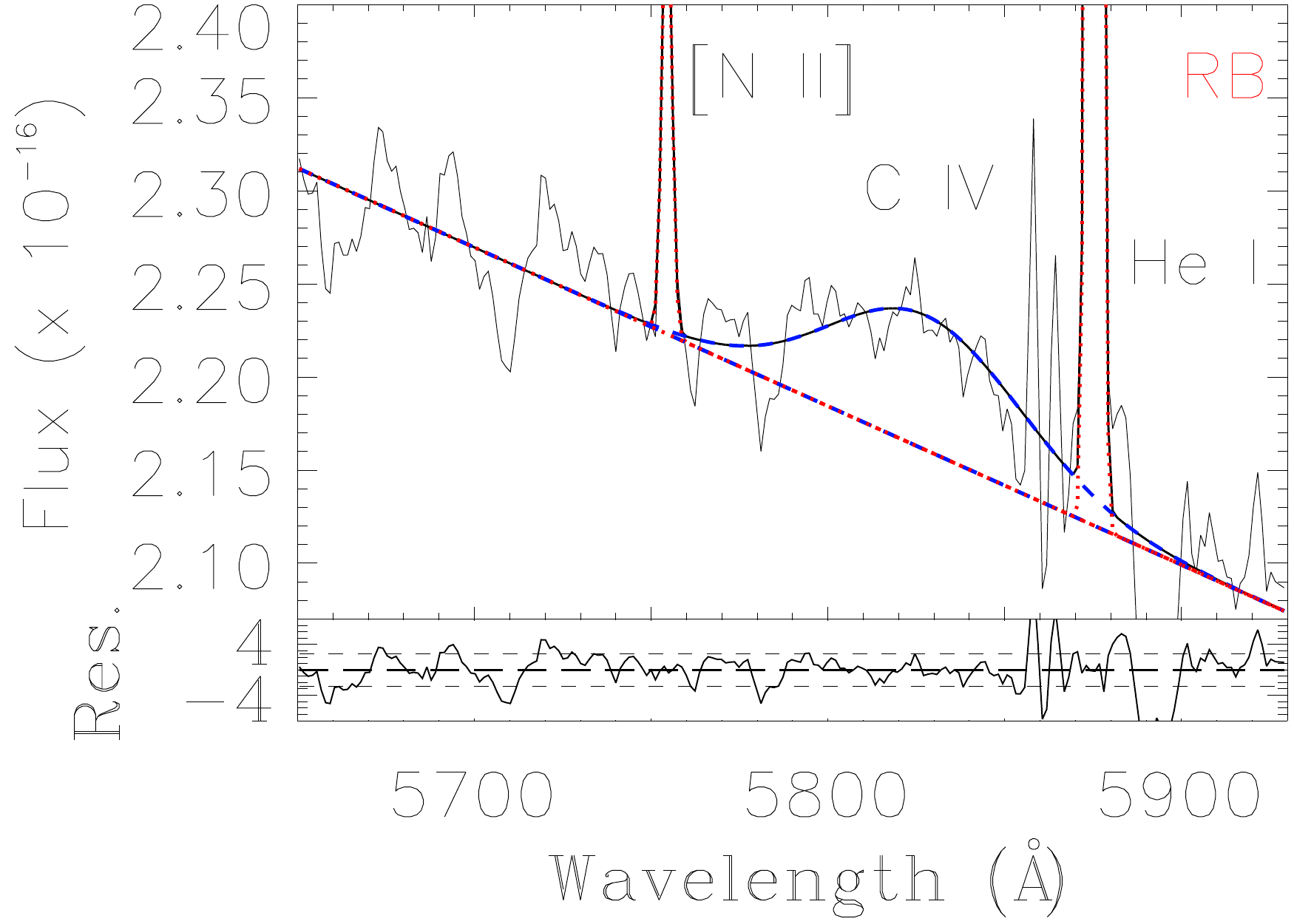}\\
\includegraphics[width=0.245\linewidth]{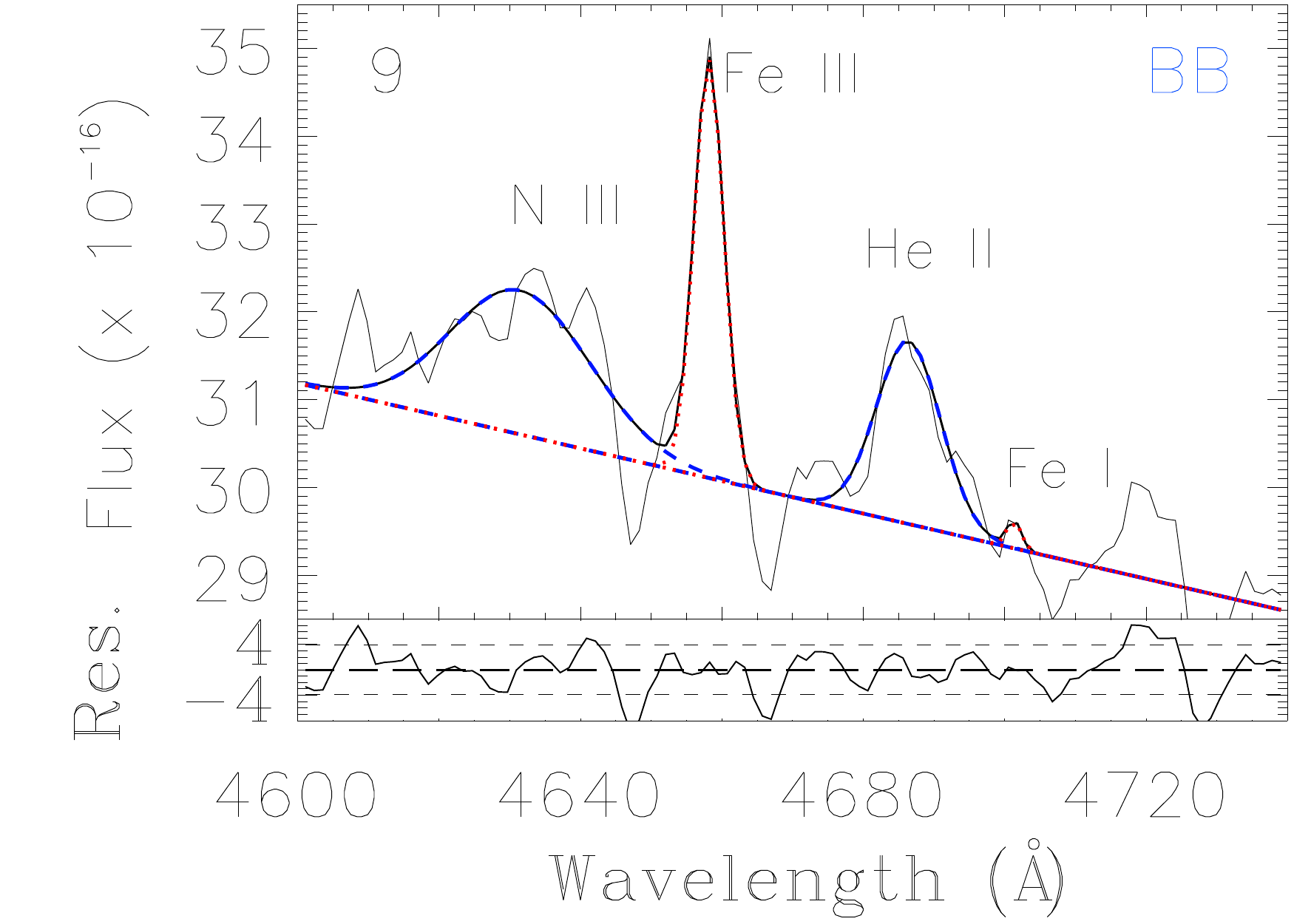}~
\includegraphics[width=0.245\linewidth]{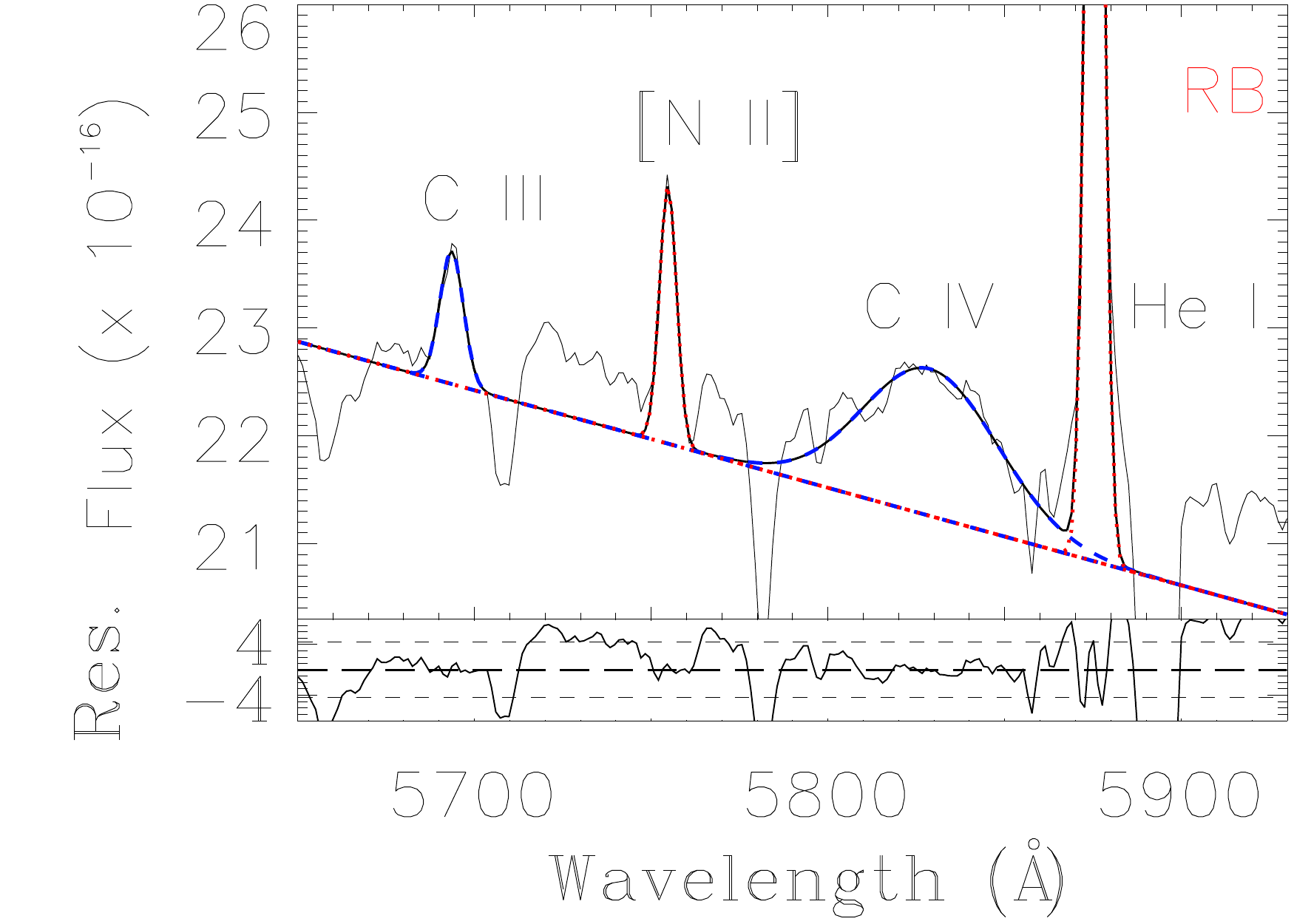}~
\includegraphics[width=0.245\linewidth]{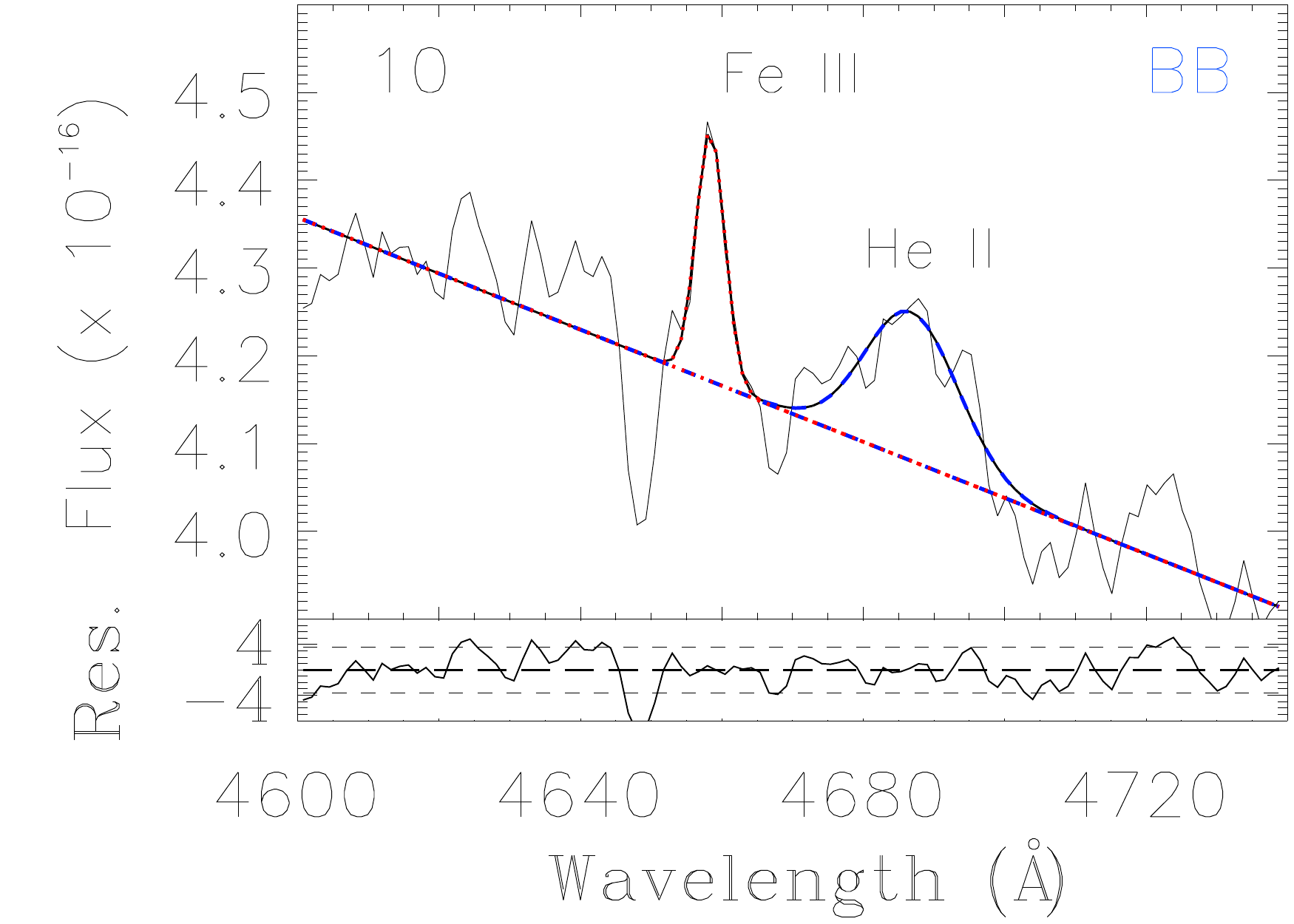}~
\includegraphics[width=0.245\linewidth]{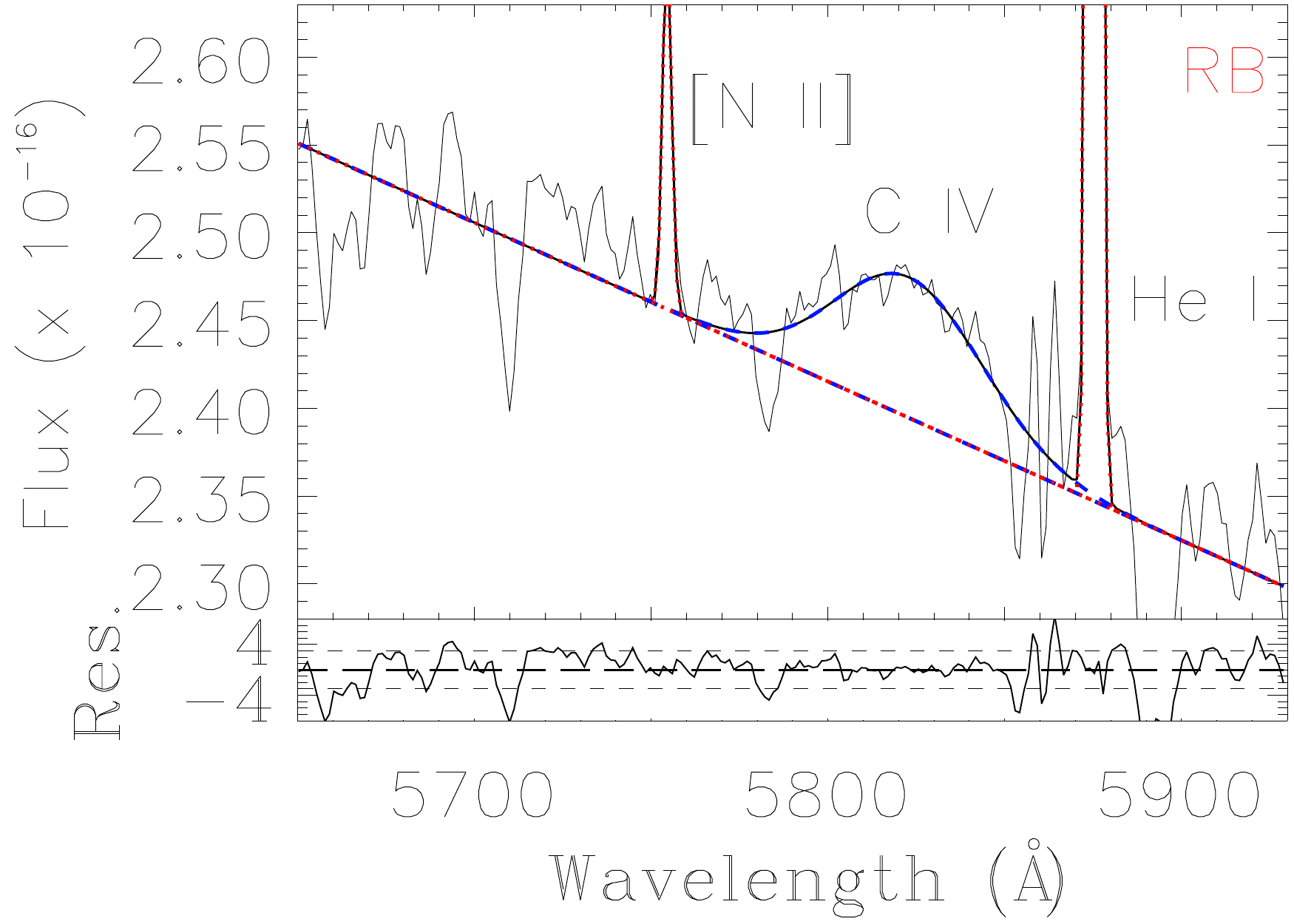}\\
\includegraphics[width=0.245\linewidth]{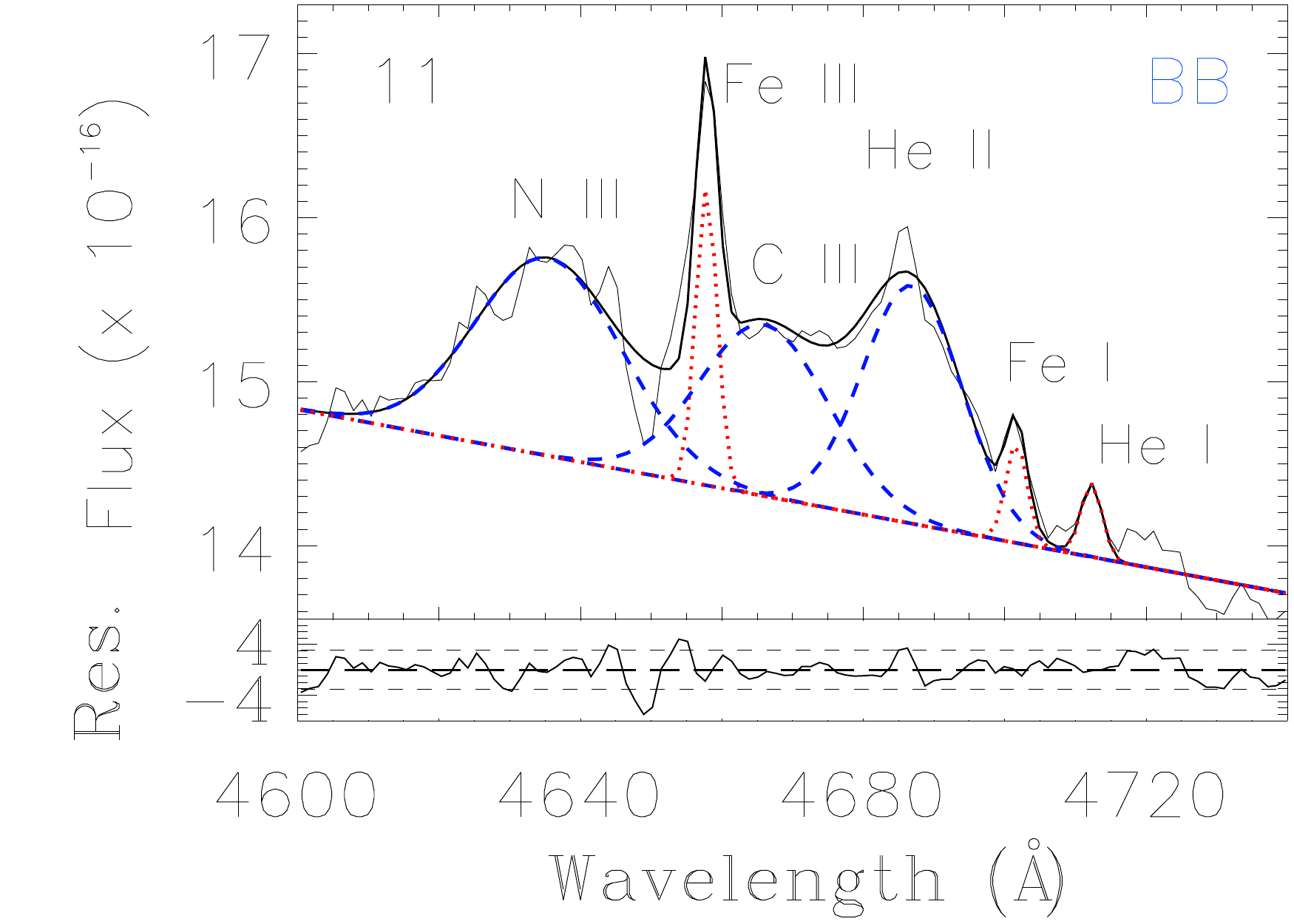}~
\includegraphics[width=0.245\linewidth]{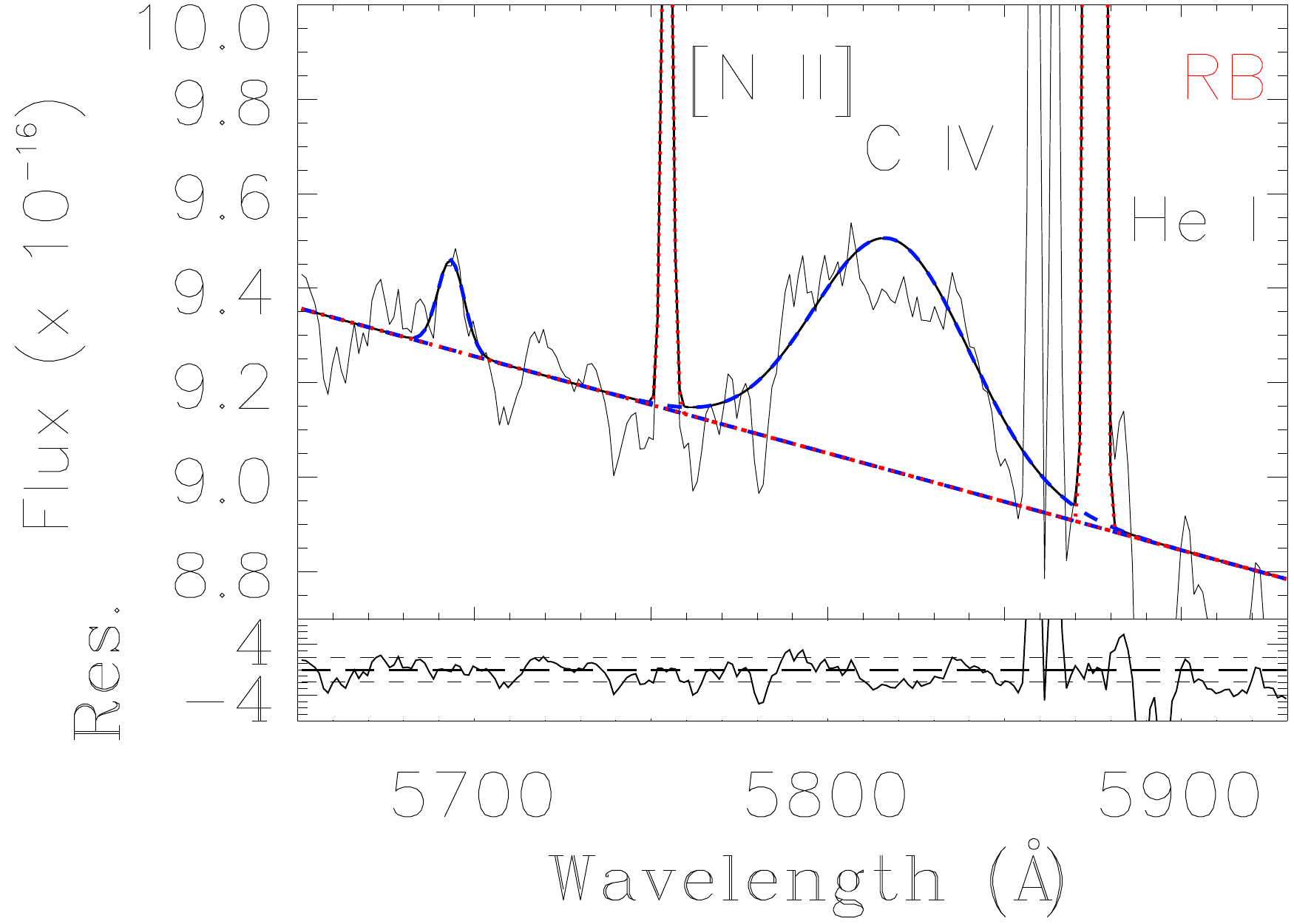}~
\includegraphics[width=0.245\linewidth]{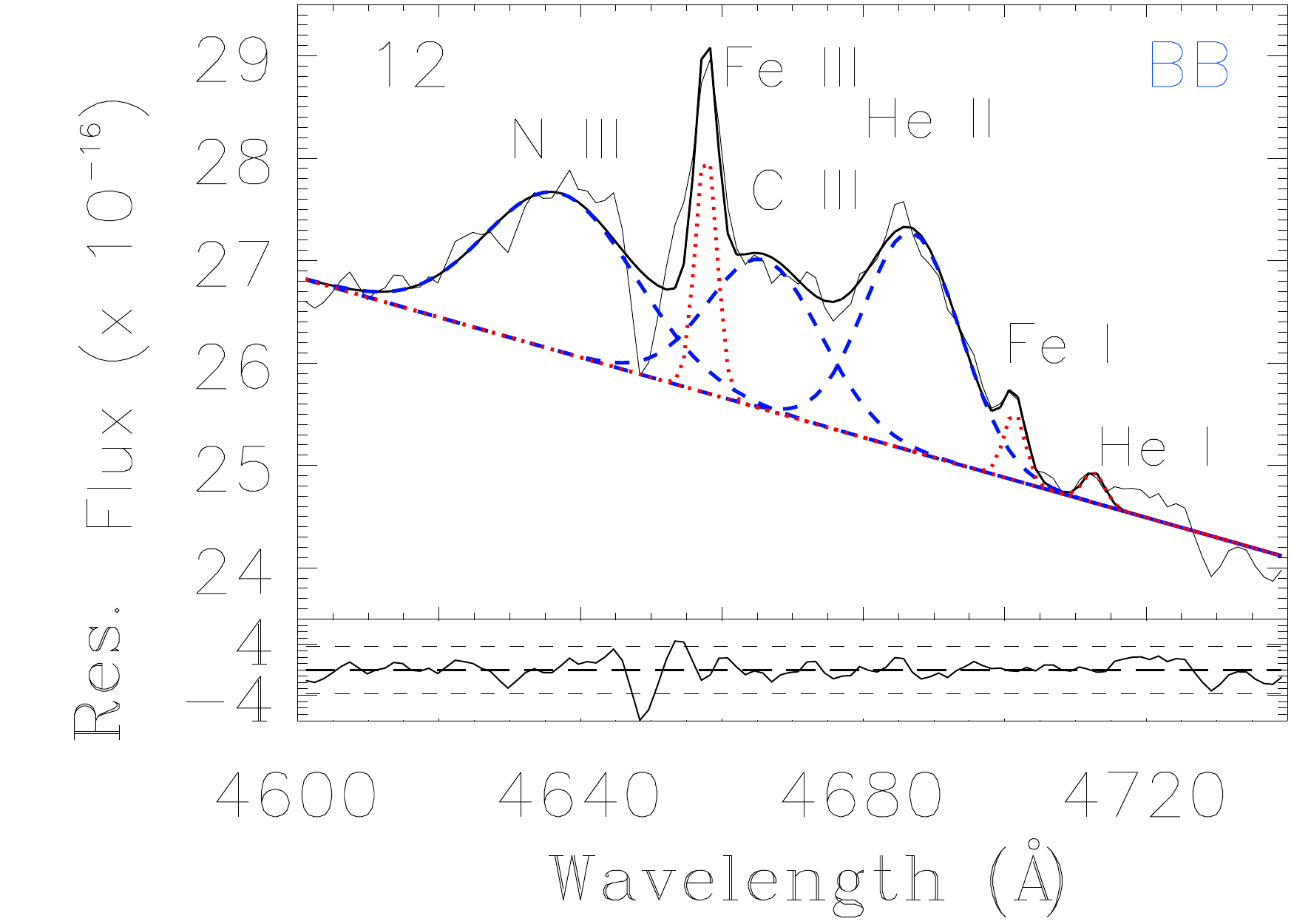}~
\includegraphics[width=0.245\linewidth]{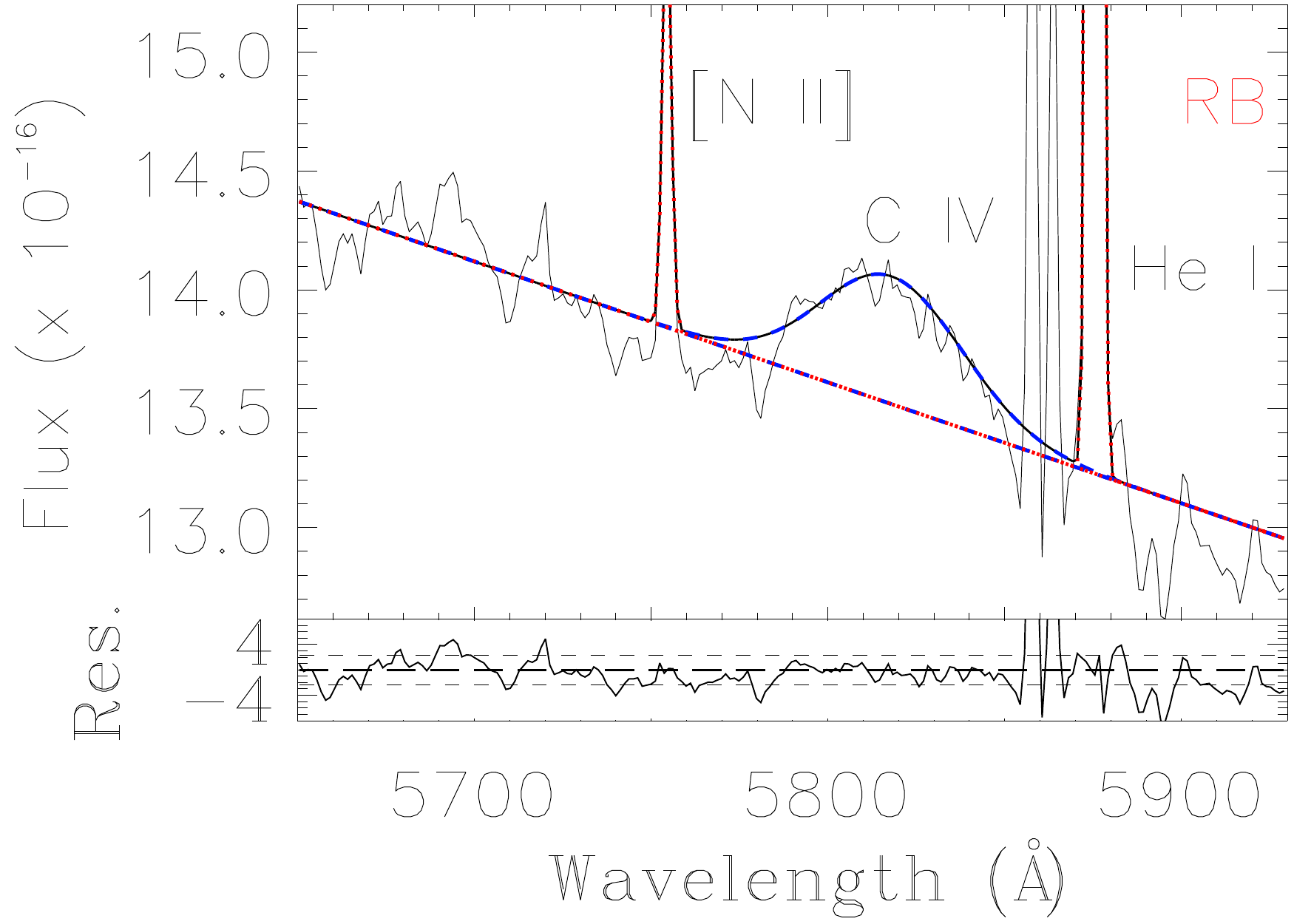}\\
\includegraphics[width=0.245\linewidth]{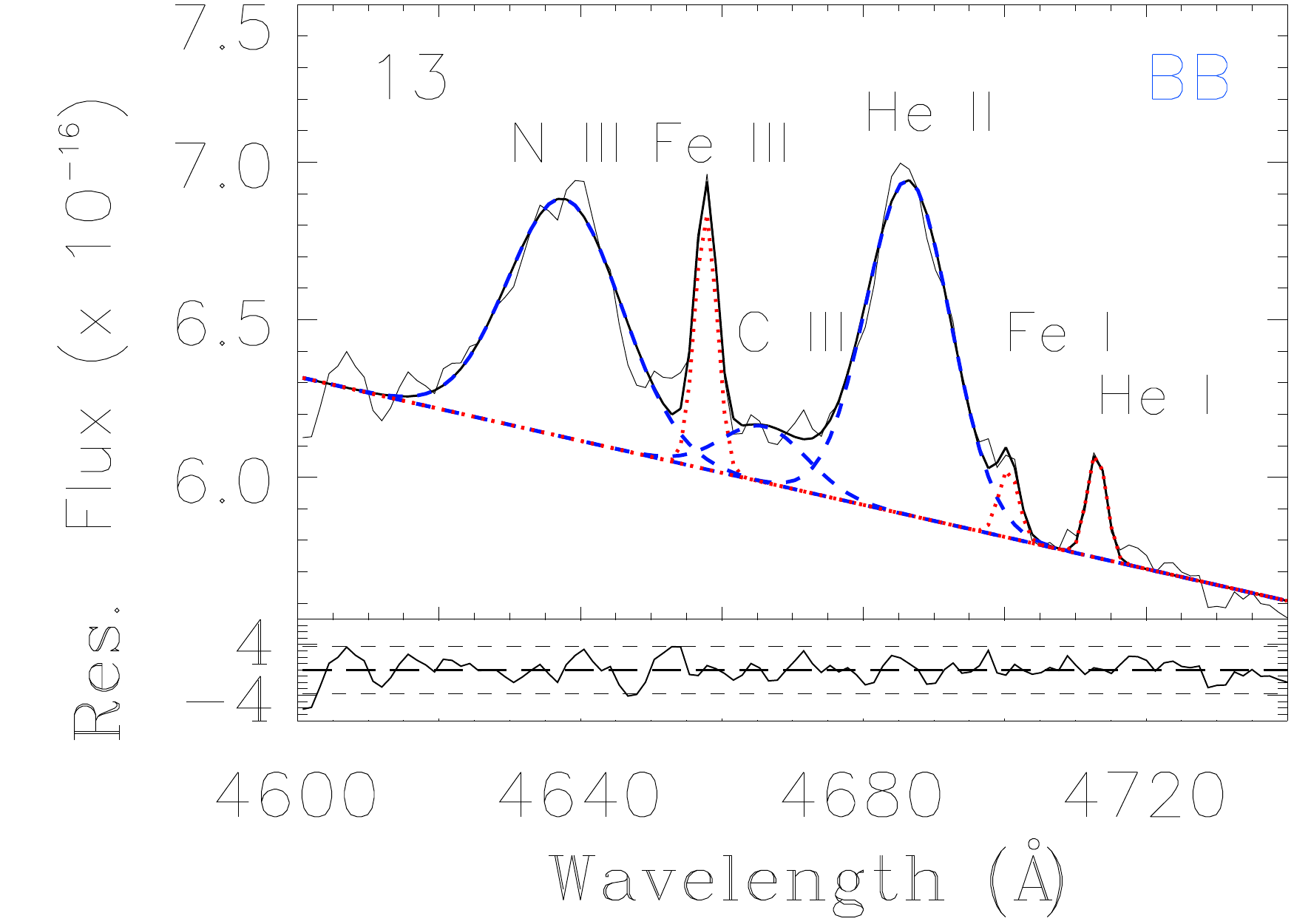}~
\includegraphics[width=0.245\linewidth]{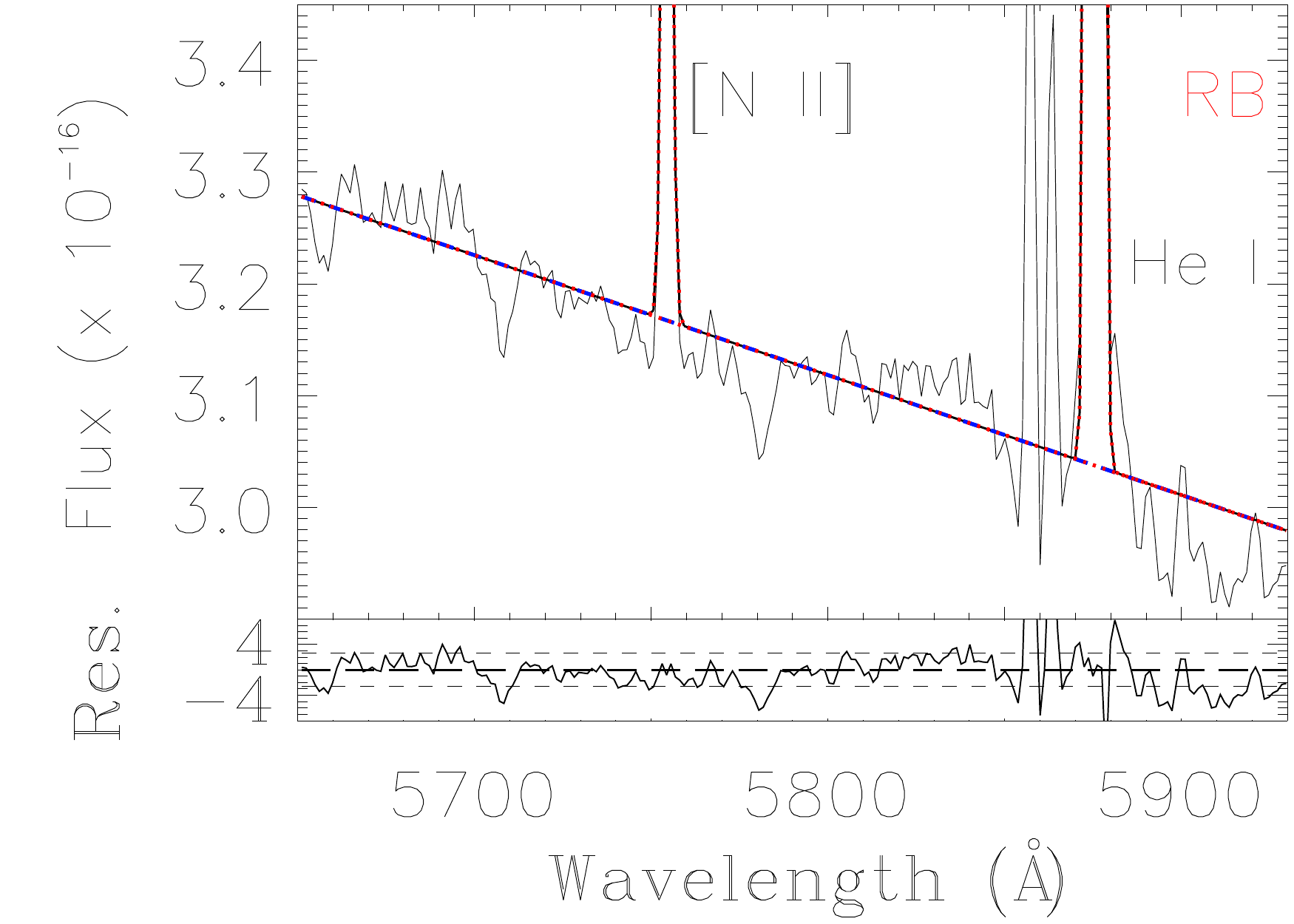}~
\includegraphics[width=0.245\linewidth]{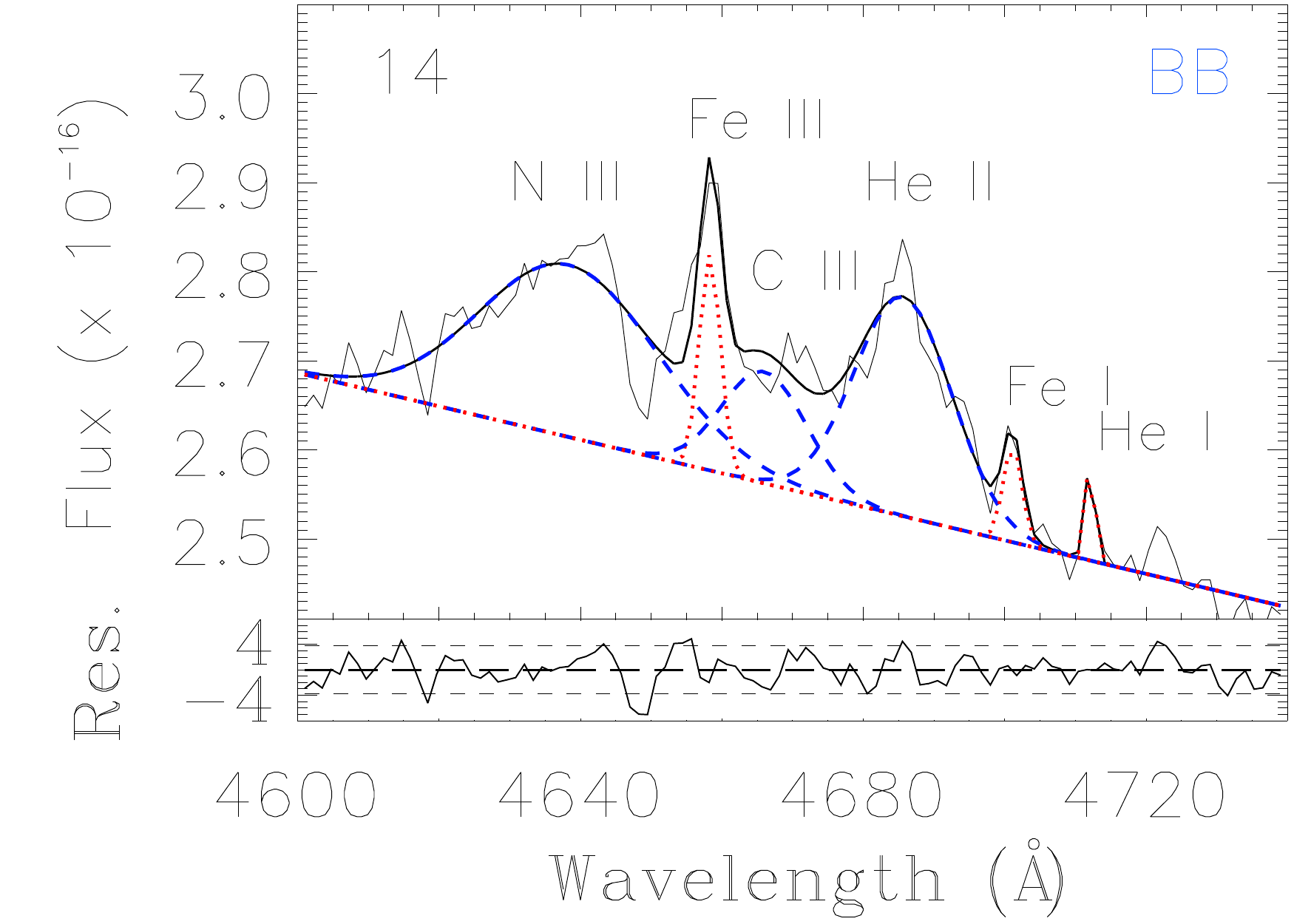}~
\includegraphics[width=0.245\linewidth]{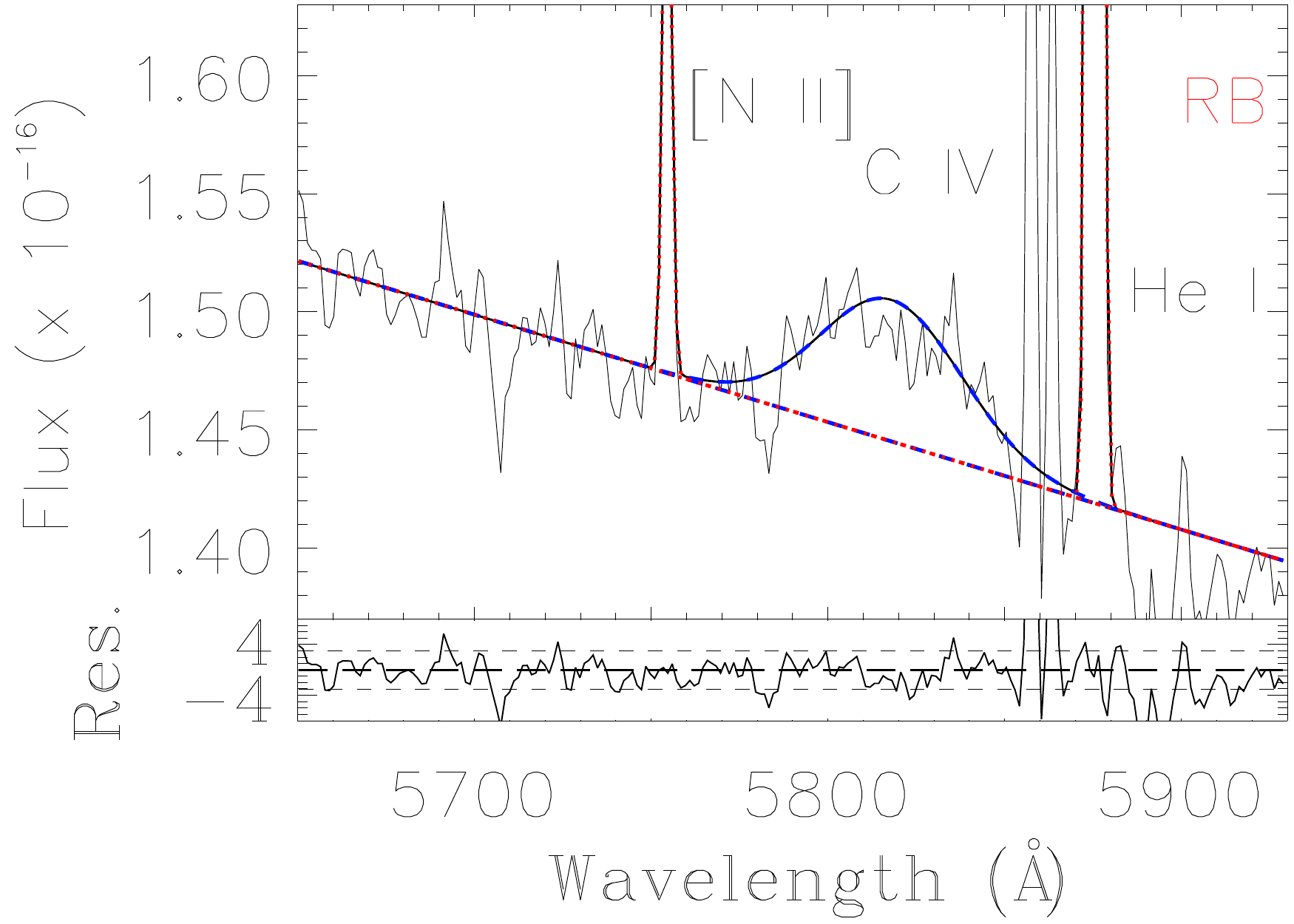}
\caption{Multi-component Gaussian fits to the {\it bumps} of the 38 star-forming complexes with WR features in the Antennae: the blue bump (BB) ({\it left})
and the red bump (RB) ({\it right}) are fitted by their broad ({\it dashed blue}) and nebular
({\it dotted red}) lines; the sum is shown in {\it black}.
The fitted continuum is shown by the
{\it dashed straight line}. The spectra are identified by the numbers
1--38 and each of the fitted components is indicated.
Residuals in per cent (\%) are shown at the bottom of each panel.
The results are listed in Table~\ref{tab:gauss}.} 
\label{fig:multi}
\end{center}
\end{figure*}

\begin{figure*}
\begin{center}
\includegraphics[width=0.245\linewidth]{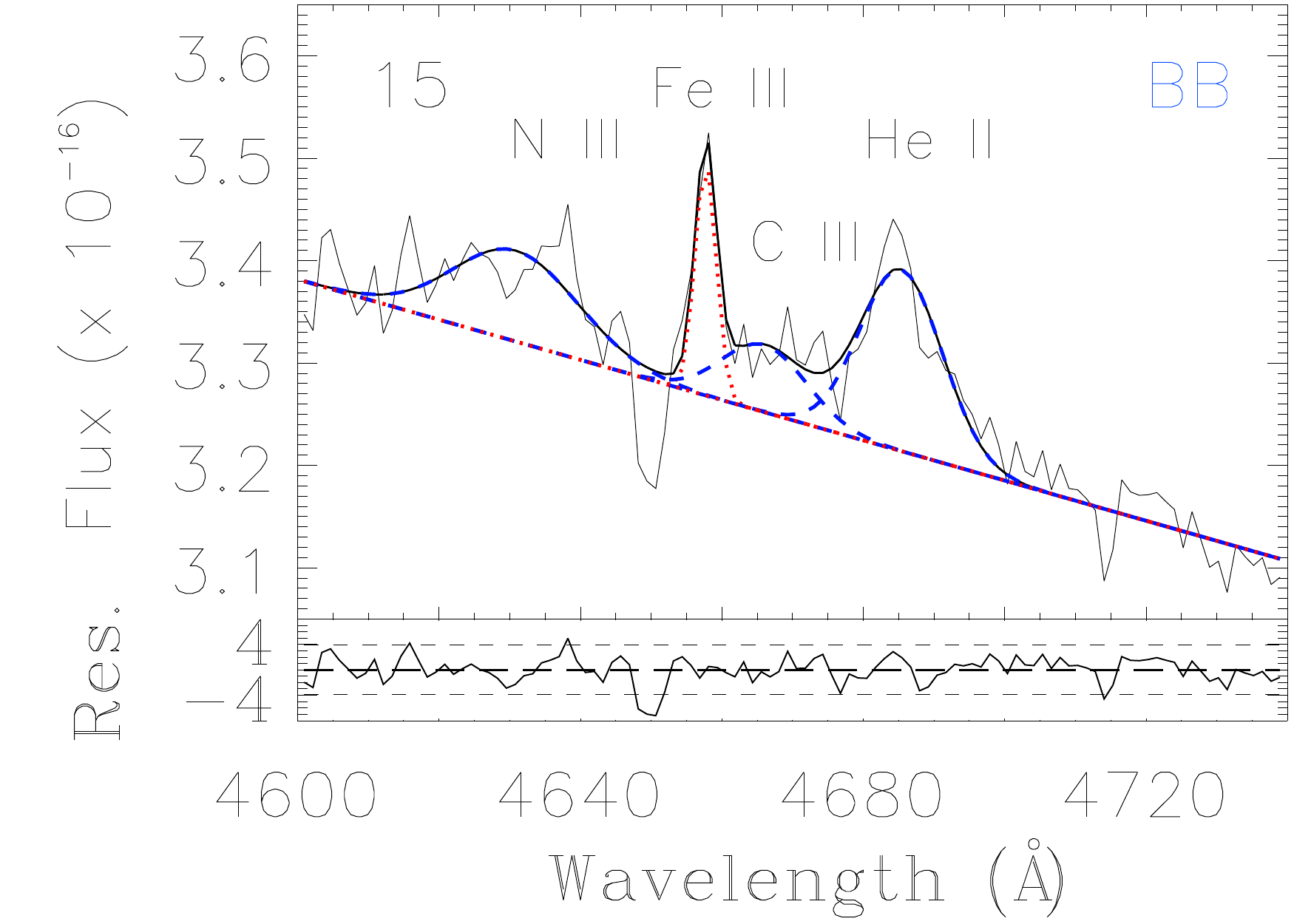}~
\includegraphics[width=0.245\linewidth]{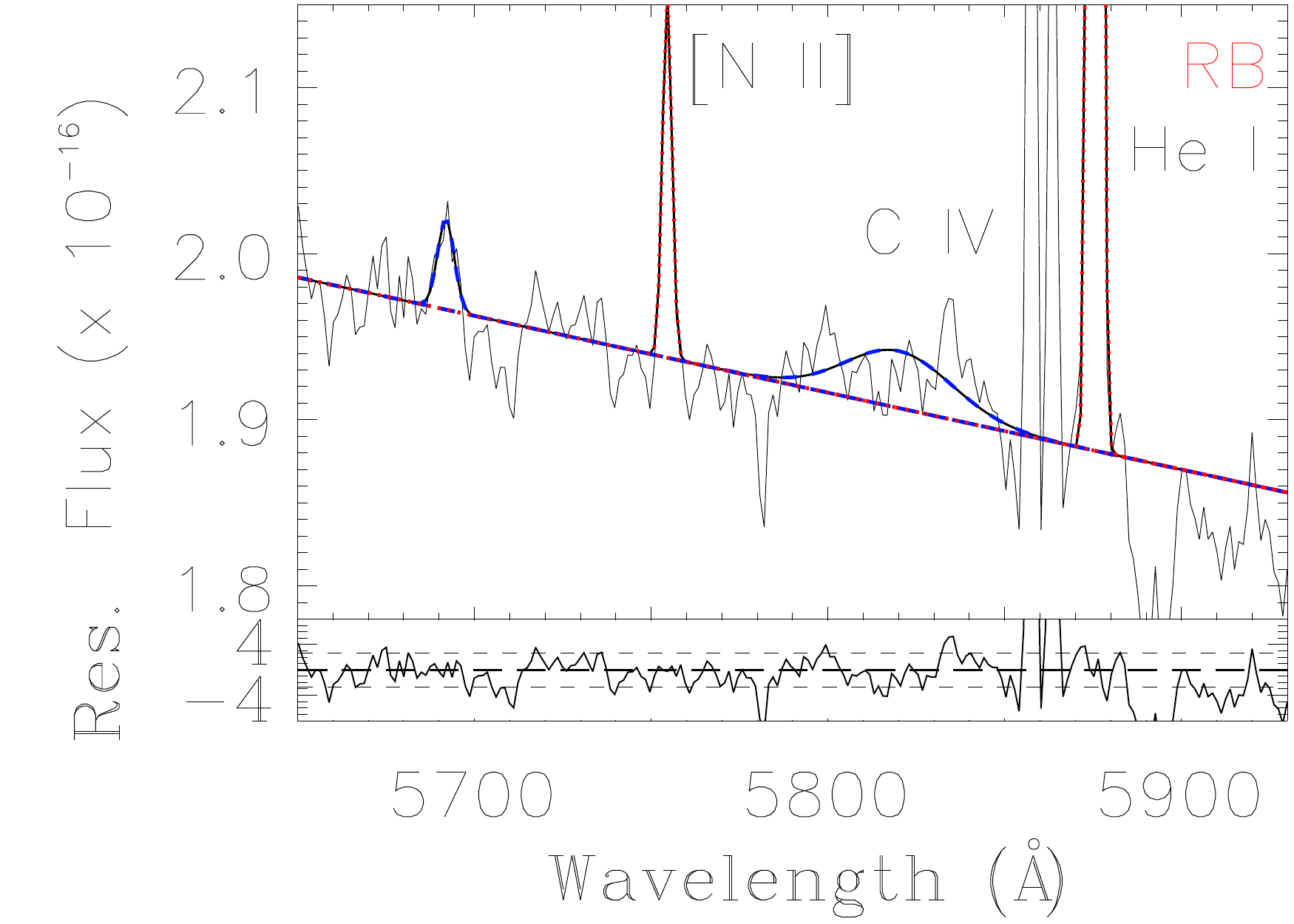}~
\includegraphics[width=0.245\linewidth]{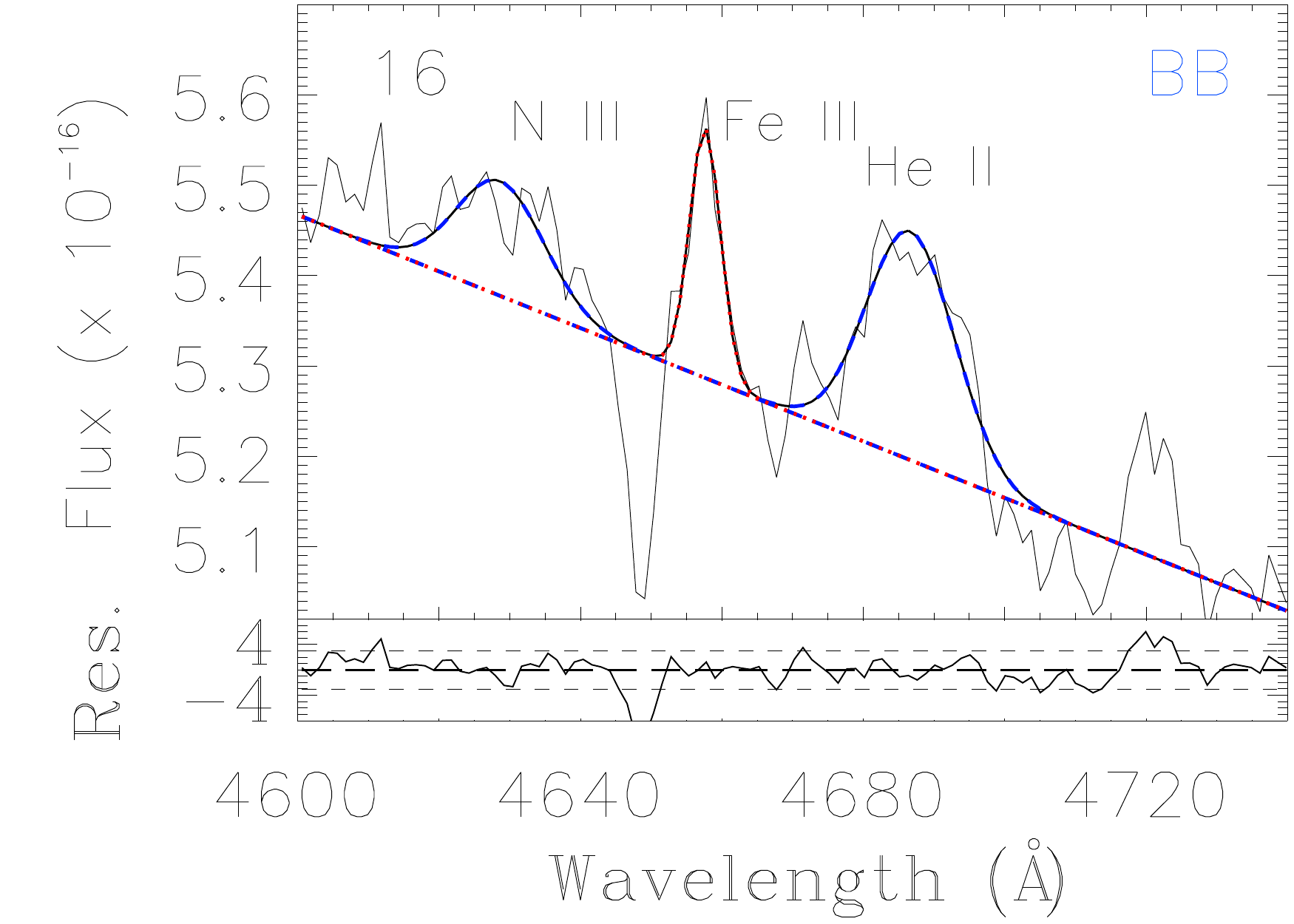}~
\includegraphics[width=0.245\linewidth]{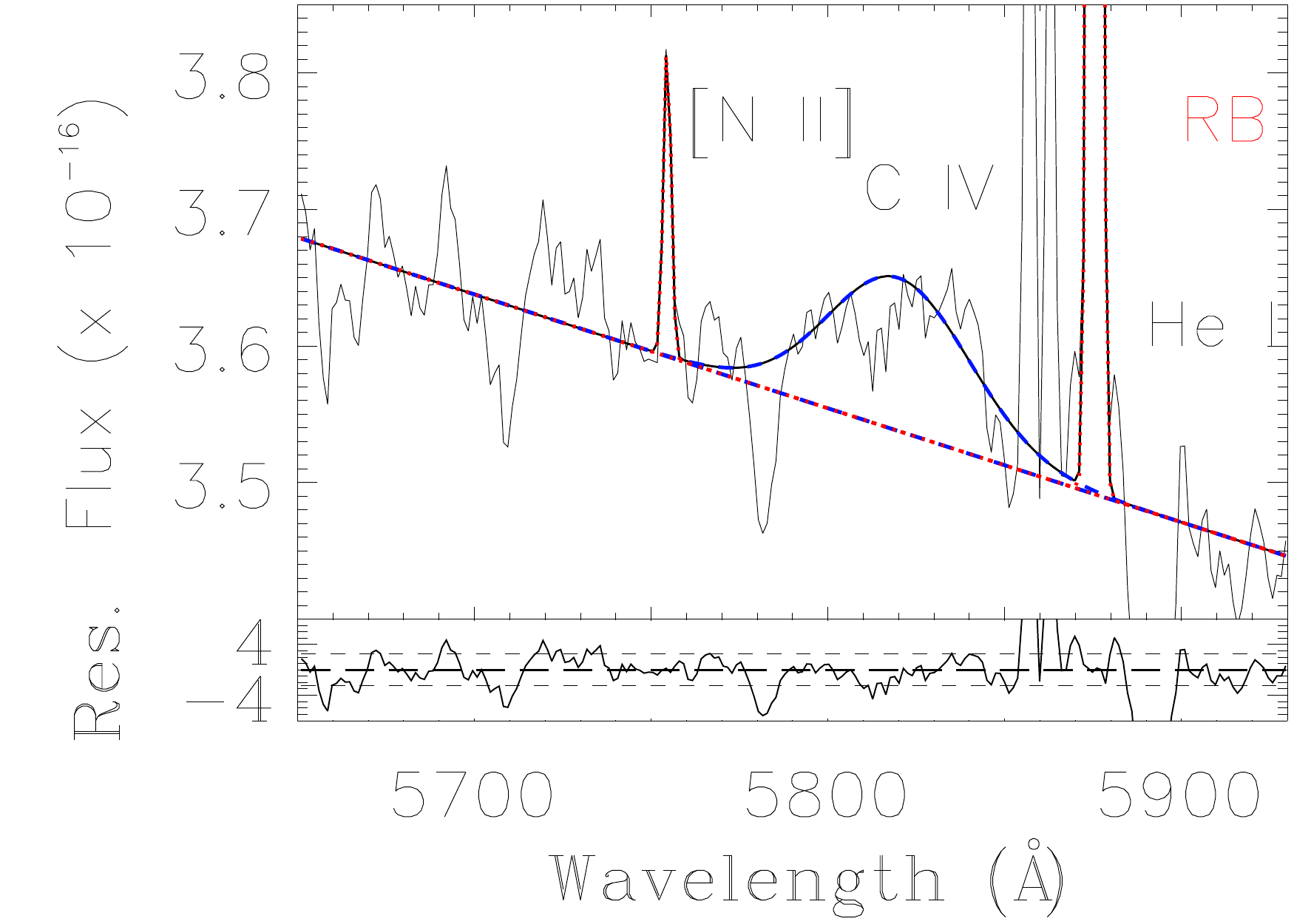}\\
\includegraphics[width=0.245\linewidth]{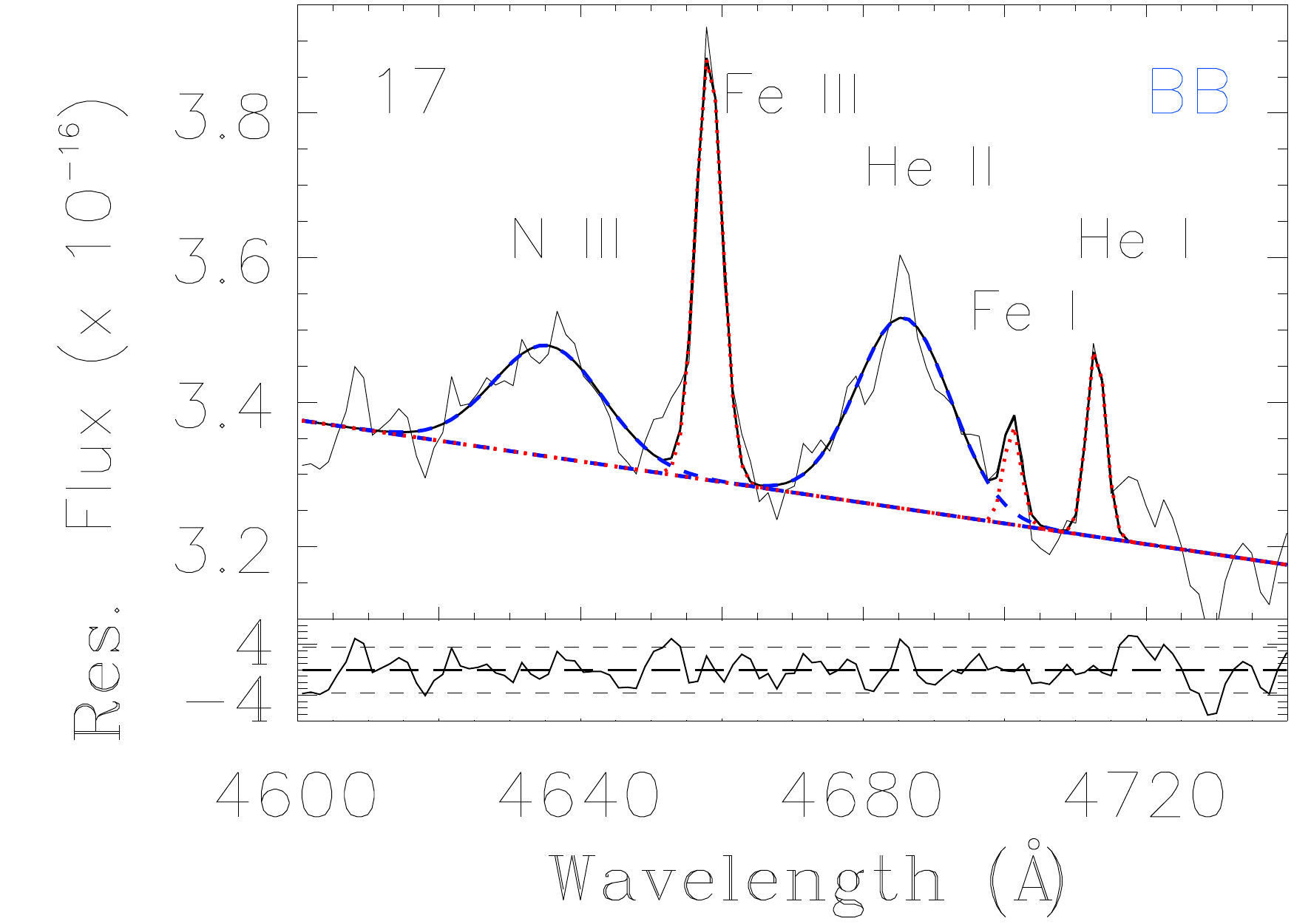}~
\includegraphics[width=0.245\linewidth]{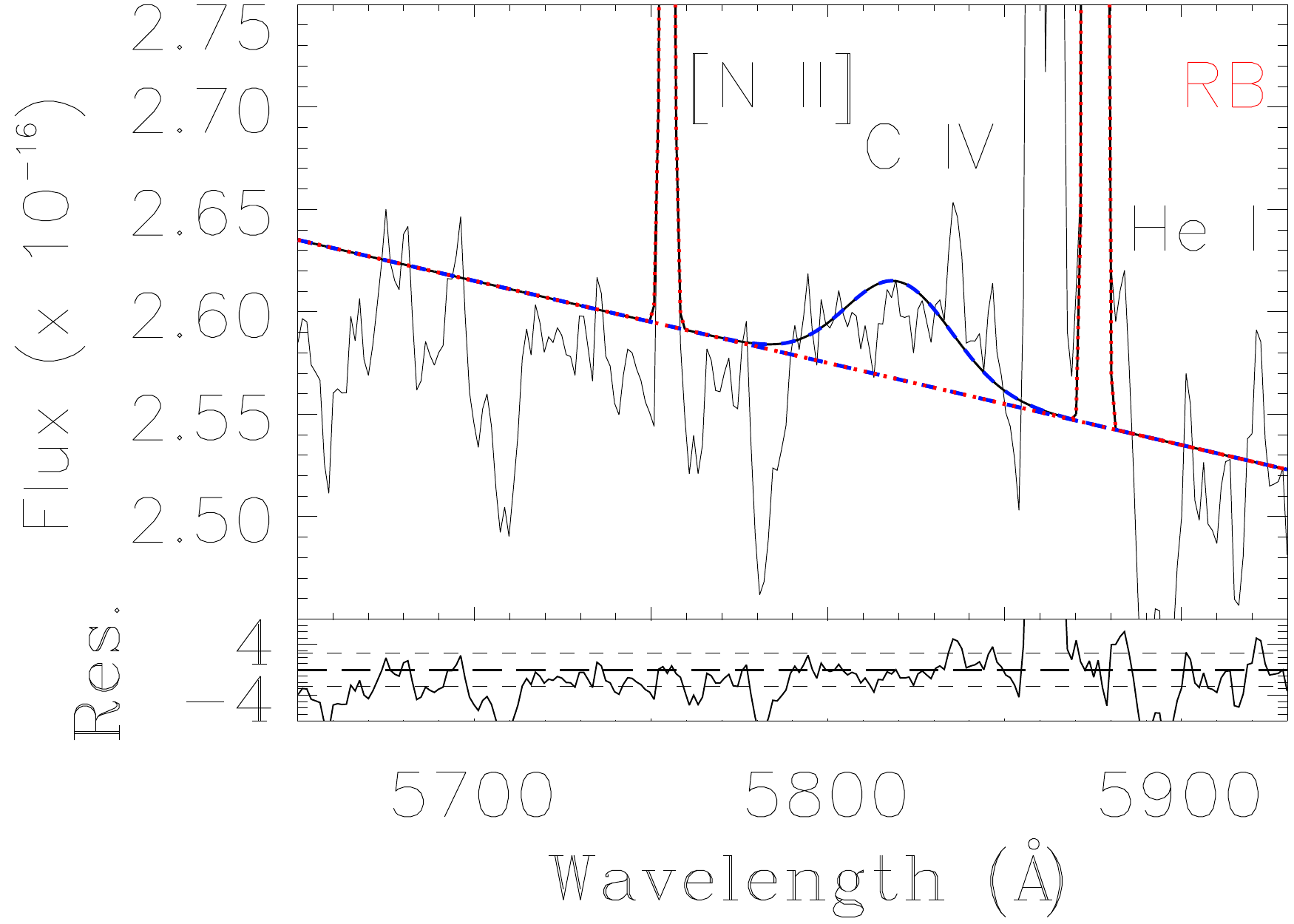}~
\includegraphics[width=0.245\linewidth]{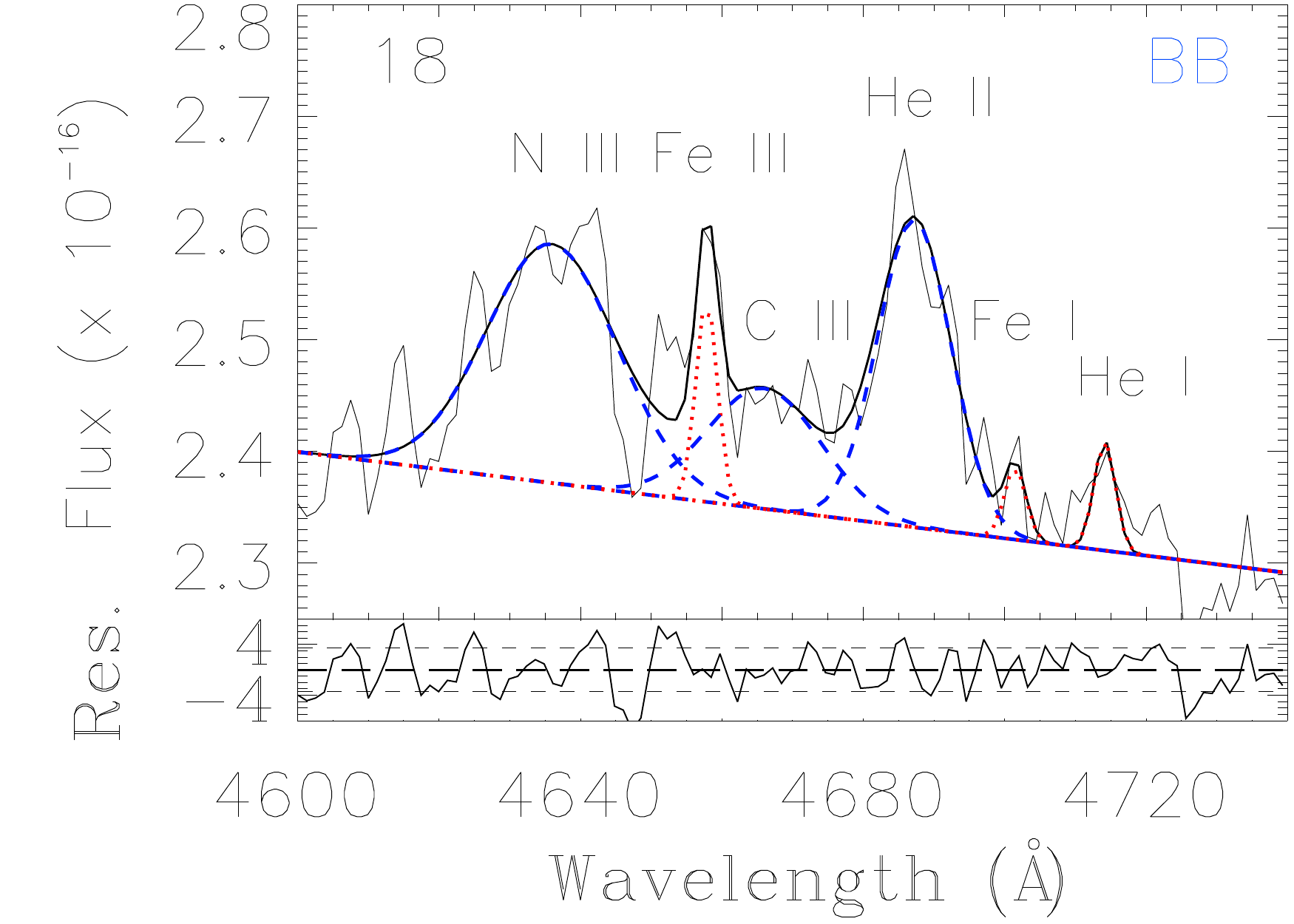}~
\includegraphics[width=0.245\linewidth]{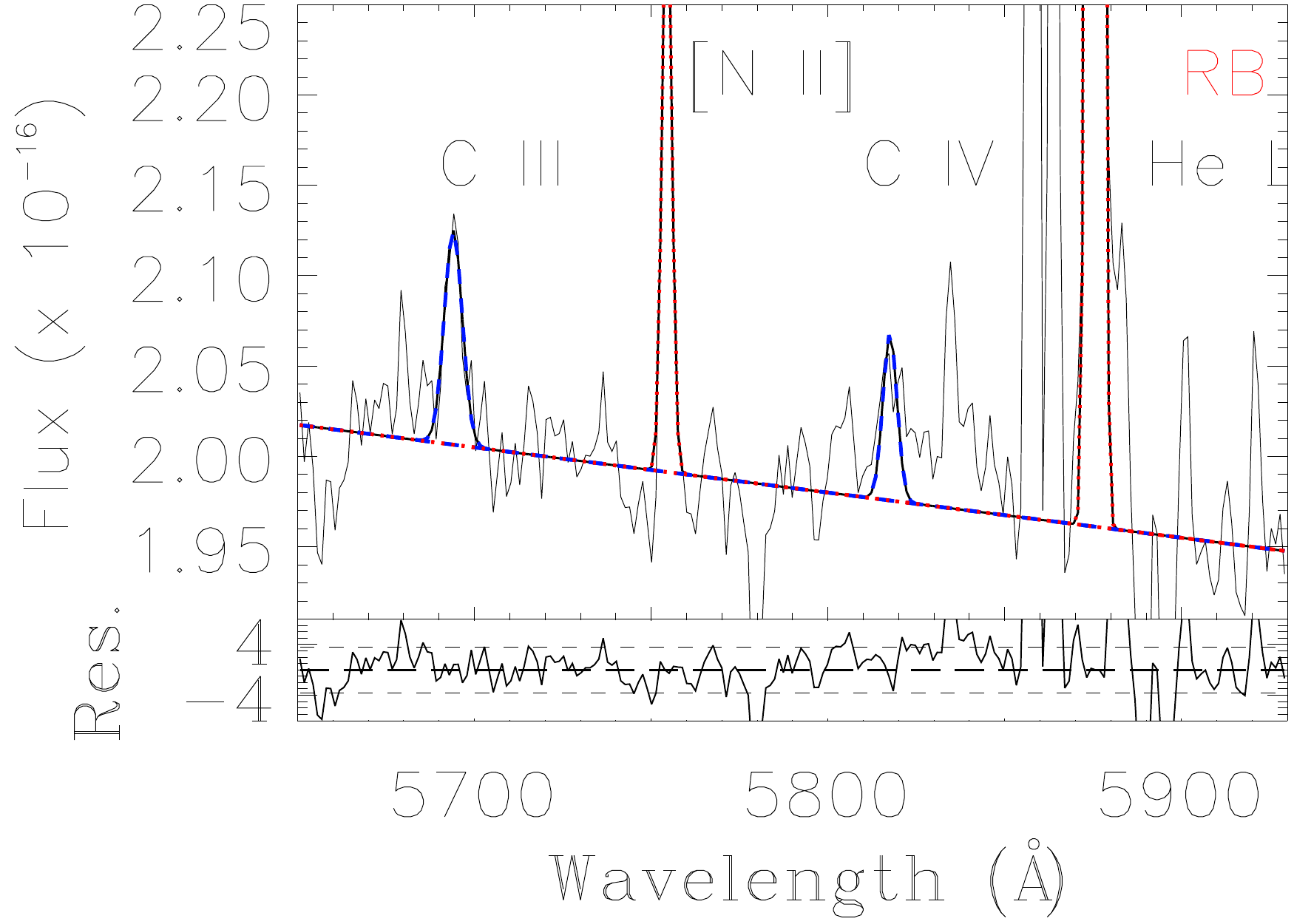}\\
\includegraphics[width=0.245\linewidth]{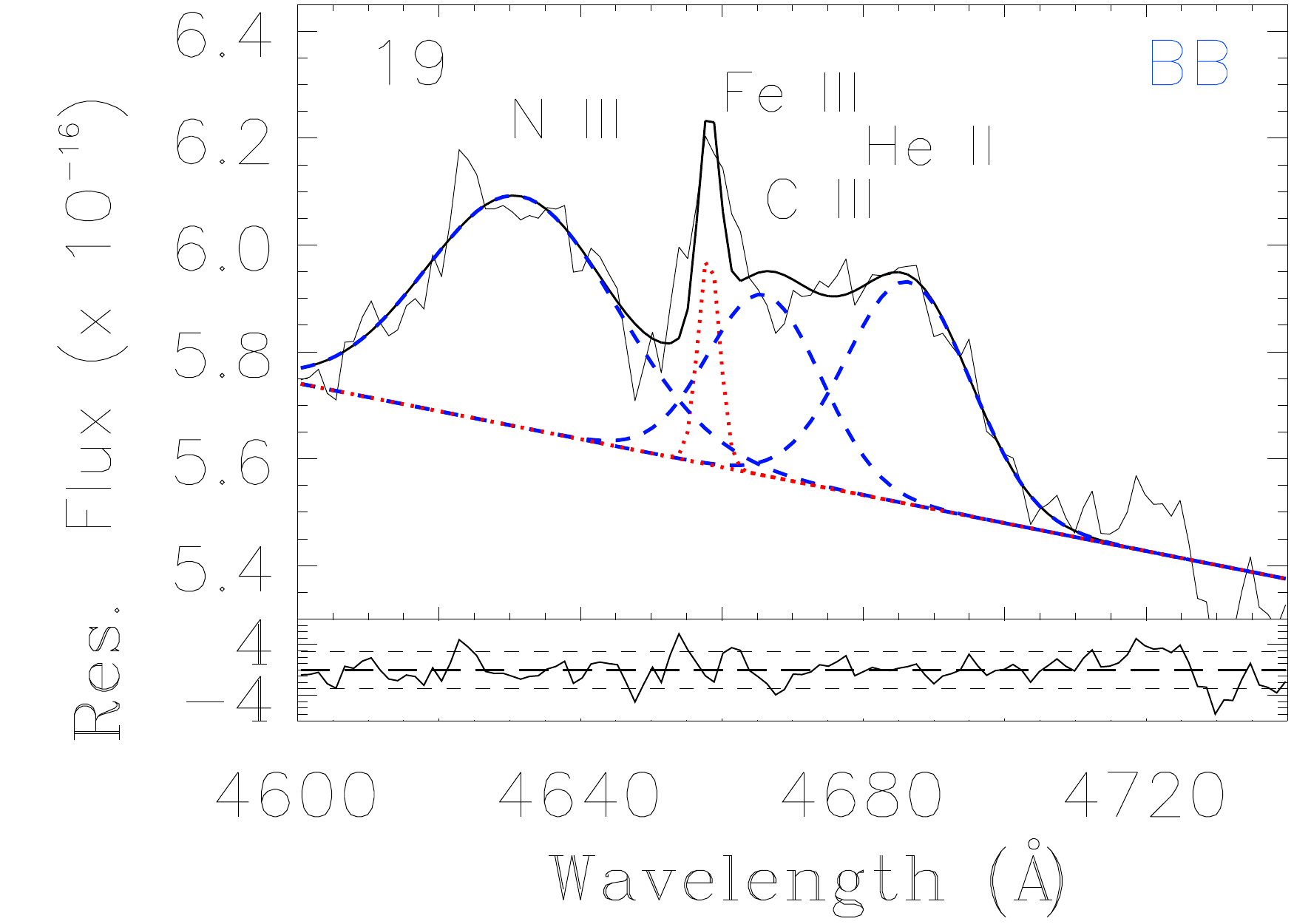}~
\includegraphics[width=0.245\linewidth]{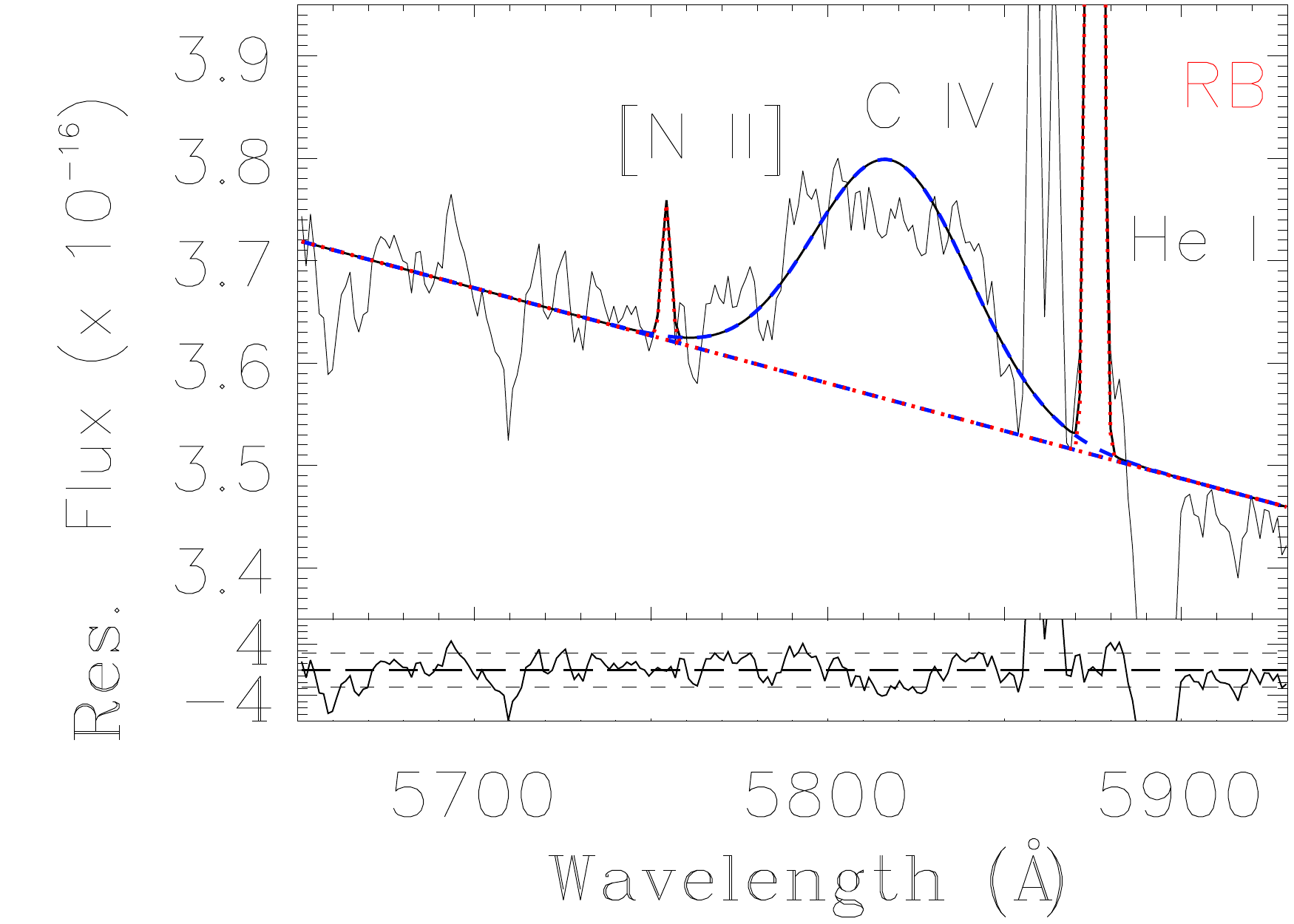}~
\includegraphics[width=0.245\linewidth]{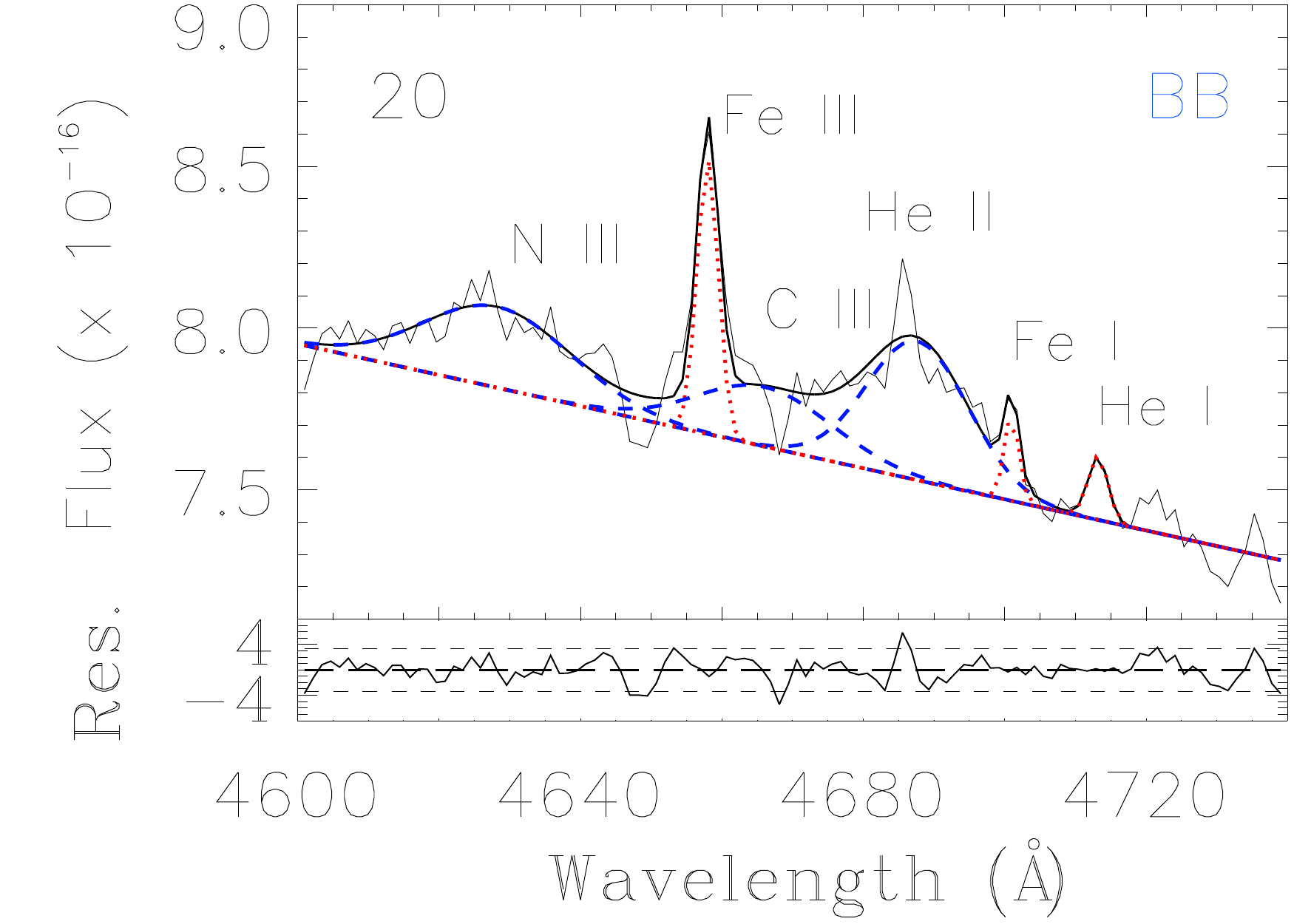}~
\includegraphics[width=0.245\linewidth]{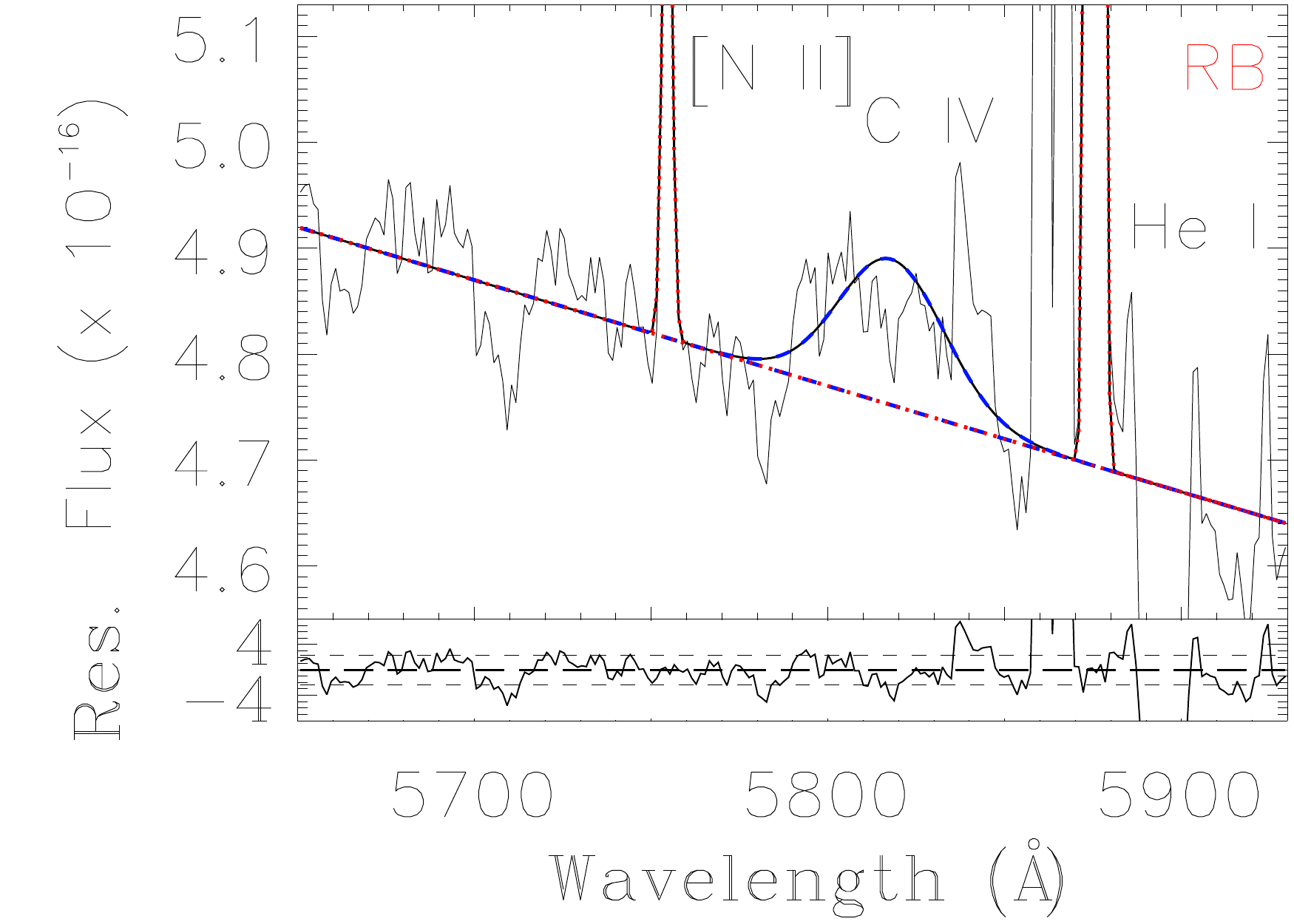}\\
\includegraphics[width=0.245\linewidth]{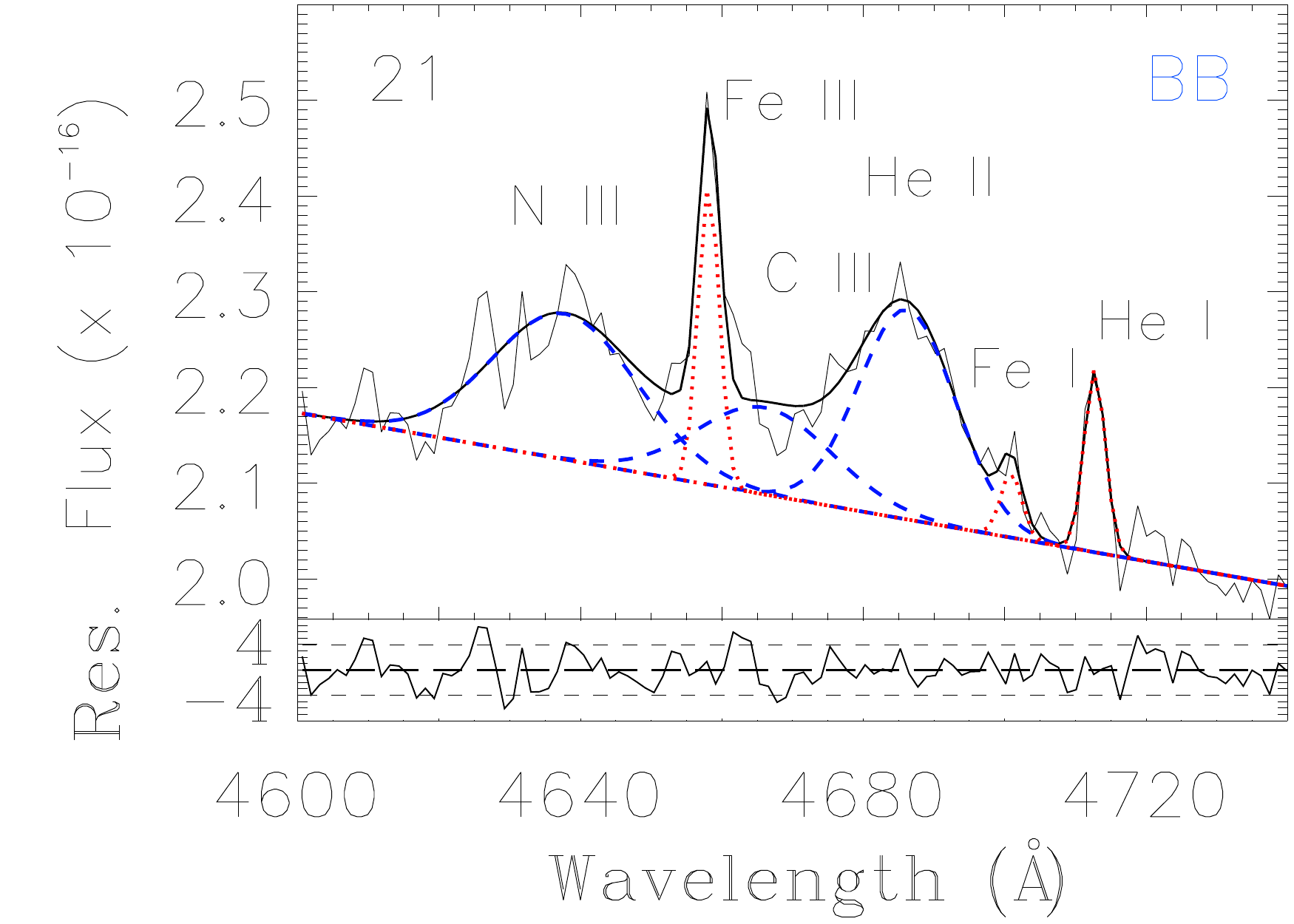}~
\includegraphics[width=0.245\linewidth]{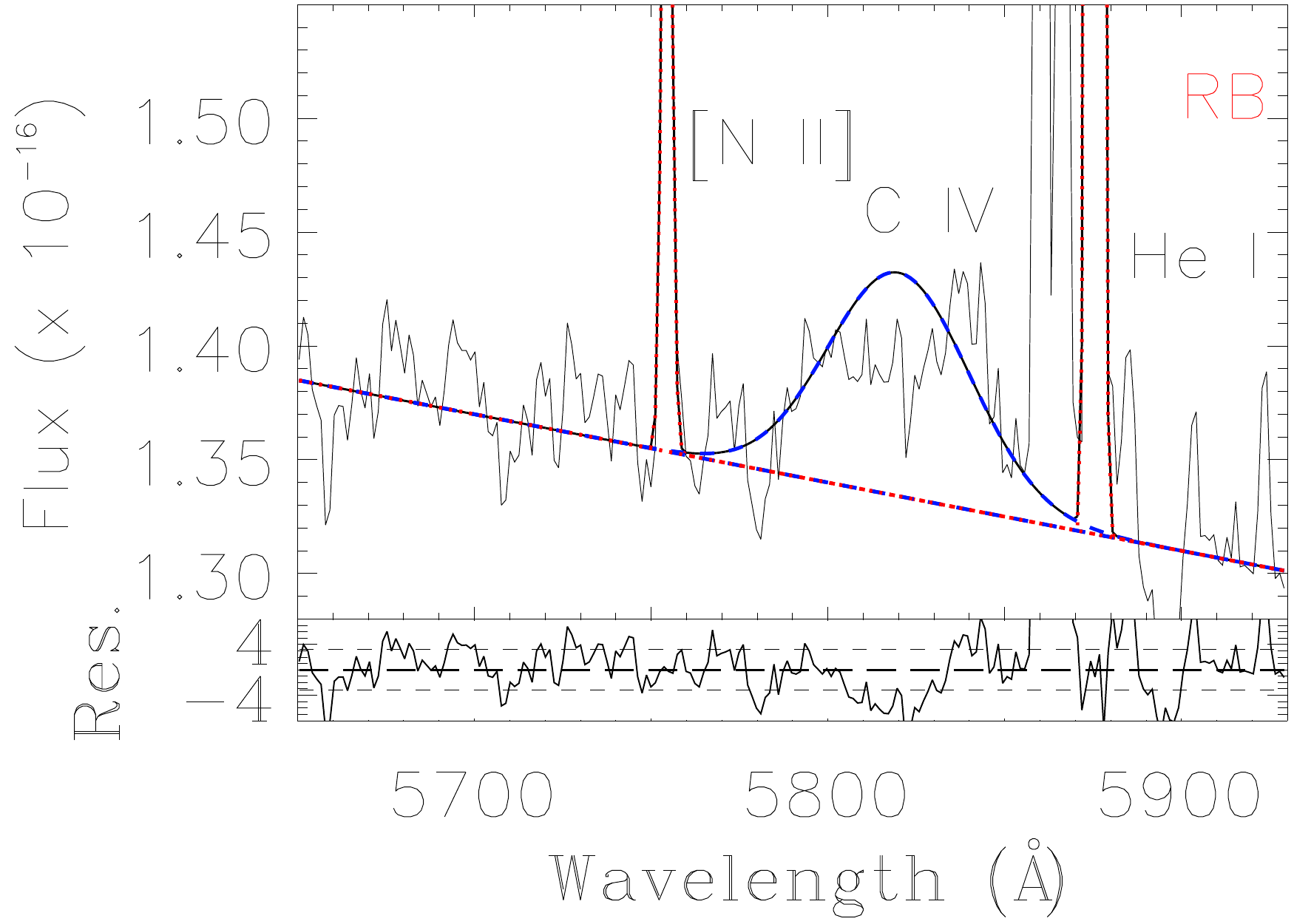}~
\includegraphics[width=0.245\linewidth]{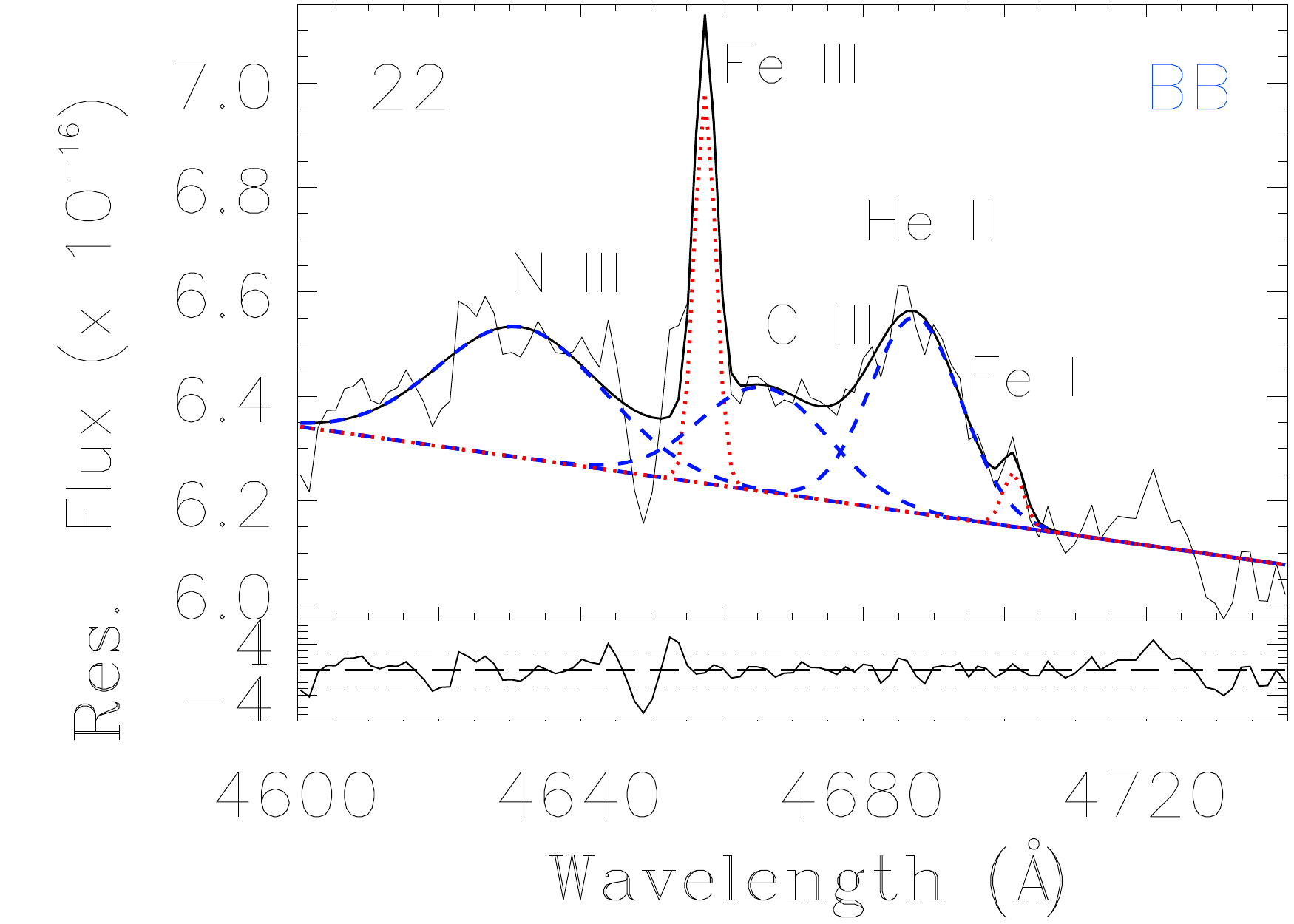}~
\includegraphics[width=0.245\linewidth]{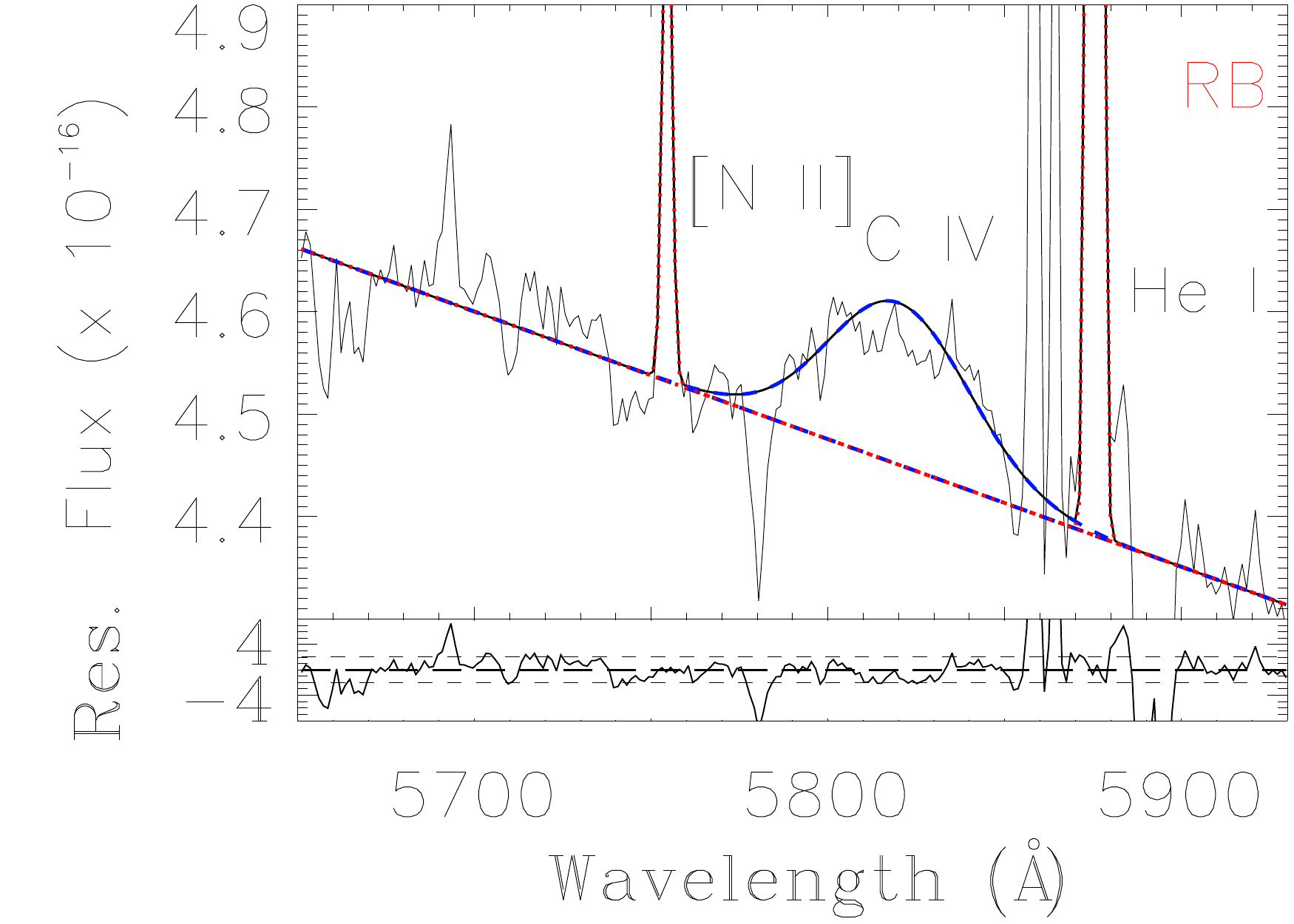}\\
\includegraphics[width=0.245\linewidth]{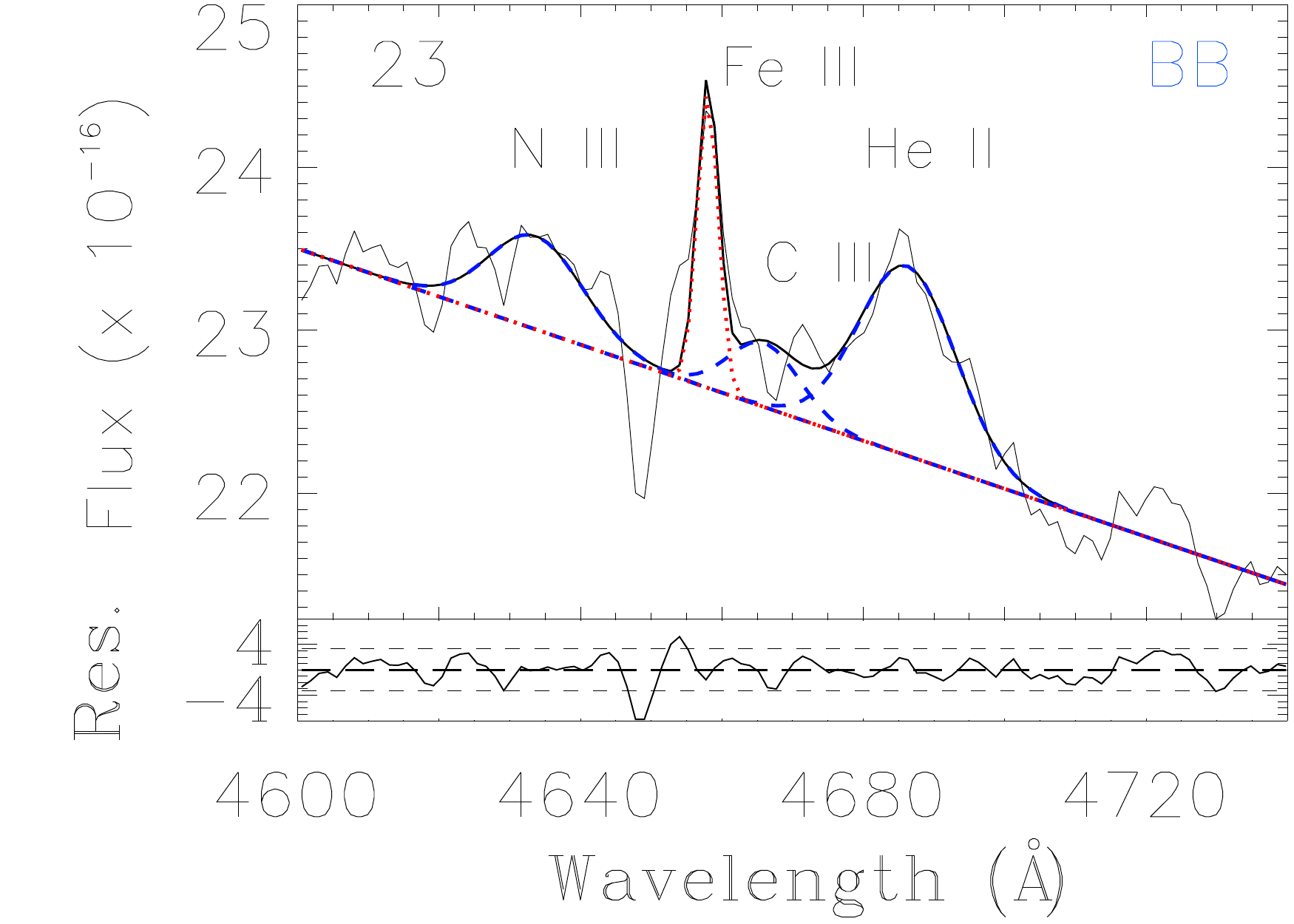}~
\includegraphics[width=0.245\linewidth]{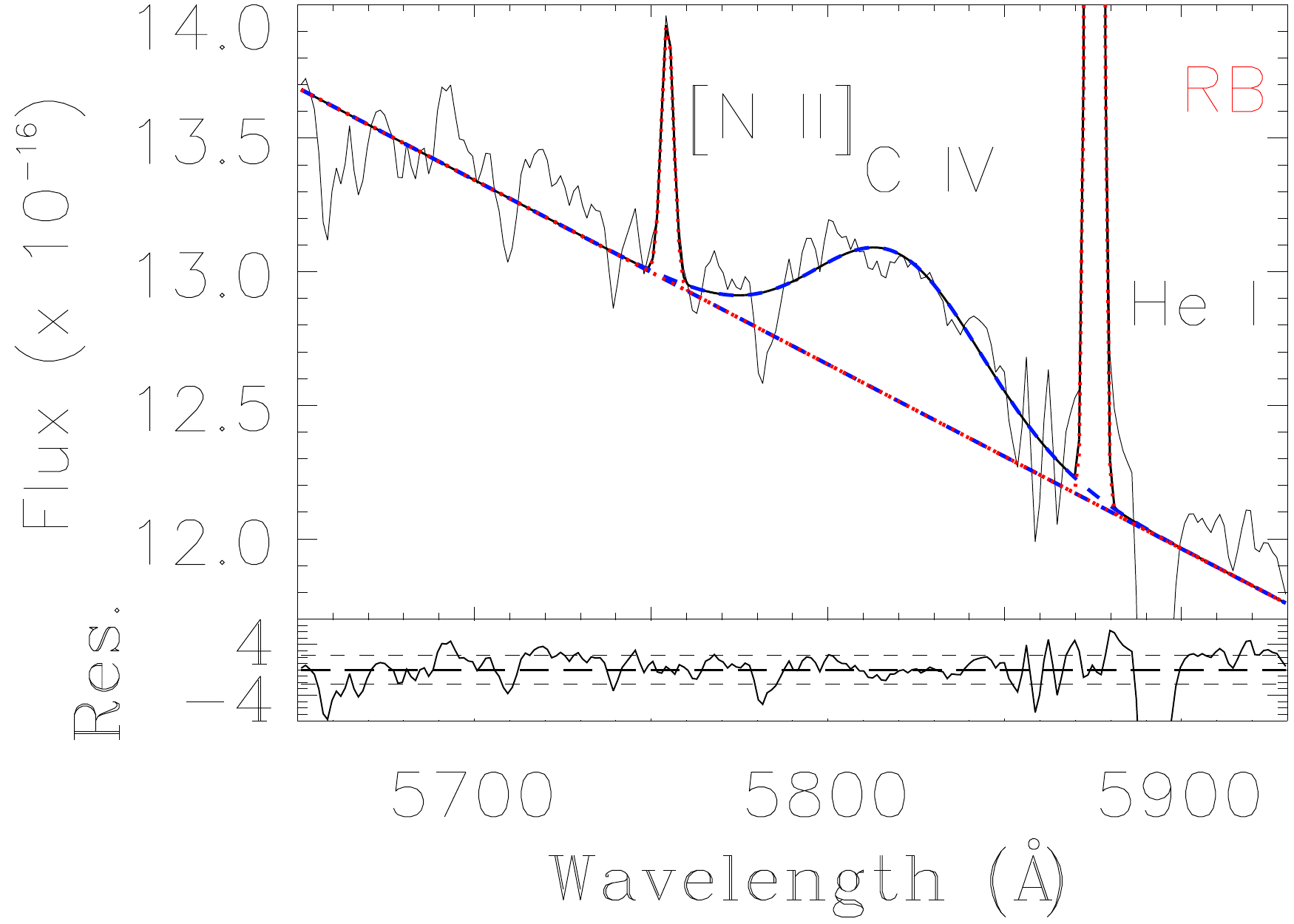}~
\includegraphics[width=0.245\linewidth]{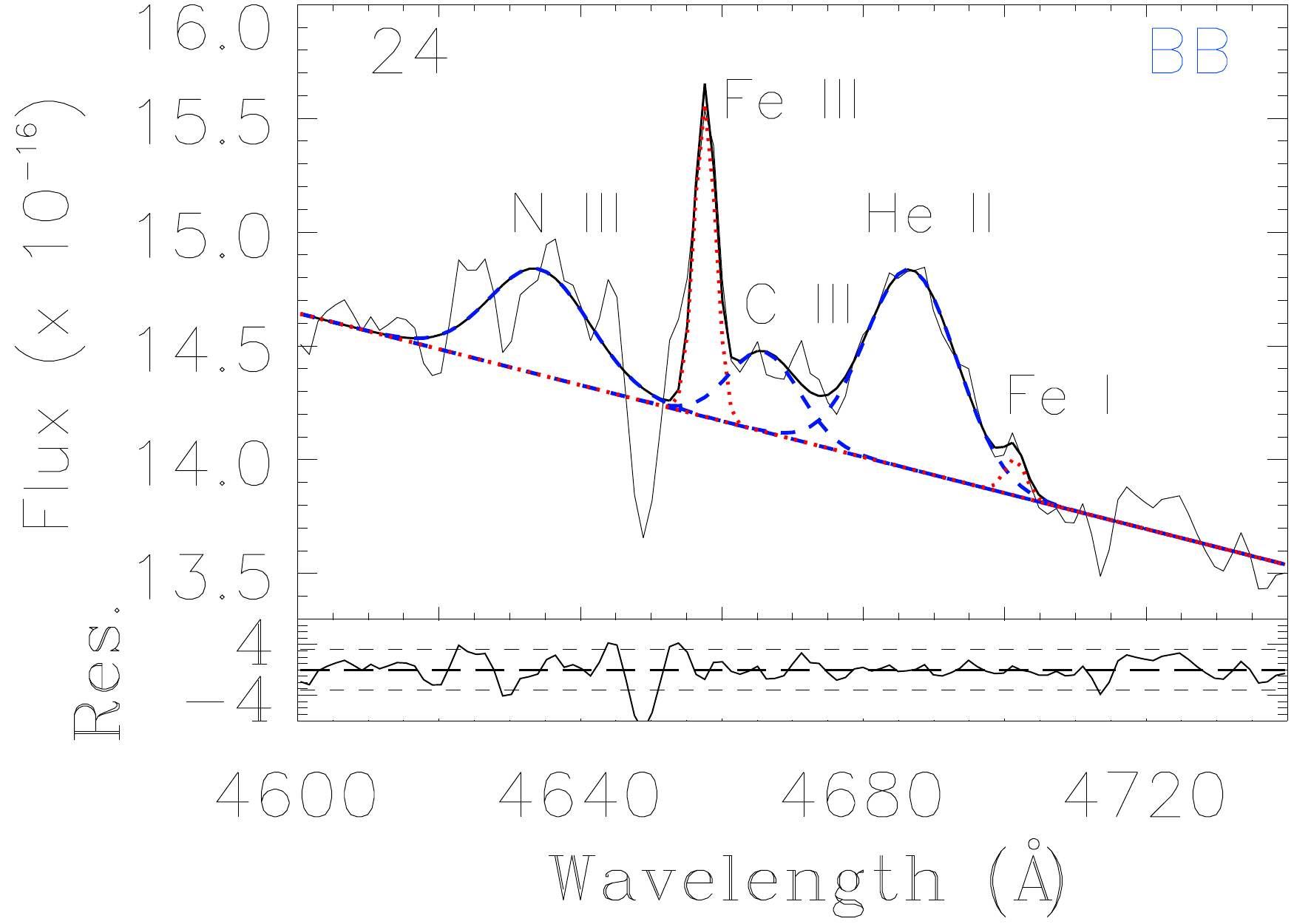}~
\includegraphics[width=0.245\linewidth]{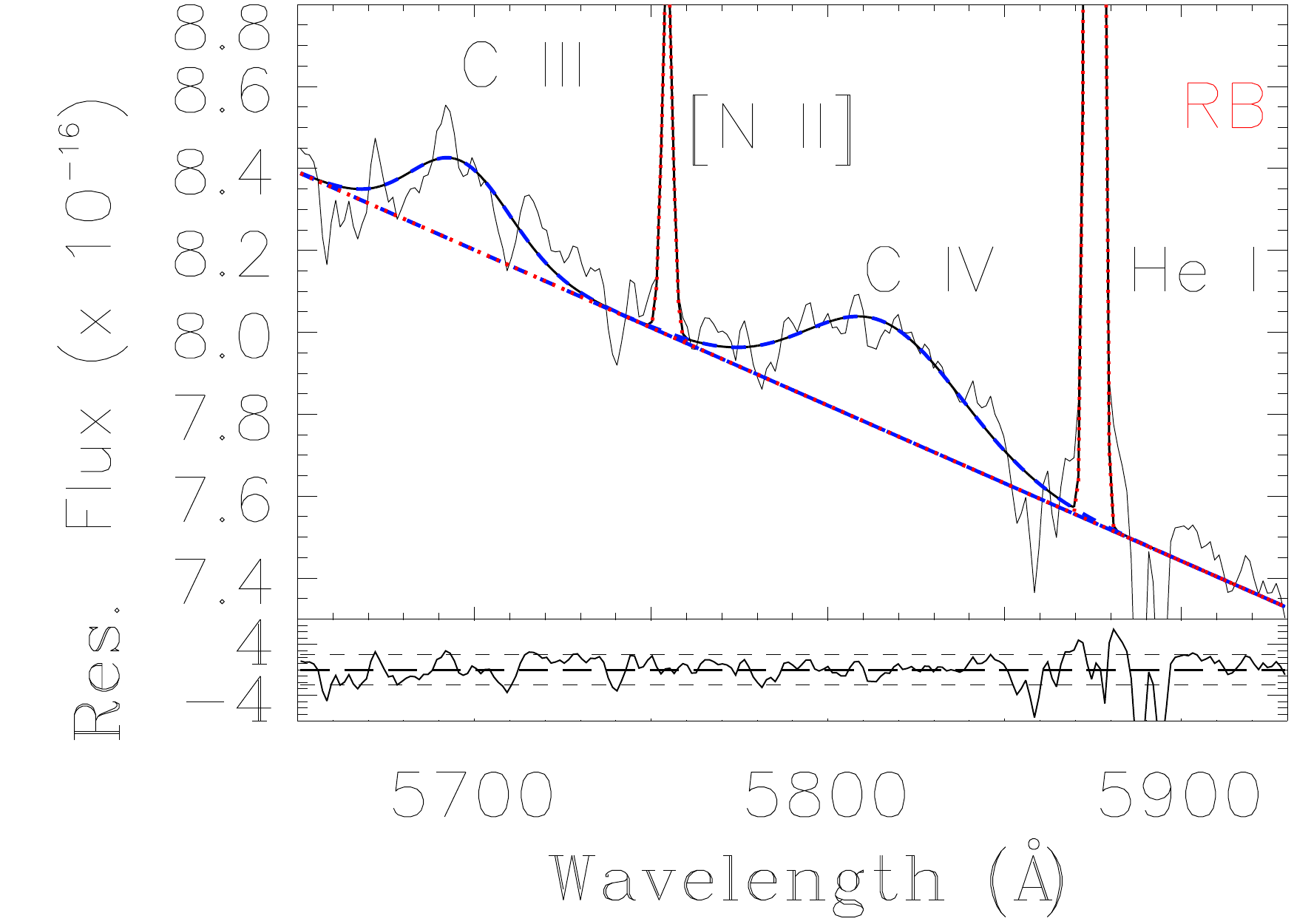}\\
\includegraphics[width=0.245\linewidth]{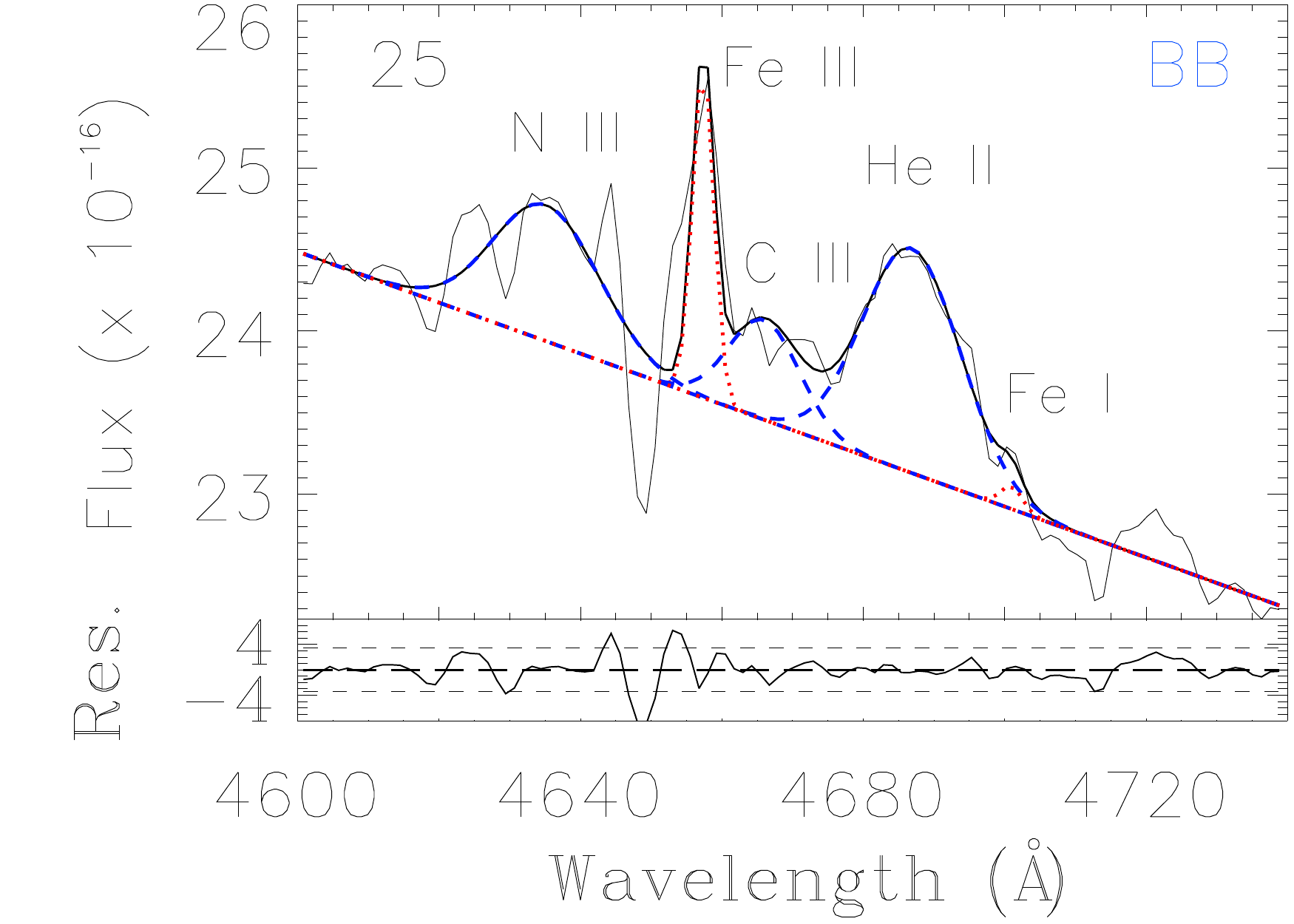}~
\includegraphics[width=0.245\linewidth]{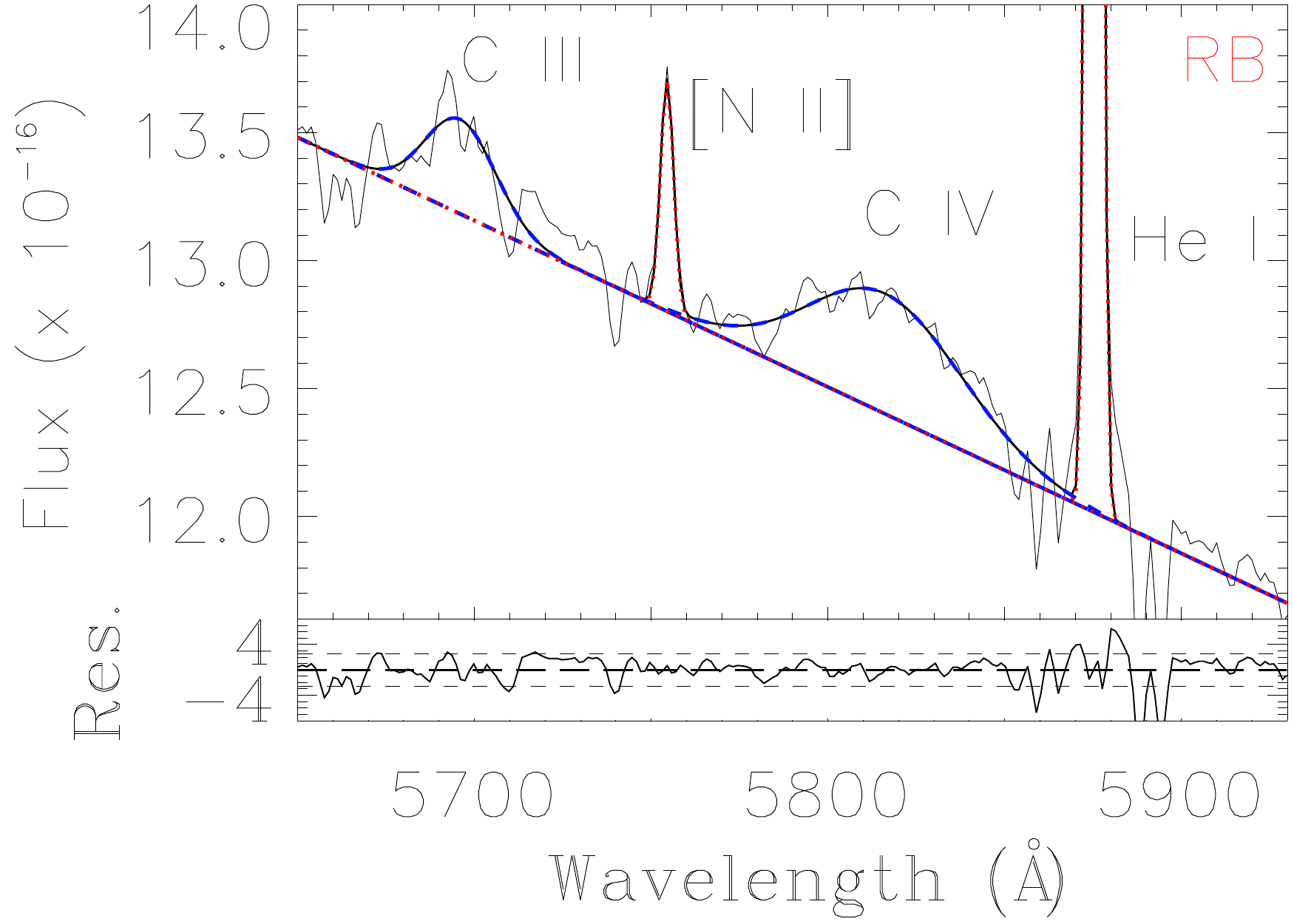}~
\includegraphics[width=0.245\linewidth]{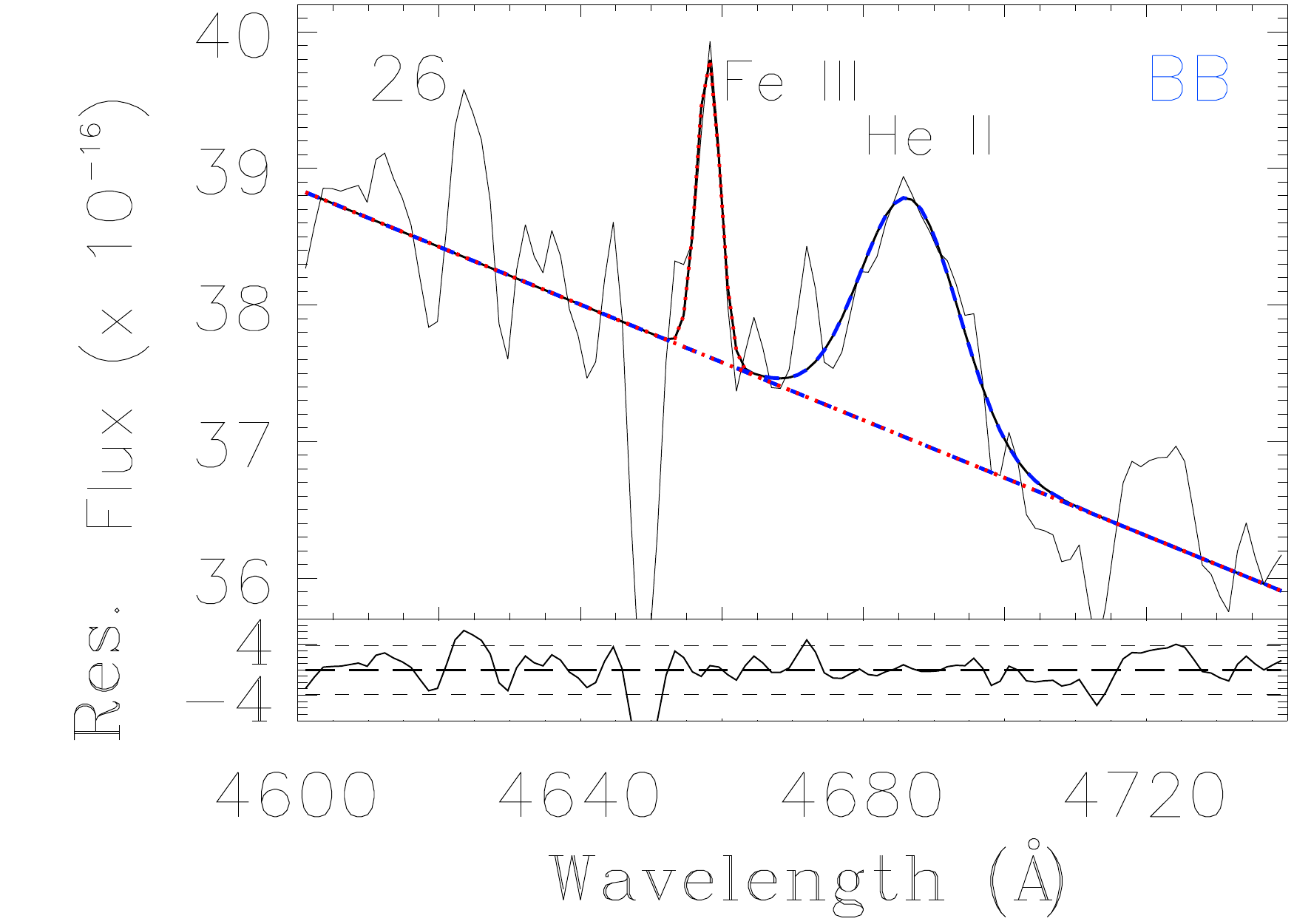}~
\includegraphics[width=0.245\linewidth]{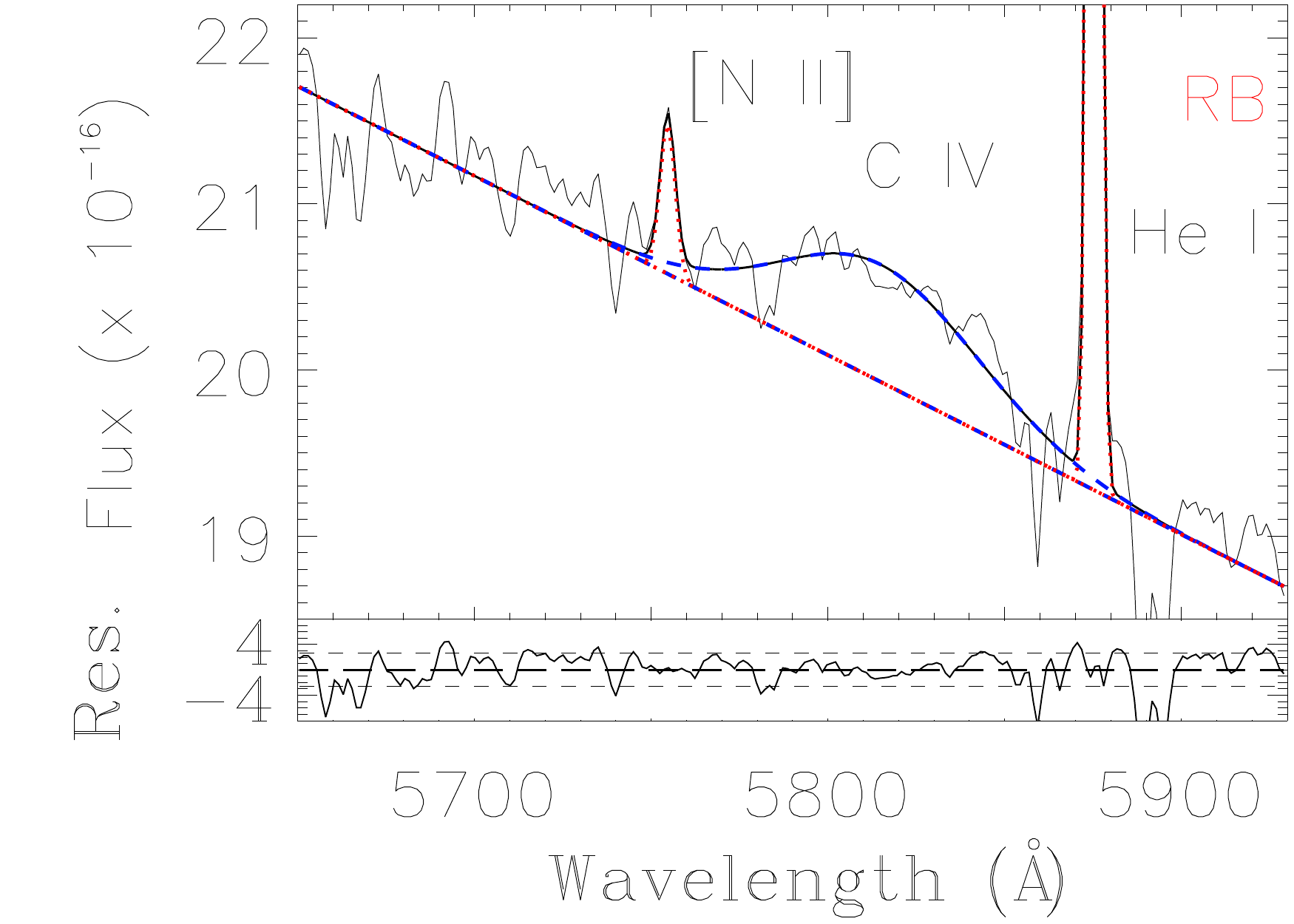}\\
\includegraphics[width=0.245\linewidth]{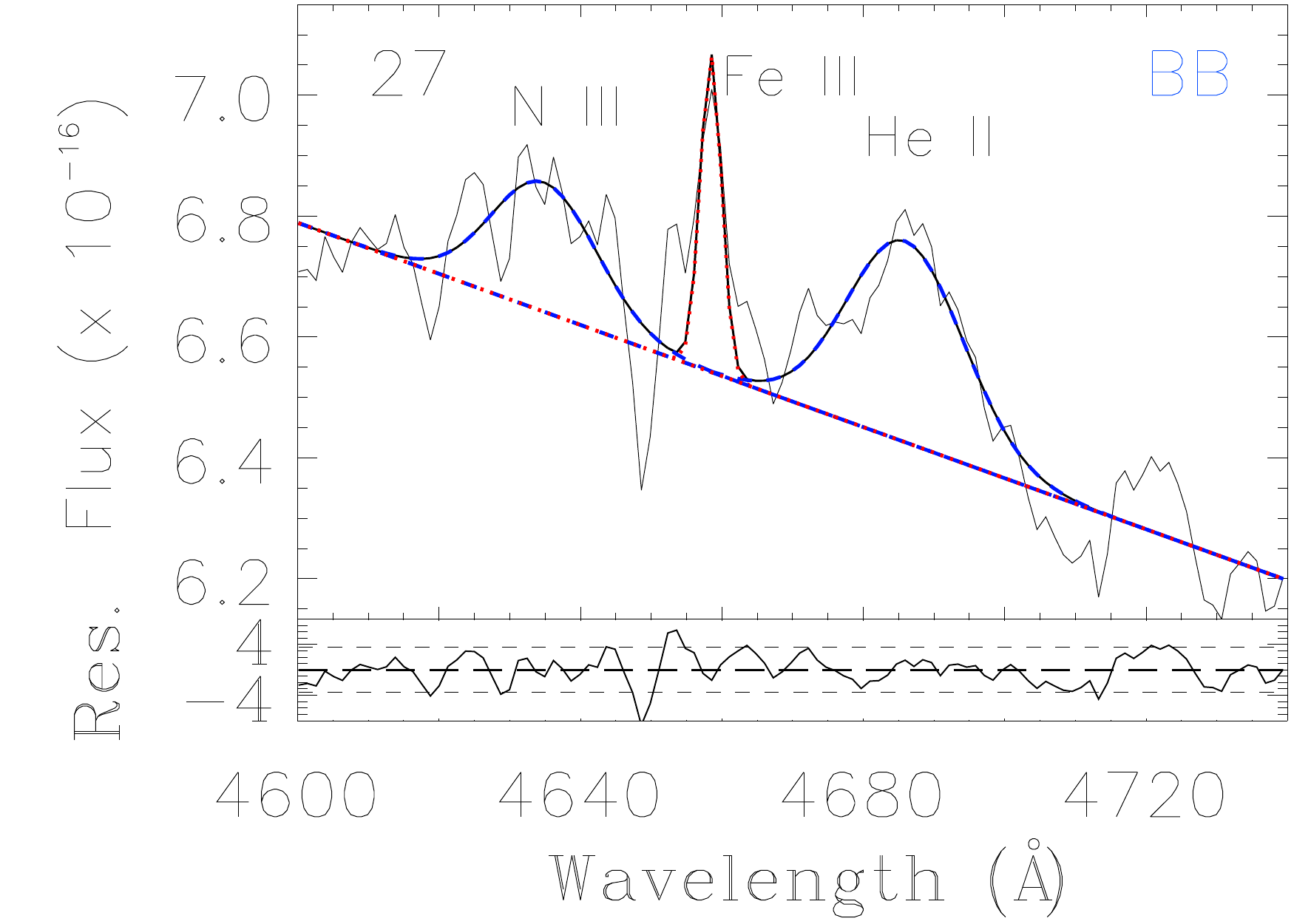}~
\includegraphics[width=0.245\linewidth]{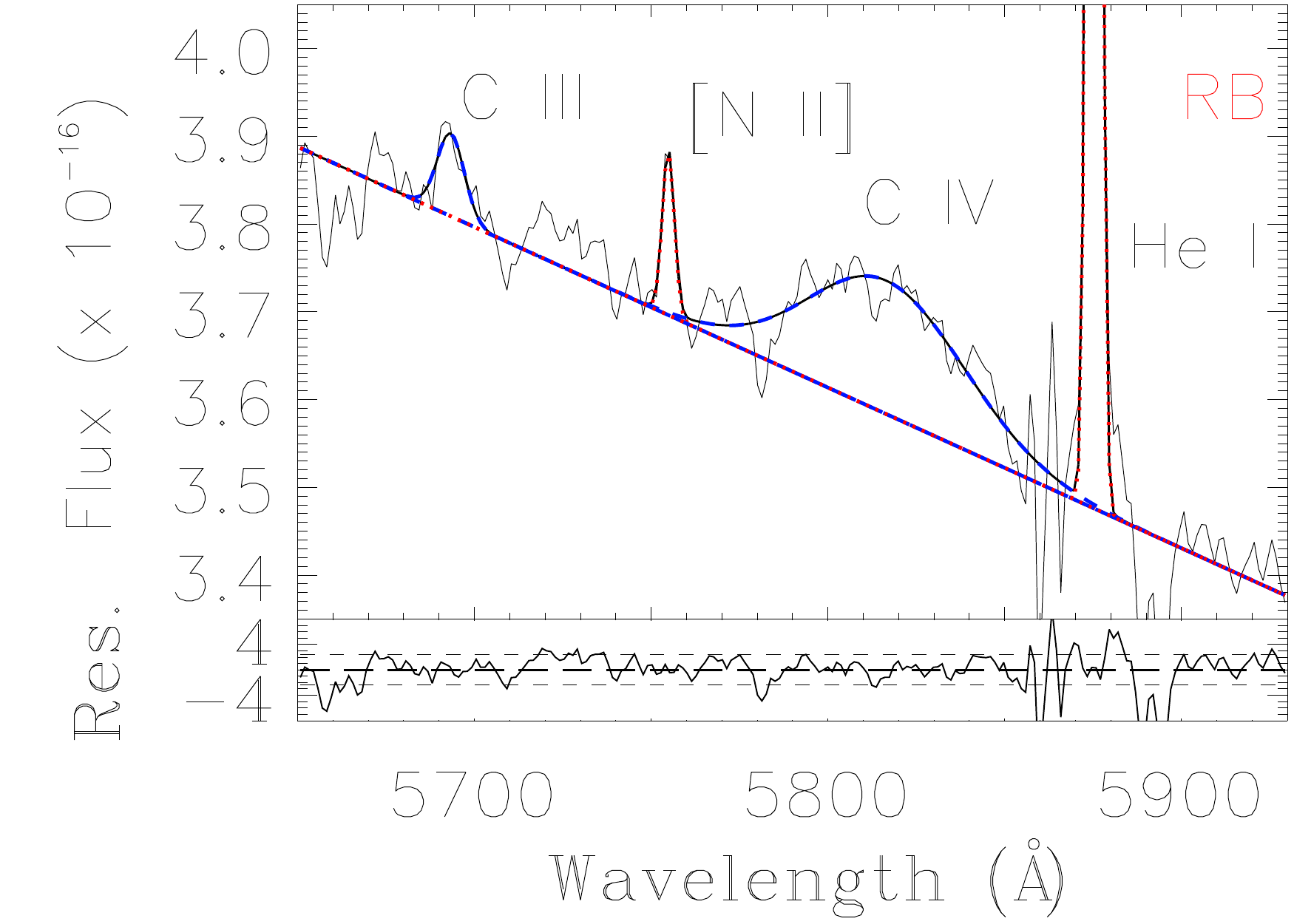}~
\includegraphics[width=0.245\linewidth]{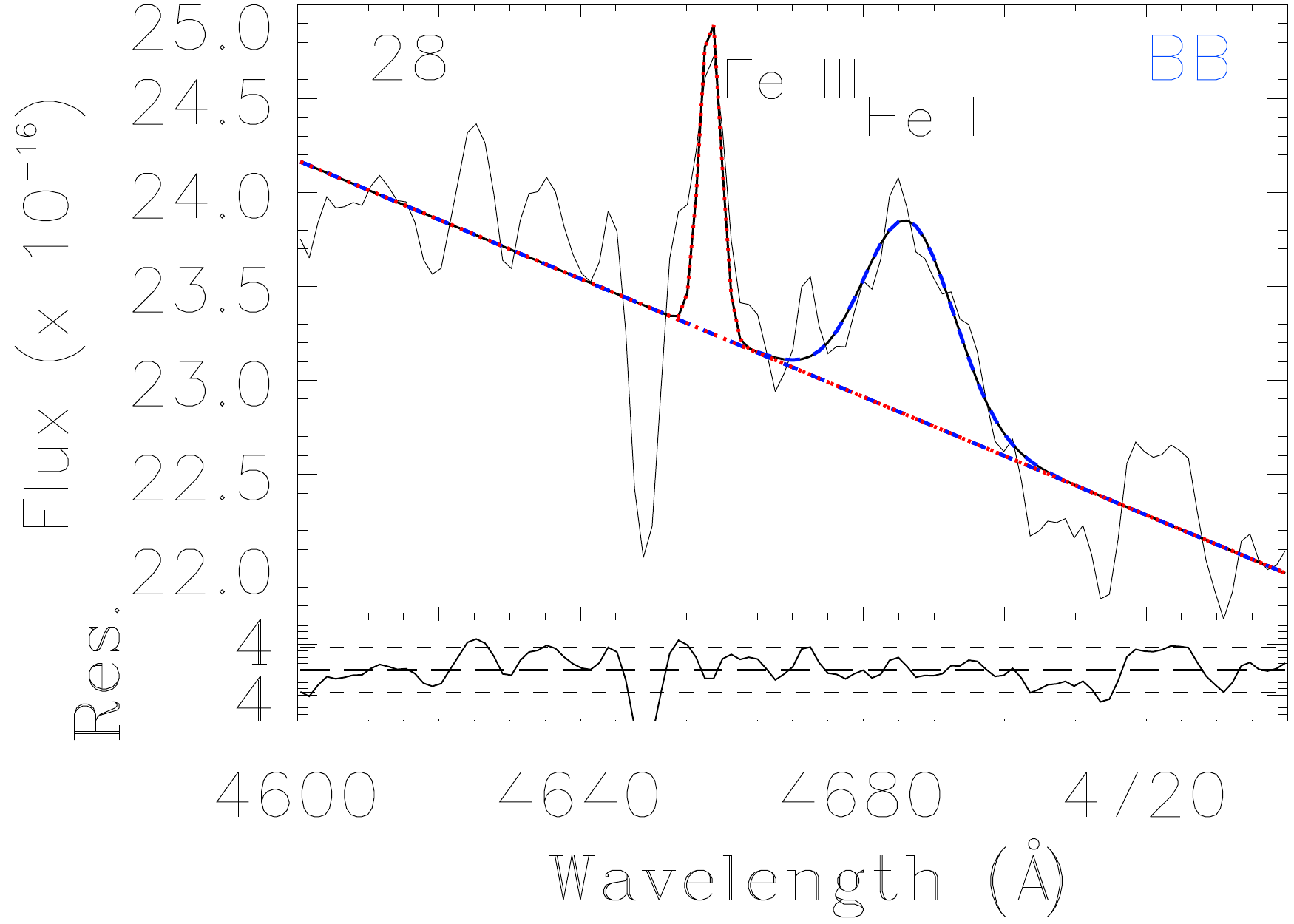}~
\includegraphics[width=0.245\linewidth]{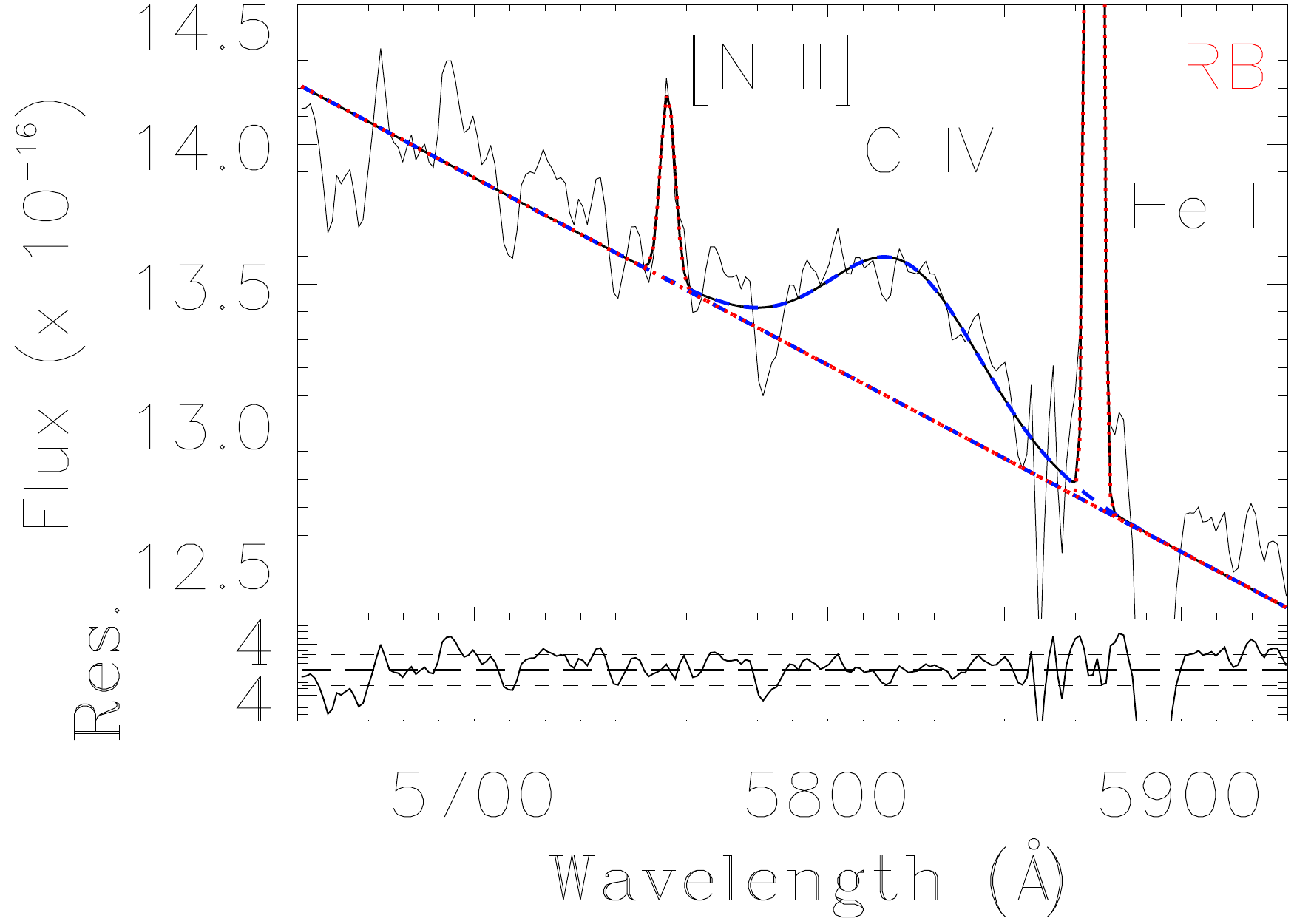}
\ContinuedFloat
\captionsetup{list=off,format=cont}
\caption{{\it -- continued}}
\end{center}
\end{figure*}

\begin{figure*}
\begin{center}
\includegraphics[width=0.245\linewidth]{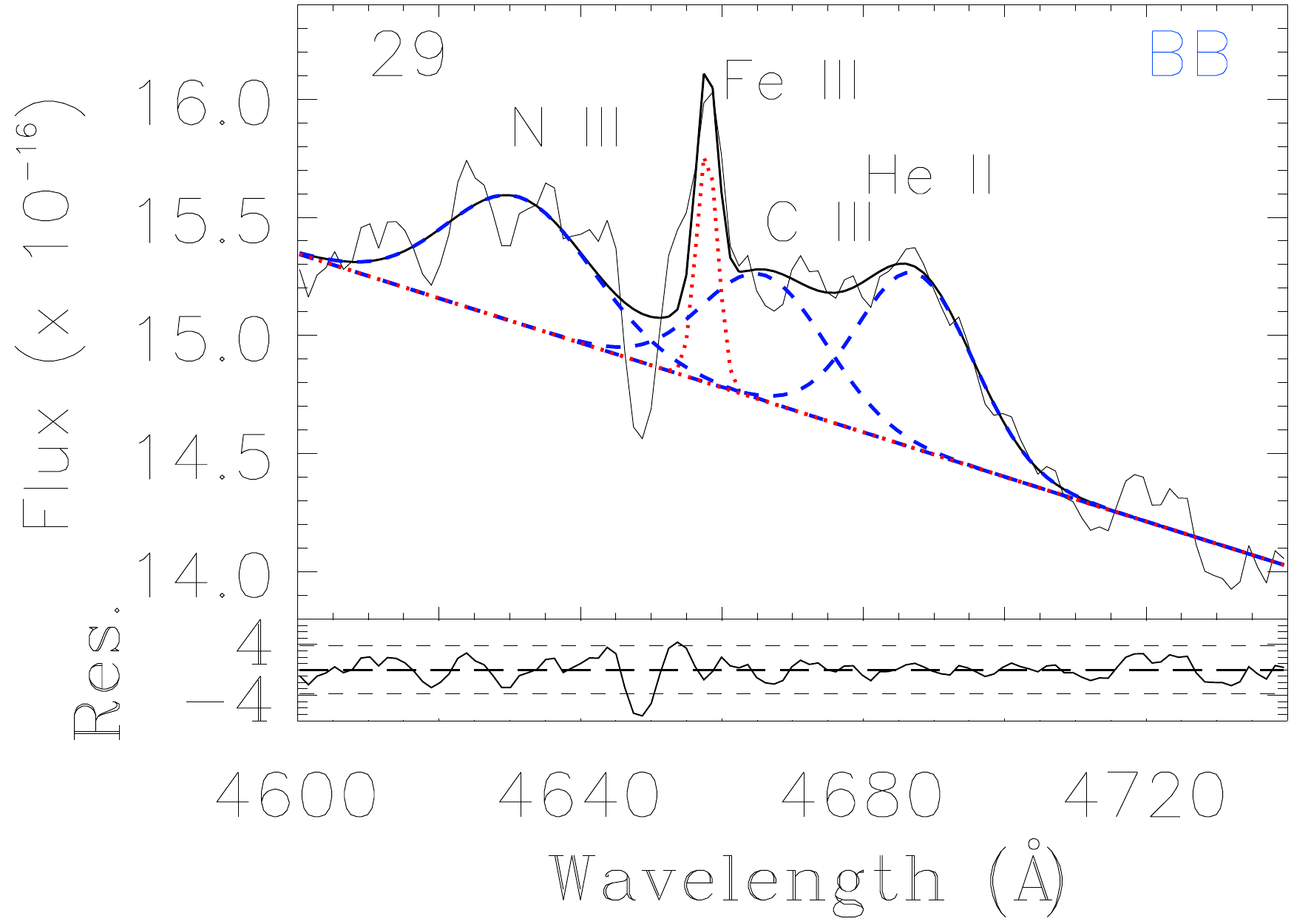}~
\includegraphics[width=0.245\linewidth]{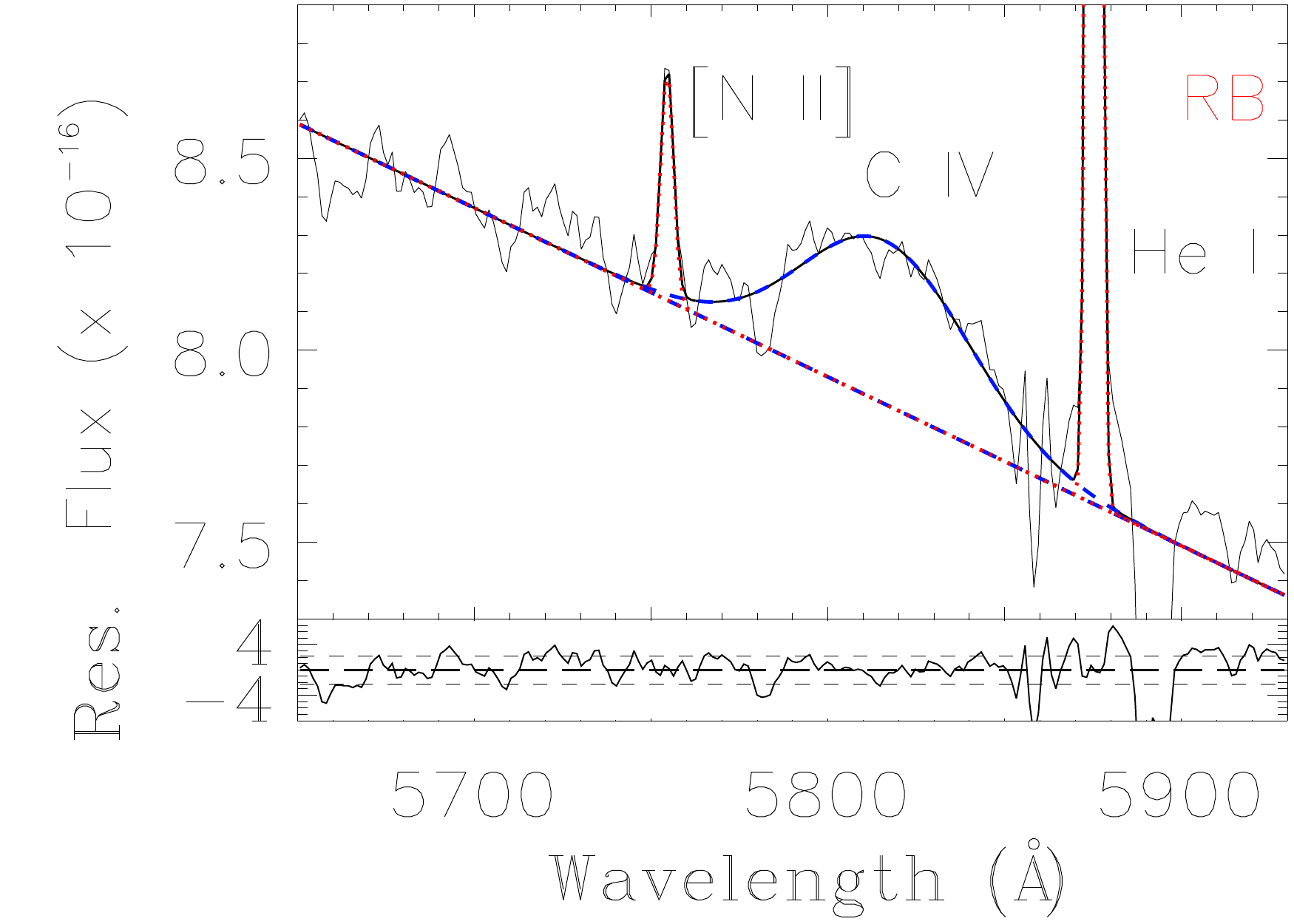}~
\includegraphics[width=0.245\linewidth]{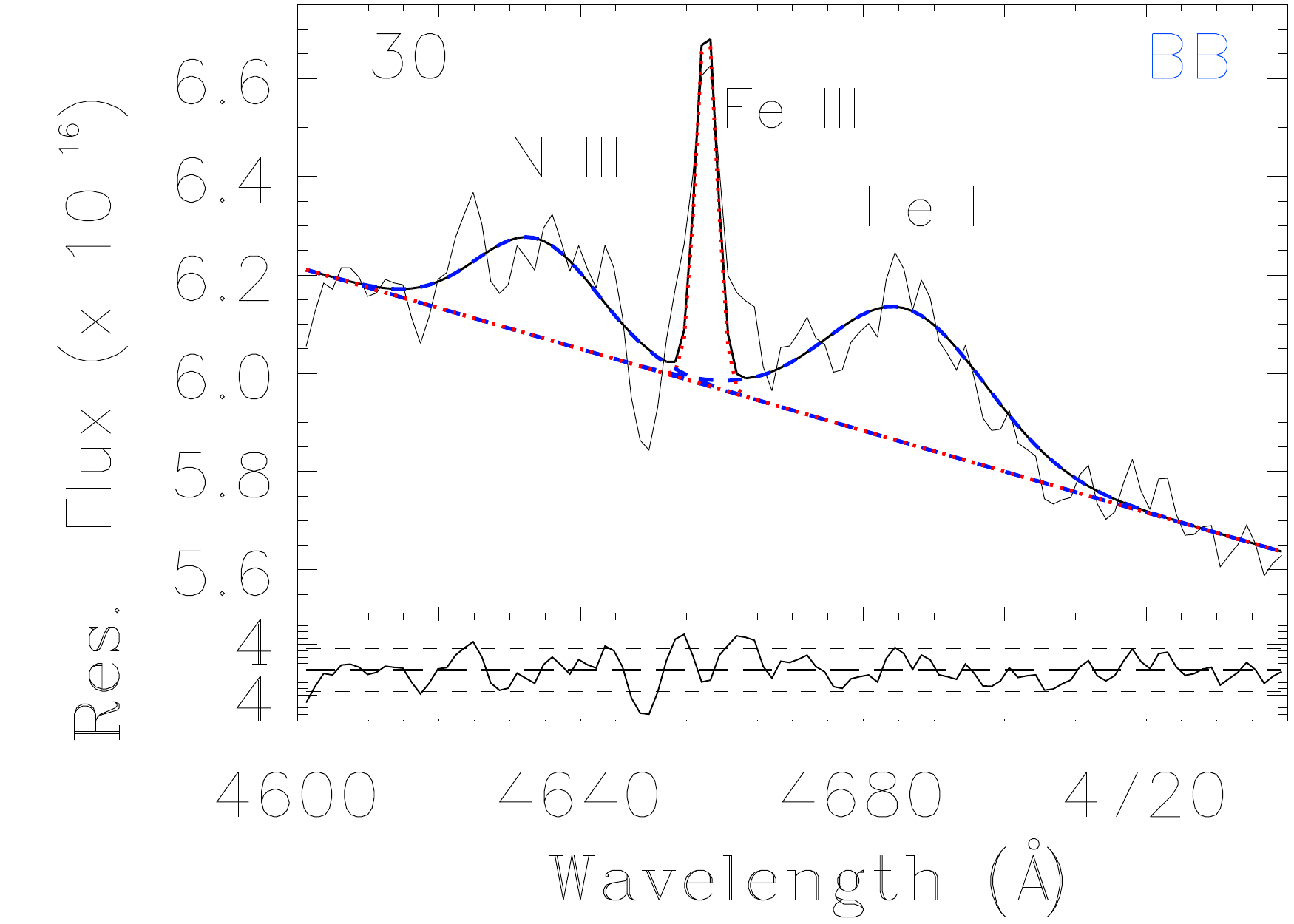}~
\includegraphics[width=0.245\linewidth]{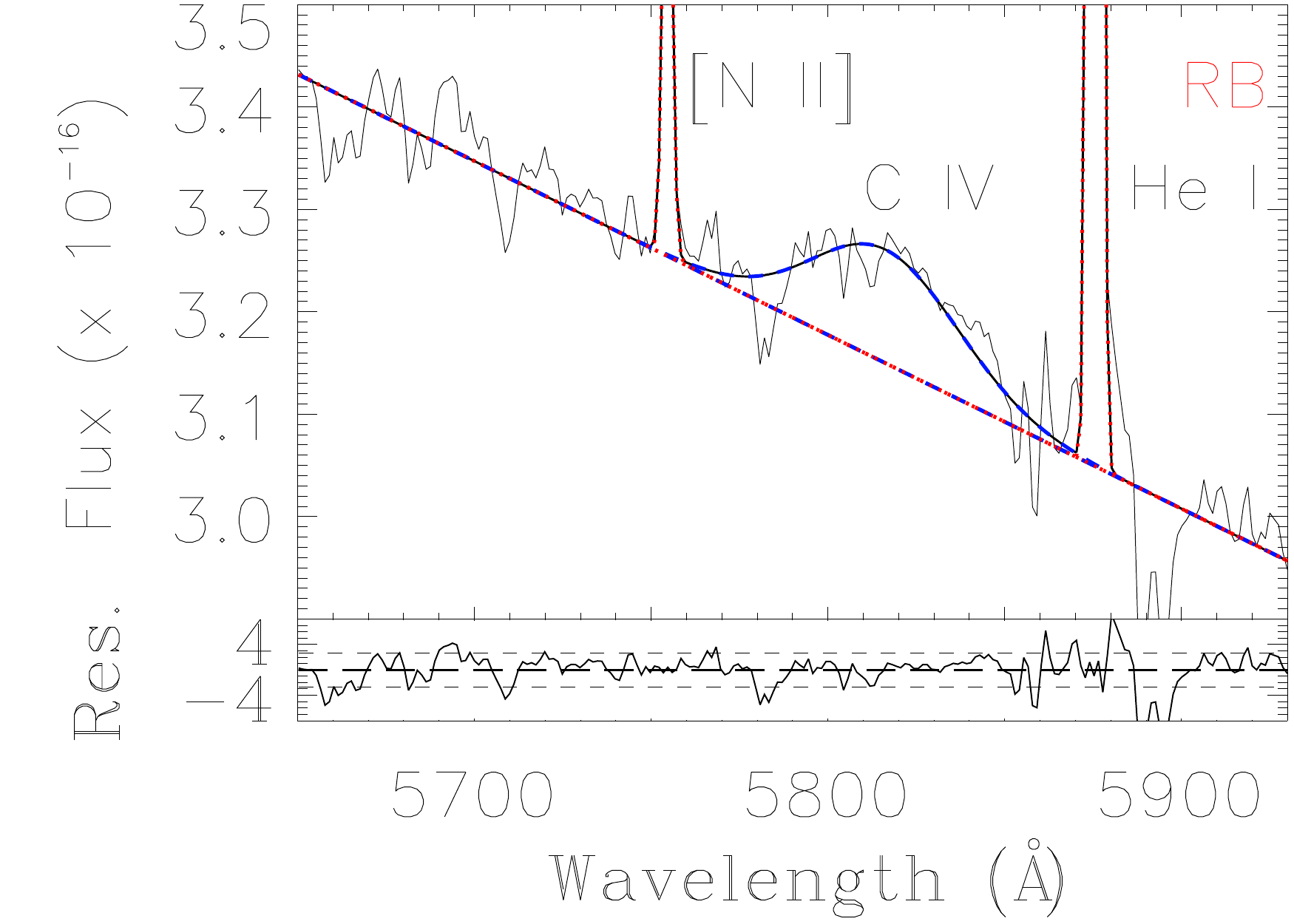}\\
\includegraphics[width=0.245\linewidth]{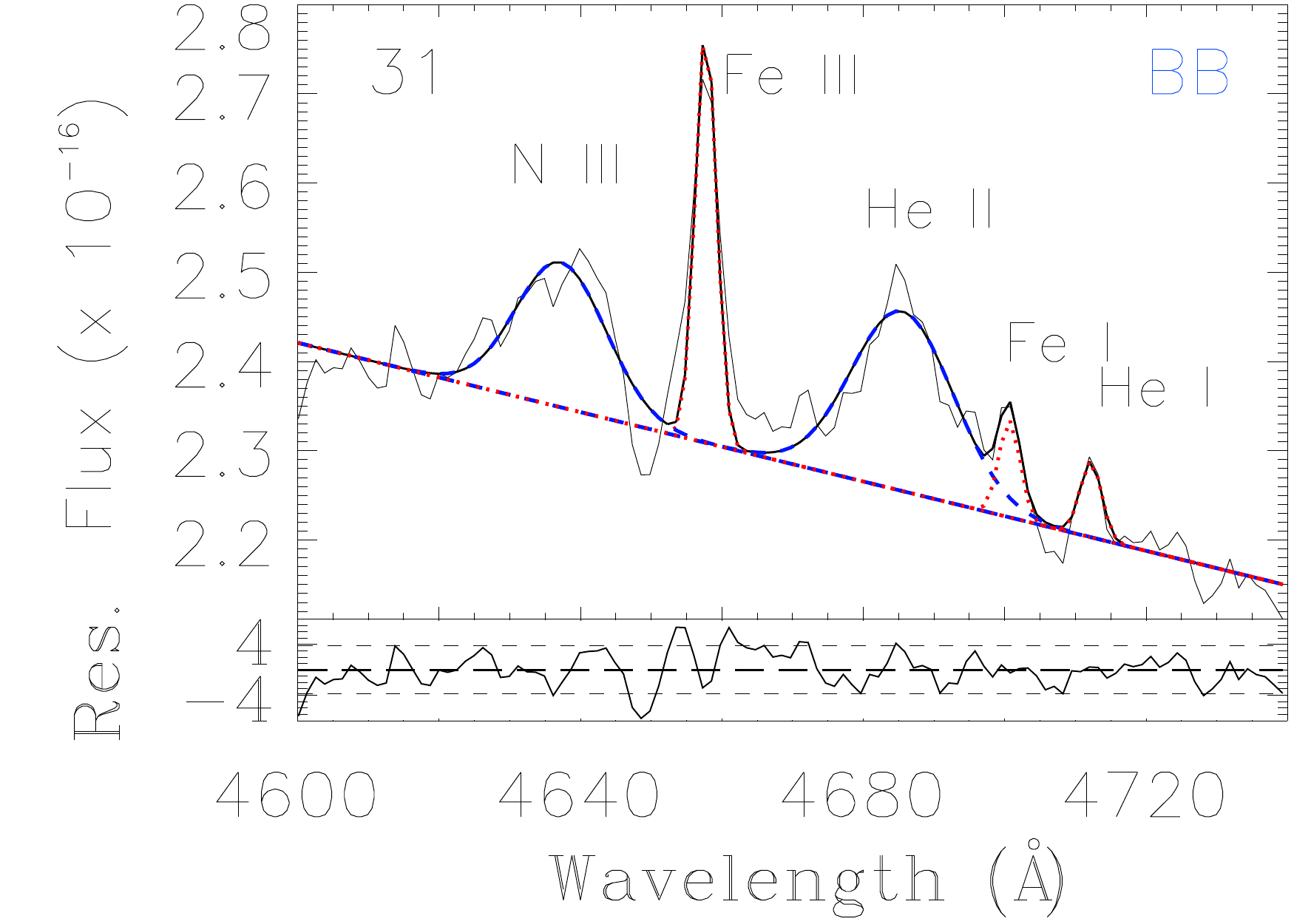}~
\includegraphics[width=0.245\linewidth]{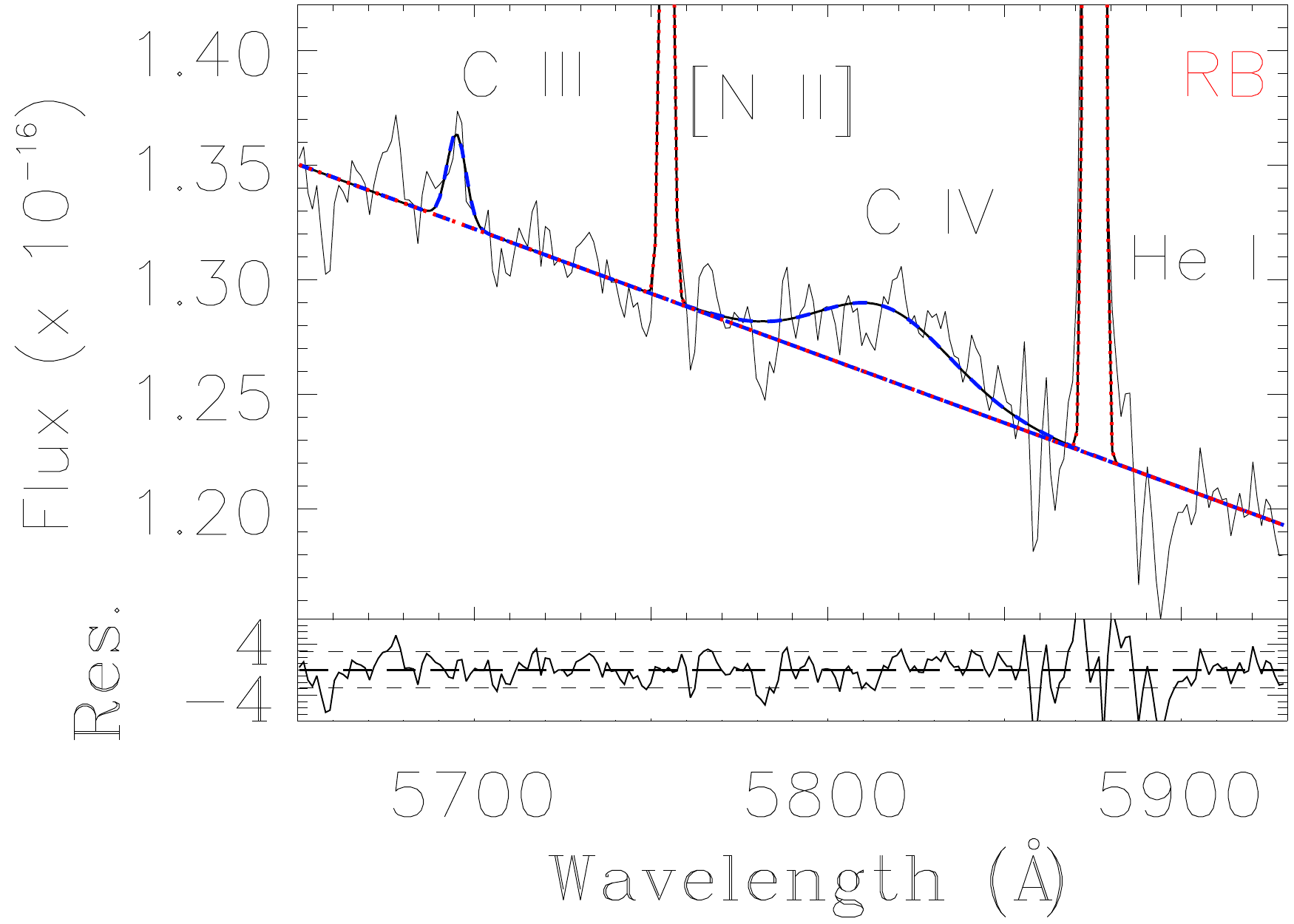}~
\includegraphics[width=0.245\linewidth]{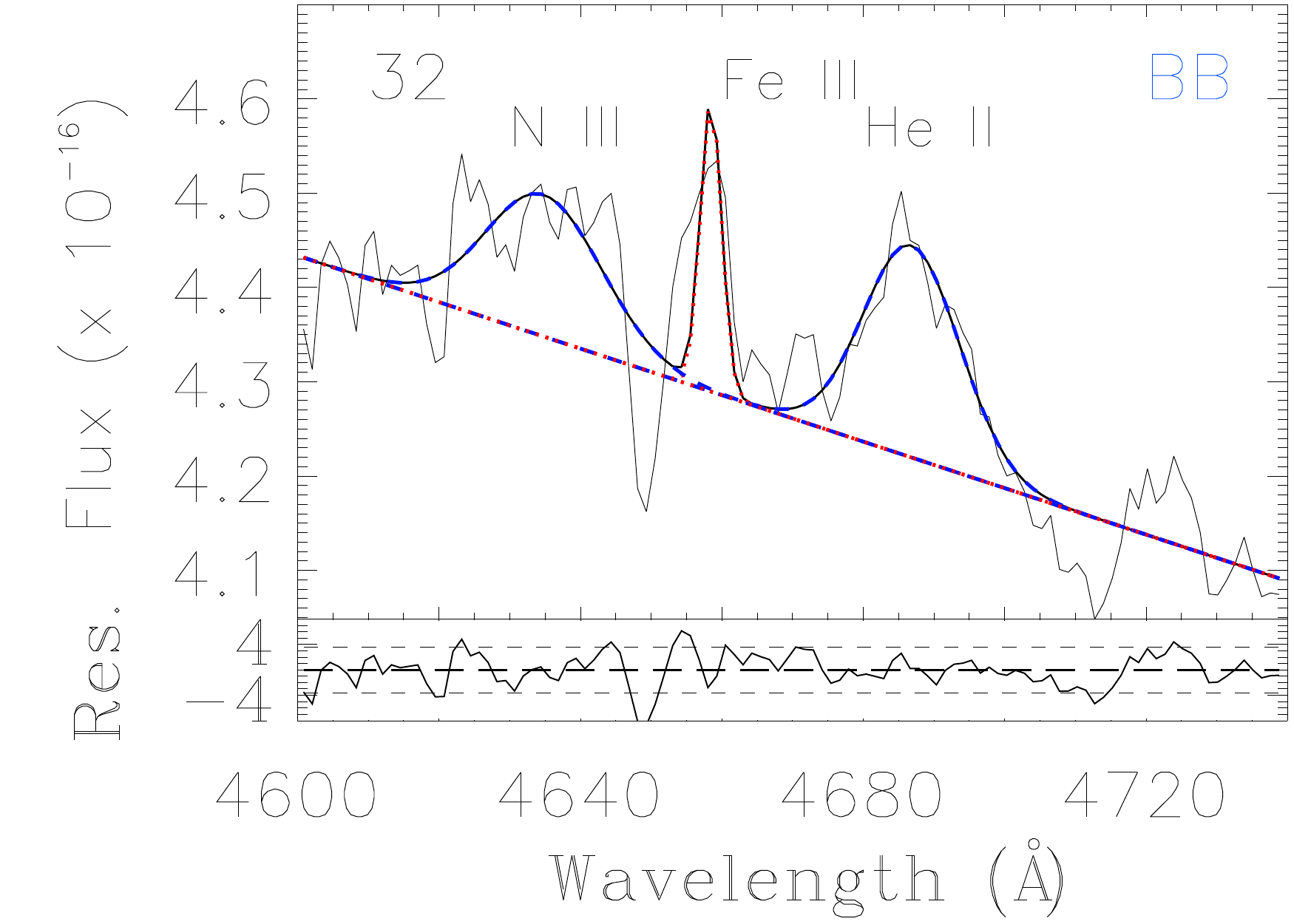}~
\includegraphics[width=0.245\linewidth]{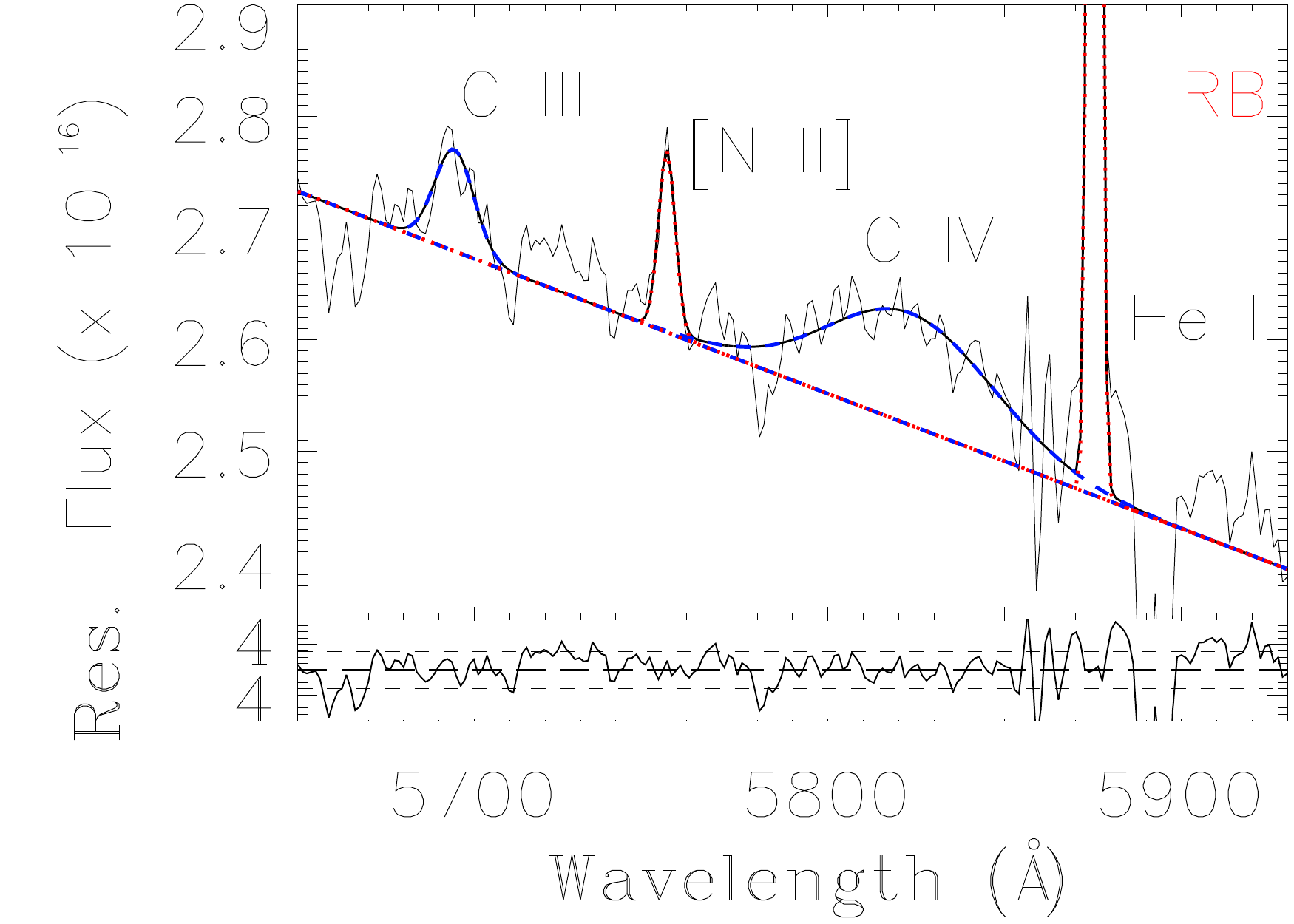}\\
\includegraphics[width=0.245\linewidth]{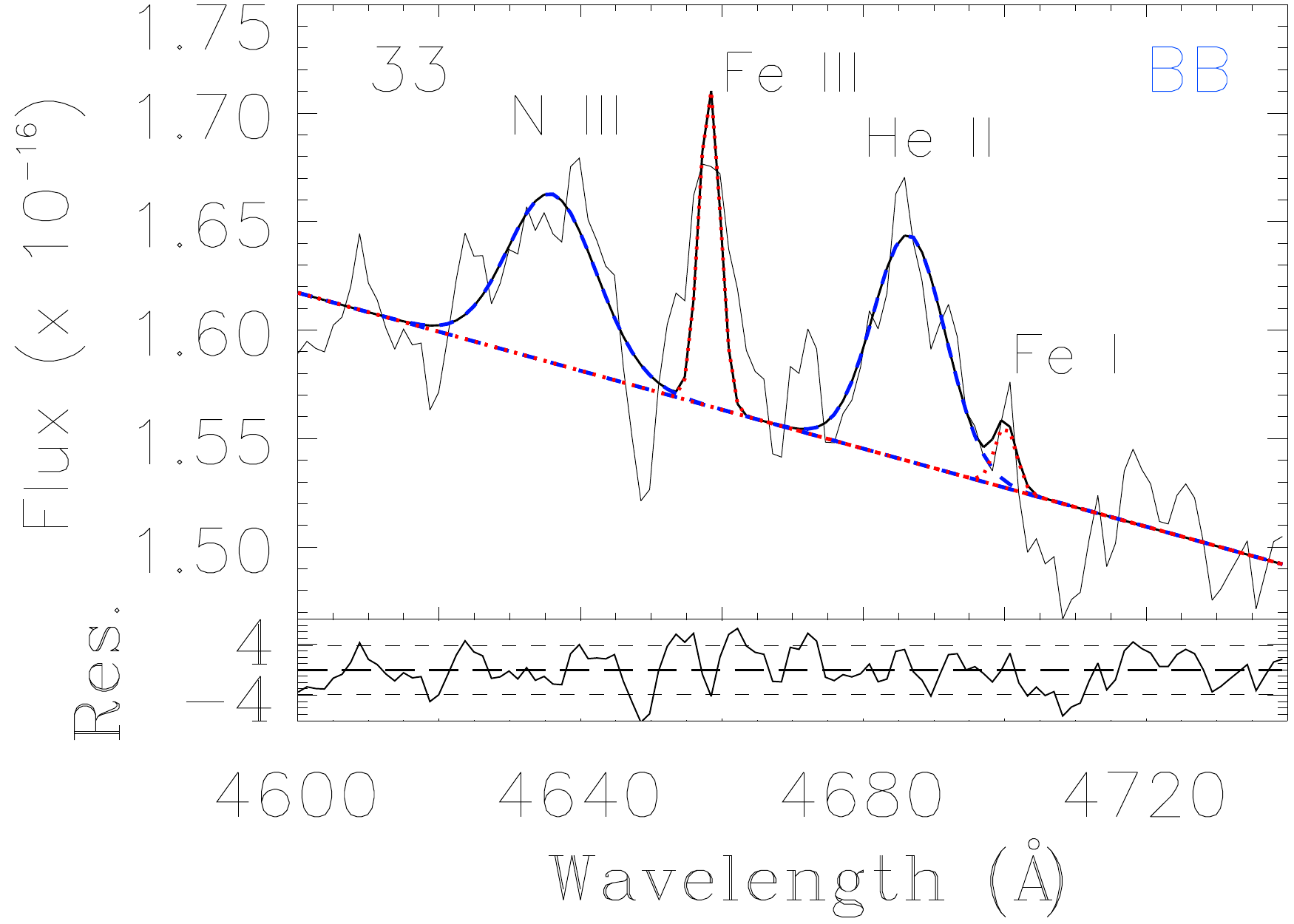}~
\includegraphics[width=0.245\linewidth]{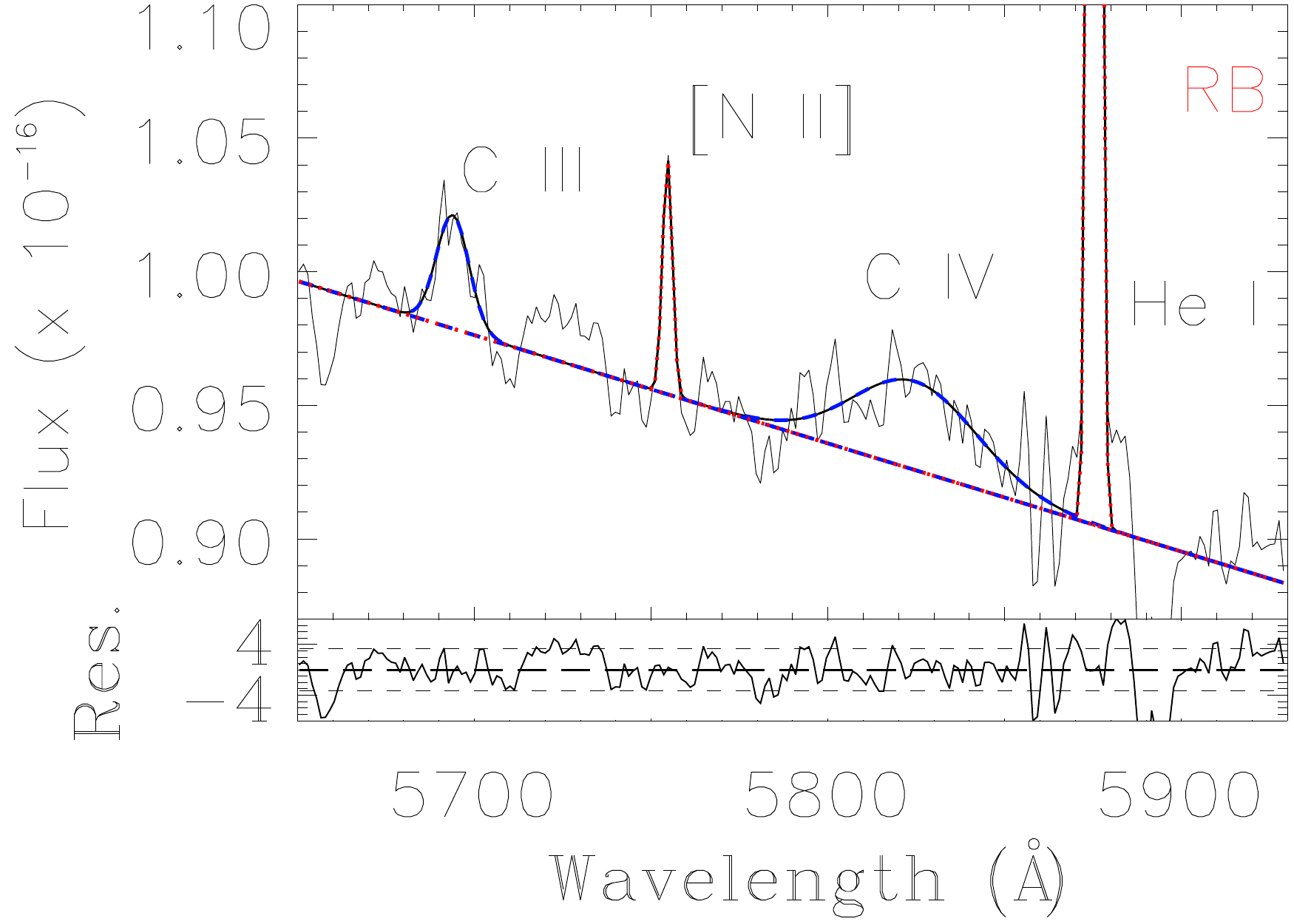}~
\includegraphics[width=0.245\linewidth]{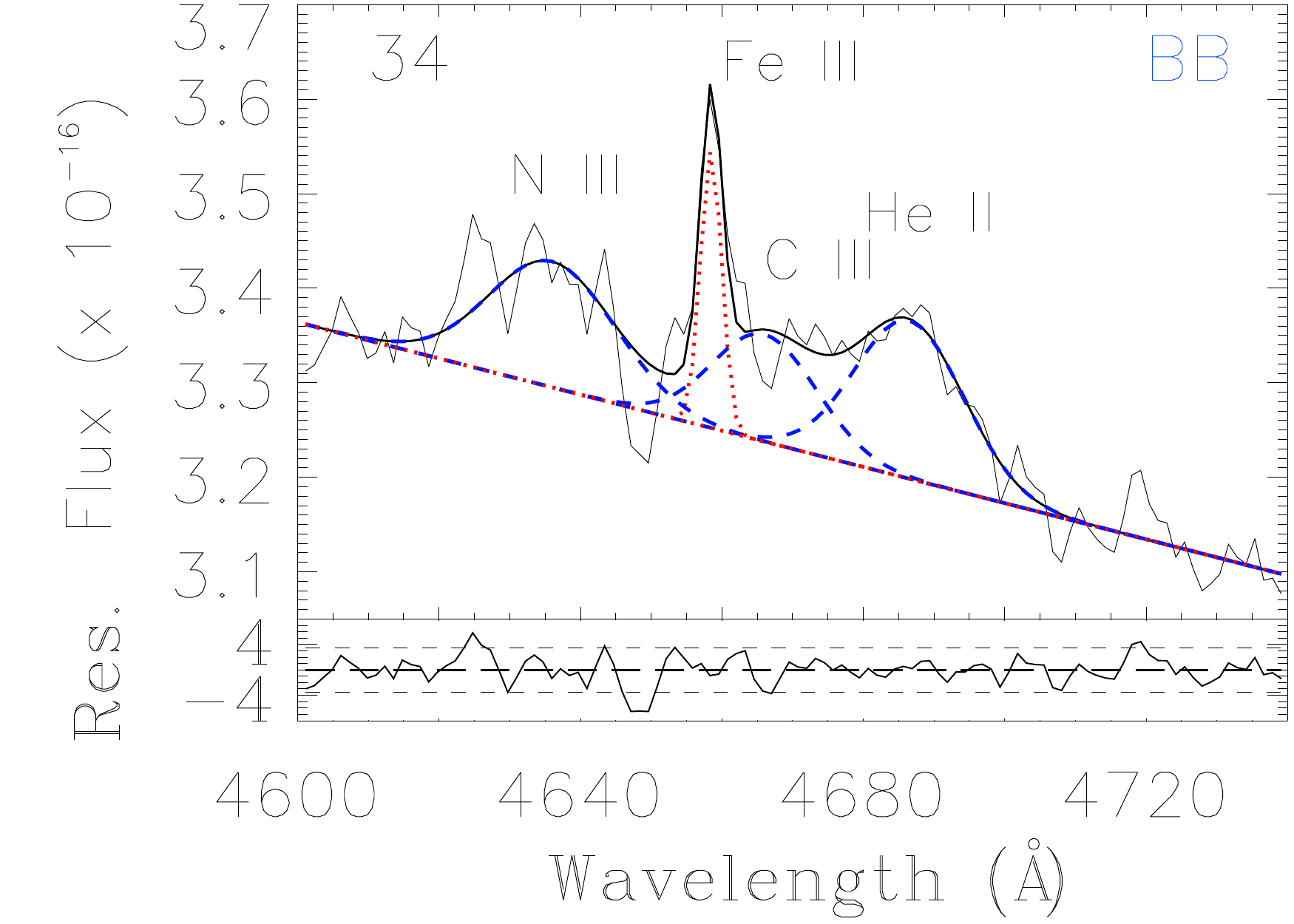}~
\includegraphics[width=0.245\linewidth]{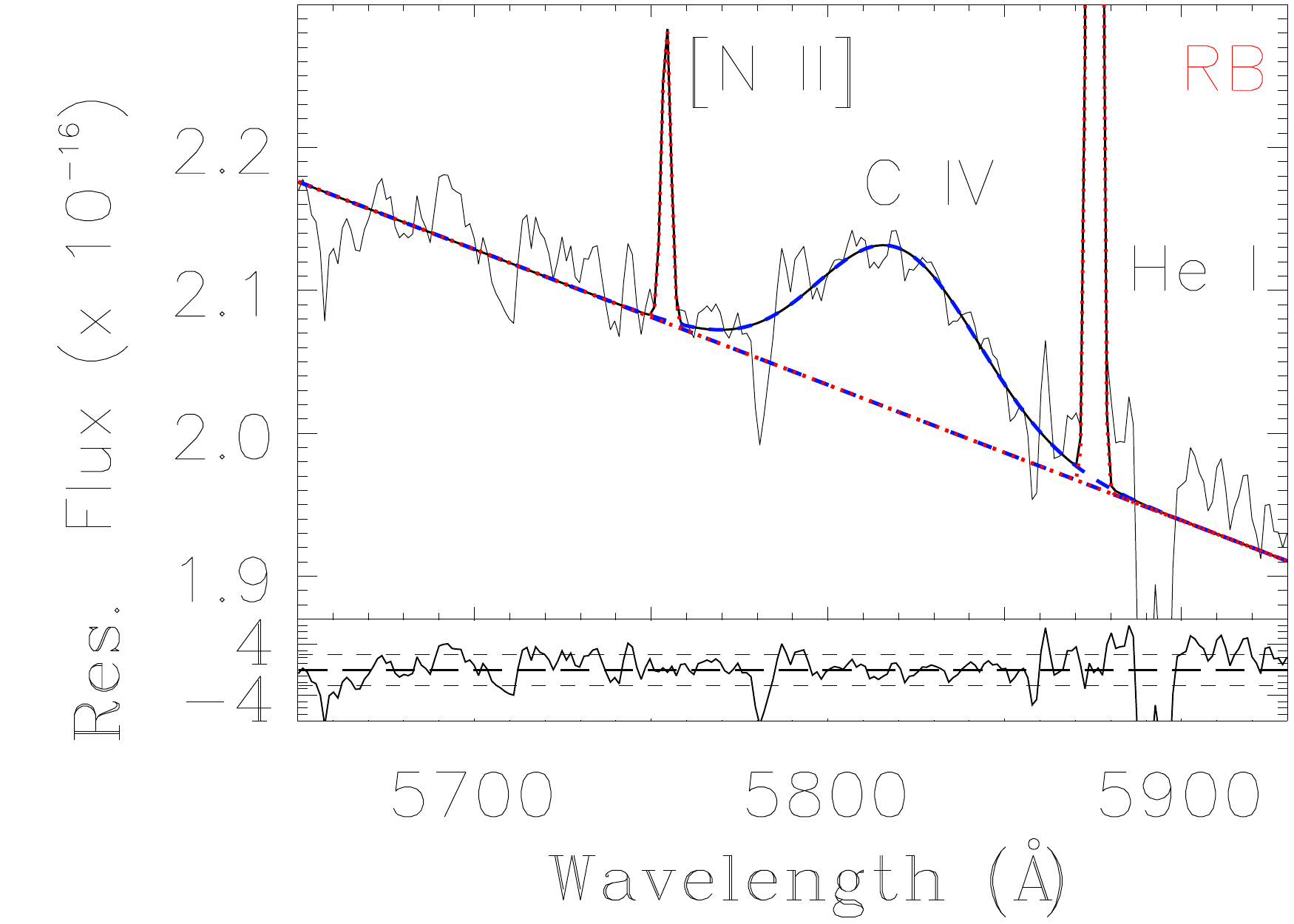}\\
\includegraphics[width=0.245\linewidth]{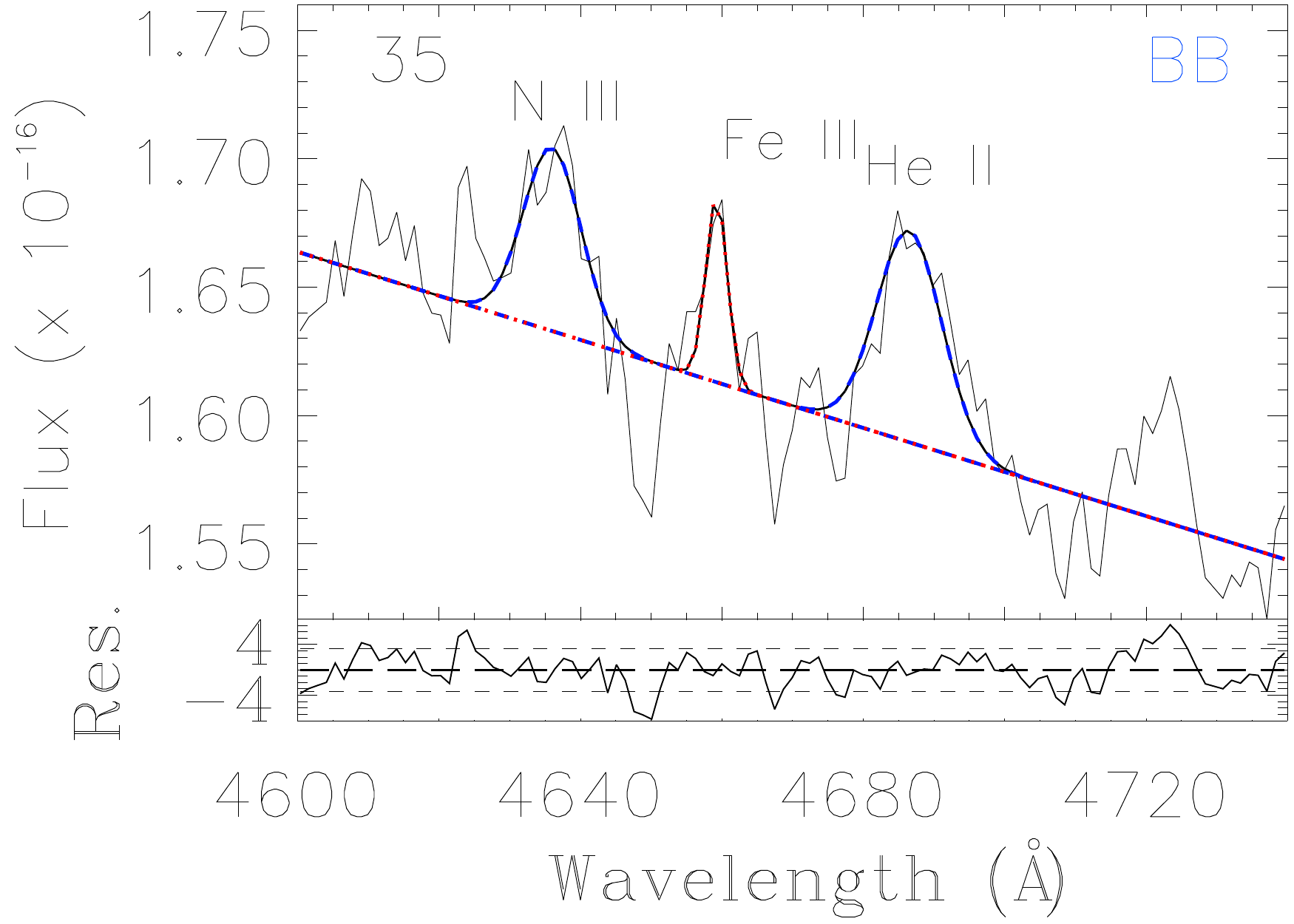}~
\includegraphics[width=0.245\linewidth]{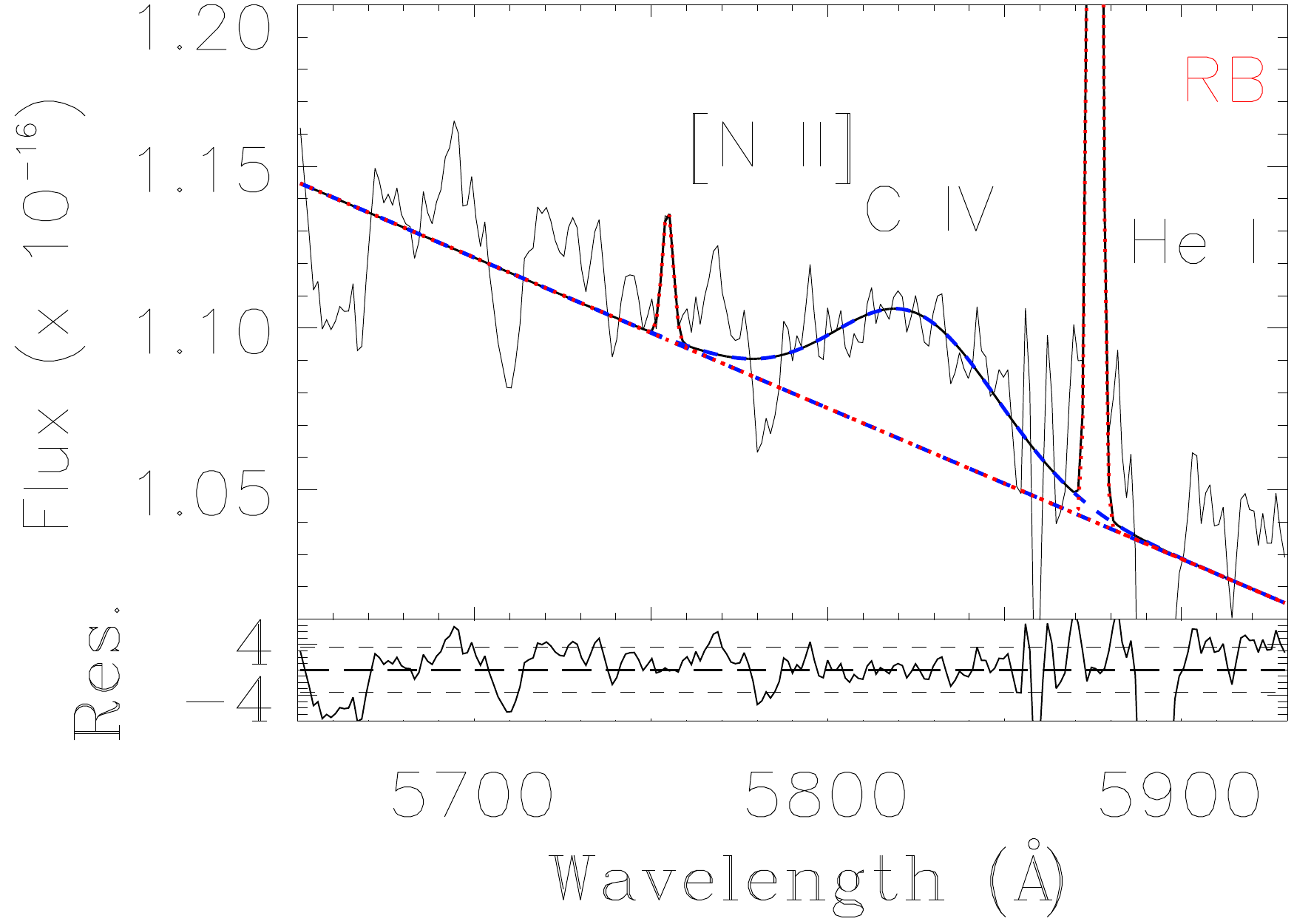}~
\includegraphics[width=0.245\linewidth]{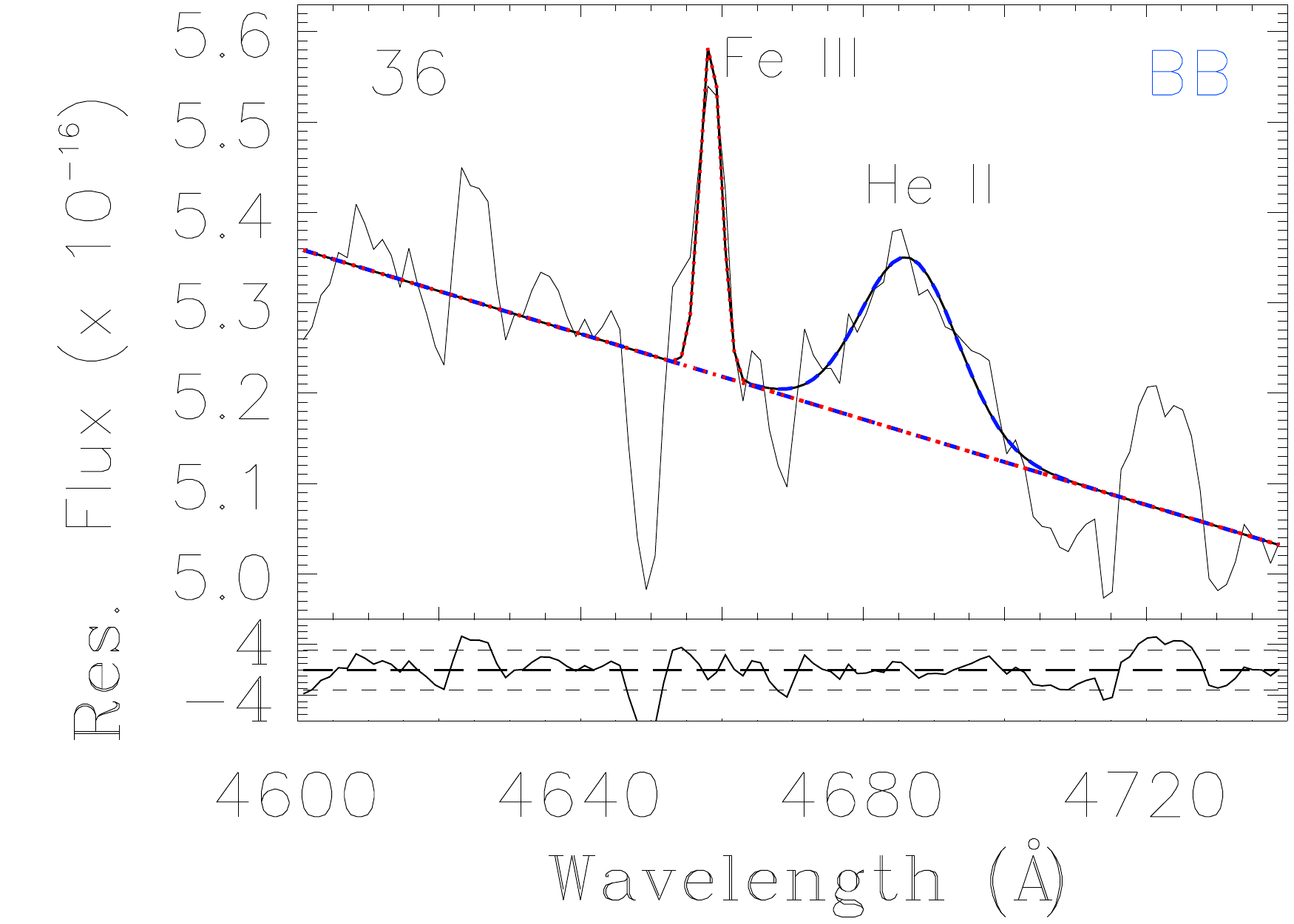}~
\includegraphics[width=0.245\linewidth]{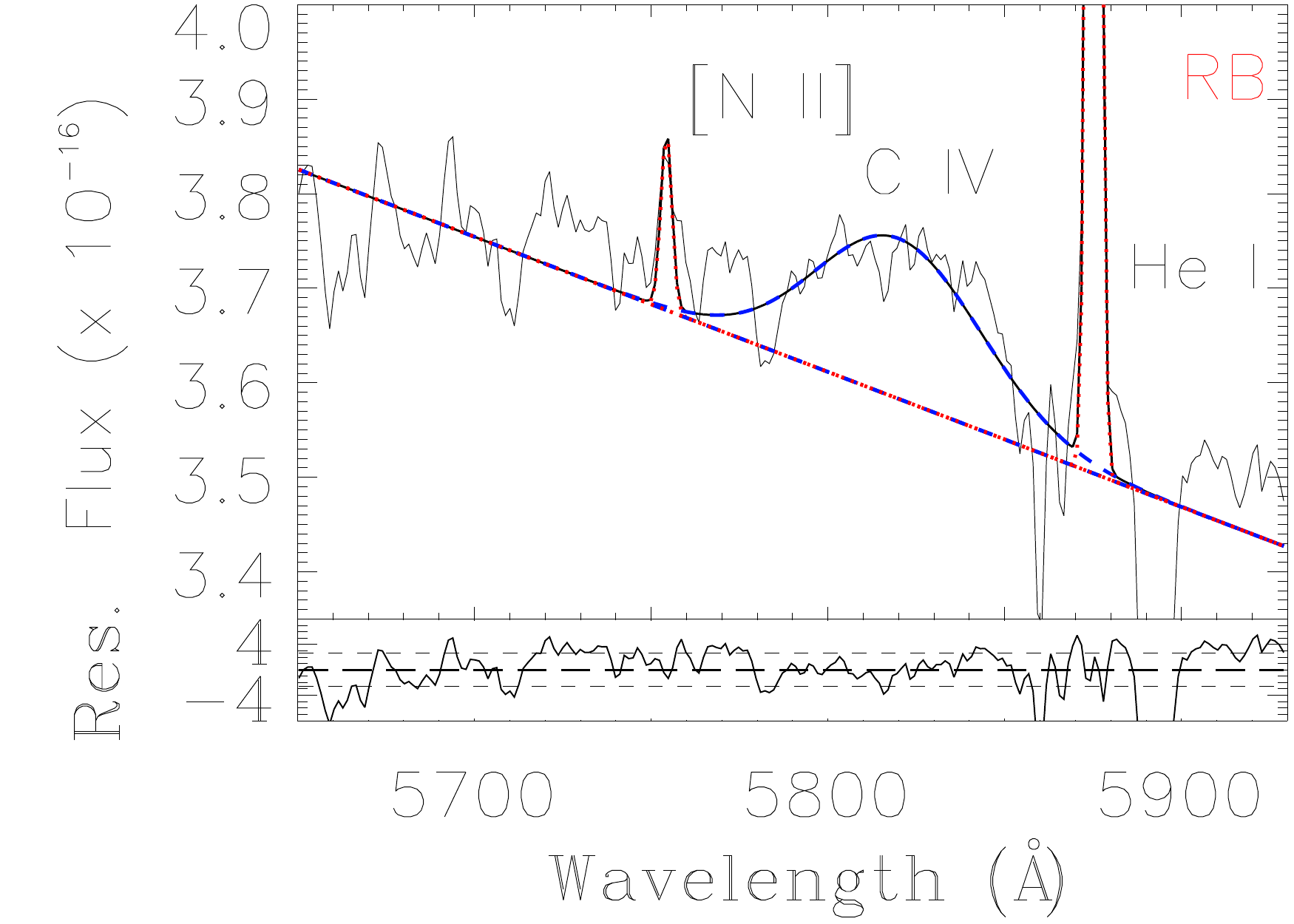}\\
\includegraphics[width=0.245\linewidth]{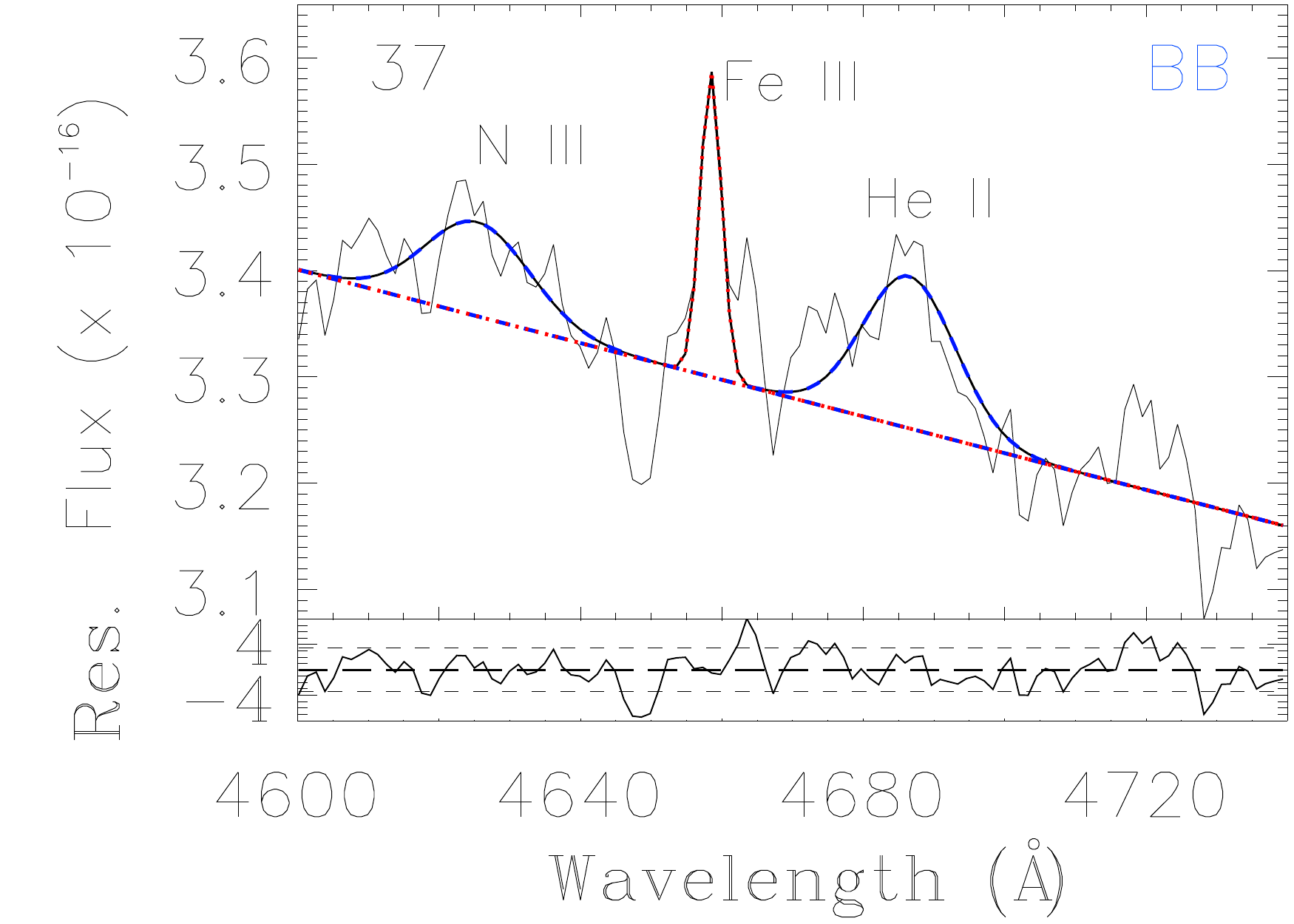}~
\includegraphics[width=0.245\linewidth]{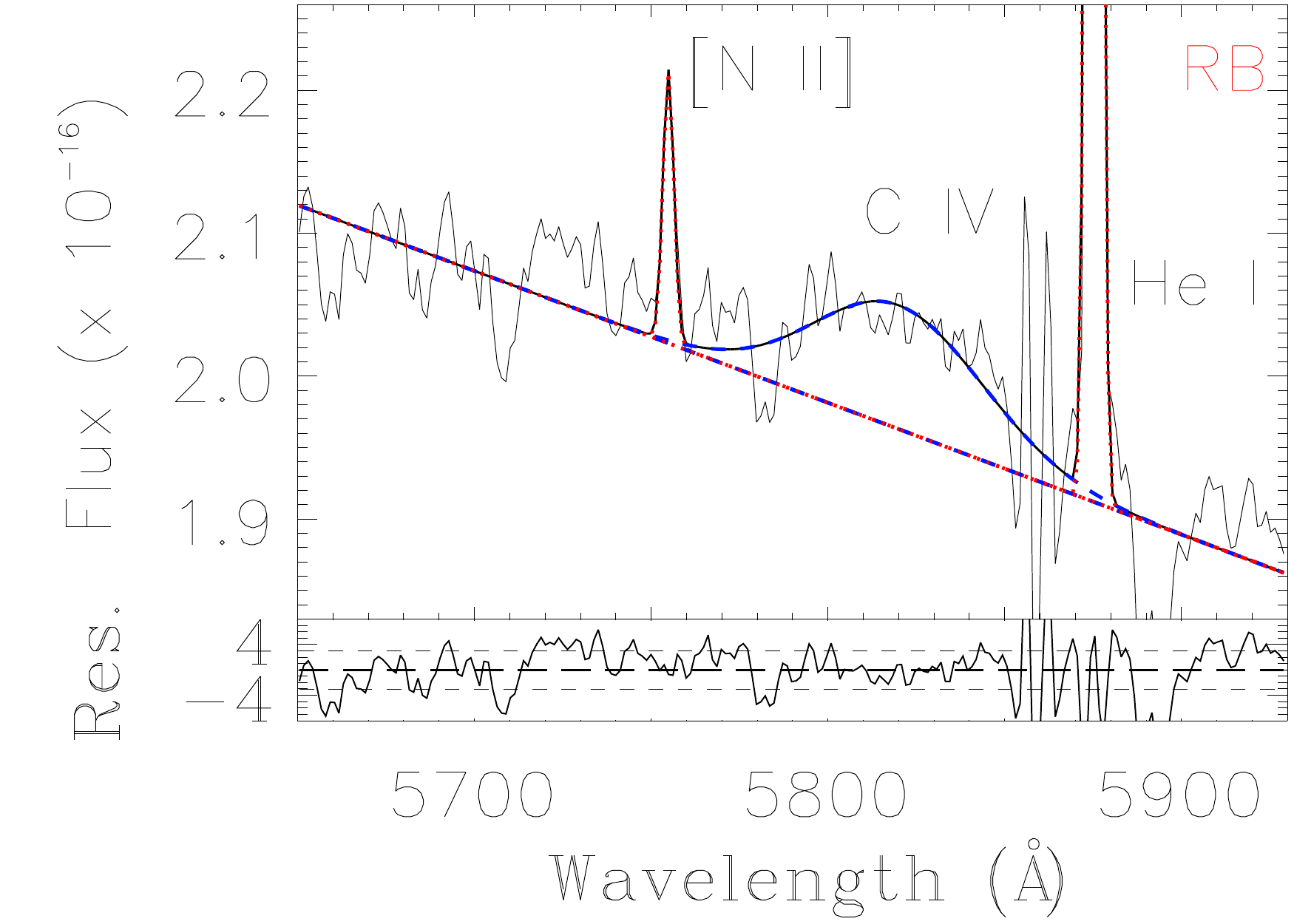}~
\includegraphics[width=0.245\linewidth]{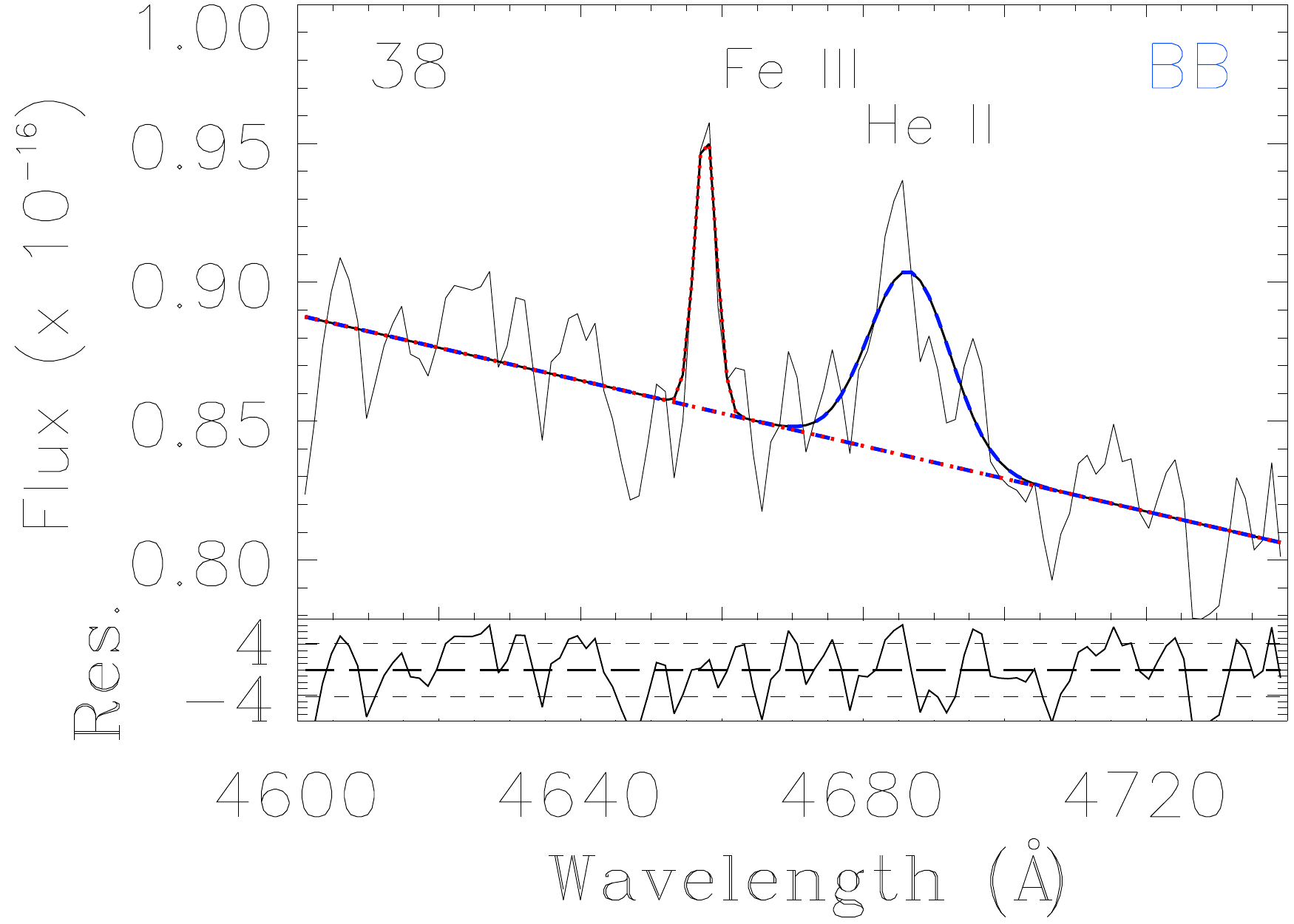}~
\includegraphics[width=0.245\linewidth]{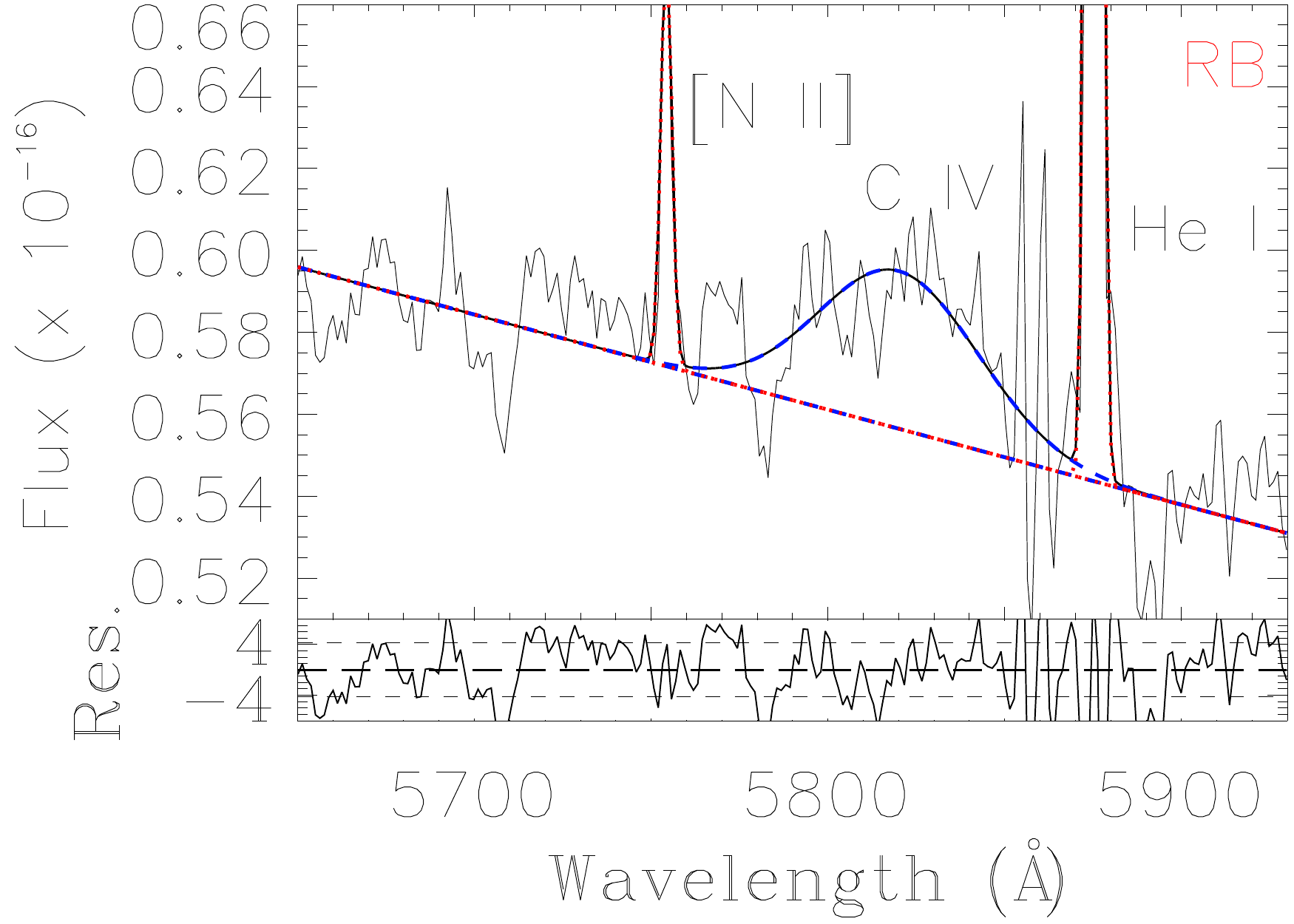}
\ContinuedFloat
\captionsetup{list=off,format=cont}
\caption{{\it -- continued}}
\end{center}
\end{figure*}

The multi-Gaussian fitting
technique help us dissect the presence and flux
from the contributing lines of the WR bumps 
\citep[e.g.,][]{2008Brinchmann,2010Lopez,2014Miralles,2017Monreal,2020Gomez}.
Furthermore, it helps us to assess the contribution from contaminant
nebular lines such as \heineb, \heiiwr, and \feiiia\, in the BB
\citep[see, e.g.,][]{2020Mayya} and \niineb\ and \civheir\ in the RB.

The BB and RB of the 38 WR spectra in the Antennae were decomposed into
their individual emission lines by following the multi-Gaussian
approach described in \cite{2020Gomez}.
This method consists of fitting the broad WR features with
multi-Gaussian components using a tailor-made code that
uses the {\sc idl} routine {\sc lmfit}\footnote{The {\sc lmfit}
function (lmfit.pro) performs a non-linear least squares fit
to a function with an arbitrary number of parameters.
It uses the Levenberg-Marquardt algorithm,
incorporated in the routine {\it mrqmin} from \citet{Press1992}.}.

\begin{table*}
\small\addtolength{\tabcolsep}{-1.5pt}
\begin{centering}
\caption{\label{tab:gauss}
Parameters of the broad emission lines from multi-component Gaussian fitting${^\dagger}$.
}
\begin{tabular}{cccccccccccccccccccc}
\hline
ID |& \multicolumn{9}{c}{Blue bump} |& \multicolumn{6}{c}{Red bump}|\\
\hline
 & \multicolumn{3}{c}{\heiiwr}|& \multicolumn{3}{c}{\niiiwr}|& \multicolumn{3}{c}{\ciiiwrb}|& \multicolumn{3}{c}{\civwrr}|& \multicolumn{3}{c}{\ciiiwrr}|& \multicolumn{2}{c}{bumps}|& \multicolumn{2}{c}{S/N}\\
   & $L$    &FWHM & EW  & $L$   & FWHM & EW  & $L$    &FWHM & EW   & $L$    & FWHM & EW  & $L$ & FWHM & EW  & $L_\mathrm{BB}$& $L_\mathrm{RB}$& $_{\rm BB}$ & $_{\rm RB}$\\
(1)& (2)  &(3)  & (4) & (5) & (6)  & (7) & (8)  &(9)  & (10) & (11) & (12) & (13) &(14)&(15) &(16)&(17)&(18)&(19)&(20)\\
\hline
1  & 56.0 & 16.6& 2.5 & 39.5& 22.3 & 1.7& 16.5 & 16.1 & 0.7 & 74.2 & 75.6 & 5.3 & 0.0 & 0.0 & 0.0 &112.0 & 74.2& 60 &127\\
2  & 20.3 & 16.9& 2.2 & 19.0& 25.5 & 2.0& 12.3 & 20.1 & 1.3 & 25.2 & 70.6 & 4.8 & 0.0 & 0.0 & 0.0 & 51.6 & 25.2& 48 &117\\
3  & 22.3 & 13.0& 4.4 & 14.0& 15.6 & 2.6& 27.9 & 60.6 & 5.3 & 14.3 & 63.6 & 4.8 & 0.0 & 0.0 & 0.0 & 64.1 & 14.3& 58 & 93\\
4  &  2.8 &  9.1& 1.0 &  1.1& 15.5 & 0.4&  0.0 &  0.0 & 0.0 &  6.2 & 58.9 & 3.6 & 0.0 & 0.0 & 0.0 &  3.9 &  6.2& 56 & 78\\
5  &  3.4 & 17.7& 2.7 &  2.8& 24.5 & 2.1&  0.9 & 16.4 & 0.7 &  5.3 & 96.4 & 6.6 & 0.0 & 0.0 & 0.0 &  7.1 &  5.3& 50 & 95\\
6  &  2.4 & 14.9& 2.3 &  1.2& 15.0 & 1.2&  0.0 &  0.0 & 0.0 &  2.8 & 82.9 & 4.5 & 0.0 & 0.0 & 0.0 &  3.6 &  2.8& 68 & 82\\
7  & 11.0 & 13.2& 1.0 & 11.5& 16.0 & 1.0&  3.1 &  8.3 & 0.3 & 12.7 & 41.7 & 1.8 & 6.1 &13.6 & 0.8 & 25.6 & 18.8& 65 & 52\\
8  &  1.3 & 12.9& 1.0 &  1.8& 22.6 & 1.3&  0.2 & 10.2 & 0.1 &  1.9 & 64.0 & 2.2 & 0.0 & 0.0 & 0.0 &  3.3 &  1.9& 64& 77\\
9  &  9.3 & 10.7& 0.9 & 15.9& 23.2 & 1.4&  0.0 &  0.0 & 0.0 & 25.6 & 45.0 & 3.0 & 4.0 & 7.8 & 0.4 & 25.2 & 29.6& 51& 46\\
10 &  1.1 & 15.6& 0.7 &  0.0&  0.0 & 0.0&  0.0 &  0.0 & 0.0 &  1.6 & 48.9 & 1.7 & 0.0 & 0.0 & 0.0 &  1.1 &  1.6& 55& 67\\
11 & 10.1 & 16.6& 1.9 & 12.3& 24.5 & 2.3&  9.3 & 21.4 & 1.8 & 10.6 & 51.8 & 3.0 & 0.7 & 8.8 & 0.2 & 31.7 & 11.3& 64&103\\
12 & 14.2 & 16.0& 1.6 & 16.1& 25.0 & 1.7& 11.5 & 18.8 & 1.2 & 10.6 & 47.1 & 2.0 & 0.0 & 0.0 & 0.0 & 41.8 & 10.7& 54& 87\\
13 &  6.3 & 14.1& 3.0 &  6.0& 18.9 & 2.7&  1.0 & 14.1 & 0.5 &  0.0 &  0.0 & 0.0 & 0.0 & 0.0 & 0.0 & 13.3 &  0.0& 53& 77\\
14 &  1.7 & 16.3& 1.8 &  2.4& 29.8 & 2.5&  0.7 & 14.1 & 0.8 &  1.2 & 47.1 & 2.0 & 0.0 & 0.0 & 0.0 &  4.8 &  1.2& 53& 66\\
15 &  1.0 & 13.7& 0.9 &  0.8& 20.8 & 0.6&  0.4 & 14.1 & 0.3 &  0.5 & 35.3 & 0.7 & 0.1 & 5.5 & 0.2 &  2.2 &  0.6& 51& 74\\
16 &  1.6 & 14.7& 0.8 &  0.8& 14.6 & 0.4&  0.0 &  0.0 & 0.0 &  2.2 & 47.1 & 1.6 & 0.0 & 0.0 & 0.0 &  2.3 &  2.2& 66& 80\\
17 &  1.7 & 15.7& 1.4 &  1.2& 18.2 & 0.9&  0.0 &  0.0 & 0.0 &  0.7 & 35.3 & 0.7 & 0.0 & 0.0 & 0.0 &  2.9 &  0.7& 56& 77\\
18 &  1.4 & 12.3& 1.6 &  1.9& 21.8 & 2.2&  0.9 & 18.8 & 1.0 &  0.2 &  4.7 & 0.2 & 0.3 & 6.0 & 0.4 &  4.2 &  0.5& 58& 55\\
19 &  3.6 & 20.6& 1.7 &  5.7& 31.7 & 2.7&  2.7 & 18.8 & 1.3 &  5.1 & 51.8 & 3.6 & 0.0 & 0.0 & 0.0 & 12.0 &  5.1& 68& 75\\
20 &  3.0 & 16.9& 1.1 &  2.6& 24.6 & 0.9&  1.7 & 22.5 & 0.6 &  2.0 & 35.3 & 1.1 & 0.0 & 0.0 & 0.0 &  7.3 &  2.0& 59& 87\\
21 &  1.5 & 16.1& 1.9 &  1.5& 24.2 & 2.0&  0.9 & 22.5 & 1.1 &  1.9 & 47.1 & 3.7 & 0.0 & 0.0 & 0.0 &  3.9 &  1.9& 51& 62\\
22 &  2.4 & 15.2& 1.0 &  2.9& 27.5 & 1.2&  1.8 & 21.2 & 0.7 &  3.1 & 47.1 & 1.8 & 0.0 & 0.0 & 0.0 &  7.0 &  3.1& 74& 99\\
23 &  7.9 & 16.3& 1.0 &  3.8& 15.8 & 0.4&  1.9 & 11.8 & 0.2 & 12.7 & 55.0 & 2.5 & 0.0 & 0.0 & 0.0 & 13.7 & 12.7& 61& 87\\
24 &  5.6 & 15.2& 1.1 &  3.2& 16.6 & 0.6&  1.7 & 11.8 & 0.3 &  5.7 & 51.8 & 1.8 & 2.8 &32.8 & 0.8 & 10.5 &  8.5& 63& 85\\
25 &  9.4 & 16.4& 1.1 &  6.2& 17.7 & 0.7&  3.0 & 11.8 & 0.4 & 10.0 & 51.0 & 2.0 & 3.6 &23.6 & 0.7 & 18.6 & 13.6& 58& 78\\
26 & 12.1 & 16.5& 0.9 &  0.0&  0.0 & 0.0&  0.0 &  0.0 & 0.0 & 19.4 & 65.8 & 2.4 & 0.0 & 0.0 & 0.0 & 12.1 & 19.4& 52& 76\\
27 &  2.7 & 19.4& 1.1 &  1.6& 17.5 & 0.6&  0.0 &  0.0 & 0.0 &  3.3 & 51.9 & 2.3 & 0.4 & 9.0 & 0.2 &  4.3 &  3.7& 56& 83\\
28 &  6.6 & 15.1& 0.8 &  0.0&  0.0 & 0.0&  0.0 &  0.0 & 0.0 & 10.6 & 49.6 & 2.0 & 0.0 & 0.0 & 0.0 &  6.6 & 10.6& 56& 83\\
29 &  5.9 & 19.1& 1.1 &  5.4& 24.2 & 1.0&  4.6 & 20.6 & 0.8 & 10.0 & 56.2 & 3.1 & 0.0 & 0.0 & 0.0 & 15.9 & 10.0& 53& 92\\
30 &  3.0 & 26.5& 1.4 &  1.7& 20.3 & 0.8&  0.0 &  0.0 & 0.0 &  2.2 & 48.5 & 1.8 & 0.0 & 0.0 & 0.0 &  4.7 &  2.3& 59& 45\\
31 &  1.4 & 16.9& 1.8 &  1.0& 15.0 & 1.2&  0.0 &  0.0 & 0.0 &  0.6 & 44.0 & 1.2 & 0.1 & 6.3 & 0.2 &  2.4 &  0.7& 53& 70\\
32 &  1.5 & 16.0& 1.0 &  1.2& 19.3 & 0.8&  0.0 &  0.0 & 0.0 &  2.3 & 55.2 & 2.3 & 0.5 &13.1 & 0.5 &  2.7 &  2.8& 55& 68\\
33 &  0.5 & 12.0& 0.9 &  0.5& 14.7 & 0.8&  0.0 &  0.0 & 0.0 &  0.6 & 41.5 & 1.5 & 0.2 &10.8 & 0.5 &  1.0 &  0.8& 52& 61\\
34 &  1.3 & 18.1& 1.1 &  1.2& 21.0 & 1.0&  0.8 & 17.4 & 0.7 &  2.6 & 53.3 & 3.2 & 0.0 & 0.0 & 0.0 &  3.2 &  2.6& 58& 82\\
35 &  0.4 & 10.9& 0.6 &  0.3&  9.6 & 0.5&  0.0 &  0.0 & 0.0 &  0.9 & 54.2 & 2.2 & 0.0 & 0.0 & 0.0 &  0.7 &  0.9& 60& 56\\
36 &  1.3 & 16.2& 0.7 &  0.0&  0.0 & 0.0&  0.0 &  0.0 & 0.0 &  3.9 & 55.1 & 2.7 & 0.0 & 0.0 & 0.0 &  1.3 &  3.0& 63& 77\\
37 &  0.9 & 15.4& 0.8 &  0.6& 17.4 & 0.5&  0.0 &  0.0 & 0.0 &  2.0 & 56.2 & 2.6 & 0.0 & 0.0 & 0.0 &  1.6 &  2.0& 58& 66\\
38 &  0.4 & 13.6& 1.2 &  0.0&  0.0 & 0.	&  0.0 &  0.0 & 0.0 &  0.9 & 53.2 & 3.8 & 0.0 & 0.0 & 0.0 &  0.4 &  0.9& 48& 47\\
\hline
\end{tabular}\\
\par\end{centering}
{\it Notes}. Brief explanation of columns:
(1) WR identification;
(2) luminosity of \heiiwr\ [$1.0\times10^{37}$~\ergs] at a distance of 18.1~Mpc;
(3) full width at half maximum (FWHM) [\AA];
(4) equivalent width (EW) [\AA];
(5) luminosity [$1.0\times10^{37}$~\ergs], (6) FWHM [\AA] and (7) EW [\AA] of \niiiwr;
(8) luminosity [$1.0\times10^{37}$~\ergs], (9) FWHM [\AA] and (10) EW [\AA] of \ciiiwrb;
(11) luminosity [$1.0\times10^{37}$~\ergs], (12) FWHM [\AA] and (13) EW [\AA] of \civwrr;
(14) luminosity [$1.0\times10^{37}$~\ergs], (15) FWHM [\AA] and (16) EW [\AA] of \ciiiwrr;
(17) BB luminosity ($L_{\rm BB}$) [$1.0\times10^{37}$~\ergs];
(18) RB luminosity ($L_{\rm RB}$) [$1.0\times10^{37}$~\ergs];
(19) continuum BB signal-to-noise ratio (S/N$_{\rm BB}$) at 4750--4830 \AA;
(20) S/N$_{\rm RB}$ at 5650--5730 \AA.
${^\dagger}$The multi-Gaussian fittings are shown in Fig.~\ref{fig:multi}.
\\
\end{table*}

\begin{table*}
\small\addtolength{\tabcolsep}{-1.5pt}
\begin{center}
\caption{Parameters of the nebular emission lines.}
\begin{tabular}{cccccccccccc}
\hline
ID&$\mathrm{log}F$(\hb)& EW(\hb) & EW(\ha)&$A_\mathrm{V}$&$T_\mathrm{e}$(\nii) &$T_\mathrm{e}$(\siii) &$n_{\rm e}$(\cliii) &$n_{\rm e}$(\sii)& \multicolumn{2}{c}{$\mathrm{12+log(O/H)}$} &$V_\mathrm{r}$ \\
& (erg\,cm$^{-2}$\,s$^{-1}$) &(\AA)&(\AA)& (mag)& ($\times10^{3}$ K)& ($\times10^{3}$ K)  & (cm$^{-3}$) &(cm$^{-3}$) &DR &R3 & (\kms)\\
(1)& (2)           & (3)& (4) & (5)  &(6)   & (7)& (8) & (9)& (10)& (11)& (12)\\
\hline
1 &$\mathrm{-12.574}\pm$0.001& 83.9 &415.7 & 0.70$\pm$0.01 & 7.96$\pm$0.08 &7.40$\pm$0.04 &200$\pm$290&229.8$\pm$2.9 & 9.01$\pm$0.02& 8.50$\pm$0.01 &1468.0$\pm$5.7\\
2 &$\mathrm{-12.721}\pm$0.001&121.2 &684.5 & 0.66$\pm$0.01 & 8.37$\pm$0.07 &8.02$\pm$0.05 &$\cdots$   &191.4$\pm$1.9 & 8.81$\pm$0.01& 8.50$\pm$0.01 &1433.0$\pm$5.9\\
3 &$\mathrm{-12.854}\pm$0.001&139.1 &831.2 & 0.81$\pm$0.01 & 8.60$\pm$0.09 &8.30$\pm$0.03 &240$\pm$150& 81.0$\pm$1.8 & 8.60$\pm$0.03& 8.50$\pm$0.01 &1452.7$\pm$2.9\\
4 &$\mathrm{-13.563}\pm$0.004& 54.5 &278.8 & 0.96$\pm$0.01 & 8.10$\pm$0.40 &8.80$\pm$0.40 & $\cdots$  & 74.6$\pm$1.6 & 8.62$\pm$0.09& 8.67$\pm$0.01 &1491.7$\pm$4.0\\
5 &$\mathrm{-13.393}\pm$0.001&112.8 &658.1 & 1.13$\pm$0.01 & 8.47$\pm$0.23 &8.56$\pm$0.10 &500$\pm$700& 60.1$\pm$2.0 & 8.65$\pm$0.05& 8.56$\pm$0.01 &1449.6$\pm$5.0\\
6 &$\mathrm{-13.445}\pm$0.002&150.6 &846.5 & 0.96$\pm$0.01 & 8.27$\pm$0.16 &8.19$\pm$0.10 &$\cdots$   & 51.4$\pm$1.1 & 8.65$\pm$0.04& 8.57$\pm$0.01 &1452.4$\pm$3.6\\
7 &$\mathrm{-12.924}\pm$0.004& 27.9 &183.1 & 1.20$\pm$0.01 & 6.70$\pm$0.70 &5.30$\pm$1.00 &$\cdots$   &134.8$\pm$3.0 & 8.70$\pm$0.70& $\cdots$      &1632.4$\pm$3.6\\
8 &$\mathrm{-13.548}\pm$0.004& 44.7 &285.8 & 0.97$\pm$0.01 & 7.15$\pm$0.32 &6.40$\pm$0.50 &$\cdots$   & 57.5$\pm$0.6 & 8.90$\pm$0.11& 8.97$\pm$0.01 &1544.1$\pm$3.8\\
9 &$\mathrm{-12.665}\pm$0.006& 39.7 &220.0 & 2.57$\pm$0.02 & 8.00$\pm$0.50 &7.20$\pm$0.80 &$\cdots$   &387.0$\pm$22  & 8.58$\pm$0.13& 9.11$\pm$0.01 &1710.8$\pm$9.5\\
10& $\mathrm{-13.764}\pm$0.006& 33.6 &206.4 & 0.77$\pm$0.02 & 7.00$\pm$0.50 &6.20$\pm$0.50 &$\cdots$   & 70.0$\pm$4.0 & 9.01$\pm$0.12& 8.91$\pm$0.01 &1555.0$\pm$2.2\\
11&$\mathrm{-12.756}\pm$0.002& 67.1 &434.9 & 1.29$\pm$0.01 & 7.62$\pm$0.18 &7.07$\pm$0.22 &$\cdots$   & 64.7$\pm$0.6 & 8.81$\pm$0.05& 8.83$\pm$0.01 &1581.5$\pm$4.3\\
12&$\mathrm{-12.942}\pm$0.003& 48.9 &316.5 & 0.69$\pm$0.01 & 7.48$\pm$0.16 &7.16$\pm$0.29 &$\cdots$   & 56.6$\pm$1.2 & 8.86$\pm$0.05& 8.87$\pm$0.01 &1616.9$\pm$0.9\\
13&$\mathrm{-13.225}\pm$0.001& 98.0 &666.6 & 0.79$\pm$0.01 & 7.55$\pm$0.13 &7.58$\pm$0.08 &$\cdots$   & 67.0$\pm$4.0 & 8.87$\pm$0.06& 8.77$\pm$0.01 &1644.5$\pm$4.1\\
14&$\mathrm{-13.764}\pm$0.003& 65.0 &400.8 & 0.79$\pm$0.01 & 7.36$\pm$0.30 &7.81$\pm$0.35 &$\cdots$   & 48.3$\pm$0.1 & 8.92$\pm$0.05& 8.88$\pm$0.01 &1629.3$\pm$6.8\\
15&$\mathrm{-13.822}\pm$0.004& 37.5 &230.9 & 0.95$\pm$0.01 & 7.80$\pm$0.40 &7.70$\pm$0.60 &300$\pm$700& 24.0$\pm$1.6 & 8.81$\pm$0.10& 8.81$\pm$0.01 &1634.0$\pm$4.8\\
16&$\mathrm{-13.693}\pm$0.007& 23.0 &131.8 & 0.93$\pm$0.02 & 7.30$\pm$0.80 &6.10$\pm$3.50 &$\cdots$   & 22.4$\pm$3.0 & 8.95$\pm$0.32& 8.94$\pm$0.01 &1652.7$\pm$3.8\\
17&$\mathrm{-12.938}\pm$0.003&109.3 &635.6 & 1.50$\pm$0.01 & 7.96$\pm$0.20 &7.66$\pm$0.15 &$\cdots$   & 87.2$\pm$2.7 & 8.83$\pm$0.04& 8.65$\pm$0.01 &1570.8$\pm$4.2\\
18&$\mathrm{-13.001}\pm$0.003& 75.4 &511.5 & 2.67$\pm$0.01 & 6.99$\pm$0.26 &6.70$\pm$1.10 &$\cdots$   & 75.2$\pm$1.6 & 8.92$\pm$0.16& $\cdots$      &1607.2$\pm$4.6\\
19&$\mathrm{-13.800}\pm$0.008& 21.9 &125.6 & 0.70$\pm$0.02 & 8.90$\pm$0.70 &6.40$\pm$1.10 &$\cdots$   & 25.6$\pm$2.0 & 8.61$\pm$0.17& 8.67$\pm$0.01 &1580.6$\pm$2.4\\
20&$\mathrm{-13.305}\pm$0.004& 53.4 &309.8 & 1.08$\pm$0.01 & 8.10$\pm$0.50 &8.00$\pm$0.60 &$\cdots$   & 28.0$\pm$4.0 & 8.82$\pm$0.16& 8.58$\pm$0.01 &1467.8$\pm$1.7\\
21&$\mathrm{-13.617}\pm$0.003&102.7 &546.0 & 1.00$\pm$0.01 & 8.30$\pm$0.28 &8.07$\pm$0.30 &100$\pm$600& 73.2$\pm$2.4 & 8.83$\pm$0.05& 8.61$\pm$0.01 &1488.1$\pm$4.6\\
22&$\mathrm{-12.994}\pm$0.004& 44.0 &313.4 & 1.87$\pm$0.01 & 7.20$\pm$0.40 &6.60$\pm$0.50 &$\cdots$   & 83.0$\pm$6.0 & 8.92$\pm$0.14& 8.95$\pm$0.01 &1583.3$\pm$3.7\\
23&$\mathrm{-12.989}\pm$0.005& 31.4 &215.5 & 1.27$\pm$0.01 & 7.40$\pm$0.60 &6.60$\pm$0.80 &$\cdots$   & 59.0$\pm$7.0 & 8.68$\pm$0.27& 9.05$\pm$0.01 &1650.8$\pm$4.7\\
24&$\mathrm{-12.989}\pm$0.003& 46.9 &336.1 & 0.96$\pm$0.01 & 6.87$\pm$0.29 &6.31$\pm$0.27 &$\cdots$   & 79.9$\pm$0.7 & 8.83$\pm$0.16& 9.13$\pm$0.01 &1655.6$\pm$4.4\\
25&$\mathrm{-12.983}\pm$0.004& 32.5 &234.5 & 0.86$\pm$0.01 & 7.40$\pm$0.50 &6.20$\pm$0.60 &$\cdots$   & 60.1$\pm$2.6 & 8.55$\pm$0.29& 9.15$\pm$0.01 &1626.8$\pm$5.2\\
26&$\mathrm{-12.944}\pm$0.007& 16.9 &132.8 & 1.16$\pm$0.02 & 8.30$\pm$0.70 &6.70$\pm$0.90 &$\cdots$   & 80.0$\pm$7.0 & 8.43$\pm$0.28& 8.92$\pm$0.01 &1608.7$\pm$3.8\\
27&$\mathrm{-13.693}\pm$0.005& 26.6 &173.1 & 0.78$\pm$0.01 & 7.20$\pm$0.60 &5.90$\pm$3.40 &$\cdots$   & 37.0$\pm$4.0 & 8.70$\pm$0.50& 9.08$\pm$0.01 &1593.7$\pm$6.0\\
28&$\mathrm{-13.047}\pm$0.006& 27.0 &178.4 & 1.12$\pm$0.01 & 7.40$\pm$0.70 &6.00$\pm$4.00 &$\cdots$   & 60.0$\pm$5.0 & 8.60$\pm$0.40& 9.10$\pm$0.01 &1574.4$\pm$4.5\\
29&$\mathrm{-13.344}\pm$0.004& 27.4 &183.2 & 1.05$\pm$0.01 & 8.00$\pm$0.60 &6.80$\pm$0.60 &$\cdots$   & 40.5$\pm$0.9 & 8.51$\pm$0.19& 8.93$\pm$0.01 &1664.1$\pm$2.0\\
30&$\mathrm{-13.465}\pm$0.003& 47.6 &329.6 & 1.09$\pm$0.01 & 7.53$\pm$0.24 &7.04$\pm$0.32 &$\cdots$   & 55.1$\pm$2.4 & 8.75$\pm$0.08& 8.95$\pm$0.01 &1690.8$\pm$8.7\\
31&$\mathrm{-13.445}\pm$0.001&100.5 &695.0 & 1.29$\pm$0.01 & 6.90$\pm$0.11 &6.50$\pm$0.13 &1800$\pm$1800&219.0$\pm$2.8& 9.06$\pm$0.03 & 9.05$\pm$0.01 &1675.9$\pm$5.7\\
32&$\mathrm{-13.603}\pm$0.005& 28.1 &201.8 & 1.56$\pm$0.01 & 7.80$\pm$0.60 &5.60$\pm$3.20 &$\cdots$   & 64.7$\pm$2.1 & 8.40$\pm$0.70& $\cdots$    &1632.2$\pm$1.6\\
33&$\mathrm{-13.868}\pm$0.004& 47.5 &315.4 & 1.42$\pm$0.01 & 6.60$\pm$0.50 &5.00$\pm$3.00 &$\cdots$   & 91.0$\pm$8.0 & 8.80$\pm$0.60& $\cdots$    &1676.9$\pm$5.8\\
34&$\mathrm{-13.651}\pm$0.005& 28.4 &204.7 & 1.69$\pm$0.01 & 7.50$\pm$0.70 &6.30$\pm$0.90 &$\cdots$   & 41.0$\pm$4.0 & 8.67$\pm$0.25& 9.01$\pm$0.01 &1691.2$\pm$3.8\\
35&$\mathrm{-14.120}\pm$0.009& 24.0 &146.5 & 1.46$\pm$0.02 & 7.60$\pm$1.00 &6.50$\pm$1.50 &$\cdots$   & 67.0$\pm$7.0 & 8.10$\pm$2.90& $\cdots$    &1674.1$\pm$14.6\\
36&$\mathrm{-13.706}\pm$0.009& 17.9 &102.8 & 1.16$\pm$0.02 & 9.20$\pm$1.30 &7.40$\pm$1.40 &100$\pm$400& 49.0$\pm$13  & 8.20$\pm$0.50& 8.81$\pm$0.01 &1547.1$\pm$4.3\\
37&$\mathrm{-13.878}\pm$0.005& 38.3 &215.0 & 0.97$\pm$0.01 & 7.90$\pm$0.50 &7.00$\pm$0.70 &$\cdots$   & 26.7$\pm$3.5 & 8.60$\pm$0.17& 8.96$\pm$0.01 &1672.6$\pm$4.5\\
38&$\mathrm{-13.947}\pm$0.007& 50.2 &330.2 & 1.78$\pm$0.02 & 8.20$\pm$0.60 &7.40$\pm$0.60 &$\cdots$   & 28.0$\pm$13  & 8.60$\pm$0.50& 8.77$\pm$0.01 &1694.5$\pm$3.2\\
\hline
\end{tabular}\\
{\it Notes}. Brief explanation of columns:
(1) WR identification (1--38);
(2) ${^\dagger}$reddening-corrected fluxes of $\mathrm{log}F$(\hb) [erg\,cm$^{-2}$\,s$^{-1}$];
(3) equivalent width (EW) of \hb\ [\AA];
(4) EW of \ha\ [\AA];
(5) visual extinction ($A_{\rm V}$) [mag];
(6) electron temperature ($T_\mathrm{e}$) of the low-ionization zone (\nii) [K];
(7) $T_\mathrm{e}$ of the medium-ionization zone (\siii) [K];
(8) electron density ($n_{\rm e}$) from \cliii\ [\cmvol];
(9) $n_{\rm e}$(\sii) [\cmvol];
(10) R3-oxygen abundance;
(11) Direct method (DM)-oxygen abundance;
(12) radial velocity ($V_\mathrm{r}$) [\kms].
\\
\label{tab:neb}
\end{center}
\end{table*}

We were able to decompose the BB 
into \heii, \niii\ and \ciii\ broad emission lines,
while the \ciii\ and \civ\ broad emission lines were used for the RB.
\ciiiwrr\ clearly indicates WCL presence,
whilst if there is only \civwrr\ it suggests WCE-subtypes.
\niiiwr\ in the BB is a WNL component.
To avoid any contamination from narrow-nebular lines, we also fitted the 
\fei, \feiii\ and \hei\ in the BB and the \nii\ and \hei\ in the RB.
This approach independently confirms what was obtained through Galactic templates,
with the benefit of also having determined
the parameters of the emission features contributing to the WR bumps. 
Details of the estimated fluxes, EWs and FWHMs are listed in Table~\ref{tab:gauss}
and the fits are illustrated in Fig.~\ref{fig:multi} for all WR spectra.

\subsection{Physical properties of the WR environments}

The rest of the nebular lines of the spectra were analysed
following {\sc iraf} standard routines \citep{Tody1993} and were corrected for
extinction by using the $c$(H$\beta$) value estimated from the
Balmer decrement method.
We assume an intrinsic Balmer decrement ratio corresponding to
a case B photoionised nebula with electron temperature
($T_\mathrm{e}$) = 10000~K and
the electron density ($n_\mathrm{e}$) = 100~cm$^{-3}$ 
\citep[see][]{Osterbrock2006} and
the reddening curve of \cite{Cardelli1989}.
Reddening-corrected line fluxes for different nebular lines
used to determine the physical conditions in the ionized zones
of the 38 WR spectra are presented in Table~\ref{tab:neb2}, in the Appendix~\ref{sec:appC}.

\subsubsection{Oxygen chemical abundance}

We calculated oxygen chemical abundance using the direct method (DM).
This implies that the determined oxygen abundance
depends on the physical conditions in the gas: 
$T_\mathrm{e}$ and $n_\mathrm{e}$.

We need $T_\mathrm{e}$ for the low and medium ionization zones to estimate 
the ionic $\rm{O^{+}}$ and $\rm{O^{++}}$ abundances.
We used $T_\mathrm{e}$(\nii) as the temperature for the low ionization region.
Due to the fact that MUSE does not observe the blue part of the spectra where \oiiia\ resides,
we cannot determine $T_\mathrm{e}$(\oiii).
We use instead the $T_\mathrm{e}$(\siii) for the medium ionization region,
which has been proven to be reliable \citep{2015Berg,2020James}.

$T_\mathrm{e}$(\nii) and $n_{\rm e}$(\sii) were calculated simultaneously
using {\sc PyNEB} \citep{2015Luridiana} 
and the \nii: I($\lambda6548$+$\lambda6584$)/I($\lambda5755$) and
\sii: I($\lambda6717$)/I($\lambda6731$) diagnostics, respectively.
We also used {\sc PyNEB} to determine simultaneously $T_\mathrm{e}$(\siii)
and $n_{\rm e}$(\sii) using the \siii: I($\lambda6312$)/I($\lambda9069$)
and \sii\ diagnostics, respectively.
For the few regions where \cliii$\rm{\lambda5517/\lambda5537}$ were detected,
$n_{\rm e}$(\cliii) was also determined simultaneously with $T_\mathrm{e}$(\nii).
We estimated the uncertainties propagating the relative
error of the line fluxes to the {\sc PyNEB} determinations.
For those regions where \siii\ was not detected,
we estimated $T_\mathrm{e}$(\siii) for the medium ionization
zone using the relations from \citet{1992Garnett}:
 \begin{equation}
  T_{\rm{e}}^{\rm{low}}=0.7 \thinspace T_{\rm{e}}^{\rm{medium}}+3000\thinspace \rm{K}
 \end{equation}
 
Ionic abundances were estimated using {\sc PyNEB}, where we used as inputs
the $T_{\rm{e}}$ corresponding to each ionization zone,
the $n_{\rm e}$(\sii) estimated previously, as well as the \oiib\ and \oiic\
lines for the $\rm{O^{+}}$ ion, while \oiiib\ and \oiiic\
lines for the $\rm{O^{2+}}$ ion. 
The relative uncertainty of the ionic abundance for each line is determined
from the quadratic sum of the relative $T_{\rm{e}}$, $n_{\rm{e}}$,
and flux ratios uncertainties. 
The ionic abundances are obtained by an error-weighted average of the
ionic abundance for each line. 
 Finally, we sum both ionic abundances to obtain the total oxygen abundance.

\begin{figure*}
\begin{center}
\includegraphics[width=\linewidth]{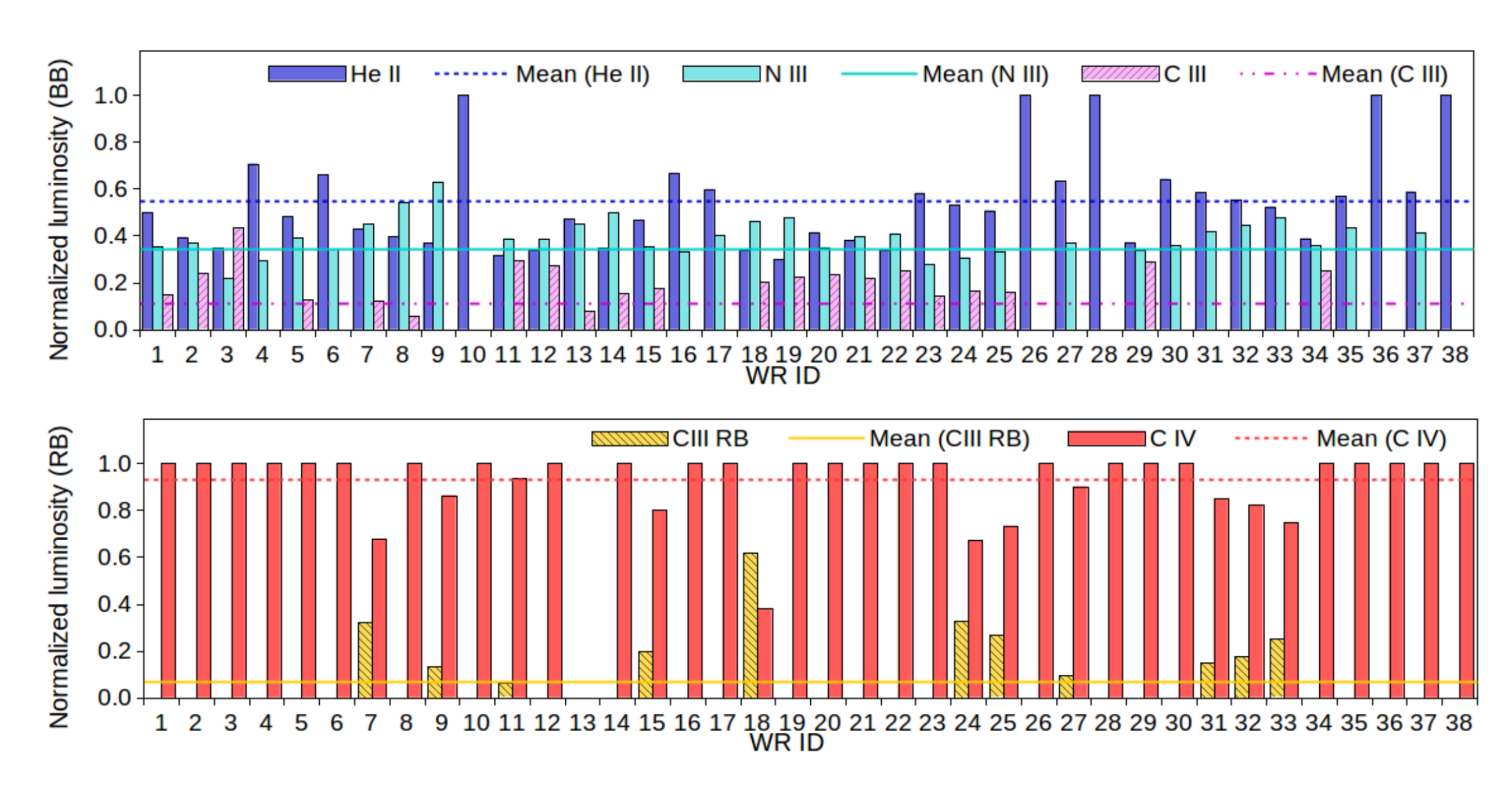}
\caption{({\it top})
The contribution from each WR broad component to the total luminosity of the BB feature, 
$L\mathrm{(BB)}$.
\heii\ is the dominating emission line, followed by
\ciii\ and \niii\ with averaged values of $\sim$55\%, $\sim$35\% and
$\sim$10\% of the total $L$(BB), respectively.
({\it bottom}) The contribution of the \ciii\, and \civ\ emission lines to the RB luminosity, $L\mathrm{(RB)}$.
On average, the \civ\, contributes to $>$85\% of $L\mathrm{(RB)}$.
The multi-Gaussian fittings are presented in Fig.~\ref{fig:multi}
and the results listed in Table~\ref{tab:gauss}.
}
\label{fig:histogram1}
\end{center}
\end{figure*}

For comparison, we also calculate the oxygen abundance 
in the emission-line regions using the observational R3 relation 
\cite[following][]{1992Vacca, 2006Bastian}
\begin{equation}
\mathrm{log(O/H)= -0.69\,log\,R_3-3.24},
\end{equation}
where $R_\mathrm{3}$ is defined as
\begin{equation}
R_3= \frac{I(\mathrm{O\,III~4959}) + I(\mathrm{O\,III~5007})}{I(\mathrm{H}\beta)} 
\end{equation}
which is valid for $-0.6 \leq\ \mathrm{log}\,R_\mathrm{3} \leq\ 1.0$.

We summarize the results in Table~\ref{tab:neb}.
The mean oxygen abundance resulted in
$\mathrm{12+log(O/H)} \approx 8.9$, which is practically Solar 
metallicity ($Z=0.02$). 
As a reference, the Solar metallicity given by
\citet{2009Asplund} is 12+log(O/H)=8.69 (Z=0.014).

\subsubsection{Kinematic information}

We considered the observed wavelengths of the
\oiiic, \hb\ and \ha\ lines, which have the highest signal-to-noise ratio,
to determine the radial velocities of the different SSCs with WR features.
The spectral resolution of the observations was high enough to enable us
to also obtain information about the kinematics.
Table~\ref{tab:neb} lists the radial velocities for the 
nebular environment of each WR SSC,
ranging from 1430 to 1700 \kms, which is in excellent agreement with
previous kinematical studies \citep[e.g.,][]{2006Bastian}.
According to NASA/NED, the heliocentric velocity for NGC\,4038 is 1642$\pm$12 \kms.

\section{Discussion}

The entire sample of WR stars we report in the Antennae are hosted in H\,{\sc ii} regions,
however, not all the H\,{\sc ii} regions in these galaxies harbour WR stars.
For example, the smallest and southernmost galaxy, NGC\,4039,
does not exhibit any hint of WR features.
Furthermore, data cubes covering the field in the outskirts of the Antennae
were found to have intermediate ages (100--300 Myr) in
\cite{2010Whitmore}, therefore, it is not rare for it not to have WR stars.

Notice that the physical sizes of the complexes found to harbour WR stars in the Antennae
galaxies are quite large (see Table~\ref{tab:class}), 
with diameters between $\sim100-350$~pc,
which are comparable to the size of the Tarantula in the LMC \citep[$\sim300$~pc; see][]{2019Crowther}.

Extragalactic WR stars are mostly detected at the location of H\,{\sc ii}
complexes, generally tracing the spiral arms of the galaxies.
This is also the case for the Antennae, particularly for NGC\,4038,
the northern and larger of this pair of galaxies
(see Fig.~\ref{fig:Antennae_RGB}).
The distribution of star-forming regions in figure~19 of
\citet{2019Gunawardhana} confirms that all the WR features
reported in the present work concur with very young H\,{\sc ii}
regions with ages $\lesssim$4~Myr.

Interestingly, WR\,1, the SSC complex with the strongest WR features, 
and therefore the largest number of estimated WR stars (800 WR), 
is located at the bridge between these merging galaxies.
It also has the strongest hydrogen lines in emission 
(see below).

We estimated a total of 4053 $\pm$ 84 WR stars in the Antennae galaxies,
which is among the highest number reported in the literature of
galaxies harbouring such stars.
Even if we adopt a foreground Galactic extinction
$A_{\rm V}$=0.127~mag \citep{2011Schlafly} as a lower limit for the 
quantification
of WR stars, the number would be $\sim2000$~WR stars.
On the other hand,
if we consider a greater distance to the Antennae, for example 
the 22~Mpc estimated by \citet{2008Schweizer} based on the type Ia supernova 2007sr,
the total number of WR stars would raise by $\sim$48\%, 
given a higher intrinsic luminosity of their WR bumps.

Through the detailed multi-Gaussian decomposition
presented in Section~3, we were able to avoid any contamination
from nebular emission lines and evaluate the contribution
from each broad emission line to the entire luminosity of both WR features.
On average, the He\,{\sc ii} emission line corresponds to $\sim$55\% of the BB luminosity 
($L\mathrm{(BB)}$), 
followed by \ciii\ and \niii\ with $\sim$35\% and $\sim$10\%, respectively.
This is illustrated in Fig.~\ref{fig:histogram1} {\it top panel}.

In particular, for WR\,10, 26, 28, 36 and 38, their BB is constituted
only by \heii, even though these are classified as WCE-subtypes.
We recall that this classification comes from their RB feature.
It does not mean that there is no carbon in the BB,
but rather that it is probably not intense enough and therefore unresolved by 
our analysis.
Thus, one must be careful to rule out the presence of WC-types in
those cases in which the BB is dominated by \heii.
The RB is determinant for any classification.
By definition \ciii\ is not present in WCE-subtypes.
\civ\ clearly dominates the RB luminosity ($L$(RB)) contributing with $>$85\%
to the total in WCL-subtypes
(see Fig.~\ref{fig:histogram1} {\it bottom panel}).

We do not find any evidence of the nebular \heii\ emission line with the multi-Gaussian
fitting decomposition, apparently only common in metal-poor environments
\citep[$\mathrm{12+log(O/H)} \leq 8.4$; see, e.g.,][]{2011Kehrig,2018Kehrig}.
Thus, we do not need to invoke other sources of 
hard ionizing radiation
\citep[see, e.g., figure 1 in][ it is expected that higher 
the metallicity, lower the intensity of nebular \heii\ emission line]{2019Schaerer}.

\begin{figure}
\begin{center}
\includegraphics[width=0.95\linewidth]{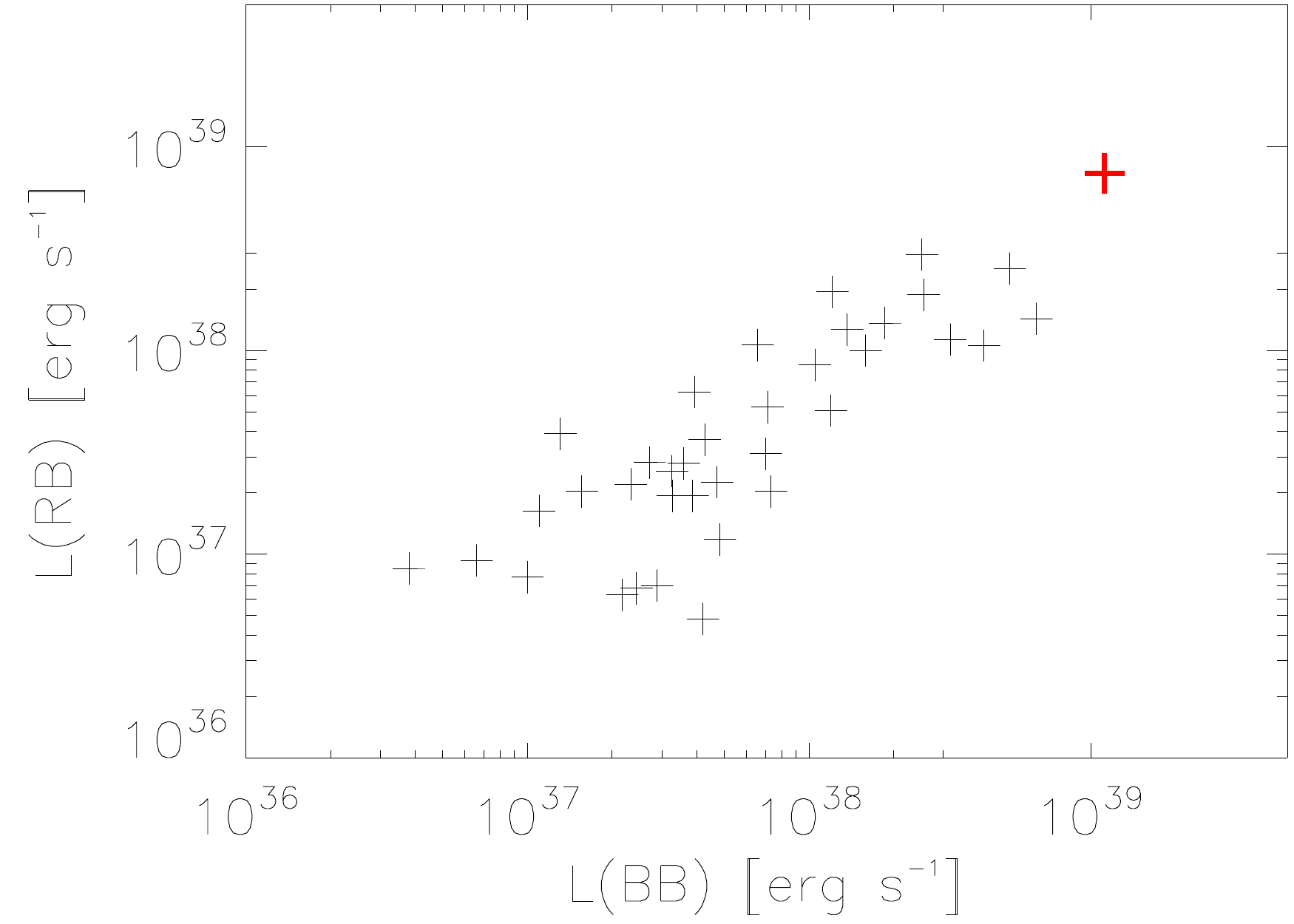}
\caption{$L\mathrm{(BB)}$ vs. $L\mathrm{(RB)}$ relation 
of the 38 WR SSC complexes in the Antennae.
For reference, the red symbol corresponds to WR\,1,
the one with the highest luminosities in its bumps.
}
\label{fig:lum1}
\end{center}
\end{figure}

\begin{figure}
\begin{center}
\includegraphics[width=0.95\linewidth]{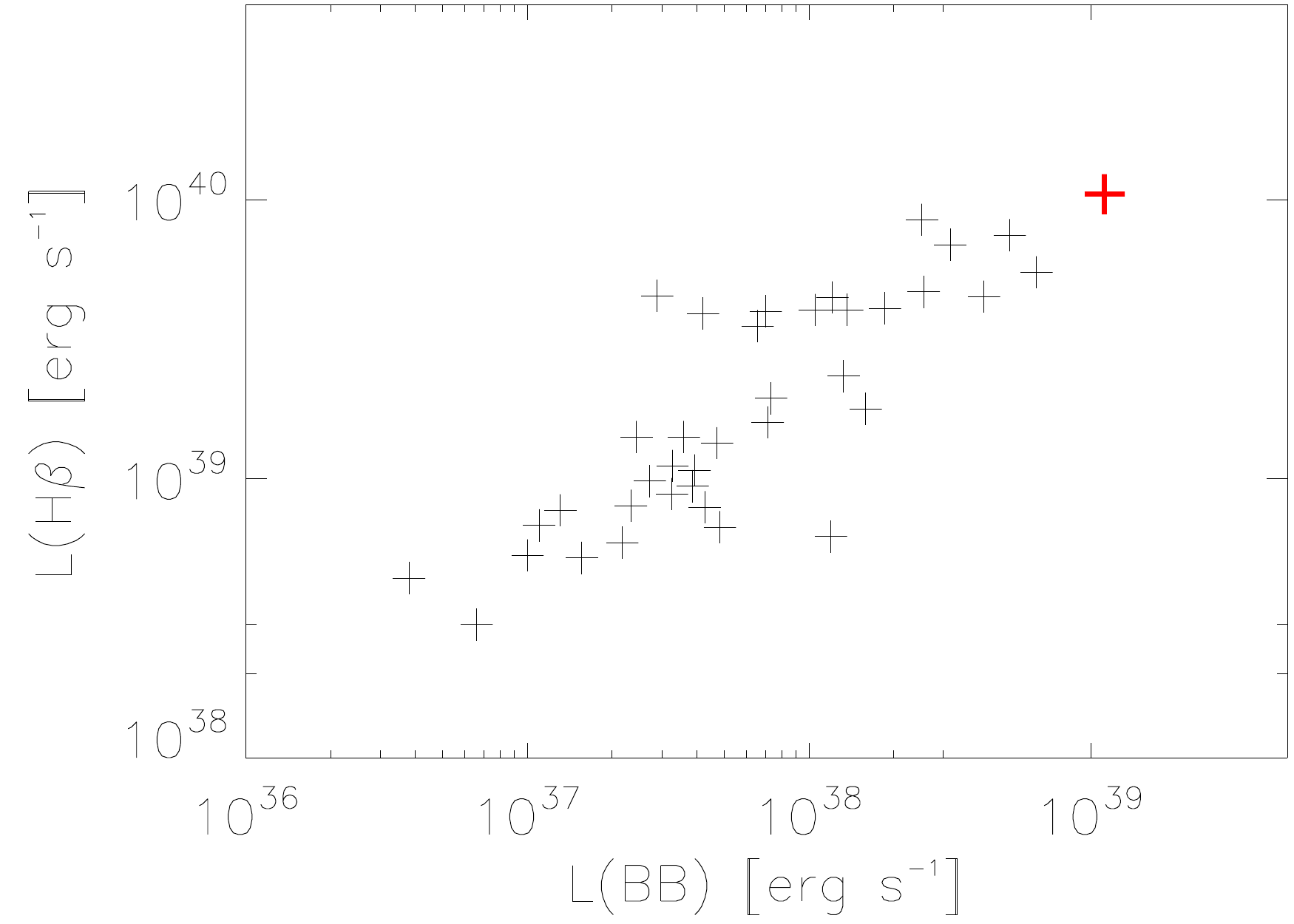}
\caption{$L$(\hb) vs. $L$(BB)
for the 38 WR SSC complexes in the Antennae.
The red symbol corresponds to WR\,1,
the one with the highest L(\hb), correlated with number of O-type stars.
}
\label{fig:lum2}
\end{center}
\end{figure}

In Fig.~\ref{fig:lum1} we illustrate the $L$(BB) vs. 
$L$(RB) relation for the 38 WR spectra. This figure shows that there
is a trend, with higher $L$(BB) corresponding to higher $L$(RB).
Furthermore, since the total \hb\ luminosity, $L$(H$\beta$), 
can be roughly associated with the population of
ionizing O-type stars, we show in Fig.~\ref{fig:lum2} the
relation between $L$(BB) and $L$(H$\beta$). This figure shows that 
those clusters with the highest
number of WR stars also show the highest values in the luminosities
of \hb, which is not unexpected since WR stars are considered descendants
of O-type stars. We note that WR\,1 is located in the most 
luminous regions in both figures.

\begin{figure}
\begin{centering}
\includegraphics[width=0.95\linewidth]{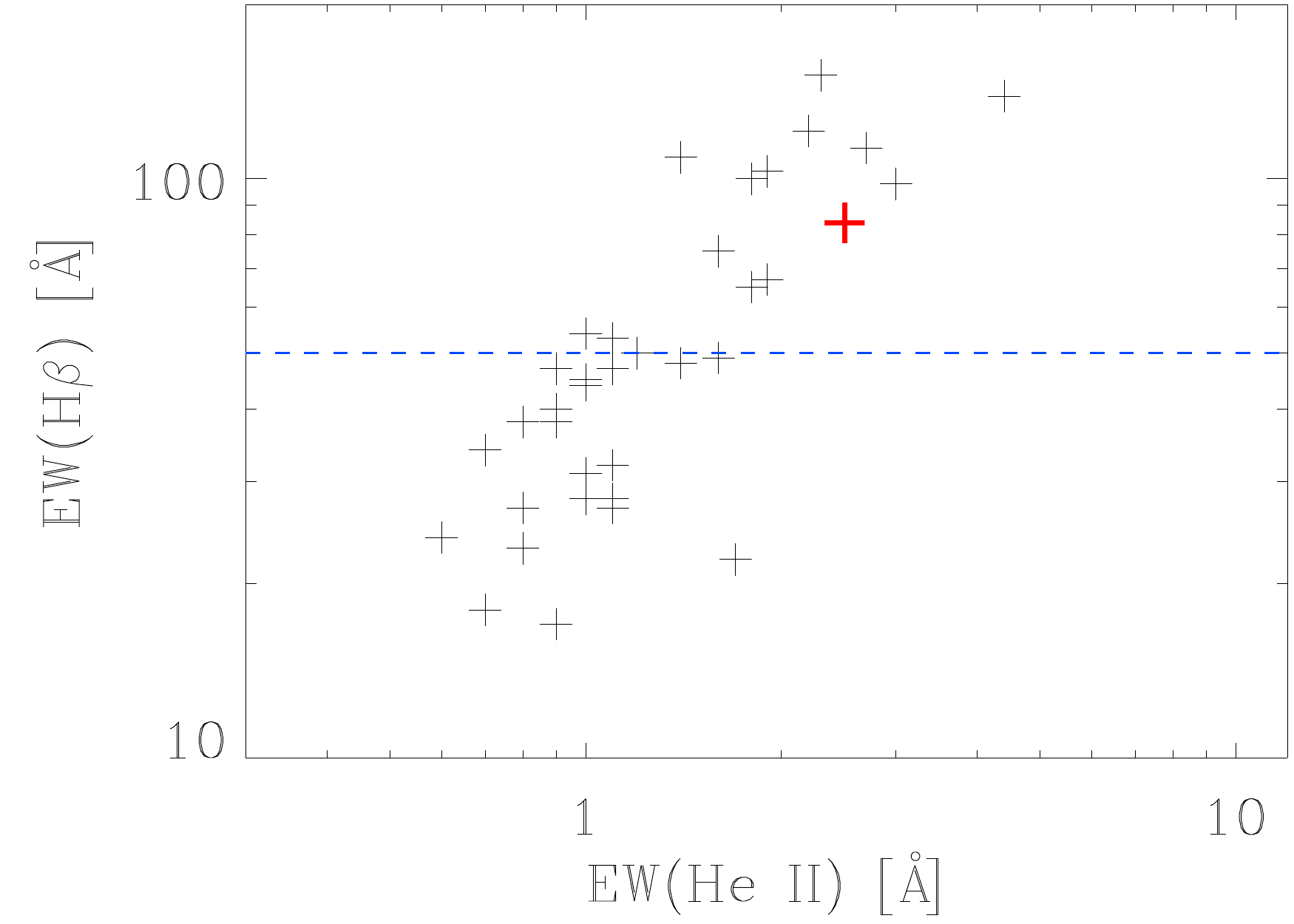}
\par\end{centering}
\caption{\label{fig:ew}
EW(\hb) vs. EW(\heii)
for the 38 WR SSC complexes in the Antennae.
The red symbol corresponds to WR\,1,
with one of the highest EW(\hb), related to the age of the starburst,
and also one of the highest EW(\heii).
EW(\hb)$>$50 \AA, above the {\it blue dashed line}, suggest a starburst age $\leq$5~Myr.
}
\end{figure}

The EW of the \hb\ emission lines can be used to assess the age of young starbursts.
In Fig.~\ref{fig:ew} we show the EW of \hb\ versus the EW of \heii\ for the 38 WR SSC 
complexes in the Antennae.
It has been suggested that an EW(\hb)$>$50 \AA\ corresponds to starburst ages 
$\leq$5~Myr \citep[see, e.g.][]{2016Chavez}.
Many of our objects are above this value,
suggesting an even younger stellar population.
Our diagram shows that the greater the EW(\hb), 
the stronger the intensity of the \heii, also meaning more WR stars.
However, it is important to emphasize that, although all the WR stars
in the Antennae are hosted in H\,{\sc ii} regions, not all the
detected H\,{\sc ii} regions harbour WR stars.

\begin{figure*}
\begin{centering}
\includegraphics[width=\linewidth]{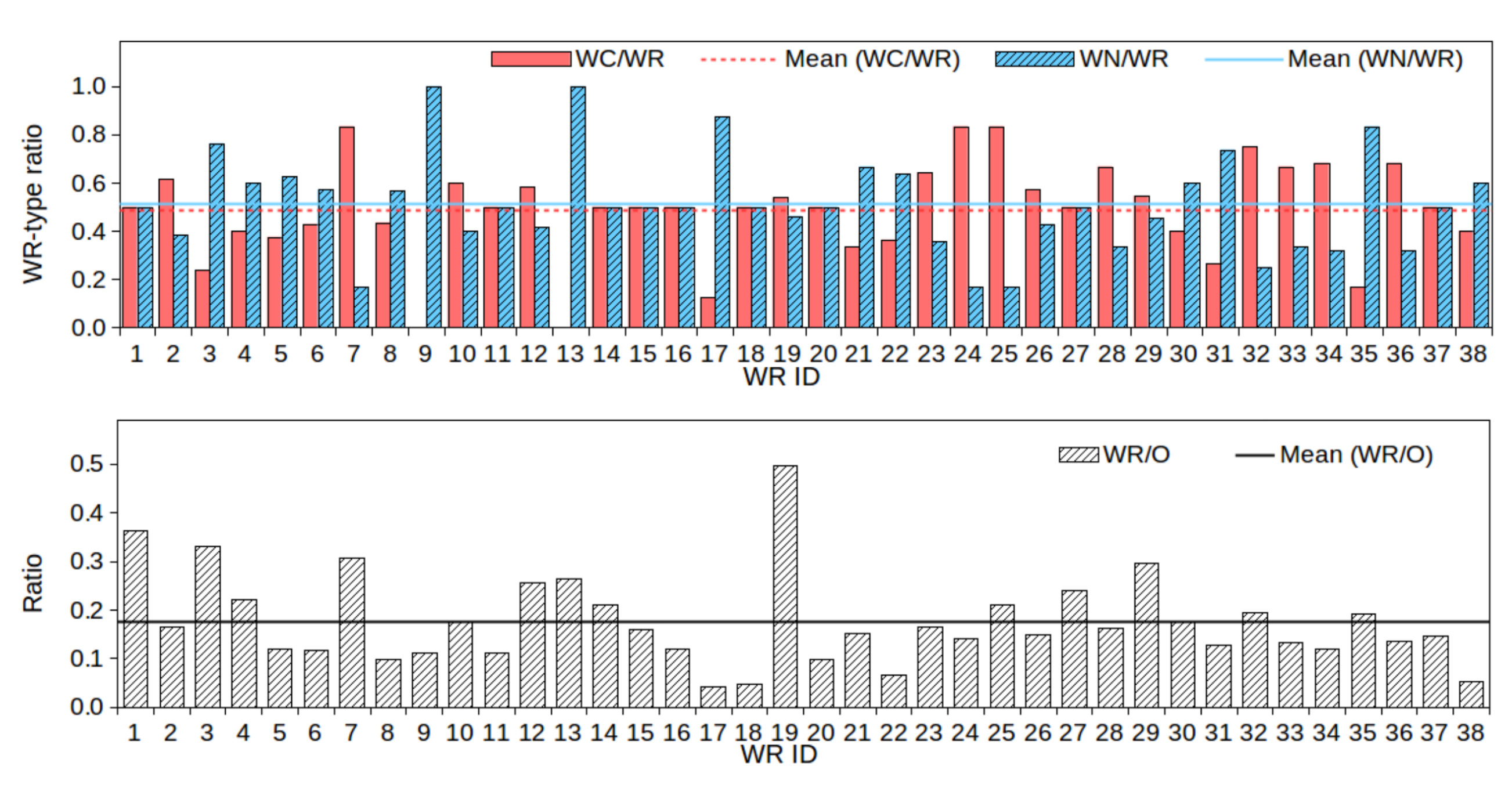}
\par\end{centering}
\caption{\label{fig:histogram2}
({\it top}) Histogram of WC/WR ({\it red}) and WN/WR
({\it blue}) ratios of WR stars;
({\it bottom}) WR/O ratio of stars.}
\end{figure*}

Reviewing the literature of the so-called WR galaxies,
there is a consensus in the massive stars community that a
high-metallicity favours WC-type stars and its later subtypes (WCE, WCL and WO),
which are otherwise rare in low-metallicity galaxies.
As previously discussed, given the wavelength range of our spectra,
we cannot inquire whether or not there are any WO-type stars present.
However, any WO-type stars would be already included in the WC-type stars
count since they, too, do display the RB feature.
Spectrographs equipped with Integral Filed Units (IFU) such as MEGARA at
the GTC covering the blue spectral range are essential to
unveil the O-rich WR population, little studied or observed
at extragalactic distances.

\begin{figure}
\begin{center}
\includegraphics[width=0.95\linewidth]{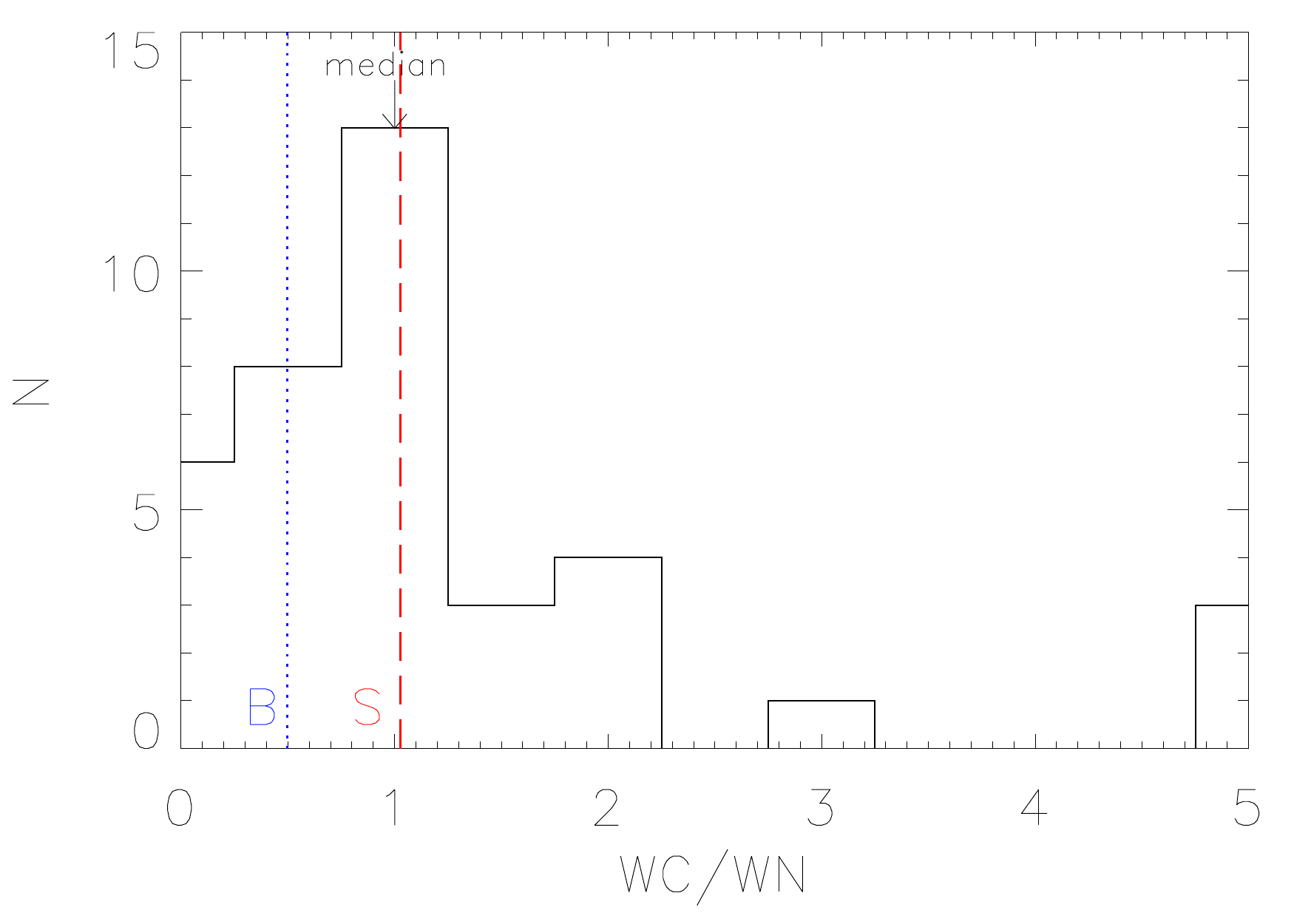}\\
\includegraphics[width=0.95\linewidth]{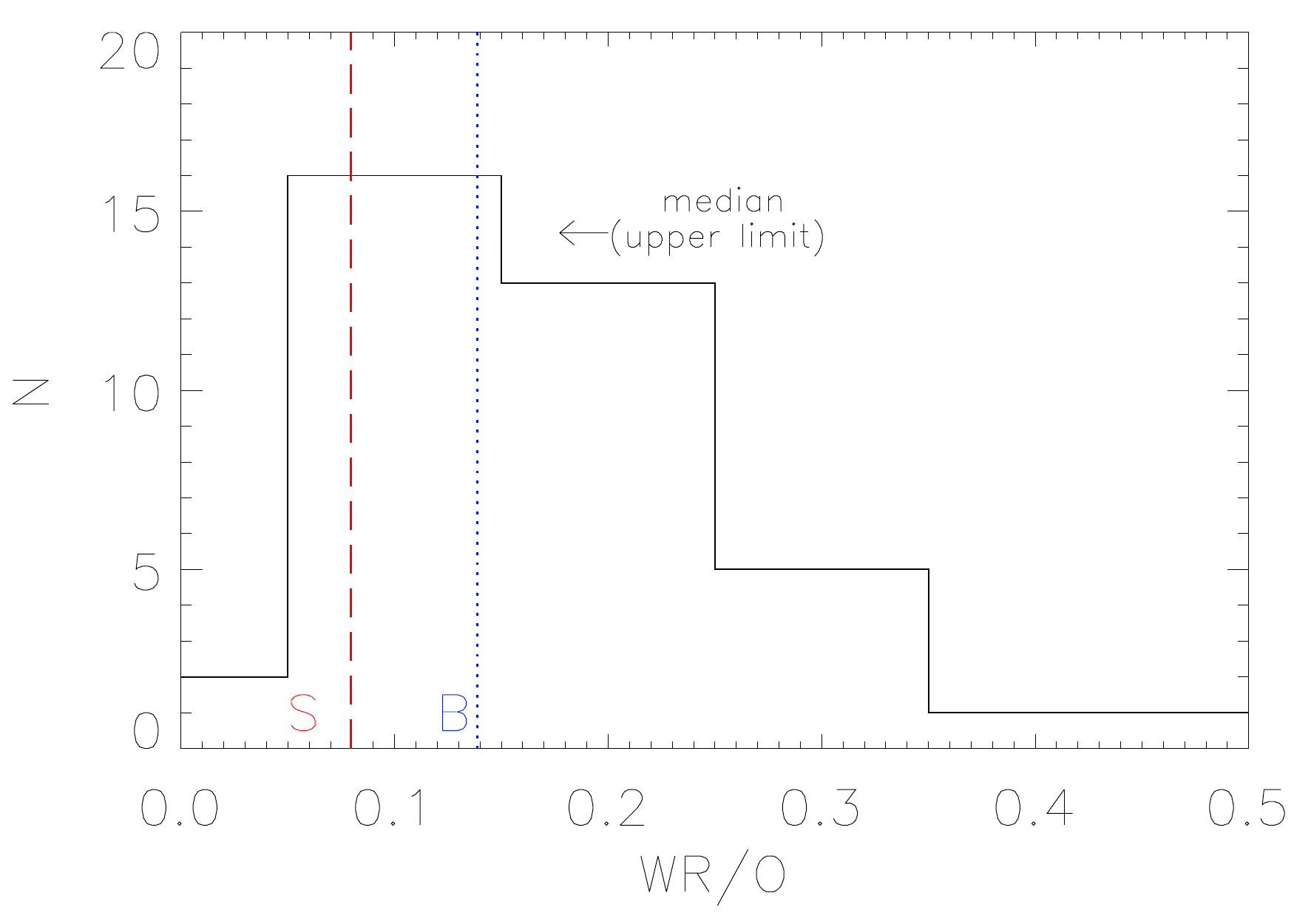}
\caption{
({\it top}) WC/WN ratio. Median value $\sim$1 is indicated with an arrow.
({\it bottom}) WR/O ratio. An upper limit for the median value $\sim$0.2 is indicated with an arrow.
Vertical {\it blue dotted} and {\it red dashed lines} indicate predicted values from BPASS binary and single models, respectively, in both panels.
}
\label{fig:ratios}
\end{center}
\end{figure}

Of the total number of WR stars found in the Antennae,
half correspond to WN-types, and the other half are WC-types
(see Fig.~\ref{fig:histogram2}, {\it top panel}).
The global WC/WN ratio in the Antennae is thus $\sim$1
(see Fig.~\ref{fig:ratios} {\it top panel}).
It is interesting to note in Table~\ref{tab:class} that we observe
the highest local WC/WN ratios when WCL-subtypes are present.
There are only two zones which have WNL stars, solely,
the rest correspond to WR cluster complexes with a combination of WNL, WCE or WCL-subtypes.
WNE templates were not needed in any region studied.
It is reasonable to discard this population given that
neither was \nv\ observed in the BB after a careful multi-Gaussian fitting
(see Fig.~\ref{fig:multi} and Table~\ref{tab:gauss}).

The averaged fraction of WR stars
over the number of O-type stars in the Antennae is $\sim$20 per cent,
which is the value obtained for the majority of the regions studied
(see Fig.~\ref{fig:histogram2}, {\it bottom panel}).
Nevertheless, \citet{2017Eldridge} warns us that
"{\it one problem with such comparisons
is uncertainty in how complete each observational sample is,
especially for the WR/O ratio where both stellar types
are hot and difficult to find in optical surveys}."
Our determination of the global number of O-type stars is necessarily
incomplete considering that we have only taken into account those in
clusters where WR features were found.
Increasing the number of O-type stars, which is to be expected, the WR/O ratio
will decrease, so our estimate represents an upper limit
(see Fig.~\ref{fig:ratios} {\it bottom}).

\begin{figure}
\begin{center}
\includegraphics[width=0.95\linewidth]{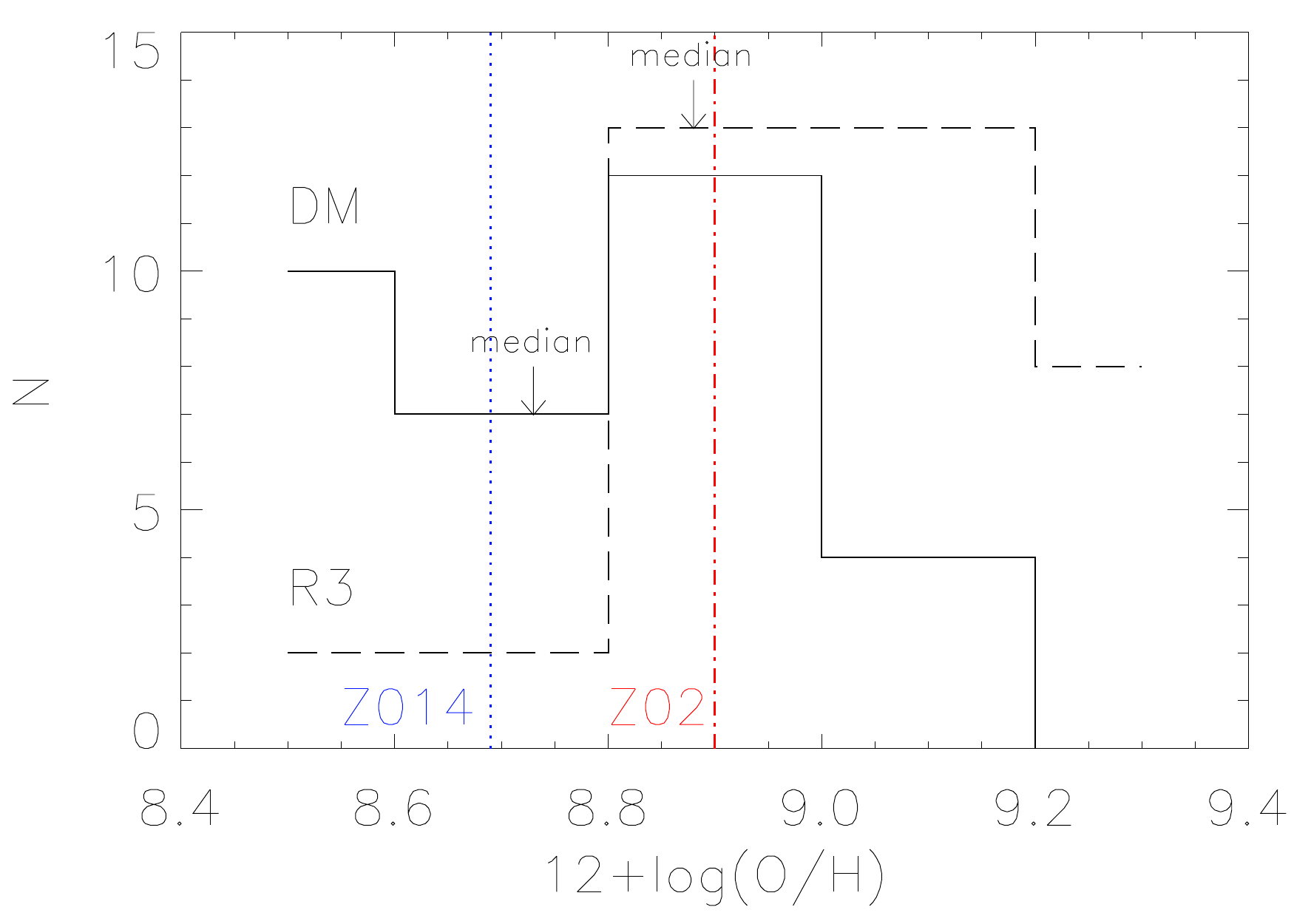}\\
\caption{
Histograms of the oxygen abundances determined with the R3 ({\it dashed line}) a DM ({\it solid line}) methods.
Median values $\sim$8.7 and $\sim$8.9, for the R3 and DM method, are indicated with an arrow.
As a reference, metallicity Z=0.014 (Solar) and Z=0.02, are indicated by a vertical {\it blue dotted} and {\it red dotted-dashed lines}, respectively.
}
\label{fig:abundance}
\end{center}
\end{figure}

In Fig.~\ref{fig:abundance} we show that the metallicity of the Antennae corresponds
with values around Solar, with and oxygen abundance
$\mathrm{12+log(O/H)} \approx 8.9$.
At this metallicity, the WC/WN ratio is expected to be $\sim$0.5 and $\sim$1,
for binary and single stars, respectively (see Fig.~\ref{fig:ratios} {\it top panel}),
according to the current Binary Population and
Spectral Synthesis (BPASS v2.2.1) models by \cite{2017Eldridge}.

\begin{figure}
\begin{center}
\includegraphics[width=0.95\linewidth]{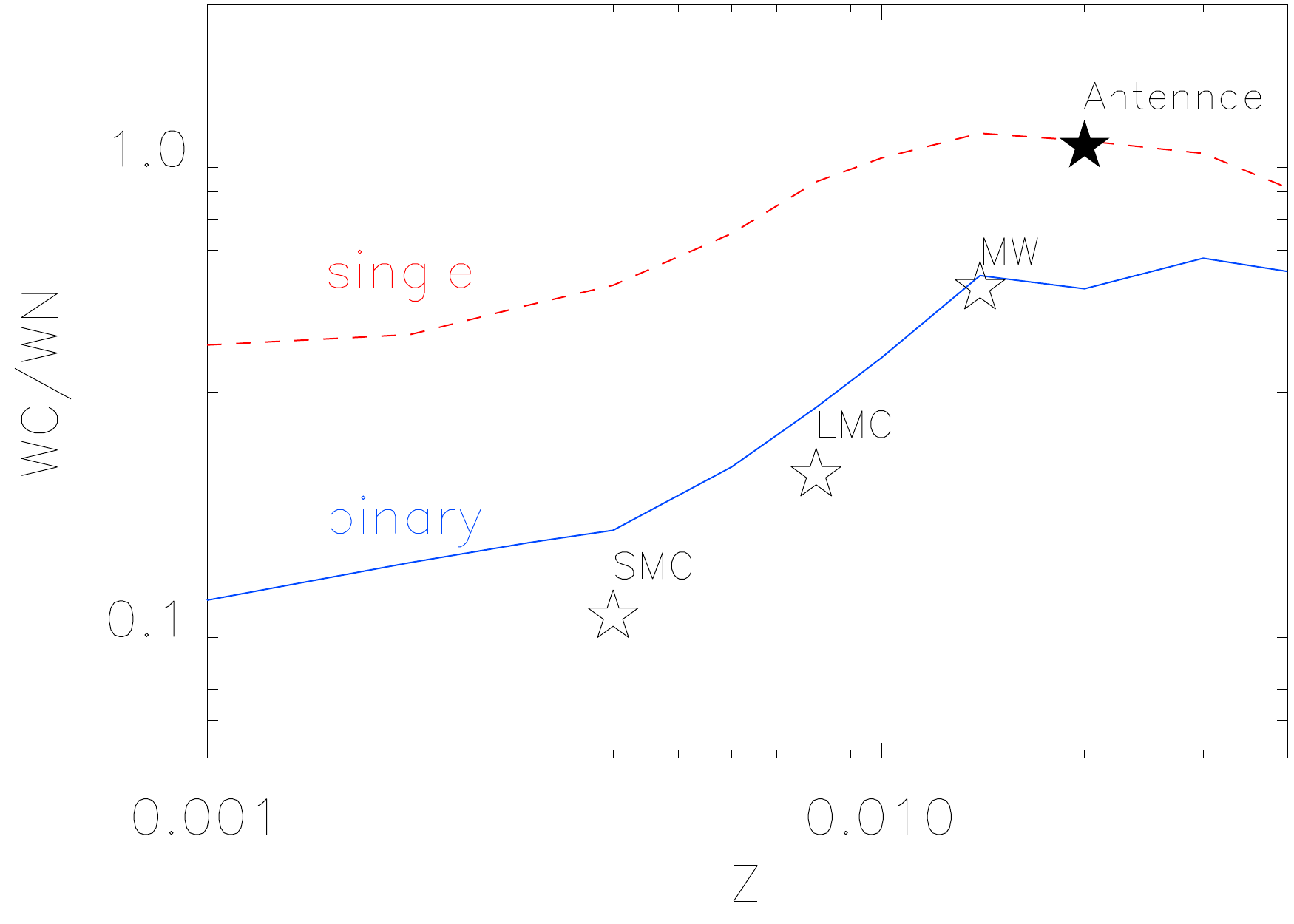}\\
\includegraphics[width=0.95\linewidth]{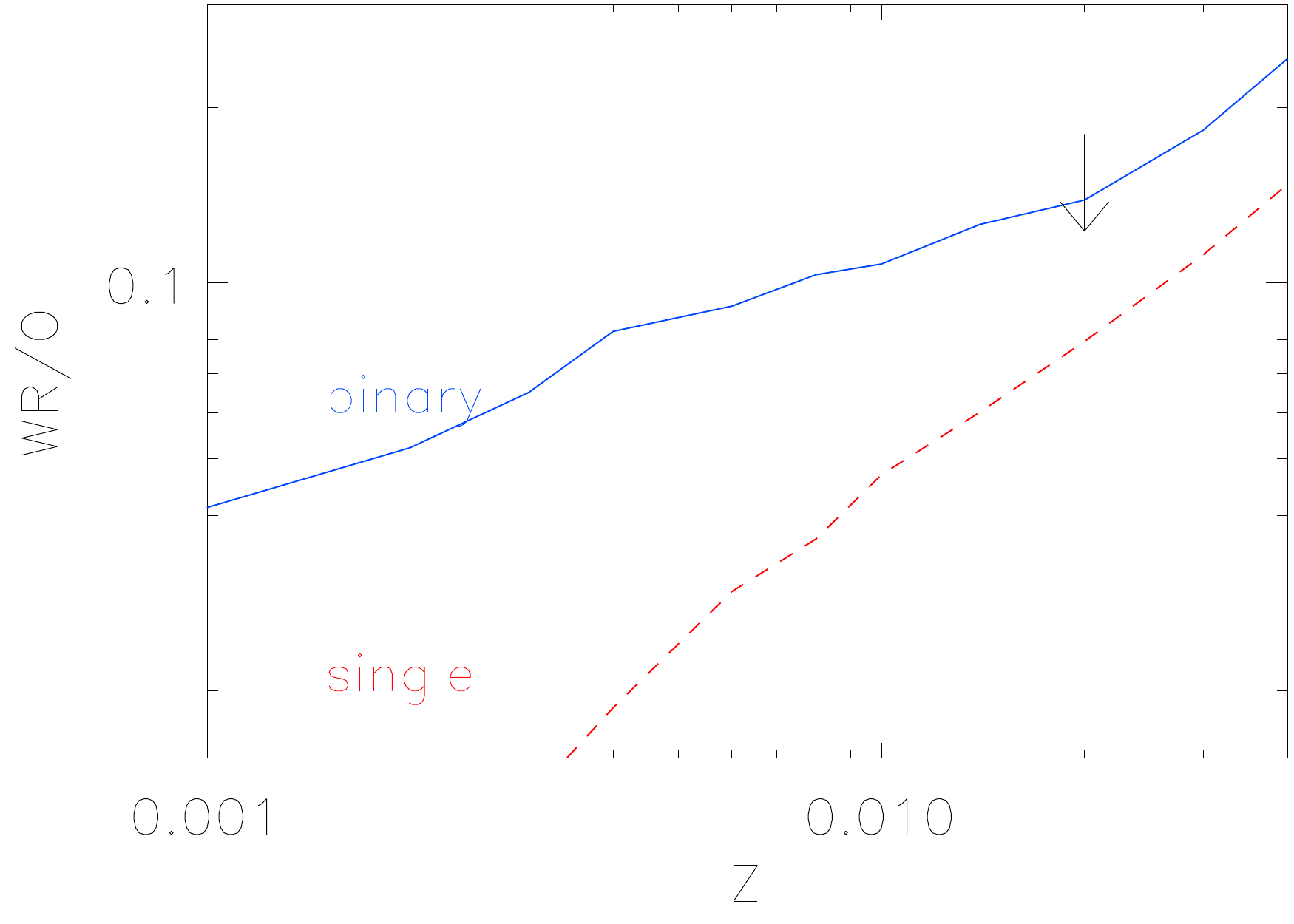}
\caption{({\it top})
WC/WN vs. metallicity.
{\it Blue lines} correspond to BPASS binary models and
{\it red dashed lines} the single models.
The {\it black star} shows the averaged WC/WN estimation
for the Antennae.
For comparison, we also show the values for the Magellanic Clouds
and the Milky Way taken from \citet{2015Rosslowe}, shown with {\it open stars}.
({\it bottom}) WR/O vs. metallicity.
The upside down arrow represents an upper limit for the Antennae,
since we only computed the number of O-type stars 
in the SSCs with WR stars.}
\label{fig:bpass}
\end{center}
\end{figure}

In Fig.~\ref{fig:bpass} we show the BPASS models for the WC/WN and WR/O ratios vs.
oxygen abundance to compare with our estimations for the Antennae.
This has been illustrated in \citet{2017Monreal} and \citet{2019Neugent}
for several galaxies with different metallicities.
The global WC/WN ratio for the Antennae is in agreement with a
population of single stars, not unusual since at Solar metallicity,
the effect of binaries is expected to be small.
On the other hand, if one assumes that each SSC complex with WR stars represents an
independent zone of star formation,
the local WR/O ratios should to be taken account.
There are 7 cases with WC/WN ratios $\lesssim$0.5:
WR\,3, 9, 13, 17, 21, 31 and 35 (see Table~\ref{tab:class}),
corresponding to those with the highest WN/WR ratios in Fig.~\ref{fig:histogram2} {\it top panel}.
In these WR complexes, WN-types dominate over WC-types by $\times$2,
in accordance with the binary scenario (see Fig.~\ref{fig:ratios} {\it top panel}).
Several studies in the literature consider global WC/WN
and WR/O ratios \citep[see e.g][and references therein]{2017Monreal},
or at specific galactocentric zones when there is a metallicity gradient
\citep[e.g.,][]{2015Rosslowe}.
This is not a trivial issue for the Antennae, seeing that
it is actually a pair of merging galaxies and thus
any galactocentric parameter may not be relevant.

Finally, the template-fitting method has been used in several works to classify WR stars
\citep[e.g.,][]{2006Hadfield,2007Hadfield,2013Kehrig,2020Gomez}.
However, it is still not clear how representative a template
spectrum is for a given subtype. 
That is, what is the dispersion in the strengths of different
lines in the individual spectra used for obtaining the template.
An in-depth study of line strengths of WR stars in the Galaxy
updated with new astrometric and photometric information from
{\it Gaia} would be required to address this question.

\section{SUMMARY AND CONCLUDING REMARKS}

We have used VLT MUSE data cubes from the ESO archive
to search for WR stars in star-forming complexes of the Antennae galaxies.
We reported their number, classification and distribution. Our results can be summarized as follows.

\begin{itemize}

\item We detected 38 WR SSC complexes
with 4053 $\pm$ 84 WR stars, out of which 
2021 $\pm$ 60 are WNL, 2032 $\pm$ 59 ~WC-types.
We cannot uncover a plausible presence of WO-type stars given the limited spectral
range covered by the observations used for this study.
Further observations that covering the VB could confirm or rule out this population,
the rarest type of WR stars.
However, thanks to the GTC spectrum of WR\,1, we can rule out the presence of WO-type stars in this SSC complex, the one with the highest number of WR stars in the Antennae.

\item Galactic WR templates of WNL, WCE and WCL-subtypes were
appropriate to classify and quantify the WR stars of each SSC complex,
given the metallicity of the Antennae, with an oxygen abundance
$\mathrm{12+log(O/H)} \approx 8.9$.

\item We analysed the observed WR blue and red bumps using multiple-component
Gaussian fitting in order to recover the ionic transitions
responsible for the bumps and report their main parameters.
In all cases, the recovered ions are consistent with those
expected for the inferred subtype using the templates.
Avoiding any nebular contamination, we evaluated the contribution
for each broad emission line to the entire luminosity of the BB and the RB
of the entire WR sample.
We do not find any evidence of the nebular \heii\ emission line.
This is explained by theoretical models, as discussed, given the high metallicity of the Antennae.

\item We estimated the main physical properties of the WR nebular environments,
oxygen abundances, and presented information regarding their kinematics.

\item We derive a global WC/WN ratio $\sim$1, which,
according to predictions of the current BPASS models
is consistent with the single-star scenario, not unusual
considering the Solar metallicity of the Antennae.
We determined the number of O-type stars in the SSC
complexes with WR features and estimated a global WR/O ratio around 20 per cent.

\item With this work, Antennae has one of the largest number of WR stars
recorded in the literature.
The detection of this number of WR stars in the
Antennae increases the sample of extragalactic WR stars, SNIbc candidates
and other post-SN by-products.

\end{itemize}

\section*{Acknowledgements}
The authors thank the referee for valuable suggestions that clarified 
our estimations on the number of WR stars in the Antennae.
VMAGG acknowledges support from the Programa de Becas 
posdoctorales funded by Direcci\'{o}n General 
de Asuntos del Personal Acad\'{e}mico (DGAPA) of the Universidad 
Nacional Aut\'{o}noma de M\'{e}xico (UNAM).
VMAGG thanks Nate Bastian for kindly sharing his WR spectra
for a preliminary analysis that prompted the interest in this study.
The authors are thankful to J.J.\,Eldridge for kindly sharing their 
models to compare with our results.
VMAGG, JAT and SJA acknowledge funding by DGAPA UNAM 
PAPIIT projects IA100720 and IN107019.
This work is based on data obtained from the ESO Science Archive Facility, program ID: 095.B-0042, 
and observations from GTC public database.
Observations made with the NASA/ESA {\it Hubble Space Telescope} were obtained
from the data archive at the Space Telescope Science Institute.
STScI is operated by the Association of Universities for Research in
Astronomy, Inc. under NASA contract NAS 5-26555.
This research has made use of the NASA/IPAC Extragalactic Database (NED),
which is funded by the National Aeronautics and Space Administration
and operated by the California Institute of Technology.

\section*{Data availability}
The data underlying this work are available in the article.
All the observations are in the public domain.
The links and observation IDs are available in the article.
The reduced OSIRIS files will be shared on request to the first author.

\appendix

\section{VLT MUSE observations}
\label{sec:datasets}

\begin{figure}
\begin{center}
\includegraphics[width=\linewidth]{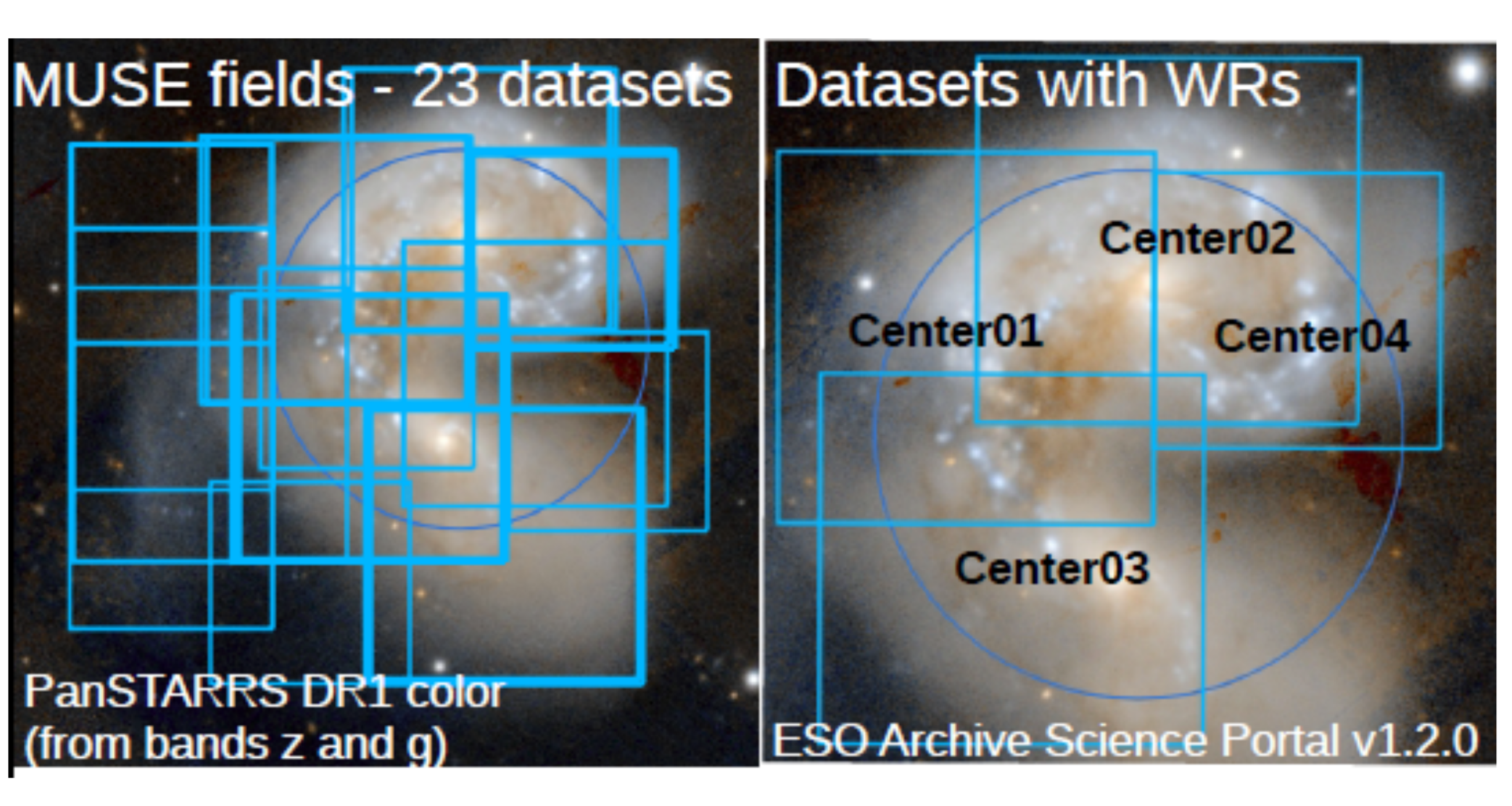}
\caption{Publicly available VLT MUSE fields
from ESO Science Archive Facility.
({\it left}) A total of 23 datasets of observations
cover the entire galaxies. ({\it right}) The FoV of the four 
data cubes used in the present work.}
\label{fig:datasets}
\end{center}
\end{figure}

Fig.~\ref{fig:datasets} shows the FoV of the 23 datasets of the Antennae
obtained with the VLT MUSE instrument. The {\it left panel} shows all available observations whilst
the {\it right panel} shows only the four data cubes that include the contribution of WR features.
These are the data used in the present work. Details of these four observations
are listed in Table~\ref{tab:muse}.

\begin{figure*}
\begin{center}
\includegraphics[width=\linewidth]{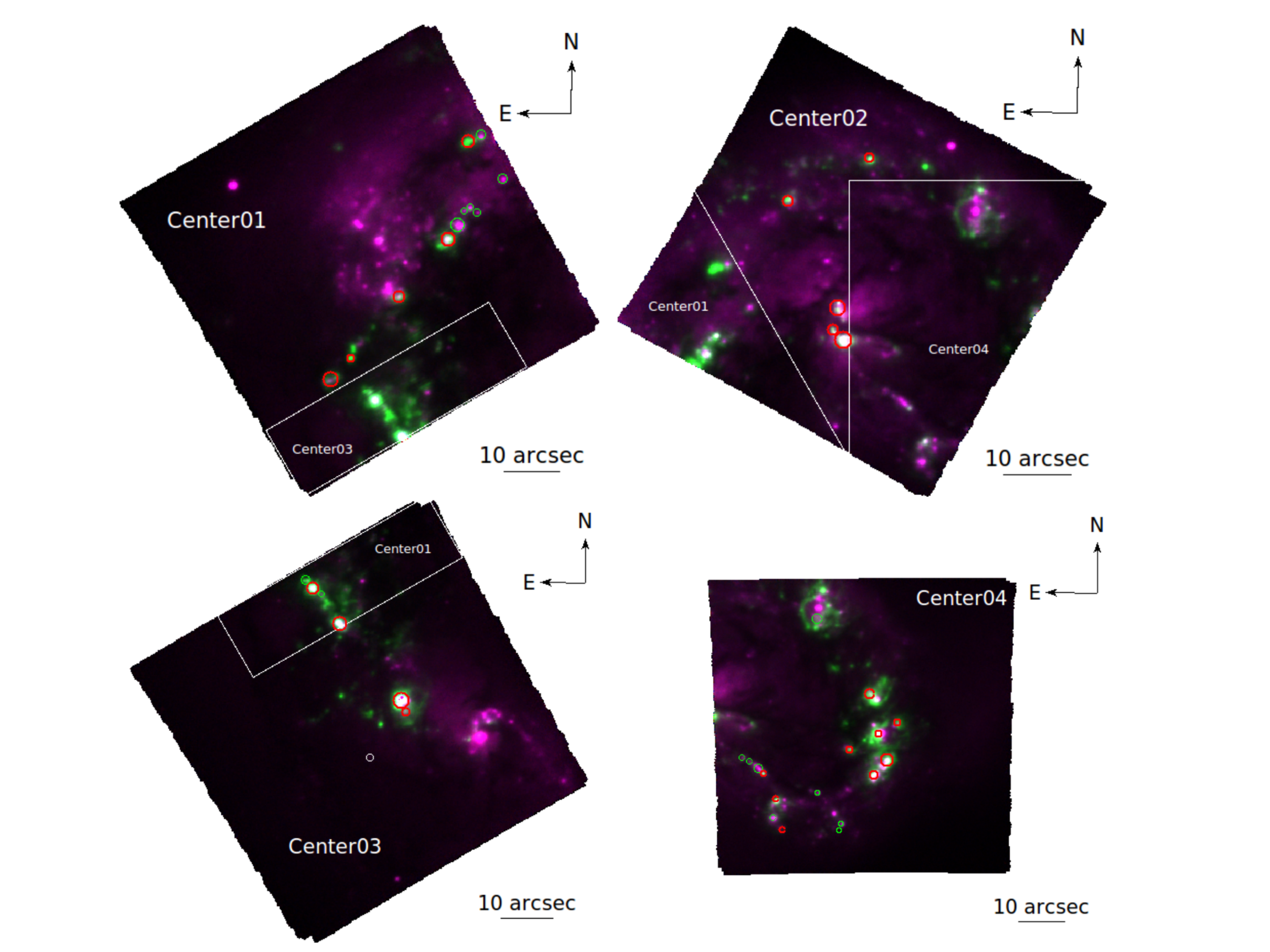}
\caption{Colour-composite images created from the four MUSE
datasets, simulating narrow-band filters,
each with 3~\AA\ of bandwidth,
centered on 4680~\AA\ (redshifted wavelength of \heii - blue), 4800~\AA\
(continuum -red) and 6596~\AA\ (redshifted wavelength of \ha - green).
The position of WR features
were identified by inspecting the images are shown with red circle. 
Subsequent search in IFU spectra revealed the presence of other
WR features, which we identify with green circles.}
\label{fig:muse_fields}
\end{center}
\end{figure*}

\begin{table*}
\small\addtolength{\tabcolsep}{-2pt}
\begin{center}
\caption{Details of the MUSE datasets${^\dagger}$ in the Antennae.}
\begin{tabular}{ccccccccc}
\hline
Pointing& Archive ID & OB ID & \multicolumn{2}{c}{Coordinates (J2000)}    &\multicolumn{2}{c}{Date of observation}& FoV &Exposure    \\
        &            &       & R.A. & Dec.                                & start    & end                        &(arcmin) &time (s)\\
 (1)    & (2)        & (3)   & (4)  & (5)                                 & (6)      & (7)                        & (8)     & (9)    \\
\hline  
Center01& ADP.2017-03-28T13:08:20.713 & 200354691 &12:01:55.93 & --18:52:16.5 &2015-04-23&2015-04-23 & 2.02 &$5400$\\
Center02& ADP.2017-03-28T13:08:20.697 & 200356231 &12:01:52.70 & --18:51:54.2 &2015-05-11&2015-05-11 & 2.00    &$5400$\\
Center03& ADP.2017-03-28T13:08:20.689 & 200356367, 200356555&12:01:55.19 & --18:53:06.7 &2015-05-11&2015-05-14 & 2.00    &$5400$\\
Center04& ADP.2017-03-28T13:08:20.681 & 200356499 &12:01:50.64 & --18:52:10.0 &2015-05-13&2015-05-21 & 1.50  &$5400$\\         
\hline
\end{tabular}\\
{\it Notes}. Brief explanation of columns: ${^\dagger}$Telescope: ESO-VLT-U4; instrument: MUSE;
technique of observation: IFU;
data type: CUBE (IFS); pixel scale = 0.2 arcsec;
number of observations per pointing = 2;
spectral range: 4600-9350 \AA; Spectral resolution (R)= 2989;
principal investigator: Weilbacher, Peter M.;
data processing certified by ESO; data Level 3; program ID: 095.B-0042;
(1) pointing ID;
(2) archive ID;
(3) Observation ID;
(4) coordinates (J2000): Right ascension (R.A.);
(5) declination (Dec.);
(6) date of the observation; start (year-month-day) and,
(7) end;
(8) field of view (FoV) (arcmin);
(9) exposure time (s).
\label{tab:muse} 
\end{center}
\end{table*}

\section{Other non-WR detection}
\label{sec:qso}
\begin{figure*}
\begin{center}
\includegraphics[width=1\linewidth]{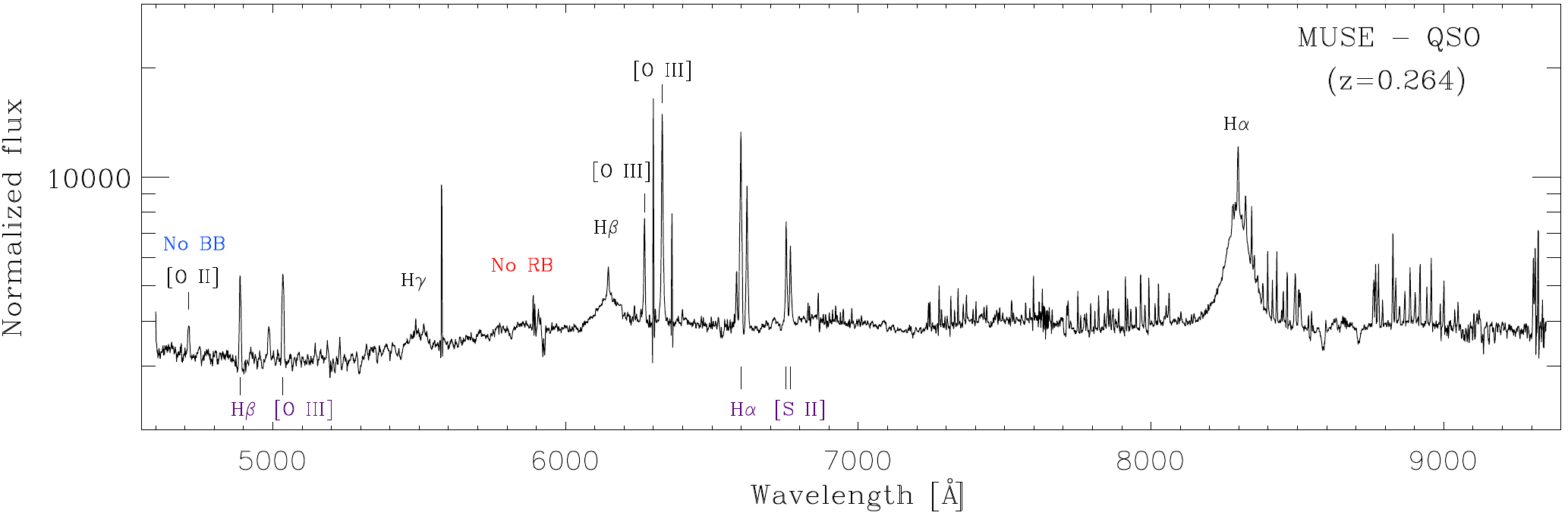}
\caption{MUSE spectrum of a spurious WR candidate in the Antennae.
The false BB and RB, among the strong emission lines
identified at z$\sim$0.264, are indicated above the continuum.
The BB turned out to be \oiia. Below the continuum some other lines
corresponding to the Antennae (z$\sim$0.005) were identified.
The spectrum contains information from astronomical objects
at different distances.
Its location is indicated in Fig.~\ref{fig:Antennae_RGB}, with a white circle labeled as qso.
Flux is normalized to 10$^{-20}$ erg\,cm$^{-2}$\,s$^{-1}$\,\AA$^{-1}$.}
\label{fig:qso}
\end{center}
\end{figure*}

During our search of WR features in the Antennae,
we found a bright object with BB emission which resulted in a probable QSO.
Its coordinates (J2000) are $(\alpha, \delta)$=(12:01:55.0528, --18:53:16.031).
Its MUSE spectrum is presented in Fig.~\ref{fig:qso}.

\section{Nebular emission lines}
\label{sec:appC}

The spectra of the SSCs with WR features studied here can be used to characterise
the physical properties of their environments. The reddening-corrected nebular
lines are listed in Table~\ref{tab:neb2}. These were used to estimate physical
parameters such as $T_\mathrm{e}$, $n_\mathrm{e}$ and oxygen abundances of the 
38 WR spectra reported in Table~\ref{tab:neb}.

\begin{table*}
\small\addtolength{\tabcolsep}{-1.5pt}
\begin{center}
\caption{Reddening-corrected line fluxes, relative to \hb=100.}
\begin{tabular}{cccccccccccccc}
\hline
ID&\oiii &\oiii  &\sii &\sii  &\nii & \nii  & \nii &\cliii &\cliii&\siii&\siii&\oii&\oii\\
 & $\lambda4959$ & $\lambda5007$ & $\lambda6717$ & $\lambda6731$ & $\lambda5755$ & $\lambda6548$ &$\lambda6584$ & $\lambda5517$  & $\lambda5537$& $\lambda6312$& $\lambda9069$& $\lambda7320$& $\lambda7330$\\
(1)&(2)& (3) & (4)& (5)& (6) & (7)   &(8)   &(9)    &(10)&(11)&(12)&(13)&(14)\\
\hline
1 &59.0$\pm$0.1&176.4$\pm$0.1&28.3$\pm$0.1&24.4$\pm$0.1&0.7$\pm$0.1&31.7$\pm$0.1& 95.6$\pm$0.1&0.4$\pm$0.1&0.3$\pm$0.1&1.4$\pm$0.1&46.8$\pm$0.3&3.2$\pm$0.1&2.6$\pm$0.1\\
2 &60.1$\pm$0.1&179.0$\pm$0.1&30.0$\pm$0.1&25.0$\pm$0.1&0.7$\pm$0.1&25.4$\pm$0.1& 76.8$\pm$0.1&0.4$\pm$0.1&0.3$\pm$0.1&1.1$\pm$0.1&29.3$\pm$0.2&2.7$\pm$0.1&2.2$\pm$0.1\\
3 &60.6$\pm$0.1&180.2$\pm$0.1&15.8$\pm$0.1&11.9$\pm$0.1&0.5$\pm$0.1&17.7$\pm$0.1& 53.1$\pm$0.1&0.4$\pm$0.1&0.3$\pm$0.1&1.0$\pm$0.1&24.1$\pm$0.1&1.6$\pm$0.1&1.3$\pm$0.1\\
4 &34.6$\pm$0.1&101.7$\pm$0.2&35.4$\pm$0.1&26.5$\pm$0.1&0.7$\pm$0.1&26.9$\pm$0.1& 81.3$\pm$0.3&0.2$\pm$0.1&0.2$\pm$0.1&0.8$\pm$0.1&17.0$\pm$0.3&1.3$\pm$0.1&1.2$\pm$0.1\\
5 &48.8$\pm$0.1&144.6$\pm$0.1&28.2$\pm$0.1&20.7$\pm$0.1&0.6$\pm$0.1&22.7$\pm$0.1& 67.8$\pm$0.1&0.3$\pm$0.1&0.2$\pm$0.1&0.9$\pm$0.1&20.4$\pm$0.2&1.9$\pm$0.1&1.6$\pm$0.1\\
6 &47.7$\pm$0.1&142.0$\pm$0.1&21.0$\pm$0.1&15.3$\pm$0.1&0.6$\pm$0.1&21.7$\pm$0.1& 65.5$\pm$0.1&0.3$\pm$0.1&0.2$\pm$0.1&0.9$\pm$0.1&22.1$\pm$0.2&1.5$\pm$0.1&1.3$\pm$0.1\\
7 & 3.5$\pm$0.2& 10.0$\pm$0.1&28.2$\pm$0.1&22.6$\pm$0.1&0.5$\pm$0.2&34.1$\pm$0.1&104.2$\pm$0.5& $\cdots$  & $\cdots$  & $\cdots$  & 8.8$\pm$0.2&0.4$\pm$0.1&0.3$\pm$0.1\\
8 &12.4$\pm$0.1& 36.6$\pm$0.1&26.9$\pm$0.1&19.8$\pm$0.1&0.5$\pm$0.1&32.7$\pm$0.1&100.6$\pm$0.4&0.3$\pm$0.1&0.2$\pm$0.1&0.4$\pm$0.1&19.6$\pm$0.1&1.0$\pm$0.1&0.9$\pm$0.1\\
9 & 7.9$\pm$0.3& 22.8$\pm$0.2&39.1$\pm$0.4&37.6$\pm$0.3&1.1$\pm$0.2&42.9$\pm$0.5&131.4$\pm$1.8&0.3$\pm$0.2& $\cdots$  & $\cdots$  &13.6$\pm$0.3&1.7$\pm$0.2&1.5$\pm$0.2\\
10&15.6$\pm$0.2& 45.3$\pm$0.1&30.7$\pm$0.1&22.9$\pm$0.1&0.5$\pm$0.1&33.1$\pm$0.2&101.5$\pm$0.7&0.4$\pm$0.2&0.2$\pm$0.1&0.3$\pm$0.1&19.9$\pm$0.1&1.1$\pm$0.1&1.0$\pm$0.1\\
11&20.0$\pm$0.1& 59.0$\pm$0.1&33.1$\pm$0.1&24.5$\pm$0.1&0.6$\pm$0.1&29.1$\pm$0.1& 88.6$\pm$0.2&0.2$\pm$0.1&0.2$\pm$0.1&0.4$\pm$0.1&15.0$\pm$0.2&1.3$\pm$0.1&1.2$\pm$0.1\\
12&17.7$\pm$0.1& 52.5$\pm$0.1&31.9$\pm$0.1&23.4$\pm$0.1&0.6$\pm$0.1&30.5$\pm$0.1& 92.8$\pm$0.2&0.2$\pm$0.1& $\cdots$  &0.4$\pm$0.1&13.9$\pm$0.2&1.2$\pm$0.1&1.3$\pm$0.1\\
13&24.6$\pm$0.1& 72.5$\pm$0.1&19.8$\pm$0.1&14.7$\pm$0.1&0.6$\pm$0.1&29.1$\pm$0.1& 88.8$\pm$0.1&0.4$\pm$0.1&0.3$\pm$0.1&0.7$\pm$0.1&22.1$\pm$0.1&1.4$\pm$0.1&1.3$\pm$0.1\\
14&16.9$\pm$0.1& 50.2$\pm$0.1&31.5$\pm$0.1&22.9$\pm$0.1&0.6$\pm$0.1&32.1$\pm$0.1& 97.3$\pm$0.2&0.3$\pm$0.1&0.2$\pm$0.1&0.5$\pm$0.1&14.6$\pm$0.2&1.3$\pm$0.1&1.3$\pm$0.1\\
15&21.2$\pm$0.1& 62.8$\pm$0.1&37.1$\pm$0.1&26.2$\pm$0.1&0.7$\pm$0.1&30.7$\pm$0.1& 92.7$\pm$0.4&0.3$\pm$0.1&0.3$\pm$0.1&0.4$\pm$0.1&12.6$\pm$0.2&1.5$\pm$0.1&1.7$\pm$0.1\\
16&14.0$\pm$0.2& 40.4$\pm$0.1&27.2$\pm$0.1&19.2$\pm$0.1&0.5$\pm$0.2&29.6$\pm$0.1& 89.3$\pm$0.6&0.4$\pm$0.2&1.6$\pm$0.1&0.2$\pm$0.1&13.8$\pm$0.3&0.9$\pm$0.2&1.4$\pm$0.2\\
17&36.0$\pm$0.1&107.0$\pm$0.2&27.4$\pm$0.1&20.8$\pm$0.1&0.7$\pm$0.1&27.7$\pm$0.1& 84.0$\pm$0.4&0.4$\pm$0.1&0.2$\pm$0.1&0.7$\pm$0.1&22.3$\pm$0.1&1.9$\pm$0.1&1.7$\pm$0.1\\
18& 5.8$\pm$0.1& 17.1$\pm$0.1&29.3$\pm$0.1&22.0$\pm$0.1&0.6$\pm$0.1&37.1$\pm$0.1&113.4$\pm$0.6&0.2$\pm$0.1& $\cdots$  &0.2$\pm$0.1&11.4$\pm$0.3&0.8$\pm$0.1&0.9$\pm$0.1\\
19&33.9$\pm$0.1& 99.2$\pm$0.4&25.5$\pm$0.1&18.1$\pm$0.1&0.7$\pm$0.2&21.6$\pm$0.1& 65.3$\pm$0.5&0.4$\pm$0.2& $\cdots$  &0.3$\pm$0.2&15.5$\pm$0.2&1.1$\pm$0.2&1.4$\pm$0.1\\
20&46.5$\pm$0.1&137.9$\pm$0.4&34.0$\pm$0.1&24.1$\pm$0.1&0.5$\pm$0.1&21.4$\pm$0.1& 64.7$\pm$0.3&0.3$\pm$0.1&0.2$\pm$0.1&0.6$\pm$0.1&16.1$\pm$0.2&1.9$\pm$0.1&2.3$\pm$0.2\\
21&41.4$\pm$0.1&122.0$\pm$0.2&28.4$\pm$0.1&21.2$\pm$0.1&0.7$\pm$0.1&25.8$\pm$0.1& 80.0$\pm$0.2&0.4$\pm$0.1&0.3$\pm$0.1&0.8$\pm$0.1&21.1$\pm$0.2&2.6$\pm$0.1&2.6$\pm$0.1\\
22&13.3$\pm$0.1& 39.5$\pm$0.1&34.0$\pm$0.2&25.7$\pm$0.1&0.6$\pm$0.1&35.0$\pm$0.2&107.0$\pm$0.7&0.3$\pm$0.1& $\cdots$  &0.3$\pm$0.1&14.5$\pm$0.3&1.0$\pm$0.1&1.0$\pm$0.1\\
23& 9.7$\pm$0.2& 28.2$\pm$0.1&41.2$\pm$0.2&30.3$\pm$0.2&0.6$\pm$0.2&33.8$\pm$0.2&103.1$\pm$0.7&0.3$\pm$0.1& $\cdots$  &0.2$\pm$0.1&10.3$\pm$0.2&0.8$\pm$0.1&0.6$\pm$0.1\\
24& 7.4$\pm$0.1& 21.9$\pm$0.1&29.9$\pm$0.1&22.6$\pm$0.1&0.5$\pm$0.1&34.7$\pm$0.1&105.9$\pm$0.3&0.3$\pm$0.1& $\cdots$  &0.3$\pm$0.1&14.8$\pm$0.1&0.7$\pm$0.1&0.5$\pm$0.1\\
25& 6.9$\pm$0.1& 20.2$\pm$0.1&32.9$\pm$0.1&24.2$\pm$0.1&0.6$\pm$0.1&32.1$\pm$0.1& 97.6$\pm$0.4&0.3$\pm$0.1& $\cdots$  &0.2$\pm$0.1&11.5$\pm$0.1&0.6$\pm$0.1&0.4$\pm$0.1\\
26&15.3$\pm$0.2& 44.0$\pm$0.1&39.4$\pm$0.2&29.6$\pm$0.1&0.8$\pm$0.2&30.9$\pm$0.1& 94.5$\pm$0.8&0.7$\pm$0.3& $\cdots$  &0.3$\pm$0.1&11.9$\pm$0.3&0.9$\pm$0.2&0.7$\pm$0.2\\
27& 8.7$\pm$0.2& 25.4$\pm$0.1&38.2$\pm$0.1&27.5$\pm$0.1&0.5$\pm$0.1&30.8$\pm$0.1& 93.9$\pm$0.5&0.3$\pm$0.1& $\cdots$  &0.1$\pm$0.1&10.0$\pm$0.2&0.6$\pm$0.1&0.2$\pm$0.1\\
28& 8.3$\pm$0.2& 23.7$\pm$0.1&40.7$\pm$0.2&30.0$\pm$0.1&0.6$\pm$0.2&32.8$\pm$0.1&100.3$\pm$0.7&0.4$\pm$0.2&0.2$\pm$0.1&0.2$\pm$0.1& 9.5$\pm$0.2&0.7$\pm$0.1&0.5$\pm$0.1\\
29&14.7$\pm$0.1& 42.8$\pm$0.1&37.8$\pm$0.1&27.2$\pm$0.1&0.7$\pm$0.2&29.9$\pm$0.1& 90.7$\pm$0.4&0.4$\pm$0.1& $\cdots$  &0.3$\pm$0.1&12.2$\pm$0.2&0.9$\pm$0.1&0.7$\pm$0.1\\
30&13.3$\pm$0.1& 39.3$\pm$0.1&40.3$\pm$0.1&29.5$\pm$0.1&0.7$\pm$0.1&34.0$\pm$0.1&103.8$\pm$0.3&0.3$\pm$0.1&0.2$\pm$0.1&0.4$\pm$0.1&14.5$\pm$0.1&1.1$\pm$0.1&0.9$\pm$0.1\\
31& 9.7$\pm$0.1& 28.8$\pm$0.1&29.2$\pm$0.1&25.2$\pm$0.1&0.7$\pm$0.1&45.6$\pm$0.1&139.7$\pm$0.1&0.2$\pm$0.1&0.2$\pm$0.1&0.4$\pm$0.1&21.6$\pm$0.1&1.4$\pm$0.1&1.1$\pm$0.1\\
32& 4.5$\pm$0.2& 12.9$\pm$0.1&31.7$\pm$0.1&23.5$\pm$0.1&0.7$\pm$0.2&34.1$\pm$0.1&104.0$\pm$0.1&0.3$\pm$0.2& $\cdots$  &0.1$\pm$0.1&10.0$\pm$0.2&0.5$\pm$0.1&0.3$\pm$0.1\\
33& 3.8$\pm$0.1& 11.3$\pm$0.1&31.3$\pm$0.2&23.9$\pm$0.1&0.4$\pm$0.1&34.5$\pm$0.1&105.6$\pm$0.5&0.3$\pm$0.2& $\cdots$  &0.1$\pm$0.1&10.0$\pm$0.1&0.5$\pm$0.1&0.3$\pm$0.1\\
34&11.3$\pm$0.1& 32.7$\pm$0.1&37.0$\pm$0.2&26.7$\pm$0.1&0.7$\pm$0.2&34.6$\pm$0.2&105.7$\pm$0.8&0.3$\pm$0.2&1.4$\pm$0.1&0.2$\pm$0.1&12.4$\pm$0.2&0.9$\pm$0.1&0.7$\pm$0.1\\
35& 2.6$\pm$0.3&  5.6$\pm$0.2&26.1$\pm$0.1&19.4$\pm$0.1&0.5$\pm$0.2&26.6$\pm$0.2& 81.8$\pm$1.0&0.4$\pm$0.3& $\cdots$  & $\cdots$  & 5.6$\pm$0.2&0.3$\pm$0.2&0.2$\pm$0.2\\
36&21.9$\pm$0.3& 62.8$\pm$0.2&44.2$\pm$0.5&32.1$\pm$0.3&1.0$\pm$0.4&28.2$\pm$0.1& 86.0$\pm$0.9&0.6$\pm$0.3&0.5$\pm$0.3&0.3$\pm$0.2&11.2$\pm$0.4&0.9$\pm$0.3&0.5$\pm$0.3\\
37&13.2$\pm$0.2& 38.9$\pm$0.1&45.6$\pm$0.2&32.3$\pm$0.1&0.7$\pm$0.1&29.0$\pm$0.1& 88.7$\pm$0.5&0.3$\pm$0.1& $\cdots$  & $\cdots$  &10.0$\pm$0.2&1.1$\pm$0.1&0.9$\pm$0.1\\
38&24.2$\pm$0.1& 71.2$\pm$0.2&42.1$\pm$0.5&29.9$\pm$0.3&0.7$\pm$0.2&28.5$\pm$0.2& 87.4$\pm$1.0&0.4$\pm$0.2&0.2$\pm$0.2&0.5$\pm$0.1&15.0$\pm$0.1&1.4$\pm$0.2&1.2$\pm$0.2\\
\hline
\end{tabular}\\
\label{tab:neb2}
\end{center}
\end{table*}

\end{document}